\documentclass[pre,preprint,aps,a4paper,floatfix]{revtex4_new}
\usepackage[T1]{fontenc}
\usepackage{color}
\usepackage{graphicx}
\usepackage{amsmath}
\usepackage{bm}
\usepackage{epsfig}
\usepackage[utf8]{inputenc} 
\usepackage{rotating}
\usepackage{natbib}
\pdfoutput=1

\begin{document}

\title{Hard-body models of bulk liquid crystals}

\author{Luis Mederos}
\email{lmederos@icmm.csic.es}
\affiliation{Instituto de Ciencia de Materiales de Madrid, CSIC, Sor Juana In\'es de la Cruz, 3, E--28049 Madrid, Spain.}
\author{Enrique Velasco}
\email{enrique.velasco@uam.es}
\affiliation{Departamento de F\'{\i}sica Te\'orica de la Materia Condensada
and Instituto de Ciencia de Materiales Nicol\'as Cabrera,
Universidad Aut\'onoma de Madrid, E-28049 Madrid, Spain}

\author{Yuri Mart\'{\i}nez-Rat\'on}
\email{yuri@math.uc3m.es}
\affiliation{Grupo Interdisciplinar de Sistemas Complejos (GISC),
Departamento de Matem\'{a}ticas,Escuela Polit\'{e}cnica Superior,
Universidad Carlos III de Madrid, Avenida de la Universidad 30, E--28911, Legan\'{e}s, Madrid, Spain}

\date{\today}

\begin{abstract}
Hard models for particle interactions have played a crucial role in the understanding of the structure of condensed matter. 
In particular, they help to explain the formation of oriented phases in liquids made of anisotropic molecules or colloidal 
particles, and continue to be of great interest in the formulation of theories for liquids in bulk, near interfaces and in 
biophysical environments. Hard models of anisotropic particles give rise to complex phase diagrams, including uniaxial and biaxial
nematic phases, discotic phases, and spatially ordered phases such as smectic, columnar or crystal. Also, their mixtures exhibit 
additional 
interesting behaviours where demixing competes with orientational order. Here we review the different models of hard particles 
used in the theory of bulk anisotropic liquids, leaving aside interfacial properties, and discuss 
the associated theoretical approaches and computer simulations, 
focusing on applications in equilibrium situations. 
The latter include one-component bulk fluids, mixtures and polydisperse 
fluids, both in two and three dimensions, and emphasis is put on 
liquid-crystal phase transitions and complex phase behaviour in general.
\end{abstract}

\maketitle
\tableofcontents 

\newpage

\section{Introduction}

\subsection{Liquid-crystalline phases}

Common states of matter are gas, liquid, and crystal. Therefore, the term `liquid crystal' seems, in principle, a contradiction in
itself. However, liquid crystals have been known since their discovery by Friedrich Reinitzer \cite{Reinitzer:1888a} 125 years ago. 
The name suggests an intermediate behaviour between liquid and crystal and, in fact, this is the case: 
Liquid crystals are states of matter that exhibit liquid and crystal properties simultaneously. They can flow and form droplets
(liquid-like properties) but, at the same time, they present some kind of long-range molecular order (orientational and 
sometimes also partial positional order) which translates into anisotropic macroscopic optical, electric, and magnetic 
properties, and also elasticity (crystal-like properties). Liquid crystals are then \emph{mesophases}, i.e. phases or states
of matter intermediate bewteen liquid and crystal phases, and their properties have been used with advantage in many
technological applications (for an introduction to the subject see e.g. 
\cite{deGennes,Chandrasekhar,Collings,Khoo,Gray,review_discotics}; for a review on applications see the excellent series of 
books edited by Bahadur \cite{Review_applications}).

\begin{figure}[h]
\includegraphics[width=15cm,angle=0]{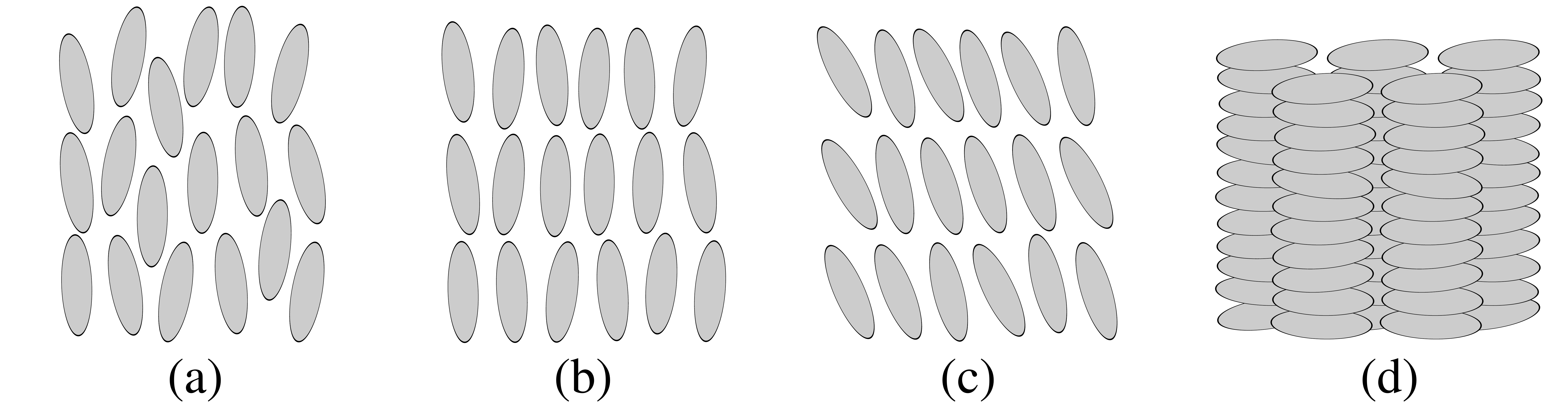}
\caption{Schematic of three liquid-crystal phases. (a) Nematic. (b) Smectic A. (c) Smectic C. (d) Columnar.}
\label{figg1}
\end{figure}

The common feature of all liquid-crystalline phases is that they exhibit orientational order. Liquid crystals are formed
by anisotropic molecules. Orientational long-ranged order means that molecules align, on average, along a particular direction, called
the \emph{director}, specified by a unit vector $\hat{\mathbf{n}}$ referred to the laboratory-fixed reference frame.
Distinction between different liquid-crystal phases comes from the occurrence of partial positional order.
A nematic (N) liquid crystal is a mesophase characterized by the presence of orientational order but the lack of positional order;
therefore, molecular centers of mass are completely disordered, see Fig. \ref{figg1}(a). In a smectic (S) phase, the centres of mass 
of the molecules are arranged in liquid-like layers, so that the system shows one-dimensional
positional order \cite{footnote1}. In the case that the layers are perpendicular to the director,
Fig. \ref{figg1}(b), the smectic
is called smectic A (S$_{\rm A}$). In a smectic C phase (S$_{\rm C}$), the liquid layers are tilted with respect
to the director, see Fig. \ref{figg1}(c). Disc-like particles usually form N phases and, at higher densities, they tend to form 
the so-called columnar (C) phase, which consists of a regular two-dimensional array of liquid-like 
columns, Fig. \ref{figg1}(d). 
A complete description of the different mesophases can be found e.g. in \cite{deGennes,Chandrasekhar,Collings}.

\subsection{The role of hard interactions in systems of isotropic particles}

In simple atomic or effectively spherical molecular systems interacting through physical interactions (no chemical
bonds), for example argon or methane at sufficiently high temperatures, the repulsive part of the two-body interaction
at short distances plays a crucial role in determining the local liquid structure. In particular, the radial distribution function
$g(r)$, which gives the probability of finding a particle at a distance $r$ from a given particle at the origin, 
can be reasonably approximated by that of a system of hard spheres (HS), a system of spheres interacting via overlap
interactions (no configurations involving overlap of spheres are allowed in the partition function).
Moreover, computer simulations first \cite{Rosenbluth,Alder,Wood} and theory later \cite{Tarazona}
showed that a HS system undergoes a first-order freezing transition when the packing fraction (fraction of volume occupied 
by spheres) of the liquid is equal to $\eta_{l}=0.494$. This liquid coexists with a face-centred-cubic (fcc) solid of
packing fraction $\eta_{s}=0.545$. This result is a bit surprising because, on a first look, it is not easy to accept 
that a system with no attractions can freeze. The explanation, of course, relies on entropy: when the packing fraction is 
high enough, spheres as a whole have more free volume available (hence entropy) in the ordered than in the liquid phase. 

In summary, hard-core interactions in simple systems can qualitatively explain most of their
equilibrium properties; the phenomenon of condensation from a vapour (liquid-vapour transition) and the dependence
of freezing with temperature are due to the attraction between particles and can be treated perturbately. 
It is therefore not surprising that perturbation theories for simple systems have given excellent results, even with quantitative
agreement, in the theory of condensed matter, not only for the liquid \cite{Hansen}, but also for the solid phase \cite{Rascon:1996a}.

\subsection{Hard interactions in systems of anisotropic particles}

One of the first questions one can ask about the equilibrium properties of liquid crystals concerns the role
of anisotropic hard repulsions between particles in determining the structure of the system. Is that role similar to that 
of HS in the case of simple liquids? The answer to this question is more complicated
than in the case of simple liquids, mainly because the coupling between orientational and positional order. 
See \cite{review_convex} for an early review about the role of hard particle models in liquid-crystalline phases.

It soon became clear, 
thanks to the seminal paper of Onsager \cite{Onsager}, that a system made solely 
of anisotropic bodies interacting through hard interactions, e.g. hard spherocylinders (HSC), undergoes an entropy-driven 
first-order isotropic-nematic transition when its packing fraction is high enough. There is an intuitive argument
to understand this transition. Imagine that you want to accommodate needles or matches into a box. If there are only a few matches, 
you can arrange them as you want, and you will have a completely disordered system of needles, i. e. an isotropic phase.
But when many needles have to go into the box they will have to be oriented in an ordered arrangement in order to fit
into the box. Therefore, packing considerations require the equilibrium phase at high packing fraction to be a nematic phase.

An equivalent argument to explain the nematic-smectic transition at yet higher packing fraction is more subtle. 
If we consider the nematic-smectic transition as due to the generation of a one-dimensional positional-order wave
in an already perfectly aligned nematic, then we have to address additional questions such as the role of the specific 
particle shape. It is not enough, as in the case of the isotropic-nematic transition, to demand a sufficiently
anisotropic particle shape since, for example, a fluid of parallel hard ellipsoids (HE) does not exhibit a smectic phase,
while one made of parallel HSC does when the length-to-breadth ratio is high enough. 

In fact, does the fact that a given model system possesses some particular stable liquid-crystal phase mean that a hard 
interaction is responsible for the stability of such a phase? Even though many features of phase behaviour can be explained by 
hard models, the general answer to this question is no. But the possibility to fabricate tailor-made colloidal particles
that behave close to size-monodisperse hard bodies with virtually any shape has revitalised hard-particle models and their 
research and applications. Also, because virus particles are completely monodisperse and their interactions are known with a 
high accuracy, they are ideally suited for testing theoretical ideas and results. Granular matter made of macroscopic grains
that interact through overlap interactions can also be modelled by hard models where particles interact elastically. Systems
of vibrated grains can in some regimes behave like hard particles that explore phase space with a Boltzmann probability, and
some ordered patterns obtained in these systems resemble liquid-crystal phases.

In summary, hard-body models in the theory of liquid crystals may play a role similar to that of the HS model
in the theory of simple fluids. Moreover, carefully prepared and stabilised colloidal particles may behave essentially 
as  hard bodies and, therefore, hard-body models for liquid crystals can be directly tested experimentally using
anisotropic colloidal particles. Vibrated granular matter made of anisotropically-shaped grains is another field of 
exploration for hard models. The applications of the statistical mechanics of mesophases for hard models are very numerous
in many fields. 

\subsection{Scope of this review}

In this paper we review hard-body models as applied to the study of the equilibrium properties of liquid
crystal phases. The scope of the review is on fluids made of convex hard bodies, with 
emphasis on liquid-crystalline phases and phase transitions. Only bulk properties are covered; interfacial properties and 
inhomogeneous systems are reviewed in separate work \cite{forthcoming}.
Theories and results for the bulk isotropic phase will not be reviewed (excellent reviews exist; see e.g. Nezbeda 
\cite{Nezbeda}). Nor will dynamical properties and lattice models be mentioned. Also, the topic of flexibility, so important for real 
liquid-crystal-forming molecules, filamentous viruses and other colloidal particles, is left aside. 
An enormous volume of literature is devoted to the theory and computer
simulation of the liquid-crystal phases of hard-model fluids; as a consequence, and in order to present a manageable 
exposition of the subject, we have referenced only our personal choice of representative literature. Experimental results
are mentioned insofar as they help to illustrate, motivate or support studies on hard-body fluids. Some
reviews partially cover this subject \cite{Frenkel_rev,Leker_rev1,Leker_rev2,Tarazona_review}. 
For more specific resources on liquid crystals we refer the reader to the works cited in the introduction. 

\section{Hard-body models}

Let us consider a collection of $N$ particles in a volume $V$.
Let ${\bm r}_n$ be the centre-of-mass position of the $n$th particle with respect to the laboratory frame, 
and $\{\hat{\bm\Omega}_i^{(n)}\}$, with $i=1,2,3$, a set of three unit vectors describing its orientation with respect to the same frame. 
If the particles are hard bodies, the potential energy of an arbitrary configuration of the particles can be strictly written as a sum 
of two-body potentials,
\begin{eqnarray}
U\left({\bm r}^{(n)},\{\hat{\bm\Omega}_i^{(n)}\}\right)=\sum_{n\ne m}\phi\left({\bm r}_{nm},\{\hat{\bm\Omega}_j^{(n)}\},
\{\hat{\bm\Omega}_k^{(m)}\}\right),\hspace{0.4cm}{\bm r}_{nm}\equiv{\bm r}_n-{\bm r}_m.
\end{eqnarray}
The two-body potential $\phi$ has a simple structure:
\begin{eqnarray}
\phi\left({\bm r}_{nm},\{\hat{\bm\Omega}_j^{(n)}\},\{\hat{\bm\Omega}_k^{(m)}\}\right)=
\left\{\begin{array}{ccc}\infty,&r<\sigma\left(\hat{\bm r}_{nm},\{\hat{\bm\Omega}_j^{(n)}\},\{\hat{\bm\Omega}_k^{(m)}\}\right)&
\hbox{($n$ and $m$ overlap)}\\\\0,&
r>\sigma\left(\hat{\bm r}_{nm},\{\hat{\bm\Omega}_j^{(n)}\},\{\hat{\bm\Omega}_k^{(m)}\}\right)
&\hbox{(}n \hbox{ and } m \hbox{ do not overlap)}\end{array}\right.
\end{eqnarray}
The function $\sigma\left(\hat{\bm r}_{nm},\{\hat{\bm\Omega}_j^{(n)}\},\{\hat{\bm\Omega}_k^{(m)}\}\right)$ is the so-called
{\it contact distance}, which gives the minimum distance between two nonoverlaping particles at 
fixed orientations and fixed relative interparticle vector. This function in fact is sufficient to define 
the particle model.

Since the probability of a given configuration is proportional to the Boltzmann weighting factor $\exp{(-\beta U)}$, where
$\beta=1/kT$, with $k$ Boltzmann's constant and $T$ temperature, all permissible configurations of the $N$-particle system
will have zero potential energy and the Boltzmann factor is independent of $T$. The average internal energy $E$ at temperature $T$ 
only contains the kinetic
contribution, $E=3NkT/2$, and the equilibrium state of the system, given by $\delta F=0$, $\delta^2 F>0$ (where $\delta$ indicates variations
with respect to particle configurations at fixed density $\rho_0$ and $T$), is independent of the thermal
energy $kT$, which appears
as an irrelevant scaling factor in the free energy $F/kT=3N/2-S/k$, where the entropy $S$ only depends on the density
$\rho_0=N/V$. The physical behaviour is solely controlled by the entropy, which is 
related to the volume of the configurational space (hence actual volume) accessible to particles. One of the fascinating
properties of hard-body systems is that all ordered stable arrangements of particles have an entropic origin. The only
relevant parameter in the canonical ensemble is the density $\rho_0$ or, equivalently, the packing fraction $\eta$, defined
as the fraction of volume occupied by particles; in a monodisperse system, $\eta=\rho_0 v$, where $v$ is the volume of the 
particles.\\
\\
\begin{figure}[h]
\includegraphics[width=14cm,angle=0]{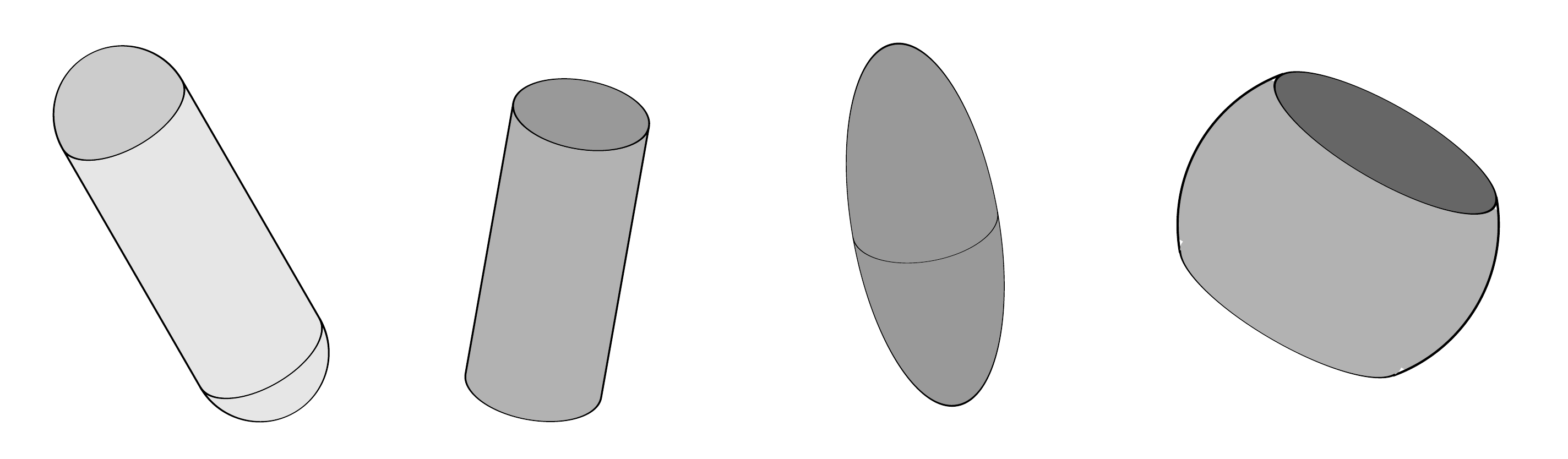}
\caption{Some popular models of hard body. From left to right: spherocylinder, cylinder, ellipsoid and cut sphere.}
\label{models}
\end{figure}

\subsection{Particles in 2D and 3D}

A virtually infinite number of hard-particle models can be defined. Here we focus on convex hard particles 
characterised by a shape and two aspect ratios. For bodies of revolution only one aspect ratio, $\kappa=l/d$, measuring the 
length ($l$) over width ($d$) ratio, is necessary. 
Prolate bodies (`rods') have $\kappa>1$, while oblate bodies (`platelets') have $\kappa<1$. In the limit $\kappa\to\infty$ all 
particles are equivalent and the virial coefficients tend to the same values. The limit $\kappa\to 0$ corresponds to 
infinitely thin platelets; here not all particles are equivalent, and the cross section of the particles is crucial in 
characterising the virial coefficients and statistical mechanics of the fluid.

A popular model for prolate uniaxial particles is the {\it hard spherocylinder} (HSC), Fig. \ref{models}: a cylindrical rod of length 
$L$ and diameter $D$ capped by a hemisphere of the same diameter at each end. The aspect ratio in this case is $\kappa=(L+D)/D$. 
This was the particle model used by Onsager \cite{Onsager} in his seminal paper on ordering of hard rods. Onsager considered 
the limit $\kappa\to\infty$ (hard needles), where the exact shape of the rod is not important. This limit is still used very often but
is not realistic. In the case of a finite and more realistic
value of $\kappa$ (comparable to that of liquid-crystal molecules or colloidal particles), the particle shape plays a role. 
The limit $L=0$ corresponds to the hard sphere (HS). 

A second popular model is the {\it hard ellipsoid} (HE), with axis lengths $a,b,c$, Fig. \ref{models}. 
If all three axes are different, $a\ne b\ne c$, the model represents a biaxial particle. If the particle has symmetry of revolution, 
two lengths are identical, say $b=c$, and only a single aspect ratio $\kappa=a/b$ is relevant and we have a model for a
uniaxial particle. A numerically convenient 
approximation for the HE model is the hard Gaussian overlap model (HGO) \cite{Berne-Pechukas}.
In the limit $\kappa\to\infty$ the physical properties of all of these fluids coincide but for even moderate aspect ratios
virial coefficients are very close \cite{Frenkel,Padilla1,Padilla2}. 
Other models for prolate particles include {\it hard cylinders} (HC, Fig. \ref{models}), 
{\it chain spheres}, {\it fused spheres}, {\it rectangular prisms} (boards), and combinations of all these. For example, 
banana-shaped particles meant to reproduce biaxial nematic phases can be constructed by attaching two rods by its ends
with a fixed angle between them \cite{banana_shaped}.
When considering rigid models for simple phases, however, the most obvious choices are HSC, HE, HGO and HC.

In the case of oblate particles, meant to reproduce molecular or colloidal discotic-forming liquid crystals,
there is also a great variety of models. In his paper, Onsager \cite{Onsager} used infinitely thin discs (i.e.
infinitely flattened cylinders) to compute
the ordering properties of model discotic particles. In real systems the particle thickness may be small compared to the
width, but in any case finite. An advantage of the HE and HGO models is that they can also be used for oblate particles
when one of the lengths is very small compared to the others ($a\ll b\simeq c$).
Apart from flattened HE or HGO models, workers have used other models such as {\it hard cut spheres}, 
HCS, i.e. spheres cut by 
two planes at the same distance from the equator, Fig. \ref{models}), HC with $L\le D$, 
and {\it hard square platelets} (rectangular prisms of side 
lengths $l_1,l_2,l_3$ with $l_1\ll l_2\simeq l_3$). 

In 2D corresponding models can be defined: {\it hard ellipses}, {\it hard rectangles} and {\it hard discorectangles}, 
which are projections of respectively ellipsoids, cylinders and spherocylinders on a plane parallel to their
uniaxial axes. Fused discs can also be used. 
Zig-zag or cross particles have been examined by attaching a number of rods at fixed angles \cite{Sabi_Mex,Gurin_Mex}. 

As we have seen, a huge number of models have been used to represent real molecular or colloidal systems. The only
requirement when defining a model is the possibility to be able to explicitly or implicitly compute, in an efficient
way, the contact distance $\sigma\left(\hat{\bm r},\{\hat{\bm\Omega}\},\{\hat{\bm\Omega}'\}\right)$ between two
particles. If particles are not identical (mixtures or size polydisperse systems), contact distances between all pairs of
species must be known. Particles made of attached units are easily handled since the total contact distance can be obtained
from the partial contact distances between the units.

\subsection{Contact with real materials}

Hard models are approximate models for systems made of anisotropic molecules, since they can provide the essential correlation
characteristics originating from the closed-shell electronic distributions of simple, non-covalent systems,
and explain ordering as an entropy effect. However, hard models are athermal, and therefore cannot
account for the temperature dependence of material properties. Even density-dependent properties are not, in general,
quantitatively described by hard models. For example, calculations of the I--N phase transition using hard-rod models 
predict too low a value for the density gap at the transition as compared to real materials.
Properties related to cohesion cannot be quantitatively described by overlap interactions and anisotropic attractive 
interactions need to be included in the theoretical treatments (either theory or simulation) of liquid crystals. 
The effect of these interactions are usually treated theoretically using perturbation theory, which require the
correlation structure of the corresponding hard model. Nevertheless, hard models are useful tools to qualitatively analyse 
the properties of molecular matter.

Hard models also find application
in stabilised colloidal suspensions of Coulomb-screened anisotropic particles in a solvent. In this case the 
agreement with real materials is more quantitative. An important field of application is in suspensions of
filamentous viruses. Early investigations on the
first virus to be isolated, the Tobacco Mosaic Virus (TMV), indicated that it formed a N phase. Onsager \cite{Onsager}
explained the formation of this phase in terms of a solution of charged hard rods of and effective diameter $D_{\rm eff}$
that depends on the ionic strength of the solution; the transition was explained solely as a competition between orientational
and correlation entropies. More recent experiments on virus suspensions are analysed
with hard models of the same type \cite{Fraden_review,Fraden_review1}.

Another type of experimental colloidal system which can be closely modelled by hard models is a suspension of mineral 
or polymeric particles. Several groups have been working on these systems for a long time, and a huge list of interesting
results concerning phases with orientational and spatial ordering have been obtained.
For example, suspensions of goethite nanorods have been shown to exhibit liquid-crystalline ordering \cite{goethite,goethite1}.
The particle interactions can be made to be approximately hard by suitable chemical treatments. 
Anisotropic particles made from sheets of the layered gibbsite mineral have been stabilised by layers of grafted polymer,
a method that, in a good solvent such as toluene, produces size polydisperse platelets which interact approximately 
through hard interactions \cite{Kooij-Lek}. Various liquid-crystalline phases have been observed in this system, as
predicted by theoretical models of hard platelets. Other particles made from layered mineral materials, e.g. ZrP,
have also been shown to exhibit liquid-crystal phases \cite{chino}. Also, prolate particles made from polymeric
materials have been synthesised \cite{PMMA}, and their interactions can be closely approximated by hard interactions.

In summary, hard models are very useful to qualitatively understand phase behaviour in real liquid-crystal-forming
molecular systems. But they can also be used to analyse colloidal suspensions of anisotropic particles made of
viruses, mineral or polymeric particles, in many cases with quantitative agreement.

\section{Theories and simulations}

In this section we make a short revision of the theories for the liquid-crystal bulk state of anisotropic hard particle models
in the language of Density-Functional Theory (DFT). This language is essential to understand inhomogeneities and phase
transitions in condensed systems made of hard particles since the structure and correlations in the fluid are core 
ingredients of DFT. For a general, recent review on DFT see \cite{Lutsko}.
The physical properties of hard bodies are governed by entropy, which is related to the volume accessible to 
particles. Therefore DFT theories, which are based on geometry and excluded volume between two particles, have played an 
essential role in the development of the field. 

The revision is not extensive but stresses the main ideas and concepts used to construct theories 
and the success, or otherwise, of the resulting theories. The application of the theories to explain the formation of the 
different LC phases is then mentioned, first for the nematic, then for the phases with spatial order. 
In accord with the spirit of this review, we do not mention theories that incorporate soft attractive interactions 
(of Maier-Saupe type or perturbation theories).
A more complete review on the fundamentals and applications of theoretical approaches to a larger variety of models for liquid crystals 
can be found in \cite{Singh_review}. 

\subsection{Density-functional theory for anisotropic hard bodies}

\subsubsection{Uniaxial particles}

The key ingredient of DFT is the ensemble average local density of particles, $\rho\left({\bm r},\hat{\bm{\Omega}}\right)$, giving the local density
of particles at position ${\bm r}$ and with orientation given by the unit vector $\hat{\bm\Omega}$; this is the orientation of the 
symmetry axis of the particle, assuming it only has one (uniaxial particle); in the case of less symmetric particles, 
more vectors are needed. One can define a local orientational distribution function 
$h\left({\bm r},\hat{\bm{\Omega}}\right)$ by extracting the density dependence, 
$\rho\left({\bm r},\hat{\bm{\Omega}}\right)=\rho\left({\bm r}\right)h\left({\bm r},\hat{\bm{\Omega}}\right)$. The
local orientational distribution function is normalised, 
$\displaystyle\int d\hat{\bm{\Omega}} h\left({\bm r},\hat{\bm{\Omega}}\right) = 1$.
In the uniform I and N phases there is no spatial dependence, $\rho\left({\bm r}\right)=\rho_0$, and one can write 
$\rho({\bm r},\hat{\bm{\Omega}})=\rho_0h(\hat{\bm\Omega})$.

In DFT the Helmholtz free energy $F$ is written as a functional of the local density, $F[\rho]$. In the grand 
canonical ensemble, where $\mu,V,T$ are fixed ($\mu$ is the chemical potential), one defines the grand-canonical functional, 
\begin{eqnarray}
\Omega[\rho]=F[\rho]-\mu\int_V d{\bm r}\int d\hat{\bm\Omega}\rho\left({\bm r},\hat{\bm{\Omega}}\right).
\end{eqnarray}
The equilibrium state of the fluid follows by functionally minimising the functional, 
\begin{eqnarray}
\left.\frac{\delta\Omega[\rho]}{\delta\rho({\bm r},\hat{\bm{\Omega}})}\right|_{\rm eq}=0
\hspace{0.8cm}\hbox{at constant $\mu$, $V$ and $T$}.
\end{eqnarray}
Once the equilibrium local density $\rho\left({\bm r},\hat{\bm{\Omega}}\right)$ is 
obtained, the equilibrium free energies follow by evaluating the corresponding functional. 

To microscopically characterise orientational ordering in a uniaxial liquid crystal a set of order parameters 
$\left\{f_{nm}\left({\bm r}\right)\right\}$ are defined in terms of a spherical-harmonic expansion of the orientational distribution function:
\begin{eqnarray}
\rho\left({\bm r},\hat{\bm{\Omega}}\right)=
\rho\left({\bm r}\right)\sum_{l=0}^{\infty}\sum_{m=-l}^lf_{lm}\left({\bm r}\right)Y_{lm}\left(\hat{\bm{\Omega}}\right),
\hspace{0.4cm}f_{lm}\left({\bm r}\right)=\int d\hat{\bm{\Omega}}h\left({\bm r},\hat{\bm{\Omega}}\right)Y^*_{lm}\left(\hat{\bm{\Omega}}\right),
\end{eqnarray}
where $Y_{lm}\left(\hat{\bm{\Omega}}\right)$ are spherical harmonics.
Since $\displaystyle\int d\hat{\bm{\Omega}}\rho\left({\bm r},\hat{\bm{\Omega}}\right)=\rho\left({\bm r}\right)$, the local number density,
one obtains $f_{00}\left({\bm r}\right)=1$. It is common to limit the expansion to the $l=2$ level, and define order parameters
$f_{1m}$, with $m=-1,0,+1$, and $f_{2m}$, with $m=-2,\cdots,2$, to quantify the orientational order. For nonpolar particles,
if one chooses the $z$ axis of the reference frame along the director $\hat{\bm n}$, the order parameters reduce to 
\begin{eqnarray}
Q\left({\bm r}\right)=\int d\hat{\bm\Omega} 
h\left({\bm r},\hat{\bm{\Omega}}\right)P_2\left(\hat{\bm n}\cdot\hat{\bm{\Omega}}\right),
\end{eqnarray}
where $P_2(x)$ is the second-order Legendre polynomial and $\hat{\bm n}\cdot\hat{\bm{\Omega}}=\cos{\theta}$, where
$\theta$ is the angle between the molecular axis and the director.

The standard practise is to split the functional in ideal, $F_{\rm id}[\rho]$, and excess (or interaction), 
$F_{\rm ex}[\rho]$, contributions, so that 
\begin{eqnarray}
F[\rho]=F_{\rm id}[\rho]+F_{\rm ex}[\rho].
\end{eqnarray}
$F_{\rm id}[\rho]$ would be the free energy of a fluid of non-interacting particles with local density 
$\rho({\bm r},\hat{\bm{\Omega}})$. It has an exact expression, and it was first derived by Onsager \cite{Onsager}
by mapping the possible particle 
orientations onto a mixture of different `chemical' species, each species corresponding to a different orientation $\hat{\bm\Omega}_i$. 
With this idea $h(\hat{\bm\Omega}_i)$ is akin to $x_i$, the composition of the 
$i$th species in a multicomponent mixture, and the ideal free energy can be built from the mixing entropy of a multicomponent 
mixture. Generalising to systems with positional order:
\begin{eqnarray}
\beta F_{\rm id}[\rho]&=&\int_V d{\bm r}\int d\hat{\bm{\Omega}} \rho({\bm r},\hat{\bm{\Omega}})
\left\{\log{\left[\rho({\bm r},\hat{\bm{\Omega}})\Lambda^3\right]}
-1\right\}
\nonumber\\\nonumber\\&=&\int_V d{\bm r}\rho\left({\bm r}\right)
\left\{\log{\left[\frac{\rho\left({\bm r}\right)\Lambda^3}{4\pi}\right]}-1\right\}+
\int_V d{\bm r}\rho\left({\bm r}\right)\left<\log{\left[4\pi h\left({\bm r},\hat{\bm{\Omega}}\right)\right]}\right>_h,
\label{fid}
\end{eqnarray}
where $\left<\cdots\right>_h$ is an angular average weighted by the distribution function 
$h\left({\bm r},\hat{\bm{\Omega}}\right)$.
In the expression above $\Lambda$ is the thermal wavelength. The first term corresponds to the positional entropy, 
whereas the second is the orientational 
entropy (the factor $4\pi$ is introduced so that the latter vanishes in an orientationally disordered fluid).
The local orientational entropy per particle (`mixing entropy') in units of the Boltzmann constant $k$ is defined as
\begin{eqnarray}
s_{\rm or}\left({\bm r}\right)\equiv -\left<\log{\left[4\pi h\left({\bm r},\hat{\bm{\Omega}}\right)\right]}\right>_h=
-\int d\hat{\bm{\Omega}} h\left({\bm r},\hat{\bm{\Omega}}\right)
\log{\left[4\pi h\left({\bm r},\hat{\bm{\Omega}}\right)\right]}.
\end{eqnarray}
For uniform phases, i.e. I or N phases, we have
\begin{eqnarray}
\frac{\beta F_{\rm id}[h]}{N}&=&\log{\left(\frac{\rho_0\Lambda^3}{4\pi}\right)}-1-s_{\rm or}[h].
\end{eqnarray}
The orientational entropy $s_{\rm or}$ is one of the key actors in the theory of liquid-crystal
phase transitions. $s_{\rm or}$ has a maximum value when particles are disordered,
i.e. the I phase, with $h(\hat{\bm{\Omega}})=(4\pi)^{-1}$. 

The other actor is the excess free-energy (excess-entropy) functional $F_{\rm ex}[\rho]$, which does not have
an exact expression. Essentially two routes have been followed to construct $F_{\rm ex}[\rho]$: the Onsager
theory, based on the concept of excluded volume, and theories based on weighted densities.

\subsubsection{Excluded volume: Onsager theory}

In Onsager theory the excess free energy is written in terms of a central quantity in hard-body systems, the
{\it excluded volume} between two particles. The excluded volume is directly related to the  
second virial coefficient. Onsager's derivation of the excess functional considers the cluster or virial
expansion of the equivalent multicomponent mixture. One obtains:
\begin{eqnarray}
\frac{\beta F_{\rm ex}[h]}{N}=B_2[h] \rho_0 + \frac{1}{2}B_3[h]\rho_0^2+\cdots
\label{virial_F}
\end{eqnarray}
where $B_n[h]$ are the virial coefficients (functionals of the orientational distribution function), 
the first of which is
\begin{eqnarray}
B_2[h] =-\frac{1}{2}\left<\left<
\int_V d{\bm r}f({\bm r},\hat{\bm{\Omega}},\hat{\bm{\Omega}}^{\prime})
\right>\right>_h,
\end{eqnarray}
where $f({\bm r},\hat{\bm{\Omega}},\hat{\bm{\Omega}}^{\prime})=
\exp{\left[-\beta\phi({\bm r},\hat{\bm{\Omega}},\hat{\bm{\Omega}}^{\prime})\right]}-1$ is the Mayer function, 
and ${\bm r}$ the relative position vector
joining the centres of mass of two particles (or the position vector of one particle, with orientation 
$\hat{\bm{\Omega}}^{\prime}$, assuming the other, oriented along $\hat{\bm{\Omega}}$, lies at the origin). 
$\left<\left<\cdots\right>\right>$ denotes double angular average over the function $h(\hat{\bm{\Omega}})$.
For two bodies that interact via a hard potential, the Mayer function is particularly simple:
\begin{eqnarray}
f({\bm r},\hat{\bm{\Omega}},\hat{\bm{\Omega}}^{\prime})=\left\{\begin{array}{ll}-1,&\hbox{particles overlap}\\\\
0,&\hbox{particles do not overlap}\end{array}\right.
\end{eqnarray}
As a consequence, the second virial coefficient is directly related with the so-called `excluded volume':
\begin{eqnarray}
B_2[h] =\frac{1}{2}\left<\left<v_{\rm excl}(\hat{\bm{\Omega}},\hat{\bm{\Omega}}^{\prime})\right>\right>_h
=\frac{1}{2}\int d\hat{\bm\Omega}\int d\hat{\bm\Omega}'
v_{\rm excl}(\hat{\bm{\Omega}},\hat{\bm{\Omega}}^{\prime})
h(\hat{\bm{\Omega}})h(\hat{\bm{\Omega}}^{\prime}).
\label{v_exc}
\end{eqnarray}
$\displaystyle v_{\hbox{\tiny excl}}(\hat{\bm{\Omega}},\hat{\bm{\Omega}}^{\prime})=-\int_Vd{\bm r}f({\bm r},\hat{\bm{\Omega}},\hat{\bm{\Omega}}^{\prime})$ is the volume excluded to one particle because of
the presence of the other. We could approximately obtain the entropy of the system by noting that, if the two particles
were isolated (low-density limit), the total volume accessible
to a particle would be $V-\frac{N-1}{2}\left<\left<{v}_{\hbox{\tiny excl}}\right>\right>_h\simeq
V-NB_2[h]$ (the factor $1/2$ accounts for the fact that two particles are
involved and the excluded volume has to be shared by the two). The partition function would then be
\begin{eqnarray}
Q=\frac{\Lambda^{3N}}{N!}\left(V-NB_2[h]\right)^N=
Q_{\hbox{\tiny id}}\left(1-\rho_0B_2[h]\right)^N,
\end{eqnarray}
and the excess free energy per thermal energy unit $\beta F_{\hbox{\tiny exc}}=-\log{\left(Q/Q{\hbox{\tiny id}}\right)}$ would 
read
\begin{eqnarray}
&&\frac{\beta F_{\hbox{\tiny exc}}}{N}=-\log{\left(1-\rho_0B_2[h]\right)}=B_2[h]\rho_0+\cdots
\label{Ons}
\end{eqnarray}
The first term of this expansion coincides with that of Eqn. (\ref{virial_F}).
Onsager qualitatively demonstrated that, for hard rods in the isotropic phase, 
the ratio $B_3/B_2^2\to (D/L)\log{(L/D)}$ so that, in the
limit $L/D\to\infty$, the third virial coefficient is indeed negligible for very long rods (hard-needle limit). 
It is then plausible that higher-order virial coefficients also vanish in the same limit. 

Now the excluded volume has to be specified. For HSC particles, the excluded volume is given by
\begin{eqnarray}
v_{\rm excl}(\hat{\bm{\Omega}},\hat{\bm{\Omega}}^{\prime})=8v_0+2L^2D
\left|\sin{({\hat{\bm\Omega},\hat{\bm\Omega}'})}\right|,
\label{vexc}
\end{eqnarray}
where $v_0=\frac{\pi}{4}LD^2+\frac{\pi}{6}D^3$ is the volume of a HSC. To simplify the calculations,
Onsager considered the excluded volume in the limit of infinite aspect ratio, $\kappa\gg 1$ (hard needles), 
i.e. $L\gg D$: $v_{\rm excl}(\hat{\bm{\Omega}},\hat{\bm{\Omega}}^{\prime})=2L^2D
\left|\sin{\left({\hat{\bm\Omega},\hat{\bm\Omega}'}\right)}\right|$.
Applying the equilibrium condition 
\begin{eqnarray}
\left.\frac{\delta F[h]}{\delta h({\bm\Omega})}\right|_{\rm eq}=\lambda,
\end{eqnarray}
where $\lambda$ is a Lagrange multiplier ensuring the normalisation condition on $h({\bm\Omega})$,
one obtains an integral, Euler-Lagrange, equation for $h(\hat{\bm{\Omega}})$,
\begin{eqnarray}
h\left(\hat{\bm{\Omega}}\right)=\frac{\displaystyle
e^{\displaystyle -\rho_0\int d\hat{\bm\Omega}_1h\left(\hat{\bm{\Omega}}_1\right)v_{\rm exc}
\left(\hat{\bm{\Omega}},\hat{\bm{\Omega}}_1\right)}}
{\displaystyle\int d\hat{\bm\Omega}_3e^{\displaystyle -\rho_0\int d\hat{\bm\Omega}_2h\left(\hat{\bm{\Omega}}_2\right)v_{\rm exc}
\left(\hat{\bm{\Omega}}_3,\hat{\bm{\Omega}}_2\right)}}.
\label{EL}
\end{eqnarray}
This is an integral equation which has to be solved numerically. In addition, Onsager used a trial function in terms of a variational
parameter, $\alpha$:
\begin{eqnarray}
h(\hat{\bm\Omega})=\frac{\displaystyle e^{\displaystyle\alpha\cos{(\hat{\bm\Omega}\cdot\hat{\bm n})}}}
{\displaystyle\int d\hat{\bm\Omega}'\hspace{0.1cm}
e^{\displaystyle\alpha\cos{(\hat{\bm\Omega}'\cdot\hat{\bm n})}}}.
\label{variational}
\end{eqnarray}
In the I phase, $\alpha=0$, whilst in the N phase $\alpha>0$. Note that, in the hard-needle limit, the solution only
depends on the scaled density $\rho^*=\rho_0L^2D$. The theory predicts a first-order phase transition between the I and
N phases for the following values of packing fraction $\eta=(\pi D/4L)\rho^*$:
\begin{eqnarray}
\eta_{\rm I}=3.340\hspace{0.1cm}\frac{D}{L},\hspace{0.6cm}\eta_{\rm N}=4.486\hspace{0.1cm}\frac{D}{L}.
\label{Onsager_res}
\end{eqnarray}
The density gap becomes smaller as the particles go to the hard-needle limit. 
As the ratio $L/D$ is reduced, end-particle effects in the excluded volume
[first term in Eqn. (\ref{vexc})] become important, there is no universal scaled density,
and the coexistence values depart from the values (\ref{Onsager_res}).
The relative density gap is $\Delta\eta/\eta_{\rm N}=(\eta_{\rm N}-\eta_{\rm I})/\eta_{\rm N}\simeq 26\%$, 
too high compared with typical experimental values. The value of the uniaxial order parameter at the transition is
$Q_{\rm IN}=0.84$, also too high with respect to experiment. Lasher \cite{Lasher} numerically solved the problem
without resorting to the variational function (\ref{variational}), but using an expansion in Legendre polynomials.
The study included the $8v_0$ factor of the 
excluded volume in the calculation of the coexistence densities, thus incorporating the dependence on aspect ratio.

Onsager theory has been more closely examined, from a numerical point of view, by Kayser and Raveche \cite{Kayser} and
Herzfeld et al. \cite{pre_Odijk}, focusing on an iterative algorithm to obtain the solution. This algorithm has been shown to be
convergent. Also, Stroobants et al. \cite{Odijk}, examined the convergence properties of a exponentiated 
truncated Legendre expansion,
\begin{eqnarray}
h(\hat{\bm\Omega})=\frac{\displaystyle
\exp{\displaystyle\left\{\sum_{n=0}^{\infty}\alpha_{2n}
P_{2n}\left(\hat{\bm\Omega}\cdot\hat{\bm n}\right)\right\}}}
{\displaystyle\int d\hat{\bm\Omega}'\hspace{0.1cm}
\exp{\displaystyle\left\{\sum_{n=0}^{\infty}\alpha_{2n}P_{2n}\left(\hat{\bm\Omega}'\cdot\hat{\bm n}\right)\right\}}}.
\end{eqnarray}
The convergence of the results was studied as a function of the number of terms included in the expansion. As a main conclusion,
it was found that the coexistence results depend very much on the form of distribution function used.

The Onsager theory was modified by Zwanzig \cite{Zwanzig} by considering an idealised model where rods can only point
along a discrete set of orientations. Despite this drastic simplification, but as a bonus, Zwanzig could extend the
theory to the seventh-order virial coefficient and study the robustness of Onsager's orientational transition. 
Zwanzig was able to confirm the transition. However, the convergence of the expansion is probably poor since 
coefficients are oscillatory. A later analysis by Runnels and Colvin \cite{Runnels} used a Pad\'e expansion to limit this problem, 
and they confirmed in turn Zwanzig's result.

Calculation of virial coefficients for various models have been done for the isotropic phase, with a view
to constructing accurate resummations of the equation of state \cite{Nezbeda,Rigby,Rigby1,Rigby2,Rigby3,Rigby4,Gubbins,Solana,deMiguel,You}. 
All of these studies are focused on short particles and therefore say nothing about the validity of Onsager theory. 
Frenkel \cite{Frenkel_revisited} has calculated the scaled third, fourth and fifth virial coefficients of hard spherocylinders 
in the isotropic phase (i.e. $B_n/B_2^{n-1}$ for $n=3,4,5$) for aspect ratios $\kappa=L/D$ in the range $1$--$10^6$, 
using Monte Carlo integration.
He concluded that all coefficients vanish in the hard-needle limit. However, the vanishing limit value is obtained
rather slowly with $\kappa$, which means that Onsager theory is not quantitatively valid for realistic values of
aspect ratio. This does not imply that excluded volume effects are not responsible for nematic formation
in hard-body fluids, but rather that it is necessary to improve Onsager theory in order to obtain quantitative
predictions. To this effect, some proposals have been considered and are reviewed later.

Calculation of virial coefficients are normally done for a constant orientational distribution $h(\hat{\bm\Omega})=1/4\pi$ 
or zero order parameter $Q=0$. In contrast, Velasco and Padilla \cite{Padilla2} used Monte Carlo integration
to calculate $B_3[h]$--$B_5[h]$ for the HGO model, in the range $\kappa=1$--$10^5$, by parameterising  $h(\hat{\bm\Omega})$ 
in terms of $Q$, thus obtaining the virial coefficients in the nematic phase as  $B_n\left(Q\right)$ in the range $Q=0$--$1$. 
The vanishing of the virial coefficients in the hard-needle limit was confirmed, and convergence was seen to be rather 
insensitive with respect to the order parameter, indicating that Onsager theory is valid in the limit $\kappa\to\infty$. 

Onsager theory has played a very important role in our qualitative understanding of orientational ordering, bringing the concept 
of particle excluded volume to the forefront. It is indeed remarkable that a simple second-order virial theory, based solely
on two-particle interactions, can explain a phase transition; this is a singular case in the theory of phase transitions.
The virial expansion is useful to connect the ideal gas to denser gases, i.e. fluids where interactions begin to be
relevant, but in standard fluids of isotropic particles there is still a long way between the low-order virial expansion 
and the real equation of state of a liquid. However, the nematic phase is a peculiar example where a phase transition
to an ordered state can be understood in terms of the first term in the virial expansion, i.e. the second virial coefficient.

By the time Onsager presented his theory, computer simulations of phase transitions for condensed matter were 
not feasible. With the advent of powerful computers in the '70s, it was possible to obtain accurate predictions and
put Onsager theory to the test. Up to now many simulations on hard bodies have been performed, using MC or MD techniques.
A good starting point to review the subject is \cite{Frenkel_rev}.
Several works have specifically focused on devising algorithms for efficiently obtaining contact distances of hard bodies,
particularly ellipsoids \cite{Perram1}. The first simulations of a liquid crystal were performed by 
Vieillard-Baron \cite{VB} (1972), who used 
Monte Carlo simulation to study a fluid of 2D hard ellipses. He observed the formation of an orientationally 
ordered phase, although the system 
studied was later revealed to be too small to obtain any quantitative conclusion. Much later, 
Frenkel et al. \cite{Frenkel} (1984)
studied a fluid of hard ellipsoids. The entire phase diagram, including I, N and crystal phases, 
was mapped out as a function
of density and aspect ratio, including prolate and oblate particles. For low aspect ratios a plastic phase
was also detected. The smectic phase was not stabilised in this fluid. 

In a more extensive simulation work involving free-energy calculations, Frenkel and Mulder \cite{Frenkel0} studied the density
stability interval of the N phase of hard ellipsoids and concluded that, for aspect ratio 3, it was very narrow. 
But the stability of this fluid was put into question
by Zarragoicoechea et al. \cite{Zarra}, who concluded from their own MC simulations that the stability of the nematic phase
was due to the small system size of the samples. In an effort to solve the issue, Allen and Mason \cite{Allen} performed MC and MD simulations 
to examine the stability of the N phase of hard ellipsoids with aspect ratio 3, using system sizes larger than previously. Although they
recognised the existence of system-size effects, the N phase was observed to be stable. The first 
simulations of a liquid crystal exploring the effect of fluctuations 
was performed in 1987 \cite{Allen-Frenkel}, where dynamical precursors of the I--N transition in hard ellipsoids 
were observed. All
analyses and simulation works have concluded that the Onsager theory is qualitatively valid, but the truncated virial 
expansion after the second coefficient is only valid for aspect ratios $L/D\agt 100$, and predictions for particles with 
realistic aspect ratios are unreliable \cite{Straley_0}.

\subsubsection{Extended Onsager theories}
\label{extensions}

There have been a number of attempts to improve Onsager theory, in an effort to make it applicable to realistic particle lengths.
Three approaches have been proposed. The first incorporates the contribution of third- and higher-order virial 
coefficients to the Onsager functional. In the second, angular correlations are treated at the Onsager level (second virial
coefficient), but spatial correlations are improved using the decoupling approximation. Mixed theories have been 
proposed in an attempt to treat the coupling of both, spatial and angular correlations, in a more quantitative fashion.
Finally, an approach based on the developments of Fundamental-Measure Theory for HS, extended to anisotropic
bodies, has been proposed. In principle, this approach improves the spatial-orientational coupling. Except for the special case of 
a mixture of freely-rotating needles and discs, it has been applied in the restricted-orientation approximation, although there have been recent
developments to extend the theory to freely-rotating particles, as reviewed later.

Straley \cite{Straley_3D} calculated $B_3$ numerically for hard rods of aspect ratios
10--100, and proposed a corresponding model which approximated its angular dependence. He estimated the correction of 
$B_3$ to the results of Onsager theory, and concluded that the theory is not accurate for $\kappa<100$. 
Tjipto-Margo and Evans \cite{Margo} considered the incorporation of the third virial 
coefficient $B_3$, using hard ellipsoids as particle model. For aspect ratios larger than 5 the extended theory
predicts the correct variation of the order parameter with density as compared with simulations. The results are
poorer in the case of shorter particles. A way to incorporate the third virial coefficient in the $y$-expansion 
of Barboy and Gelbart \cite{Barboy} was presented and the corresponding results for the transition densities were seen to
agree with the MC simulations of Frenkel and Mulder \cite{Frenkel1}. Padilla and Velasco \cite{Padilla1} obtained the virial 
coefficients $B_3$ and $B_4$ of HGO particles from Monte Carlo integration, both as functions of the uniaxial
nematic order parameter, and used them to study the convergence of the virial series in terms of the results for 
the I--N transition. The series was seen to converge quite fast in the case of $\kappa=5$, but
not for aspect ratio equal to 3. Although theoretically interesting, all of these approaches are not practical since the 
calculation of high-order virial coefficients is intractable for most hard-particle models.

More recently, You et al. \cite{You} 
calculated virial coefficients up to seventh order for the isotropic phases of hard ellipsoids, spherocylinders
and truncated hard spheres for different aspect ratios, with a view to studying the convergence properties of the 
virial series. The radius of convergence was seen to be close to the close-packing limit for low aspect ratio, and
to be considerably less than the close-packed density for higher aspect ratios.

In the '70 Cotter and coworkers \cite{Cotter1,Cotter2} derived equations of state for a fluid of perfectly aligned rods, 
applying a generalisation of the scaled-particle theory (SPT) of Reiss et al. for HS \cite{Reiss} to parallel anisotropic 
particles. The theory was then generalised to a set of 
restricted orientations and results for the equation of state and the 
I--N transition were obtained \cite{Cotter3}. Lasher \cite{Lasher} modified SPT theory and obtained an extended Onsager expression
that corrected Onsager theory. The results of the generalised SPT theory, in the low-density limit, 
were compared to the then existing theories, including that of Flory \cite{Flory} and Alben \cite{Alben} 
(formulated on a lattice), by Straley \cite{Straley_0} (a review of these early theories can be found in this reference).
A problem with these theories is that angular correlations are still considered at the level of two particles but, 
for short particles, it is clear that higher-order angular correlations, represented by virial coefficients beyond the 
second, are important. Despite recent efforts in this direction, this question is as yet unsolved.

SPT and similar theories can be regarded as resummations of the virial expansion. 
Barboy and Gelbart \cite{Barboy} introduced the $y$-expansion for general hard bodies, a type of truncated Pad\'e-approximant
resummation based on the variable $y=\eta/(1-\eta)$, where $\eta$ is the packing fraction. This expansion is closely related
with the SPT theory, exhibits better 
convergence properties than the usual virial expansion based on the density, and incorporates higher-order virial
coefficients in a systematic way (providing these are known for the model at hand). The method was applied, in particular, to 
dumbells and spherocylinders in the isotropic phase \cite{Barboy}, but oriented fluids (hard parallelepipeds) were studied later,
using the restricted-orientation (Zwanzig) approximation, in Ref. \cite{Barboy1}. Comparison with previous virial expansions and 
Pad\'e approximants was made, but no clear superiority of the theory, both quantitatively and in the practical implementation, 
was inferred.

In the '80 some further proposals in the theory of nematic ordering were made, using a seemingly different perspective
which at the end was interpreted as yet another resummation theory providing qualitatively, if not quantitatively, similar
results as previous theories. Therefore, in all of these approaches angular
correlations are kept at the level of the second-order virial coefficient.
Already in 1979 Parsons \cite{Parsons} made an interesting proposal, considering potentials of the form 
$\phi\left(r/\sigma(\hat{\bm r},\hat{\bm\Omega},\hat{\bm\Omega}')\right)$, 
where $\sigma(\hat{\bm r},\hat{\bm\Omega},\hat{\bm\Omega}')$ is the contact distance function. 
He showed that, if the radial distribution function $g({\bm r},\hat{\bm\Omega},\hat{\bm\Omega}')$ can be scaled as 
\begin{eqnarray}
g({\bm r},\hat{\bm\Omega},\hat{\bm\Omega}')=g_0\left(\frac{r}{\sigma(\hat{\bm r},\hat{\bm\Omega},\hat{\bm\Omega}')}\right),
\label{PL1}
\end{eqnarray}
where $g_0$ is the corresponding function for a reference isotropic fluid, then the orientational and 
translational degrees of freedom decouple to all orders in the density expansion of the free energy.
This approximation was proposed for the first time by Pynn \cite{Pynn,Pynn1} and later used by Wulf \cite{Wulf}
for the direct correlation function in the context of the solution of the Orstein-Zernike equation for the I--N transition. 
Interestingly, as $\rho_0\to 0$, the resulting free energy reduces to that of Onsager. 
By using a HS reference system and the corresponding equation of state, Parsons
predicted a I--N phase transition at a packing fraction which decreased with rod aspect ratio. Also, using a perturbation
theory in the softness $m$ of the potential $\phi(r)=\epsilon/r^m$, predictions for soft potentials were made and seen to agree with
experimental results for $m=12$. Finally, Parsons concluded that most of the features of the I--N 
transition are due to the repulsive part of the interactions. 

Later, Lee \cite{Lee} rederived Parsons' theory from a different but equivalent point of
view, giving rise to the now known as the {\it Parsons-Lee} (PL) theory. In Lee's approach the emphasis is on the equation of state.
The approach starts from the Carnahan-Starling equation of state for HS and generalises it to the nematic case using a simple (but
a priori purely heuristic) functional scaling for the excess free energy, as 
\begin{eqnarray}
\frac{\beta F_{\rm ex}[h]}{N}=\Psi_{\rm HS}(\eta)\frac{B_2[h]}
{B_2^{\rm (HS)}},
\label{PL2}
\end{eqnarray}
where $B_2[h]$ is the angle-averaged second virial coefficient, which was defined in (\ref{v_exc}), 
and $B_2^{\rm (HS)}=4v_0$ is the second-order virial coefficient of HS, equal to half the excluded volume of
two HS ($v_{\rm exc}^{\rm (HS)}=8v_0$, $\displaystyle v_0=\frac{\pi}{6}\sigma^3$ being the HS volume).
$\eta$ is the packing fraction of the actual fluid {\it and} of the reference HS fluid; therefore, the fluid is mapped onto a HS
fluid of equal packing fraction and density consisting of hard spheres of a volume equal to that of the actual anisotropic particles,
and any density-dependent function is scaled with the ratio of excluded volumes between the actual fluid and the reference HS fluid.
$\Psi_{\hbox{\tiny HS}}(\eta)$ is the excess free-energy per unit thermal energy $kT$ and per particle of the HS fluid, 
obtained from the Carnahan-Starling theory,
$\Psi_{\rm  HS}(\eta)=(4-3\eta)\eta/(1-\eta)^2$,
or from any other theory for the equation of state of the HS fluid.
This approximation can be seen to be completely equivalent to the decoupling approximation of Parsons \cite{Parsons}. 
The theory was applied to HSC and good agreement with simulation as concerns the I--N transition was reported, 
even for fairly short particles. 
Lee also applied the theory to hard ellipsoids of revolution \cite{Lee1} and compared with computer simulation. 
Again, as the aspect ratio of the particles becomes higher, the agreement is seen to improve. 
The theory again reduces to that of Onsager in the limit of low density. 

The PL scaling, Eqn.  (\ref{PL2}), can be obtained by approximating the virial 
coefficients of the fluid $B_n[h]$, which are functionals of the orientational distribution function $h(\hat{\bm\Omega})$, 
in terms of those of the HS fluid, $B_n^{(\hbox{\tiny HS})}$, as
\begin{eqnarray}
B_n[h]=B_n^{(\rm HS)}\frac{B_2[h]}{B_2^{(\rm HS)}}.
\label{PL3}
\end{eqnarray}
The virial expansion of the excess free-energy functional $F_{\hbox{\tiny ex}}[h]$ can then be written as
\begin{eqnarray}
&&\frac{\beta F_{\rm ex}[h]}{N}=\rho_0B_2[h]+\frac{1}{2}\rho_0^2B_3[h]+\cdots\simeq 
\left(\rho_0+\frac{1}{2}\rho_0^2\frac{B_3^{(\hbox{\tiny HS})}}
{B_2^{\rm (HS)}}+\cdots\right)B_2[h]\nonumber\\\nonumber\\&&\hspace{9cm}=\Psi_{\rm HS}(\eta)\frac{B_2[h]}{B_2^{\rm (HS)}}.
\label{PL}
\end{eqnarray}
which coincides with Eqn. (\ref{PL2}).
Since $\Psi_{\rm HS}(\eta)\to 4\eta$ as $\eta=\rho_0v_0\to 0$, and $B_2^{\rm (HS)}=4v_0$, we see that 
$\beta F_{\rm ex}[h]/N\to \rho_0B_2[h]$ and therefore the PL theory recovers the
Onsager theory (\ref{Ons}) in the low-density limit.

By about the same time as Lee but independently, a close version of the PL theory was proposed by Baus and coworkers \cite{Baus,Baus1},
using yet another perspective. These authors started from the exact excess free-energy functional for a general nonuniform
fluid in terms of the direct correlation functional,
\begin{eqnarray}
c^{(2)}({\bm r},{\bm r}^{\prime},\hat{\bm\Omega},\hat{\bm\Omega}^{\prime})=-\frac{\delta^2\beta F_{\rm ex}[\rho]}
{\delta\rho({\bm r},\hat{\bm\Omega})\delta\rho({\bm r}',\hat{\bm\Omega}')}.
\end{eqnarray}
Double functional integration of this equation along a path joining a reference uniform fluid of density $\rho_0$ and
the actual nonuniform fluid, $\rho_{\lambda}({\bm r},\hat{\bm\Omega})=\rho_0+\lambda
\left[\rho({\bm r},\hat{\bm\Omega})-\rho_0\right]$, where $\lambda$ is a coupling parameter, leads to
\begin{eqnarray}
&&F_{\rm ex}[\rho]=F_{\rm ex}^{(0)}(\rho_0)-\int_V d{\bm r}\int_V d{\bm r}'\int d\hat{\bm\Omega}\int 
d\hat{\bm\Omega}^{\prime}\int_0^1d\lambda (1-\lambda)\nonumber\\\nonumber\\&&\hspace{2cm}\times
\left[\rho({\bm r},\hat{\bm\Omega})-\rho_0\right]\left[\rho({\bm r}',\hat{\bm\Omega}')-\rho_0\right]
c^{(2)}({\bm r},{\bm r}^{\prime},\hat{\bm\Omega},\hat{\bm\Omega}^{\prime};[\rho_{\lambda}]).
\label{coupling_exact}
\end{eqnarray}
Baus et al. consider the uniform phases $\rho({\bm r},\hat{\bm\Omega})=\rho_0h(\hat{\bm\Omega})$, of
a fluid of HE particles of lengths $\sigma_{\parallel}$ (along the symmetry axis) and
$\sigma_{\perp}$ (in the directions perpendicular to the symmetry axis), aspect ratio 
$\kappa=\sigma_{\parallel}/\sigma_{\perp}$, and volume $v_0=\pi\sigma_{\parallel}\sigma_{\perp}^2/6$.
Now one approximates the direct correlation function by introducing a decoupling approximation:
\begin{eqnarray}
c^{(2)}({\bm r},{\bm r}',\hat{\bm\Omega},\hat{\bm\Omega}^{\prime})\simeq 
c_{0}^{(2)}\left(\frac{\left|{\bm r}-{\bm r}'\right|}{\sigma_0};\bar{\eta}\right)
\frac{v_{\rm exc}(\hat{\bm\Omega},\hat{\bm\Omega}^{\prime})}{v_0},
\end{eqnarray}
where the reference system is a fluid of HS of diameter $\sigma_0$ and the same volume, $v_0=\frac{\pi}{6}\sigma_0^3$.  
This expression has a structure reminiscent of that in PL theory, see Eqns. (\ref{PL1})-(\ref{PL3}).
The angular factor $v_{\rm exc}(\hat{\bm\Omega},\hat{\bm\Omega}^{\prime})$ is the excluded volume of two HE particles,
which is further approximated by that of two hard Gaussian overlap particles, i.e.
\begin{eqnarray}
v_{\rm exc}(\hat{\bm\Omega},\hat{\bm\Omega}^{\prime})=
\left\{\frac{1-\chi^2(\hat{\bm\Omega}\cdot\hat{\bm\Omega}^{\prime})^2}
{1-\chi^2}\right\}^{1/2},
\end{eqnarray}
with $\chi=(\kappa^2-1)/(\kappa^2+1)$. The direct correlation function of the reference HS fluid is taken from the
Percus-Yevick approximation. All that remains to specify is the criterion to calculate the effective packing fraction $\bar{\eta}$
where the correlation function is to be evaluated. In the I phase they take $\bar{\eta}=\eta=\rho_0v_0$. However, in the
N phase orientational order leads to an effective reduction of interactions. This is similar to the situation in the HS
crystal: the highly oscillating spatial structure of the distribution functions comes from the highly-peaked periodic local
densities, and the proper correlation is a smooth function of the particle positions similar to that of a low-density fluid.
In the N phase, the high anisotropy of the orientational correlations comes mainly from the distribution function 
$h(\hat{\bm\Omega})$ and proper orientational correlations are very accurately given by the two-particle 
low-density limit. Consequently, in the N phase we expect $\bar{\eta}<\eta$.
To establish the precise relationship between $\eta$ and $\bar{\eta}$, Baus et al. proposed a criterion based on a 
structural scaling condition between the HE and the reference HS which takes into account the  geometric constraints of the 
nematic phase. To do that, they assumed that, at contact, the direct correlation functions of the  HS  system,  
evaluated at the real and at the effective density, are related by,
\begin{eqnarray}
c_{\rm PY}\left(\frac{\left|{\bm r}\right|}{\sigma_0}=1;\eta\right)=
c_{\rm PY}\left(\frac{\left|{\bm r}\right|}{\sigma_0}=x(\kappa);\bar{\eta}\right).
\end{eqnarray}
$\sigma_0$, the reference HS diameter, is the average contact distance of the HE in the isotropic phase, whereas 
$\sigma_0x$ (with $x < 1$)
is the  average contact  distance  of the  HE  in the  nematic  phase.  Since $\kappa$ is the natural length scale of the 
problem, they put $x = \kappa$ when $\kappa< 1$ and $x = 1/\kappa$  when $\kappa> 1$. 
The Percus-Yevick result for the direct correlation function of HS was used in the practical implementation of the criterion. 
It is then easy to realize that the theory of Baus et al. reduces to the PL theory: it is a decoupling approximation with
the same orientational factor, a different density prefactor, the same reference fluid but evaluated at a 
different effective density in the N phase. The theory was applied to the study of the I and N
equations of state and the I--N transition in systems of HE of different aspect ratios. Results in fair agreement with 
the computer simulation studies were obtained (see original papers for details).

The PL theory and other resummations incorporating higher-order virial coefficients have been contrasted
with computer simulations by Padilla and Velasco \cite{Padilla1} for moderately long particles in the case of the 
hard Gaussian overlap fluid. They used DFT to obtain the pressure, order parameter and I--N transition point
and compared with computer simulation results for aspect ratios 3--5.
Different theories were used. First, a decoupling approximation with an isotropic 
hard Gaussian overlap fluid instead of a hard-sphere fluid as a reference
fluid was used. The former was seen to give a closer agreement with the simulation,
but both overestimate the transition density since angular correlations are
represented at the same level in both theories. Second, extended decoupling approximations 
incorporating $B_3$ and $B_4$, obtained using a technique similar to that in Ref. \cite{Margo}, were seen to 
give improved results over the standard decoupling approximation since they incorporate higher-order angular correlations.

Williamson \cite{Williamson} made a comparison of the results from PL scaling and related
theories with simulation, using an annealing technique to minimise the functional that does not assume any particular
form for the orientational distribution function.
The original PL scaling was found to give the best results. Further comparison between Parsons-Lee 
and MC simulation was made in \cite{Jackson_HSPC}, where excellent agreement for the isotropic and nematic branches and
reasonably accurate coexistence densities and pressures were found. Again, the corresponding 
orientational distribution function at the phase transition was not accurately reproduced. 

Some of the previous works on calamitic (rodlike) nematics have been paralleled by corresponding studies on discotic systems.
These systems are made of disclike particles, which can also form nematic phases. In fact, many of the studies have  
been motivated by, or have motivated, experimental studies on colloidal suspensions
of platelike particles. Many of these particles can be synthesised and prepared such that the particles interact in solution
like (or close to) hard particles, which made it possible to test the predictions of theories for oblate hard bodies. 
However, except for the early observation of nematic ordering in solutions of platelets by 
Langmuir \cite{Langmuir}, the nematic phase of these colloids has been elusive until relatively recently 
\cite{Kooij-Lek,Mejia}, due to the large tendency of these systems to form gel phases.
 
In fact, Onsager already made predictions for a model of oblate particle consisting of infinitely thin circular platelets.
The first simulation of nematic ordering was made precisely on a system of infinitely thin platelets \cite{Frenkel-Eppenga0}.
Comparing with Onsager theory, this study evidenced the poorer predictions made by Onsager theory in the case of platelets.
This is no surprise: Onsager theory neglects virial coefficients beyond the second, which are very important in 
the case of platelets. Discotic liquid crystals will be discussed in more detail in Sec. \ref{further}.

\subsubsection{Nonuniform phases: Extended Onsager and others}
\label{nonuniform}

Since entropy alone can explain freezing of the HS fluid into a stable periodic face-centre-cubic structure, it is legitimate 
to wonder if anisotropic hard interactions can stabilise the S phase, the more symmetric liquid-crystalline phase 
beyond the N phase. The first attempts to formulate statistical-mechanical theories to deal with nonuniform liquid-crystalline
phases date back to the 70' \cite{Hosino,Hosino1} and '80 \cite{Meyer}. Hosino et al. \cite{Hosino} and Wen et al. \cite{Meyer} 
investigated a fluid of aligned rods using a second-order virial approximation and a bifurcation analysis. 
The first group also studied the effect of orientational fluctuations \cite{Hosino1} using a 
discrete-orientation approximations. 
These works were the first evidence that hard rods and, consequently, excluded-volume interactions,
could induce the formation of liquid-crystalline phases with spatial order. The origin lies in the more efficient 
packing of rods in a layered configuration. Specifically, the entropy reduction associated with ordering along one
dimension is more than compensated, at sufficiently high density, by the increased entropy due to an optimised free 
volume due to the arrangement of particles into layers.

Some time after the work of Hosino et al., Stroobants et al. \cite{Stroobants1,Stroobants2} studied a fluid of perfectly parallel HSC by MC 
simulations and observed a N--S continuous transition, followed by a first-order S-crystal transition. The formation of
a layered phase in a fluid of parallel HSC is a nontrivial result, considering that
the parallel HE model cannot stabilise into a layered structure (since its free energy can be exactly mapped onto
that of a HS fluid, which does not possess order intermediate between the fluid and the crystal).
Evidence that S ordering was robust when free particle orientations were included came a few years later 
\cite{Frenkel_nature}.  Confirmation of the stability of the smectic phase in colloidal fluids of approximately 
hard rods came from the work of Wen et al. \cite{Meyer1}, who observed layered phases in colloidal suspensions of the tobacco mosaic 
virus and in-layer undulational fluctuations. 

A few years before, Mulder \cite{Mulder_smec} proposed a virial expansion for the excess free energy of a fluid of 
parallel hard rods, including an Onsager-like, second-order term, and also the third- and fourth-order virial 
coefficients, and assumed smectic symmetry. Since particle orientations are assumed to be frozen along the director, taken 
along the $\hat{\bm z}$ unit vector, the full local particle density can be simplified, 
$\rho({\bm r},\hat{\bm\Omega})=\rho(z)\delta(\hat{\bm\Omega}-\hat{\bm z})$.
As usual, the Helmholtz free-energy functional is split in ideal and excess terms, 
$F[\rho]=F_{\rm id}[\rho]+F_{\rm exc}[\rho]$. The ideal part is
\begin{eqnarray}
\beta F_{\rm id}[\rho]&=&\int_Vd{\bm r}\int d\hat{\bm\Omega}
\rho({\bm r},\hat{\bm\Omega})\left[\log{\left(\rho({\bm r},\hat{\bm\Omega})\Lambda^3\right)}
-1\right]\nonumber\\\nonumber\\&=&A\int_L dz\rho\left(z\right)\left[\log{\left(\rho(z)\Lambda^3\right)}-1\right],
\label{eq:mul0_z}
\end{eqnarray}
where $L$ is the linear dimension of the system along the $z$ direction and $A$ the system area in the other
two perpendicular directions. The excess part can be written as a virial expansion:
\begin{eqnarray}
\beta F_{\rm ex}[\rho]&=&-\frac{1}{2}\int_Vd{\bm r}\int_Vd{\bm r}'\int d\hat{\bm\Omega}\int d\hat{\bm\Omega}'
\rho({\bm r},\hat{\bm\Omega})\rho({\bm r}',\hat{\bm\Omega}')
f({\bm r}-{\bm r}',\hat{\bm\Omega},\hat{\bm\Omega}')
\nonumber\\\nonumber\\&-&\frac{1}{6}
\int_Vd{\bm r}\int_Vd{\bm r}'\int_Vd{\bm r}''\int d\hat{\bm\Omega}\int d\hat{\bm\Omega}'\int d\hat{\bm\Omega}''
\rho({\bm r},\hat{\bm\Omega})\rho({\bm r}',\hat{\bm\Omega}')
\rho({\bm r}'',\hat{\bm\Omega}'')\nonumber\\\nonumber\\&&\hspace{1cm}\times
f({\bm r}-{\bm r}',\hat{\bm\Omega},\hat{\bm\Omega}')
f({\bm r}'-{\bm r}'',\hat{\bm\Omega}',\hat{\bm\Omega}'')
f({\bm r}-{\bm r}'',\hat{\bm\Omega},\hat{\bm\Omega}'')
+\cdots
\label{eq:mul_z}
\end{eqnarray}
Invoking the translational symmetry of the fluid in the $xy$ plane, the functional can be written as
\begin{eqnarray}
\frac{\beta F[\rho]}{A}&=&\int_L dz\rho\left(z\right)\left[\log{\left(\rho(z)\Lambda^3\right)}-1\right]
\nonumber\\\nonumber\\&+&
\int_Ldz\int_Ldz'
\rho\left(z\right)\rho\left(z'\right)B_2^{\parallel}\left(z,z'\right)
\nonumber\\\nonumber\\&+&\frac{1}{2}
\int_Ldz\int_Ldz'\int_Ldz''
\rho\left(z\right)\rho\left(z'\right)\rho\left(z''\right)
B_3^{\parallel}\left(z,z',z''\right)+\cdots
\label{eq:mul_z1}
\end{eqnarray}
where $B_n^{\parallel}$ are projected virial coefficients, the first two of which are given by 
\begin{eqnarray}
&&B_2^{\parallel}(z,z')=-\frac{1}{2}\int_A d{\bm r}_{\perp}'f_{\parallel}\left({\bm r}'_{\perp},z-z'\right),\nonumber\\\nonumber\\
&&B_3^{\parallel}(z,z',z'')=-\frac{1}{3}\int_A d{\bm r}_{\perp}'\int_A d{\bm r}_{\perp}''
f_{\parallel}\left({\bm r}'_{\perp},z-z'\right)
f_{\parallel}\left({\bm r}'_{\perp}-{\bm r}''_{\perp},z'-z''\right)
f_{\parallel}\left({\bm r}''_{\perp},z-z''\right),
\end{eqnarray}
and $f_{\parallel}$ is the Mayer function of parallel particles.
Expressions for higher-order virial coefficients can be written from the standard virial expansion (see e.g.
\cite{Hoover,Hoover1}). The frozen-orientational approximation facilitates enormously the calculation of the
first virial coefficient, and Mulder \cite{Mulder_smec} managed to calculate virial coefficients for parallel
HC up to the fourth. The occurrence of S order was analysed using a bifurcation analysis 
of the free-energy functional truncated at second order, and the effect of the higher-order terms in the
bifurcation point was discussed. A continuous transition from the nematic to the smectic phase was found, at a density
and smectic period in quite good agreement with computer simulations of the same model.

\begin{figure}
\includegraphics[width=13cm]{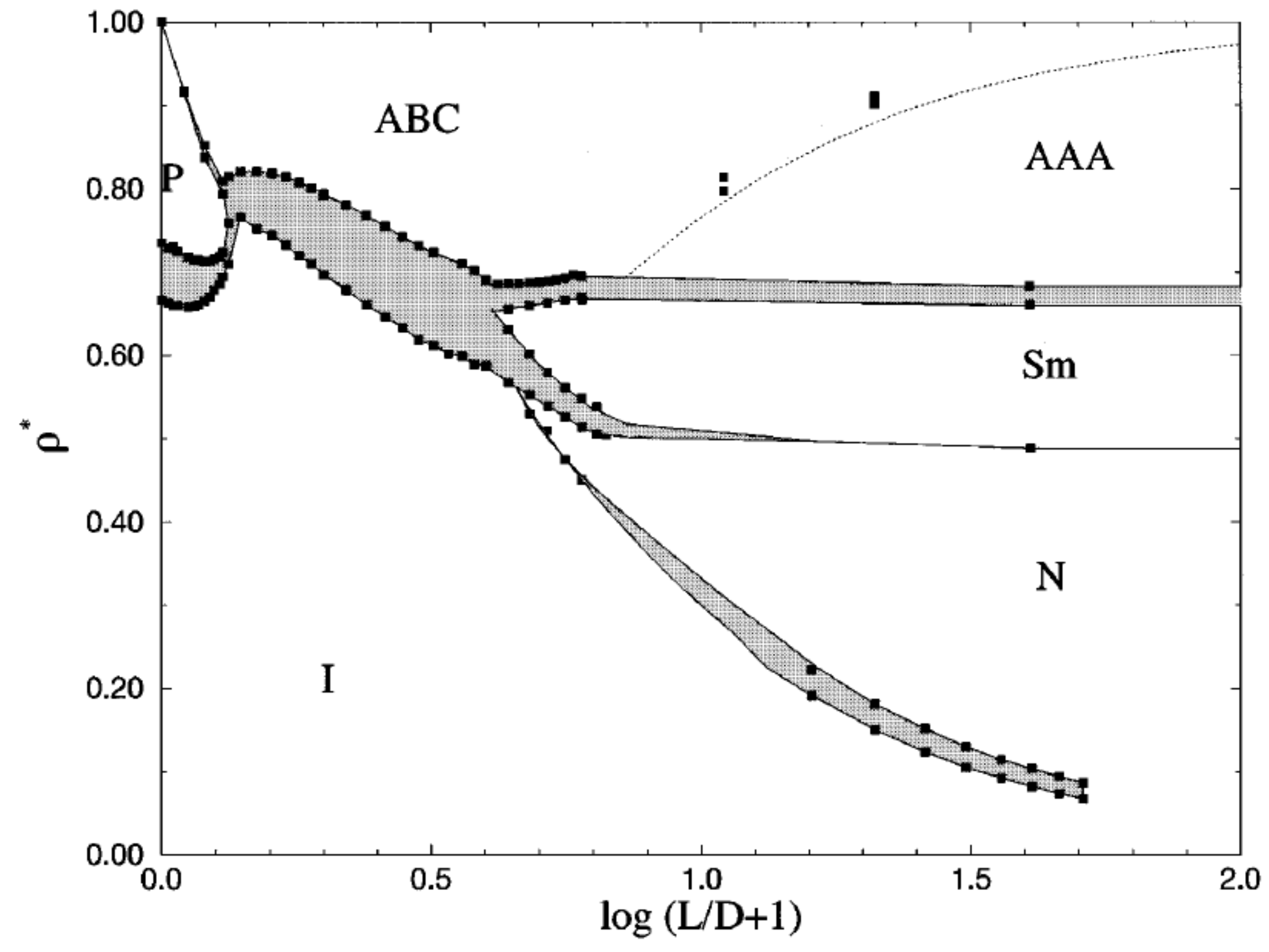}\caption{Phase diagram of a system of freely-rotating HSC, as
calculated from computer simulation \cite{FLS}, in the density-aspect ratio plane (density is $\rho^*=\rho/\rho_{\rm CP}$, where
$\rho_{\rm CP}$ is the close-packing density). Symbols indicate region of stability
of the phases: I, isotropic; N, nematic; Sm, smectic; P, plastic solid; AAA and ABC, different stackings of
crystal phases. Shaded regions correspond to two-phase regions. Reprinted with permission from \cite{FLS}. Copyright (1997), AIP Publishing LLC.}
\label{Fig_Frenkel_HSC} 
\end{figure}

The complete phase diagram of hard spherocylinders, in the density vs. aspect ratio $\kappa=L/D$ plane, has now been 
mapped out in detail by simulation \cite{Jackson_HSPC,FLS} for the whole range of aspect ratios $\kappa=0-\infty$
and for the freely-rotating model (see Fig. \ref{Fig_Frenkel_HSC}). There exists a terminal aspect ratio,
below which the N phase ceases to be stable, and there is direct 
coexistence between I and S$_A$ phases. For low values of $\kappa$ a plastic phase is formed. The crystal phase has 
AAA or ABC stacking depending on the aspect ratio.

Esposito and Evans \cite{Esposito} presented an Onsager-like density-functional theory for the nematic-smectic bifurcation point 
which was in fair agreement with the simulation data. The theory incorporated third- and higher-order correlations in an ad hoc
manner, and included orientational freedom, which allowed these workers to present a density-aspect ratio phase diagram including
isotropic, nematic and smectic-A phases. The triple point location coincided with that from the then existing theories, but
the nematic-to-smectic transition was always found to be of first order, regardless of the value of aspect ratio. This is in
contrast with predictions from more sophisticated theories, which found either a continuous transition or
a tricritical point at which the character of the transition changed from first-order to continuous. Therefore, this theory 
indicated that the N--S transition is of first order and becomes continuous only in the strict limit of parallel particles.

The reason why a fluid of hard parallel spherocylinders induces a layered phase is a subtle one, since a 
corresponding system of ellipsoid does not. This point has been investigated by Evans \cite{GTEvans}, who considered
a PL theory on particles of different shapes: ellipsoids, spherocylinders and ellipo-cylinders (a particle that
interpolates between the latter two). Nematic fluids of particles with both frozen and free orientations were 
considered. All particles formed a S phase, except ellipsoids, which were revealed as pathological in that respect.

\begin{figure}
\epsfig{file=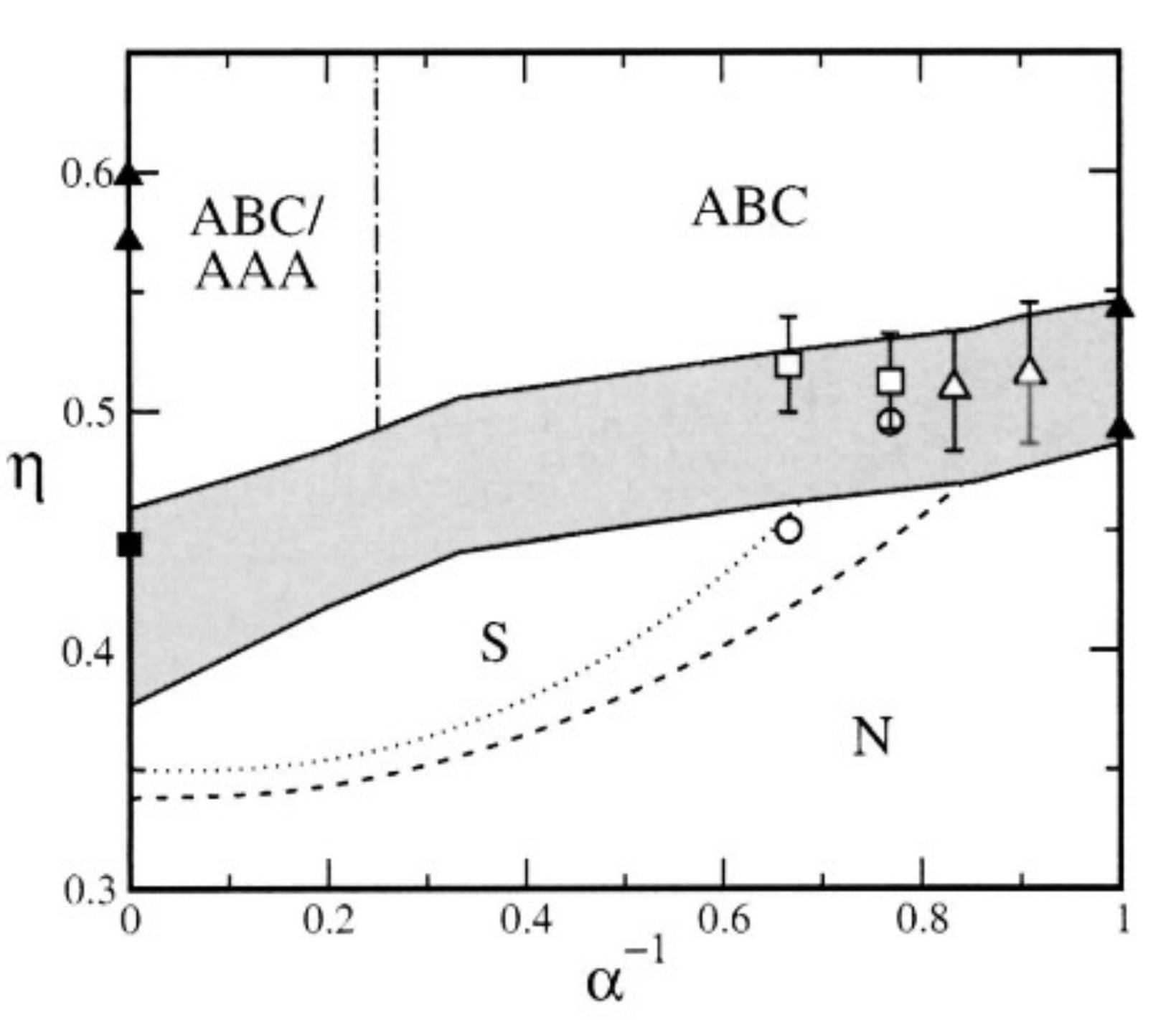,width=4.0in}
\caption{Phase diagram packing fraction-inverse shape parameter $\alpha^{-1}$ of parallel hard superellipsoids as obtained in
\cite{superellipsoids}. Continuous lines: coexistence boundaries for the smectic-solid transition using PL theory for the 
smectic and free-volume theory for the solid. Dashed line: nematic-smectic spinodal line
from PL theory; dotted line: nematic-smectic spinodal line from the theory of Mulder \cite{Mulder_smec}.
Filled triangles: simulation results for the smectic-solid transition of parallel cylinders \cite{Veerman}
and for liquid-solid coexistence in hard spheres \cite{Hoover3}. Filled square: nematic-smectic spinodal from computer simulation of 
Veerman and Frenkel \cite{Veerman}. Open symbols: simulation estimates for nematic-smectic spinodal (circles), 
first-order smectic-solid transition (squares) and for first-order nematic-solid transition (triangles) \cite{superellipsoids}. 
Labels indicate stable phases: N, nematic; S, smectic; ABC and AAA, crystalline solids with corresponding symmetries. 
Vertical dot-dashed line: approximate limit of degeneracy of ABC and AAA structures within FV theory.
Reprinted with permission from \cite{superellipsoids}. Copyright (2008), AIP Publishing LLC.}
\label{MV}
\end{figure}

This conclusion was further supported by Mart\'{\i}nez-Rat\'on and Velasco \cite{superellipsoids} in 2008, who considered
systems of parallel hard superellipsoids of revolution, which can be viewed as an interpolation between ellipsoids of 
revolution and cylinders. The shape of the particles is given by the equation:
\begin{eqnarray}
\left(\frac{R}{a}\right)^{2\alpha}+\left(\frac{z}{b}\right)^{2\alpha}=1,\hspace{0.4cm}R=\sqrt{x^2+y^2}.
\end{eqnarray}
Particles are characterized by a shape parameter $\alpha$ (with $\alpha=1$ corresponding to ellipsoids of 
revolution, while $\alpha=\infty$ is the limit of cylinders. Using computer simulation, it was shown that above
a critical value $\alpha>1.2$, the smectic phase is stabilised. This value is surprisingly close to that of
ellipsoids. Also, Mart\'{\i}nez-Rat\'on and E. Velasco used a PL approach combined with free-volume theory for the 
solid phases to obtain a complete phase diagram exhibiting regions of smectic, columnar and crystalline ordering (see Fig. \ref{MV}).
Comparison with simulation results of Veerman and Frenkel \cite{Veerman} for cylinders and superellipsoids of different
shapes \cite{superellipsoids} was reasonable.

In 1989 Taylor et al. \cite{Taylor0} presented a theory for a fluid of parallel HSC that combined ideas from scaled-particle
theory and cell theory. The idea was to decouple the spatial directions associated with order from those where the system is 
disordered, and apply a cell theory to the first and scaled-particle theory to the second. Correspondingly the total free energy 
was written as a sum of two contributions. The theory was the first to give a complete phase diagram of the model, and good
qualitative agreement with existing simulations \cite{Stroobants2} was shown. The layer spacing at the N--S transition, in particular,
was quantitatively correct. However, the theory was limited in that all
phase transitions were required to be discontinuous by construction, which is incorrect in most instances, in particular
in the case of the N--S transition. Also, density gaps are overestimated for first-order phase transitions.

In 1992 Baus and coworkers \cite{Baus2} studied the N--S transition
of the same model using their version of the scaled Onsager theory, together
with ideas from the density-functional theory for HS freezing. They
start from the exact expression for the excess free-energy functional
in terms of the direct correlation function, given by Eqn. (\ref{coupling_exact}).
The direct correlation function is then approximated as that of a spatially
uniform fluid evaluated at an effective density $\bar{\rho}$, 
\begin{eqnarray}
c^{(2)}({\bm{r}},{\bm{r}}',\hat{\bm{\Omega}},\hat{\bm{\Omega}}';[\rho])=
c_{0}^{(2)}({\bm{r}}-{\bm{r}}',\hat{\bm{\Omega}},\hat{\bm{\Omega}}';\bar{\rho}[\rho]).
\end{eqnarray}
Since the determination of the function 
$c_{0}^{(2)}({\bm{r}},\hat{\bm{\Omega}},\hat{\bm{\Omega}}';\bar{\rho})$
in the case of nonspherical particles is a rather complicated problem
by itself, Baus et al. proposed a further simplification by approximating
$c_{0}^{(2)}({\bm{r}},\hat{\bm{\Omega}},\hat{\bm{\Omega}}';\bar{\rho})$
in the case of parallel HSC in terms of the Percus-Yevick direct correlation
function of hard spheres of diameter $\sigma$ equal to the contact
distance between two (parallel) HSC. To complete the approximation,
a recipe to determine $\bar{\rho}$ in terms of the density profile
$\rho(\mathbf{r})$ in the different phases is needed. In the N phase
the density profile is constant, $\rho(\mathbf{r})=\rho$, and, therefore,
$\bar{\rho}$ is also constant and equal to $\rho$. In the S phase,
however, the density profile has to reproduce the layered structure.
Taking the layers to be perpendicular to the $z$ direction, the density
profile in the S phase only depends on $z$ and it was parameterised
as $\rho(z)=\rho\left[1+\epsilon\cos(2\pi z/z_{0})\right]$, $\rho$
being now the mean smectic density, $z_{0}$ the layer spacing, and
$\epsilon$, with $0\leq\epsilon\leq1$, the amplitude of the smectic-like
oscillation. In the work of Baus et al. only a bifurcation analysis
was carried out, studying the stability of the N phase against smectic-like
fluctuations with $\epsilon\ll1$. In that case the density profile
$\rho(z)$ only departs slightly from the mean smectic density $\rho$
and, thefore, the same approximation $\bar{\rho}(z)=\rho$ was used
for the S phase. The N--S transition of parallel HSC of aspect ratios
up to $\kappa=5$ was studied, and good agreement with computer
simulations was obtained.

\subsubsection{Weighted-density theories}
\label{WDA_sec}

More sophisticated and accurate theories, incorporating the orientational degrees of freedom, have been proposed to explain 
the transition to the smectic phase and other nonuniform phases such as the columnar. 
One of the problems with all existing theories was that they assumed the parallel particle
approximation, which is qualitatively valid in the N and S regimes but limits the possibility of quantitative analysis.
Also, density correlations, crucial in understanding phases with spatial order, are poorly accounted for in all theories
deriving from virial expansions. Some of the theories have been
formulated for general nonuniform fluids and have also been applied to interfacial phenomena. 
The new theories grew out from developments made for HS in an effort to
improve the local-density approximation (which is useless to describe crystallisation in the HS model) and used ideas from the 
corresponding density-functional approximations for HS. The central quantity in these theories is the
{\it average density}, a local density that takes into account the density distribution in the neighbourhood of a point. 
The average density is a convolution of the real density and a weighting function which describes the properties of the 
environment and whose details are optimised to give the correct bulk properties of the
HS fluid. A review of these {\it weighted-density approximation} (WDA) functionals can be found, among other sources, in
\cite{Evans}.

In WDA theory for the HS system, the excess free energy is written as
\begin{eqnarray}
\beta F_{\rm ex}[\rho]=\int_Vd{\bm r}\rho({\bm r})\Psi\left(\bar{\rho}({\bm r})\right),
\label{Fexc}
\end{eqnarray}
where $\Psi(\bar{\rho})$ is the excess free energy per particle of a uniform isotropic fluid evaluated at the local weighted 
density $\bar{\rho}({\bm r})$, which averages the fluid structure within a neighbourhood of the point ${\bm r}$. The 
weighted density is calculated as a convolution $\displaystyle\bar{\rho}({\bm r})=\int_V d{\bm r}'
\omega({\bm r}-{\bm r}')\rho({\bm r}')$, and the weight function $\omega({\bm r})$ is obtained by imposing
the functional to recover a particular approximation for the equation of state and direct correlation function of the uniform 
phase. This function, which has a range of the order of the sphere diameter $\sigma_0$, 
can be made to depend on the averaged density, which gives a powerful self-consistency to the theory. 
Two extensions of the WDA for HS to hard-rod fluids have been proposed to account for the effects of particle anisotropy
and free orientations. 

Poniewierski and
Ho\l{}yst \cite{Ponie} proposed a weighted-density functional (called the {\it smoothed density approximation}, SDA) for general
convex anisotropic bodies, focusing on a fluid of HSC. This theory used ideas from the
corresponding density-functional approximations for hard spheres. 
Let $\rho({\bm r})$ be the angle-averaged one-particle distribution function $\rho({\bm r},\hat{\bm\Omega})$, i.e.
$\displaystyle\rho({\bm r})=\int d\hat{\bm\Omega} \rho({\bm r},\hat{\bm\Omega})$. 
The orientational dependence in the model is taken care of by assuming the following form for the average density:
\begin{eqnarray}
\bar{\rho}({\bm r})=\frac{1}{\rho({\bm r})}\int_V d{\bm r}\int d\hat{\bm\Omega} \int d\hat{\bm\Omega}^{\prime}
\omega({\bm r}-{\bm r}^{\prime},\hat{\bm\Omega},
\hat{\bm\Omega}^{\prime})\rho({\bm r},\hat{\bm\Omega})\rho({\bm r}^{\prime},\hat{\bm\Omega}^{\prime}),
\end{eqnarray}
The model is specified by giving $\Psi(\rho)$ and $\omega({\bm r},\hat{\bm\Omega},
\hat{\bm\Omega}^{\prime})$. The latter was taken as
\begin{eqnarray}
\omega({\bm r},\hat{\bm\Omega},\hat{\bm\Omega}^{\prime})=
-\frac{1}{2B_2^{\rm (iso)}}f\left({\bm r},\hat{\bm\Omega},\hat{\bm\Omega}^{\prime}\right),
\label{weight}
\end{eqnarray}
where $f$ is the Mayer function of two spherocylinders, and $B_2^{\rm (iso)}$ 
is the second virial coefficient for the isotropic fluid. The excess free energy density
is chosen as
\begin{eqnarray}
\beta\Psi(\rho)=\rho_0B_2^{\rm (iso)}+\left[\beta\Psi^{(\hbox{\tiny CS})}(\eta)-4\eta\right].
\label{psi}
\end{eqnarray}
Here $\Psi^{(\hbox{\tiny CS})}(\eta)$ is the Carnahan-Starling excess free energy per particle of a fluid of HS, and the packing 
fraction is $\eta=\rho_0 v_0$, with $v_0$ the volume of a spherocylinder. This choice for $\Psi(\rho)$ guarantees that the correct first term of the
virial expansion is recovered when $\rho_0\to 0$ (and hence the Onsager limit when $L/D\to\infty$).
Interestingly, in the opposite limit $L/D\to 0$ (HS limit), $\Psi(\rho)\to\Psi^{(\hbox{\tiny CS})}(\eta)$ and Tarazona's first 
version of WDA for HS \cite{Tarazona0} is recovered. The theory was applied to a fluid of hard spherocylinders with full orientational freedom. 
The isotropic-nematic transition was obtained, using the parameterised form (\ref{EL}). As usual,
the order parameter at the transition is overestimated and the pressure underestimated). A lower bound
$L/D=2.46$ for the existence of a nematic phase was found (to be compared with $3.7$ from the simulations \cite{FLS}). Also, the nematic-smectic
transition was located using a bifurcation analysis, assuming the transition was continuous. Lifting this assumption, the nature
of the transition was found to change from first-order to continuous at $L/D=6$, which is then a tricritical point.
For $L/D=5$ the results were compared with simulations, and very good agreement
was found for the spinodal density and the smectic spacing at the transition (although the latter exhibits an unphysical
maximum near the transition). More details on the theory and of its implementation were given later 
\cite{Ponie_more}. Ho\l{}yst and Poniewierski \cite{Holyst_HPC} also applied their theory to study a simpler system where orientations are frozen,
i.e. a system of hard parallel cylinders. At the level of a 
bifurcation analysis, they obtained the spinodal line for the nematic-smectic A transition and argued that, at even higher densities,
the system exhibits smectic-B, columnar and solid phases.

\begin{figure}
\epsfig{file=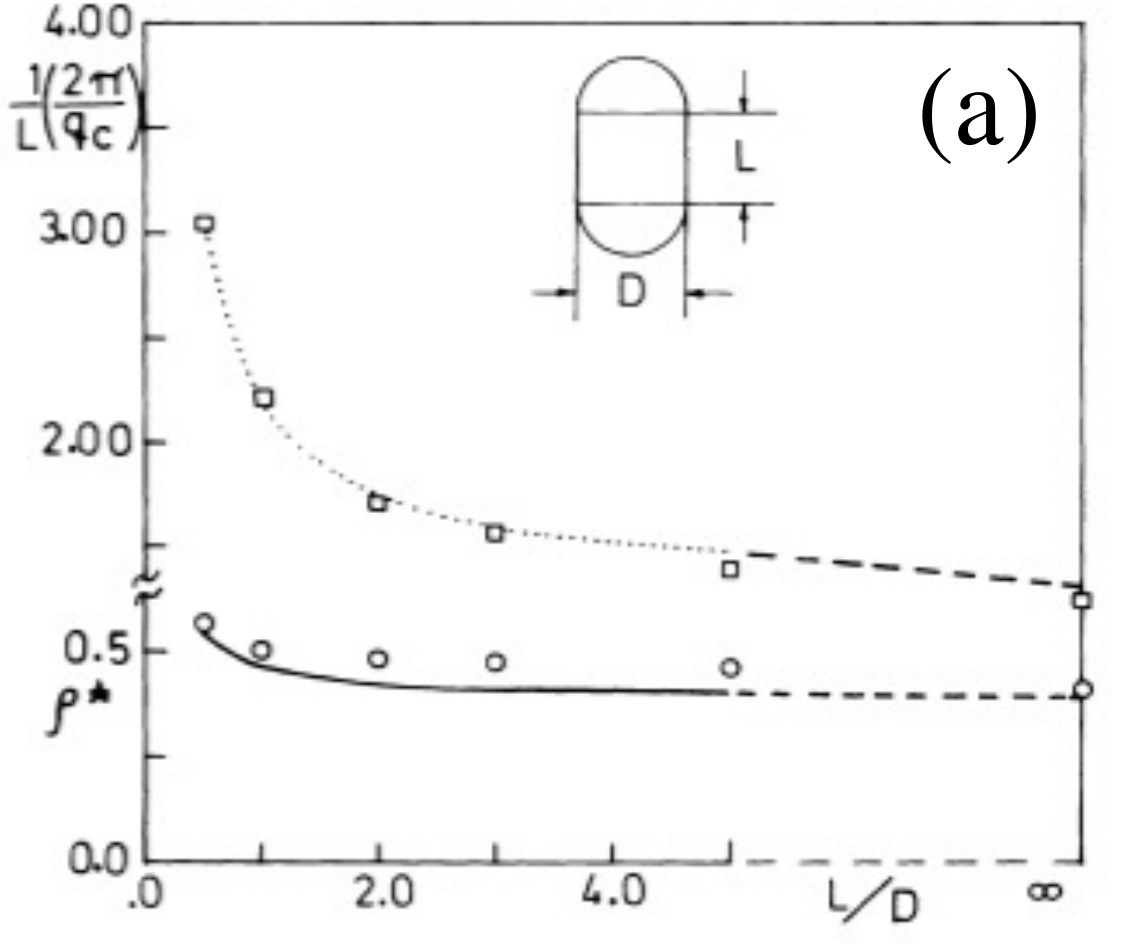,width=3.2in}
\epsfig{file=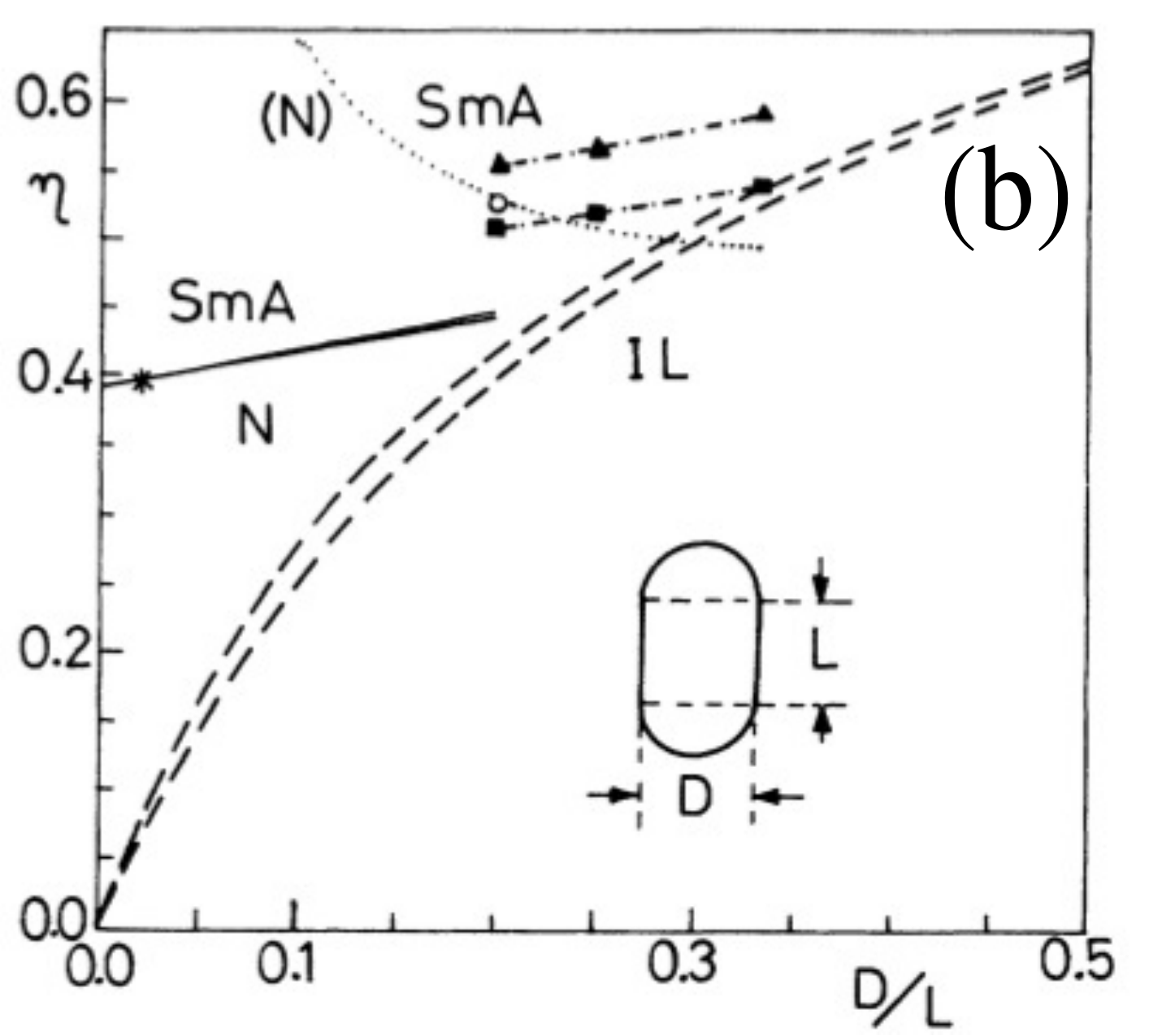,width=3.0in}
\caption{Results for the phase diagram of hard spherocylinders from the Somoza-Tarazona theory. (a) Parallel particles
\cite{ST}. (b) Freely oriented particles \cite{ST2}. Reprinted with permission from \cite{ST,ST2}. 
Copyright (1988,1990) by the American Physical Society.}
\label{ST_fig}
\end{figure}

At about the same time as Ho\l{}yst and Poniewierski but independently, Somoza and Tarazona (ST) \cite{ST,ST1,ST2} proposed an extension of 
Onsager theory in the line of the PL approach but valid for inhomogeneous fluids.  
The basic idea is that the mapping onto a HS fluid is inaccurate in a fluid with orientational 
order since the correlation structure along the director is very different from that in perpendicular directions. 
To correct this, Somoza and Tarazona proposed a reference fluid consisting of parallel HE of lengths $\sigma_{\parallel}$ and
$\sigma_{\perp}$, whose thermodynamics in turn can be mapped onto an 
equivalent fluid of HS of diameter $\sigma_0$. The expression for the excess free energy is
\begin{eqnarray}
\beta F_{\rm ex}[\rho]=\int_V d{\bm r}\int d\hat{\bm\Omega}\rho({\bm r},\hat{\bm\Omega})\Psi\left(\bar{\rho}({\bm r})\right)
\left\{\frac{\displaystyle\int_V d{\bm r}'\int d\hat{\bm\Omega}'\rho({\bm r}',\hat{\bm\Omega}')
f({\bm r}-{\bm r}',\hat{\bm\Omega},\hat{\bm\Omega}')}
{\displaystyle\int_V d{\bm r}'\rho({\bm r}')f_{\hbox{\tiny PHE}}\left({\bm r}-{\bm r}'\right)}\right\},
\label{STar}
\end{eqnarray}
where the weighted density $\bar{\rho}({\bm r})$ is calculated from $\rho({\bm r})=\int d\hat{\bm\Omega}
\rho({\bm r},\hat{\bm\Omega})$ as in the WDA theory for HS, but with a weight
function $\omega({\bm r})$ scaled by the factors $\sigma_{\parallel}/\sigma_0$ and 
$\sigma_{\perp}/\sigma_0$ in the directions parallel and perpendicular to the ellipsoids, respectively.
In the N phase, $\rho({\bm r},\hat{\bm\Omega})=\rho_0h(\hat{\bm\Omega})$ and $\bar{\rho}({\bm r})=\rho_0$,
Eqn. (\ref{STar}) coincides with that from Parsons-Lee theory (\ref{PL}), since the thermodynamics of a fluid of 
parallel HE is identical to that of HS:
\begin{eqnarray}
\beta F_{\rm ex}[\rho]=\frac{\rho_0\Psi\left(\rho_0\right)}
{\displaystyle\int d{\bm r}'f_{\hbox{\tiny PHE}}\left({\bm r}-{\bm r}'\right)}
\int d{\bm r}\int d\hat{\bm\Omega}\int d\hat{\bm\Omega}'h(\hat{\bm\Omega})h(\hat{\bm\Omega}')
v_{\rm exc}(\hat{\bm\Omega},\hat{\bm\Omega}').
\end{eqnarray}
Note that the factor between curly brackets in (\ref{STar}) is the ratio between excluded volumes
of the actual fluid and that of a reference fluid of parallel hard ellipsoids with the same spatial structure $\rho({\bm r})$.
The choice of reference ellipsoid can be made with different recipes (e.g.
by demanding equal packing fraction and equal length-to-breadth ratio of the actual and ellipsoidal particles).
A limited analysis of the theory was made in the first two papers \cite{ST,ST1}, assuming parallel particles. 
The phase diagram for parallel hard spherocylinders was calculated, and remarkable agreement with computer simulations 
\cite{Stroobants2} was obtained, see Fig. \ref{ST_fig}(a). The ST theory was also applied to a fluid of parallel oblique 
cylinders, and a full phase diagram with respect to the obliqueness parameter was obtained, containing nematic, smectic A, smectic C and
biaxial (see Fig. \ref{Somoza_fig} in Sec. \ref{biaxial_section}) phases. This work was the first to show that a transition between 
smectic-A and smectic-C phases was possible in a system of hard particles. A full implementation of the ST theory was made in \cite{ST2},
where the complete phase diagram of the hard-spherocyinder model was presented, Fig. \ref{ST_fig}(b). The theory included
the isotropic-nematic transition, which gives the same results as the PL theory. An important feature of the theory is that it predicts
a first-order nematic-smectic transition for $L/D<50$ that changes over to continuous for $L/D>50$, with $L/D=50$ a tricritical
point. In contrast, the SDA theory of Ho\l{}yst and Poniewierski predicted a tricritical point at $L/D=6$. Although
the results for the two transitions at $L/D=5$ were a bit worse than those of Ho\l{}yst and Poniewierski, the ST
theory corrected the unphysical feature of the former in the predicted smectic spacing, and also the slope of the nematic-smectic
boundary with respect to the aspect ratio $L/D$ and the limit $L/D\to\infty$. Overall the ST seems to work better for large
aspect ratios, whereas the SDA theory is slightly more accurate for shorter particles.
A discussion of the comparison between the SDA and ST theories was made by Poniewierski \cite{Poniewierski_almost_parallel},
who obtained a density functional based on the virial expansion, including the third-order term, valid in the asymptotic limit
$L/D\to\infty$ but including corrections for finite $L/D$. However, the question on the existence and location of the tricritical
point was not settled, since this point is very sensitive to higher-order particle correlations. 

A more accurate implementation of the ST theory was done 
in \cite{Us_Canada}, where the MC results of \cite{FLS} were contrasted with the theory. Very good agreement with the 
simulations was found for both I--N and N--S transitions and for all the aspect ratios considered ($L/D=3.5$--$10$). 

Graf and L\"owen \cite{Lowen0} made another proposal for a weighted-density functional, adopting an extension of the 
modified weighted-density approximation (MWDA) for the HS system. They write $F_{\rm ex}[\rho]$ as 
\begin{eqnarray}
\beta F_{\rm ex}[\rho]=N\Psi(\hat{\rho}),
\end{eqnarray}
where $\hat{\rho}$ is a global density obtained by averaging the average density 
\begin{eqnarray}
\bar{\rho}({\bm r},\hat{\bm\Omega})=\int d{\bm r}'\int d\hat{\bm\Omega}' \omega({\bm r}-{\bm r}',\hat{\bm\Omega},\hat{\bm\Omega}')
\rho({\bm r}',\hat{\bm\Omega}')
\end{eqnarray} 
in the whole volume:
\begin{eqnarray}
\hat{\rho}=\frac{1}{N}\int d{\bm r}\int d\hat{\bm\Omega}\bar{\rho}({\bm r},\hat{\bm\Omega})\rho({\bm r},\hat{\bm\Omega}).
\end{eqnarray}
The excess free energy $\Psi(\rho)$ and weighting function $w({\bm r},\hat{\bm\Omega},\hat{\bm\Omega}')$ are chosen as in 
Poniewierski and Ho\l{}yst theory, i.e. using (\ref{psi}) and (\ref{weight}). 
The results of the ST theory are superior to those obtained by Graf and L\"owen \cite{Lowen0}, since the latter
theory again predicts a wrong positive slope of the nematic-smectic phase boundary in the
density--$L/D$ phase diagram. The theory was later modified \cite{Lowen} to incorporate scaled-particle and cell-theory 
concepts,  which results in correct slope and better predictions.

\subsubsection{Fundamental-measure theories}

\label{FMT}

The Fundamental-Measure Theory (FMT) is a WDA approximation, in that
the free energy is built in terms of weighted densities, but has a
deeper root based on geometrical grounds. The theory makes use of
the concept of weighted densities, but here weighting functions are
measures of one particle (and possibly more particles), whereas WDA approximations are built on
measures of two particles (with the range of the HS diameter $\sigma_{0}$).

The FMT was first developed by Rosenfeld \cite{Yasha1,Yasha2} for hard spheres. It is naturally
formulated for a general mixture of hard particles. Rosenfeld proposed an excess free-energy functional
$F_{{\rm ex}}[\{\rho_{\nu}\}]$ (where species are labelled
with the index $\nu$) that depends on a finite set of weighted
densities $n_{\alpha}({\bm{r}})$, the latter being a sum of convolutions
of the local densities $\rho_{\nu}({\bm{r}})$ and some weighting
functions $\omega_{\nu}^{(\alpha)}({\bm{r}})$, i.e. 
\begin{eqnarray}
n_{\alpha}({\bm{r}})=\sum_{\nu}\int d{\bm{r}}'\rho_{\nu}({\bm{r}}')\omega_{\nu}^{(\alpha)}({\bm{r}}-{\bm{r}}').
\label{weighted}
\end{eqnarray}
The weights, which can be vectors or scalars, depend on the geometry
of a single particle, and their integrals result in the fundamental
geometric measures of the sphere: the mean Gaussian curvature, the
surface area and its volume. Let $\Phi({\bm{r}})$ be the local excess
free-energy density in thermal energy units, so that 
\begin{eqnarray}
{\displaystyle \beta F_{{\rm ex}}[\{\rho_{\nu}\}]=\int_{V}d{\bm{r}}\Phi({\bm{r}}).\label{Phi}}
\end{eqnarray}
To determine the dependence of the function $\Phi({\bm{r}})$ on the weighted
densities $\{n_{\alpha}\}$, two requirements were imposed. The first
is related to the uniform limit, which was chosen to be obtained from
the SPT theory \cite{Reiss}. The second requirement
was related to
the virial expansion of the direct correlation function, $c_{\mu\nu}^{(2)}({\bm{r}},{\bm{r}}')=-\delta^{2}\beta F_{{\rm ex}}[\{\rho_{\nu}\}]/\delta\rho_{\mu}({\bm{r}})\delta\rho_{\nu}({\bm{r}}')$,
which was demanded to recover the exact first (Mayer function) and
second terms of the uniform limit.

A modified version of FMT for HS was subsequently proposed \cite{Schmidt1,Schmidt2}
to correct a serious drawback of the first version, namely, the divergence
of $F_{{\rm ex}}[\{\rho_{\nu}\}]$ as the density profiles $\rho_{\nu}({\bm{r}})$
become more and more localised. This problem caused the theory not
to be able to predict crystallisation in the one-component fluid of
HS. In the new version the singularity was removed by redefining the
dependence of $\Phi({\bm{r}})$ on the weighted densities while maintaining
the bulk equation of state and the correct low-density expansion of
the direct correlation function. For a recent review of FMT for HS mixtures, see 
\cite{Roth_review}.

Finally, the modern versions of the FMT for HS were constructed from
first principles using cavity theory \cite{Tarazona1}. The idea is
to impose the fulfillment of an important requirement, namely the
dimensional crossover property: When the density profile of a $D$-dimensional
system is extremely constrained along one spatial dimension by freezing
the degrees of freedom of particles in that dimension (for example
by making $\rho_{{\rm D}}(x_{1},\dots,x_{D})=\rho_{{\rm D-1}}(x_{1},\dots,x_{D-1})\delta(x_{D})$,
where $(x_{1},\dots,x_{D})$ are the components of a $D$-dimensional
position vector), the $D$-dimensional density functional should reduce
to the $D-1$ functional. Also, the same principle holds for a collection of overlapping cavities
with sizes such that only one particle fits in each cavity. 
Using this important property, density functionals
for one-component HS \cite{Tarazona2} and HS mixtures \cite{Cuesta1}
were obtained, and it was realised that, apart from scalar and vectorial
weighted densities, tensorial densities are also needed \cite{book}.
The resulting functional gives very accurate predictions for the HS
crystal and, in particular, it was shown that the correct cell-theory
in the high density limit is recovered \cite{Tarazona2}.

Recently it has been shown that the Rosenfeld functional, which was originally
derived semi-heuristicaly, can be systematically calculated from the
virial diagrammatic expansion including the clusters that represent
particle overlap in one centre \cite{Korden1}. Further, this
formulation has been extended to any particle geometry in any dimension and for
any number of intersections \cite{Korden2}. Finally, the
relation between the structures of the density functionals obtained from
the dimensional crossover to 0D and from a resummation
of a diagrammatic expansion that only considers certain classes of
diagrams, was analysed \cite{Korden3}. As a result, the 
zero-dimensional limit was reconciled, in an elegant way, with the virial
approach.

In principle, the original ideas of Rosenfeld for HS mixtures could be applied to formulate a FMT functional for general hard 
anisotropic convex bodies. The procedure would involve first to exactly deconvolute the Mayer function in terms of 
one-particle weights, and then use dimensional analysis to obtain a free energy as a sum of products of
weighted densities up to third order in density, ensuring that the expansion captures the exact first- and second-order terms of the
direct correlation function. Since the free energy can be made to satisfy the same differential equation as 
in SPT, the packing-fraction-dependent coefficients of this sum can be extracted, except for some numerical constants. These
constants could be fixed by demanding the free energy to give the correct zeroth-dimensional (cavity) limit.
In practice, this programme cannot be followed exactly for general convex bodies. Therefore, a version of FMT was
proposed in the restricted-orientation or Zwanzig approximation (used before in some extended Onsager theories, see 
Sec. \ref{extensions}) to treat fluids made of hard parallelepipeds (HP) \cite{Cuesta2,Cuesta3,Cuesta4}.
An alternative procedure to obtain the density functional was proposed by Cuesta and Mart\'{\i}nez-Rat\'on \cite{Cuesta3}.
They realised that the free-energy density of a mixture of HP can be obtained, in {\it any} dimensionality, starting solely 
from the exact zeroth-dimensional functional by applying a differential operator to it. 
For dimensionality D$=3$ the result is
\begin{eqnarray}
\Phi^{\rm(3D)}=-n_{0}\ln(1-n_{3})+\frac{{\bm{n}}_{1}\cdot{\bm{n}}_{2}}{1-n_{3}}+\frac{n_{2x}n_{2y}n_{2z}}{(1-n_{3})^{2}},\label{3D}
\end{eqnarray}
with the usual definitions of weighted densities (\ref{weighted}).
The weights $\omega_{\nu}^{(\alpha)}({\bm{r}})$ are given by 
\begin{eqnarray}
\omega_{\nu}^{(0)}({\bm{r}}) & = & \frac{1}{8}\prod_{\tau}\delta\left(\frac{\sigma_{\nu}^{(\tau)}}{2}-|x_{\tau}|\right),\nonumber \\
\omega_{\nu}^{(1\alpha)}({\bm{r}}) & = & \frac{1}{4}\Theta\left(\frac{\sigma_{\nu}^{(\alpha)}}{2}-|x_{\alpha}|\right)
\prod_{\tau\neq\alpha}\delta\left(\frac{\sigma_{\nu}^{(\tau)}}{2}-|x_{\tau}|\right),\nonumber \\
\omega_{\nu}^{(2\alpha)}({\bm{r}}) & = & \frac{1}{2}\delta\left(\frac{\sigma_{\nu}^{(\alpha)}}{2}-|x_{\alpha}|\right)
\prod_{\tau\neq\alpha}\Theta\left(\frac{\sigma_{\nu}^{(\tau)}}{2}-|x_{\tau}|\right),\nonumber \\
\omega_{\nu}^{(3)}({\bm{r}}) & = & \prod_{\tau}\Theta\left(\frac{\sigma_{\nu}^{(\tau)}}{2}-|x_{\tau}|\right),\label{weigth_3d}
\end{eqnarray}
where $\sigma_{\nu}^{(\alpha)}=\sigma+(L-\sigma)\delta_{\alpha\nu}$, with $L$ the length of the parallelepipeds and
$\sigma^2$ their cross-sectional area. $\Theta(x)$ and $\delta(x)$ are the Heaviside and Dirac-delta functions, respectively.
A particle of species $\nu$ is defined as having its longest axis along the Cartesian axis $\nu$ (with 
$\nu=x,y$ or $z$). $\sigma_{\nu}^{(\alpha)}$ is the length of species $\nu$
in the direction of the axis $\alpha$.
Integrals of the weights give the fundamental measures, $M_{\nu}^{(\alpha)}\equiv
\int d{\bm r} \omega_{\nu}^{(\alpha)}({\bm r})$, of species $\nu$: $M_{\nu}^{(0)}=1$,
$M_{\nu}^{(3)}=v$ (with $v=L\sigma^2$ the particle volume), $M_{\nu}^{(1\tau)}=\sigma_{\nu}^{(\tau)}$
(the edge-length of species $\nu$ parallel to $\tau$), and $M^{(2\tau)}_{\nu}=L\sigma^2/\sigma^{(\tau)}_{\nu}$
(the surface area of the sides of species $\nu$ perpendicular to $\tau$). Note that within the same
free-energy density (\ref{3D}), it is possible to describe prolate (rod-like with $L>\sigma$) and oblate (plate-like
with $L<\sigma$) particles.

\begin{figure}[h]
\includegraphics[width=8.2cm,angle=0]{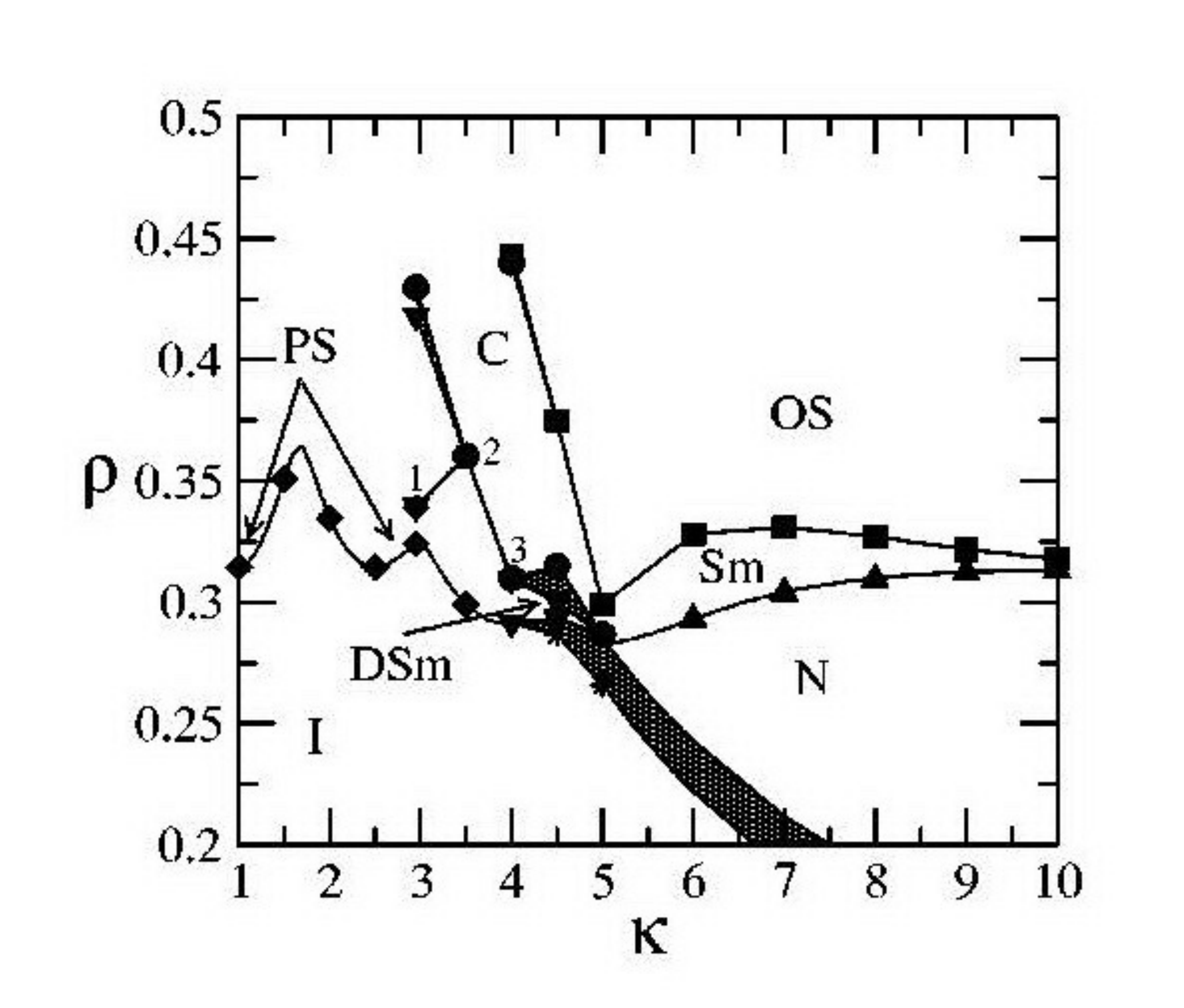}
\includegraphics[width=8cm,angle=0]{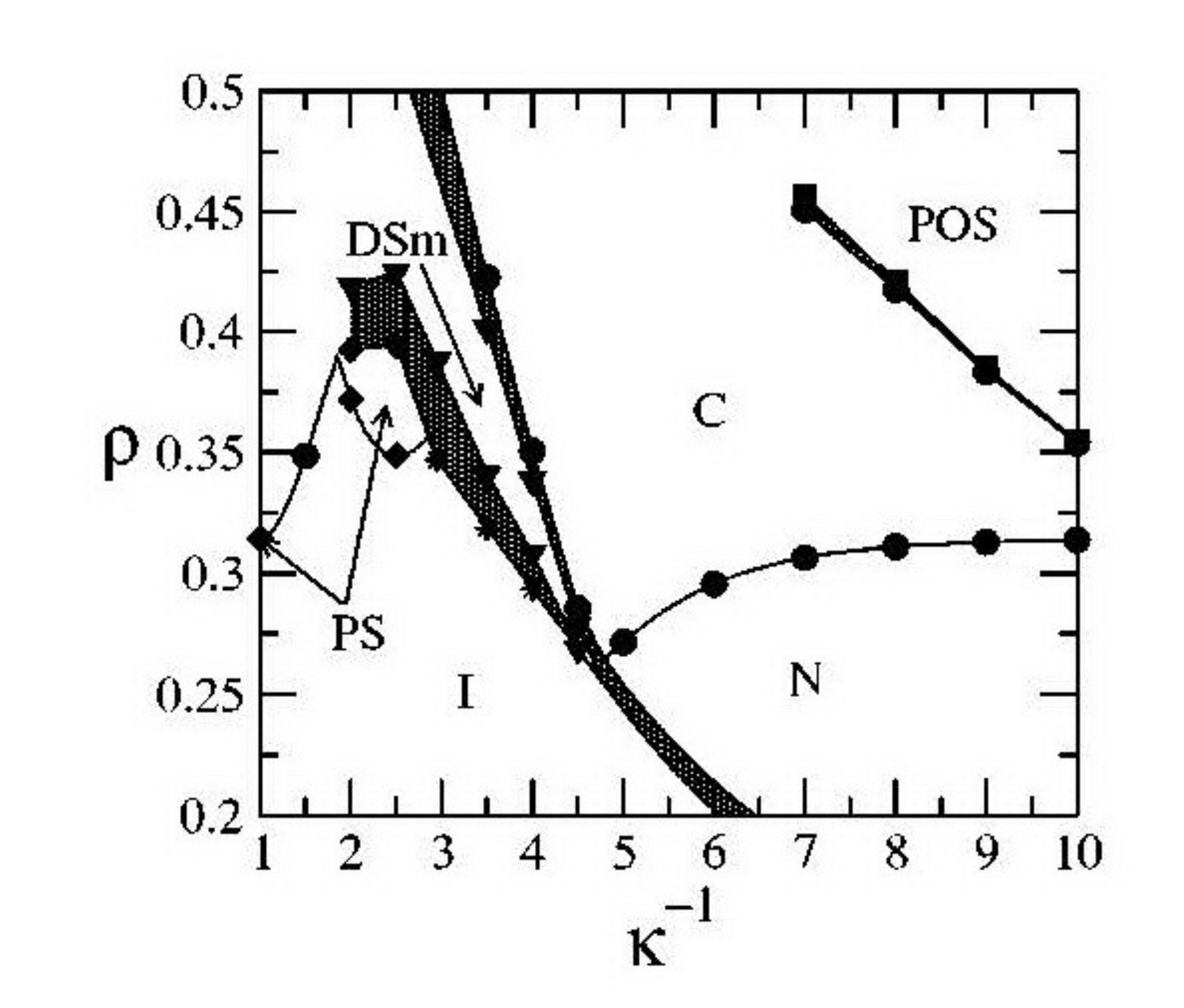}
\caption{Phase diagrams in the density-aspect ratio plane for prolate (left)
and oblate (right) hard parallelepipeds as obtained from a FMT using
the restricted orientation approximation, from \cite{Yuri5a}. Labels correspond to
isotropic (I), nematic (N), columnar (C), smectic (Sm), discotic smectic
(DSm), oriented solid (OS), plastic solid (PS), and perfectly oriented
solid (POS) phases. Shaded areas correspond to two-phase coexistence
regions. Reprinted with permission from \cite{Yuri5a}. 
Copyright (2004) by the American Physical Society.}
\label{Yuri_FMT}
\end{figure}

\begin{figure}
\includegraphics[width=4.5in,angle=0]{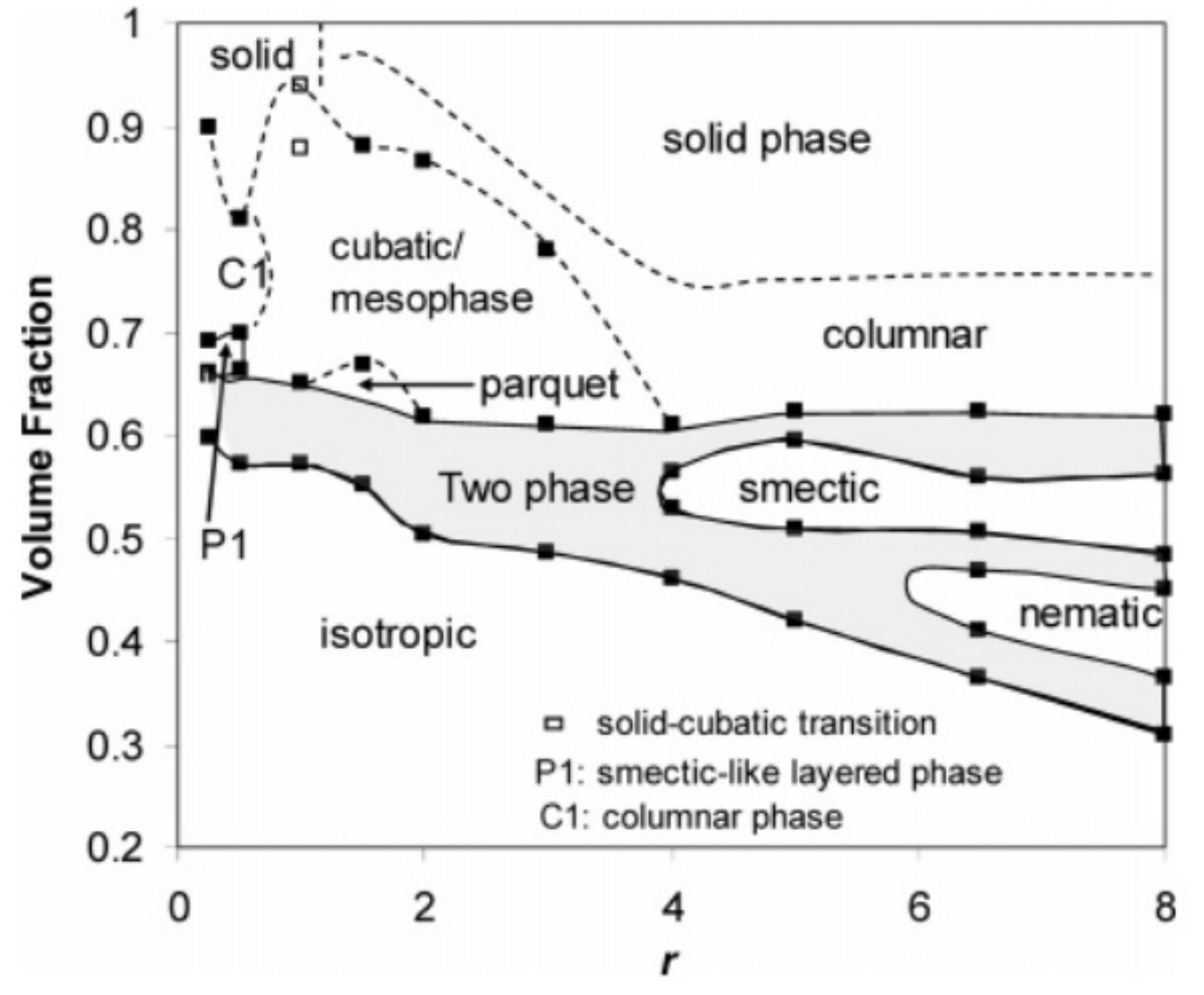}
\caption{Phase diagram of freely-rotating cuboids with aspect ratio $r$ in the range 0.25--8, from Ref. \cite{Escobedo}. 
Dashed lines indicate approximate phase boundaries. P1 is a smectic-like phase. C1 is a columnar
Open squares correspond to approximate boundaries of the solid-cubatic coexistence region. 
Reprinted with permission from Ref. \cite{Escobedo}. Copyright (2005) American Chemical Society.}
\label{Escobedo_fig}
\end{figure}

The above theory was used in 2004 by Mart\'{\i}nez-Rat\'on \cite{Yuri5a} to investigate the
phase diagram of prolate and oblate HP. Transitions to different nonuniform phases
(smectic, columnar, oriented or plastic solid) were considered. Results are shown in Fig. \ref{Yuri_FMT}. On the left side,
the phase diagram of prolate parallelepipeds is shown, while in the right side the results for oblate ones
are presented. As can be inferred from the figure, for $\kappa=L/\sigma>5$ (prolate) and $\kappa^{-1}>4.5$ (oblate)
the sequence of phase transitions from low to high densities is I$\to$N$\to$ S$\to$ OS (prolate) and
I$\to$N$\to$ C$\to$ OS (oblate), similar to what is found for freely rotating particles of other geometries like HSC, HC
and platelets. However, for low values of $\kappa$ (prolate) or $\kappa^{-1}$ (oblate), the phase diagram topology is quite
different from that of freely-rotating particles, reflecting the effect of the restriction of orientations on the stability of
nonuniform phases. For example, the model predicts the stability of a so-called {\it discotic smectic} phase (labelled DSm in the
figure). This peculiar phase is a layered structure, like the smectic phase but with the long (prolate) or short (oblate)
axis of the particles lying in the layers. Because there is no orientational order of this axis in the layers, the nematic order
parameter takes negative values at the positions of the density maxima (which correspond to the position of the layers).
FMT predicts a first-order phase transition from the I phase to the discotic smectic phase. This is in qualitative agreement with
computer simulations of the Zwanzig model with $\kappa=5$ on a lattice \cite{Casey-1}, which showed an I--DSm transition
at a density between 0.47 and 0.55. However, it should be taken into account that the parallelepipedic geometry, especially for
cuboids with $1<\kappa\alt 4$, might stabilize the cubatic and the so-called {\it parquet} (a smectic with 
in-layer tetratic order) phases, as found by John and Escobedo \cite{Escobedo} in recent MC simulations on freely rotating 
cuboids. The phase diagram predicted by the simulations is rather complex (see Fig. \ref{Escobedo_fig}).

\begin{figure}[h]
\includegraphics[width=4.5in,angle=0]{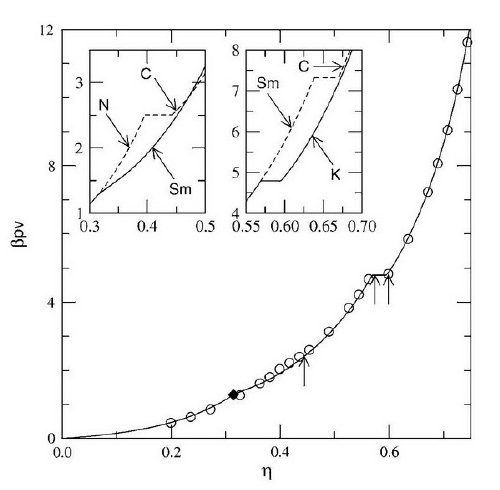}
\caption{Equation of state (pressure in reduced units vs. packing fraction) for all the
stable phases obtained from the FMT for parallel HC \cite{Capitan-1}. These phases are nematic
N (for $\eta$ up to the point indicated by full rhombus), smectic Sm (from that
point up to the discontinuity), and crystal K (from the
discontinuity up to close packing). The open circles are MC simulation
results from \cite{Veerman}. The arrows represent the N--Sm and Sm--K phase
transitions as obtained from simulations. The two insets show the
EOS for the metastable C phase in the neighborhood of the
N--C (left inset) and Sm--C (right inset) phase transitions. 
Reprinted with permission from \cite{Capitan-1}. Copyright (2008), AIP Publishing LLC.}
\label{HC-Capitan}
\end{figure}

Mart\'{\i}nez-Rat\'on et al. \cite{Yuri1} devised a differential procedure to generate functionals in higher 
dimensionalities from low-dimensional functionals for parallel particles with constant cross sections. Using the dimensional crossover 
property mentioned above, these authors obtained a functional for a mixture of parallel cylinders from the corresponding functional for a 
mixture of hard discs. The theory was used by Capit\'an et al. \cite{Capitan-1} to study the phase behaviour 
of a fluid of parallel HC, and all the possible liquid-crystalline phases, as well as the crystalline phase, 
were considered (see Fig. \ref{HC-Capitan}). The functional was numerically minimized using a Gaussian parameterisation 
for the density profiles, and very good agreement with Monte Carlo simulations \cite{Stroobants2,Veerman} was obtained 
for the equation of state, particularly for the nonuniform phases. The main result was that the C phase was found to be 
metastable with respect to the S or K (crystal) phases, which explains the observation in the simulations of a region of stability 
of the C phase which disappears with system size. Since the present functional reduces to the SPT in the 
uniform-phase limit, the description of the N phase was not as accurate as that of the nonuniform phases, and a
deviation between theory and simulations was evident in the location of the N--S transition.

An early attempt to use ideas from FMT for freely-rotating anisotropic bodies was due to
Cinacchi and Schmid \cite{Cinacchi}, who proposed a density functional
that interpolates between the FMT for HS and the Onsager theory for hard needles. They applied their 
approximation to hard ellipsoids and hard spherocylinders. The location of the isotropic-nematic transition
was calculated and compared with simulation, and much better agreement was found than in the case
of Onsager theory. In fact, the theory was found to be almost as accurate as the PL theory.

Later, Schmidt extended the FMT 
formalism to a mixture of freely rotating particles, proposing a 
numerically tractable functional for a mixture of spheres and rods of vanishing thickness (needles) \cite{Schmidt3}.
He showed that the Mayer function of spheres and needles can be exactly expressed as a sum of convolution products, 
\begin{eqnarray}
f_{\rm sn}({\bm r},\hat{\boldsymbol{\Omega}})=\left[\omega_s^{(3)}\ast \omega^{(0)}_{\rm n}\right]
({\bm r},\hat{\boldsymbol{\Omega}})+\left[\omega^{(2)}_{\rm sn}\ast 
\omega^{(1)}_{\rm n}\right]({\bm r},\hat{\boldsymbol{\Omega}}),
\label{ee}
\end{eqnarray}
where $\rm s$ and $\rm n$ stand for spheres and needles, respectively, and $*$ is a convolution product,
$[a*b]({\bm r})=\int d{\bm r}'a({\bm r}')b({\bm r}-{\bm r}')$.
The HS scalar and vectorial weights are those of the original FMT \cite{Schmidt3}, 
\begin{eqnarray}
\omega^{(3)}_{\rm s}({\bm r})=\Theta(R-r),\quad 
\boldsymbol{\omega}^{(2)}_{\rm s}({\bm r})=\delta(R-r)\frac{{\bm r}}{R},
\end{eqnarray}
with $R$ the HS radius. For hard needles the weights are   
\begin{eqnarray}
\omega^{(1)}_{\rm n}({\bm r},\hat{\boldsymbol{\Omega}})&=&
\frac{1}{4}\int_{-L/2}^{L/2}dl \delta({\bm r}+\hat{\boldsymbol{\Omega}}l),\nonumber\\\nonumber\\
\omega^{(0)}_{\rm n}({\bm r},\hat{\boldsymbol{\Omega}})&=&
\frac{1}{2}\left[\delta\left({\bm r}+\frac{\hat{\boldsymbol{\Omega}}L}{2}\right)+
\delta\left({\bm r}-\frac{\hat{\boldsymbol{\Omega}}L}{2}\right)\right],
\end{eqnarray}
with $L$ the length of the needles. Note that both weights depend on the spatial and 
orientational coordinates of a single needle. However, recovering the exact 
Mayer function requires a new weight:
\begin{eqnarray}
\omega^{(2)}_{\rm sn}({\bm r},\hat{\boldsymbol{\Omega}})=2|\boldsymbol{\omega}_{\rm s}^{(2)}({\bm r})\cdot 
\hat{\boldsymbol{\Omega}}|,
\end{eqnarray}
which depends on the position ${\bm r}$ of the HS and the needle 
orientation $\hat{\boldsymbol{\Omega}}$. Therefore, strictly speaking it is not a one-body weight. Within 
this approximation, the free-energy density is obtained as 
$\Phi=\Phi_{\rm s}+\Phi_{\rm sn}$ (i.e. the sum 
of the one-component HS free-energy density and the contribution 
coming from the interaction between spheres and needles \cite{Schmidt3}). 
The excess part of the free-energy functional is then
obtained by integration, $\beta {\cal F}[\{\rho_{\mu}\}]=\displaystyle
\int d{\bm r} \int d\hat{\boldsymbol{\Omega}} \ \Phi(\{n_{\mu}^{(\alpha)}\})$. 

To improve the theory, Schmidt et al. took a step forward by including rod-rod interactions \cite{Schmidt4}. 
In order to do that, the residual surface of rods was taken into account, in the asymptotic limit 
of high aspect ratios, by approximating the Mayer function between two rods. A new weight function is required
which depends on the orientations of both needles, a fact that turns practical applications 
of this approach into a demanding numerical task.  

In the case of HSC, the correction of order $D/L$ to the Mayer-function deconvolution
was taken into account by introducing four new geometric weights which, in turn, define a new set
of weighted densities \cite{Schmidt5}. 
The same formalism was also applied to a ternary mixture of HS, hard rods and 
hard platelets, with both anisotropic particles having vanishing thickness \cite{Schmidt6}.
In particular, the I--N transition of a one-component hard-platelet fluid was calculated with 
this functional, and the results compared with MC simulations and Onsager theory. The 
results compared well with simulations, as shown in Table \ref{table_Schmidt}.

\begin{table}
\begin{center}
\begin{tabular}{ccccccc}
\hline\hline
&& \hbox{FMT}&& \hbox{Simulation} & &\hbox{Onsager}\\
\hline\hline
$\rho^{\rm(iso)}R^3$ & & 0.418 && 0.473 && 0.667\\
$\rho^{\rm(nem)}R^3$ & & 0.460 && 0.509 && 0.845\\
$\Delta\rho R^3$ & & 0.041 && 0.036 && 0.178\\
$S_{\rm nem}$ & & 0.492 && 0.370 && 0.781\\
\hline\hline
\end{tabular}
\end{center}
\caption{Isotropic-nematic coexistence data for hard platelets of radius $R$, as obtained from
FMT \cite{Schmidt6}, simulation \cite{Frenkel-Eppenga0} and Onsager theory \cite{Schmidt6}.}
\label{table_Schmidt}
\end{table}

\begin{figure}
\epsfig{file=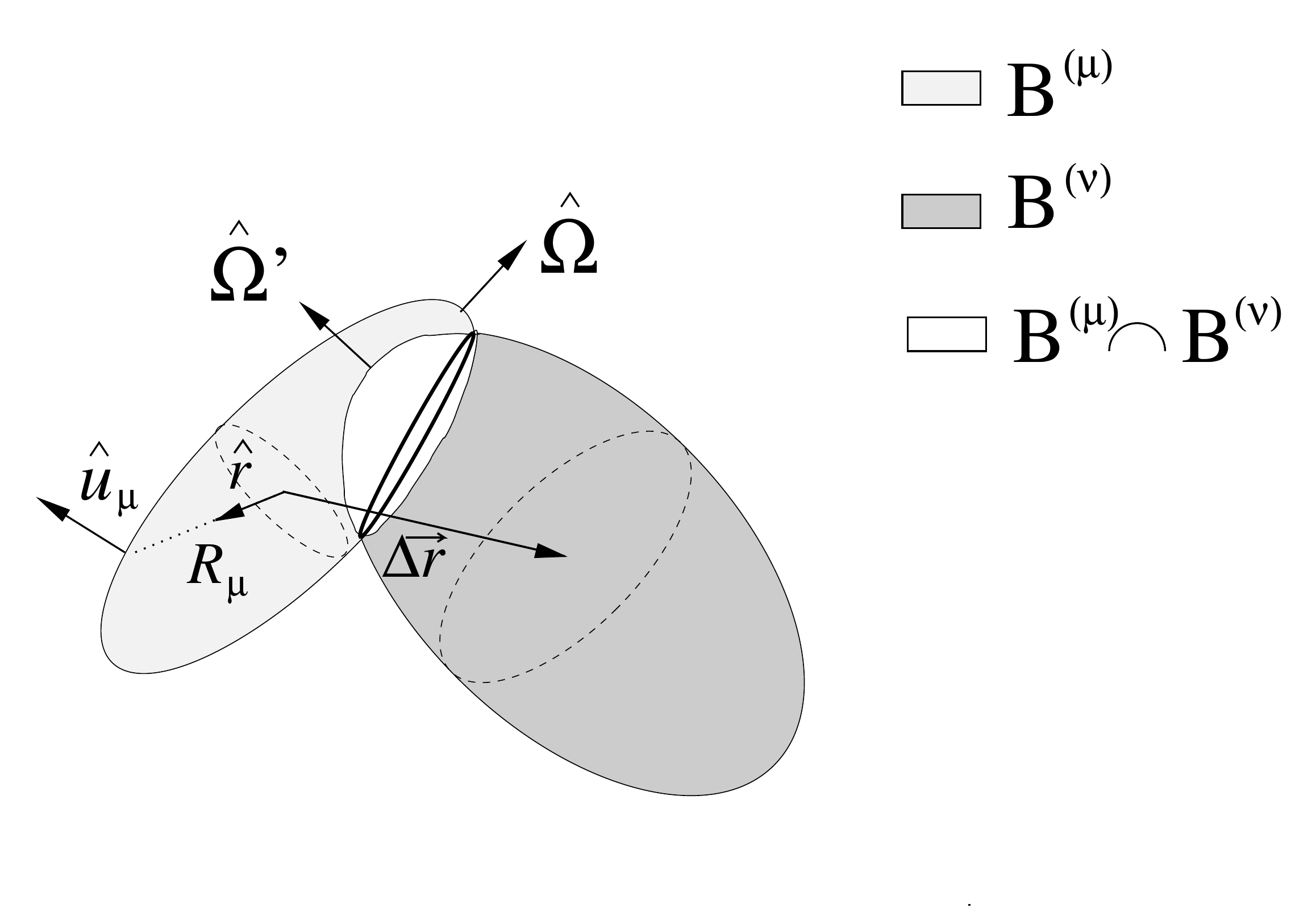,width=5.in}
\caption{Sketch of two overlapping bodies ${\cal B}^{(\mu)}$ and ${\cal B}^{(\nu)}$ with fixed orientations
given by the unit vectors $\hat{\boldsymbol{\Omega}}$ and $\hat{\boldsymbol{\Omega}}'$,
respectively. Their common intersection volume ${\cal B}^{(\mu)}\cap{\cal B}^{(\nu)}$ is shown. Also,
$\hat{\bm R}$ is a generic unit vector at the centre of the particle,
and $R_{\mu}(\hat{\bm R})$ gives the distance from the centre to its surface $\partial{\cal B}^{(\mu)}$.
$\hat{\bm u}_{\mu}$ is the corresponding normal unit vector. The thick solid line represents the curve 
$\partial {\cal B}^{(\mu)}\cap \partial {\cal B}^{(\nu)}$.}
\label{fig2}
\end{figure} 
A new development, along the lines originally proposed by Rosenfeld \cite{Yasha3}, has
recently been carried out in an attempt to formulate a general FMT approximation for
freely-rotating convex bodies. Again the theory is formulated for a mixture of hard particles with arbitrary number of
species, labelled by index $\mu=1,2,\cdots$. The main idea is based on the observation made by Rosenfeld
that, in general, the deconvolution of the Mayer function as a product of single particle weights is
intimately related to the Gauss-Bonnet theorem of differential geometry \cite{Yasha3}.
This idea has been followed by Hansen-Goos and Mecke \cite{Hansen1,Hansen2}, who applied
the Gauss-Bonnet theorem to approximately deconvolute the Mayer function of two convex particles.
Fig. \ref{fig2} is a sketch of two overlapping convex bodies ${\cal B}^{(\mu)}$ and ${\cal B}^{(\nu)}$ which, in general,
will be of different species $\mu$ and $\nu$. Let the orientations of the bodies be given by the unit vectors
$\hat{\boldsymbol{\Omega}}$ and $\hat{\boldsymbol{\Omega}}'$, and let their centre-of-mass relative position vector with respect to a fixed reference 
frame be $\Delta{\bm r}$. Also, let $\partial{\cal B}^{(\mu)}$ be the surface of body
${\cal B}^{(\mu)}$, and $\partial{\cal B}^{(\mu)}\cap{\cal B}^{(\nu)}$ the surface of this body which is inside the body
${\cal B}^{(\nu)}$. Finally, $\partial{\cal B}^{(\mu)}\cap \partial {\cal B}^{(\nu)}$ denotes the curve generated by the 
intersection  between the surfaces $\partial {\cal B}^{(\mu)}$ and $\partial {\cal B}^{(\nu)}$. Further, let $\hat{\bm r}$ be 
a generic unit vector from the centre of particle $\mu$, and $R_{\mu}(\hat{\bm r})$ the distance from this centre to a point
on the surface of the same particle. The Gaussian and mean curvatures at a point 
on the surface $\partial{\cal B}^{(\mu)}$ are
$K_{\mu}(\hat{\bm r})=k_{\mu}^{(1)}(\hat{\bm r}) k_{\mu}^{(2)}(\hat{\bm r})$ and
$\displaystyle{H_{\mu}(\hat{\bm r})=\frac{1}{2}\left[k_{\mu}^{(1)}(\hat{\bm r})+k_{\mu}^{(2)}(\hat{\bm r})\right]}$,  with
$k_{\mu}^{(\alpha)}(\hat{\bm r})$ ($\alpha=1,2$) the principal curvatures of the surface $\partial{\cal B}^{(\mu)}$.
These curvatures
characterize the intrinsic geometry of the surface $\partial {\cal B}^{(\mu)}$.
The curve $\partial {\cal B}^{(\mu)}\cap \partial {\cal B}^{(\nu)}$ at point on the surface
$\partial {\cal B}^{(\mu)}$ is characterized by its geodesic curvature $k_{g\mu}(\hat{\bm r})$.

In the present geometrical construction, the Gauss-Bonnet theorem can be expressed as
\begin{eqnarray}
&&\int_{\partial {\cal B}_{\mu}\cap{\cal B}_{\nu}} \frac{K_{\mu}(\hat{\bm r})}{4\pi} dA+\int_{\partial {\cal B}_{\nu}\cap {\cal B}_{\mu}} 
\frac{K_{\nu}(\hat{\bm r})}{4\pi} dA+\int_{\partial {\cal B}_{\mu}\cap \partial {\cal B}_{\nu}} 
\left(\frac{k_{g\mu}(\hat{\bm r})+k_{g\nu}(\hat{\bm r})}{4\pi}\right)dl\nonumber\\\nonumber\\&&\hspace{7cm}
=-f_{\mu\nu}(\Delta{\bm r},\hat{\boldsymbol{\Omega}},\hat{\boldsymbol{\Omega}}'),
\label{deconvoluted}
\end{eqnarray}
where $f_{\mu\nu}(\Delta{\bm r},\hat{\boldsymbol{\Omega}},\hat{\boldsymbol{\Omega}}')$ is the Mayer function of the two particles. 
The first two surface integrals can be expressed as 
the spatial convolutions $\omega_{\mu}^{(0)}\ast \omega_{\nu}^{(3)}$ and $\omega_{\mu}^{(3)}\ast \omega_{\nu}^{(0)}$, respectively, with
the one-particle weights defined as
\begin{eqnarray}
\omega^{(0)}_{\mu}({\bm r})&=&\frac{K_{\mu}(\hat{\bm r})}{4\pi}\delta\left(R_{\mu}(\hat{\bm r})- r\right),\nonumber\\\nonumber\\
\omega^{(3)}_{\mu}({\bm r})&=&\Theta\left(R_{\mu}(\hat{\bm r})- r\right).
\end{eqnarray}
However, the third integral can only be exactly deconvoluted in the case of two particle geometries: spheres (or parallel ellipsoids) 
and parallel or mutually perpendicular parallelepipeds. 
The last integral is a contour integral. The integral of the first term, $k_{g\mu}(\hat{\bm r})/4\pi$,  can be split into two
contributions. The first is the integral of a term proportional to $H_{\mu}(\hat{\bm r})$. The second is the integral of a term proportional  
to $\displaystyle{\frac{k_{\mu}^{(1)}(\hat{\bm r})-k_{\mu}^{(2)}(\hat{\bm r})}{2(1+{\bm u}_{\mu}\cdot{\bm u}_{\nu})}}$.
The presence of the denominator 
$1+{\bm u}_{\mu}\cdot{\bm u}_{\nu}$ is the reason why the last integral in (\ref{deconvoluted}) cannot be exactly deconvoluted. 
Approximating this denominator by unity (the lowest order in the Taylor expansion with respect to 
${\bm u}_{\mu}\cdot{\bm u}_{\nu}$) allows for the deconvolution of this contribution in terms of one-particle weights. The result is:
\begin{eqnarray}
-f_{\mu\nu}=\omega^{(0)}_{\mu}\ast\omega^{(3)}_{\nu}+\omega^{(1)}_{\mu}\ast\omega^{(2)}_{\nu}
-\boldsymbol{\omega}^{(1)}_{\mu}\ast\boldsymbol{\omega}^{(2)}_{\nu}
-\zeta \overset\leftrightarrow{\omega}_{\mu}^{(1)}\ast
\overset\leftrightarrow{\omega}_{\nu}^{(2)}+(\mu\leftrightarrow\nu),
\end{eqnarray} 
with $\omega^{(1,2)}_{\mu}$, $\boldsymbol{\omega}^{(1,2)}_{\mu}$, and 
$\overset\leftrightarrow{\omega}_{\mu}^{(1,2)}$, the 
scalar, vectorial and tensorial weighted densities (expressions for them can be found in Ref. \cite{Hansen1,Hansen2}). $\zeta$ 
is an adjustable parameter that allows for the improvement of the theory (in particular, it has been used to improve the results
for the I--N transition as compared to simulations). A free-energy density which leads to this deconvolution of the Mayer 
function and, also, which recovers Tarazona's FMT version for HS \cite{Tarazona2} (and, therefore, fulfills the dimensional crossover 
to 0D) is 
\begin{eqnarray}
\Phi&=&-n_0\ln(1-n_3)+\frac{n_1n_2-{\bm n}_1{\bm n}_2-\zeta \text{Tr}\left[
\overset\leftrightarrow{n}_1\overset\leftrightarrow{n}_2\right]}{1-n_3}
\nonumber\\
&+&\frac{3}{16\pi}\frac{{\bm n}_2^{\text{T}}\overset\leftrightarrow{n}_2{\bm n}_2
-n_2{\bm n}_2{\bm n}_2-\text{Tr}\left[\overset\leftrightarrow{n}_2^3\right]+
n_2\text{Tr}\left[\overset\leftrightarrow{n}_2^2\right]}{(1-n_3)^2}.
\label{laphi}
\end{eqnarray}
The weighted densities are calculated as
\begin{eqnarray}
n_{\alpha}({\bm r})=\sum_{\mu} \int d{\bm r}'\int d\hat{\boldsymbol{\Omega}}
\rho_{\mu}({\bm r}',\hat{\boldsymbol{\Omega}})\omega^{(\alpha)}_{\mu}({\bm r}-{\bm r}',
\hat{\boldsymbol{\Omega}}).
\end{eqnarray}
The free-energy density 
(\ref{laphi}) has the merit of being a function of one-particle densities, and consequently
its implementation is numerically cheaper than those based on the Mayer 
function. Also, the functional can be improved by adding more tensorial weighted densities 
in the second term of (\ref{laphi}) as obtained from the expansion of 
$(1+{\bm u}_{\mu}\cdot{\bm u}_{\nu})^{-1}$.   
Note that the term proportional to $\zeta$ is necessary to generate an orientational 
phase transition; its absence destabilises the N phase. The value of the 
parameter $\zeta$ can be chosen so as to
improve the results of the model compared with computer simulations, as shown by Hansen-Goos and Mecke  
\cite{Hansen1,Hansen2}.
These authors applied the present functional to study 
the I--N transition of HSC for general aspect ratio. The coexisting densities resulting from a 
particular value of $\zeta$ compare well with the simulation results.
H\"artel and L\"owen \cite{Hartel} applied the same theory to study the static and 
dynamical response of a fluid of HSC subject to the presence of aligning fields. 

\begin{figure}[h]
\includegraphics[width=4.in,angle=0]{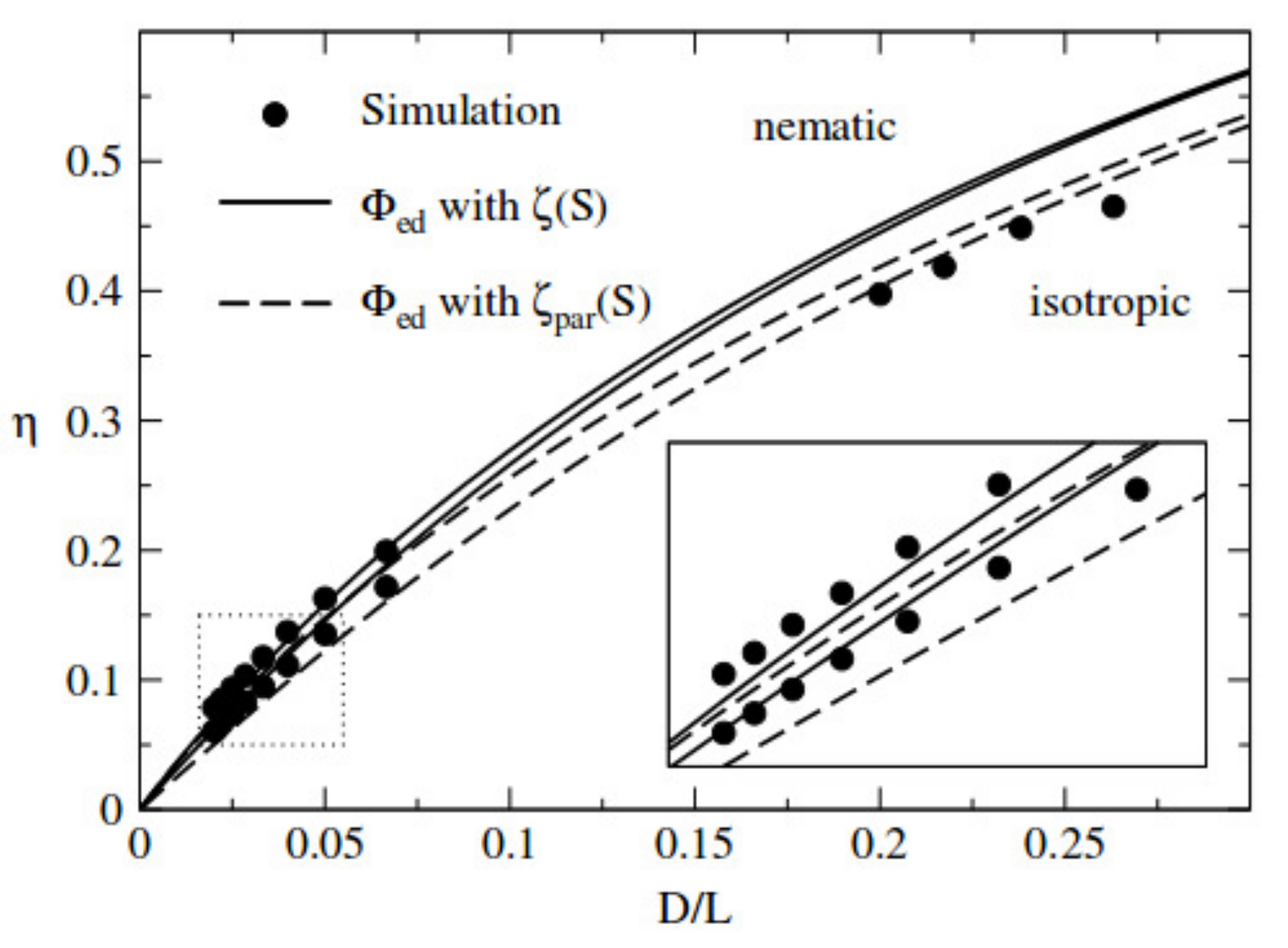}
\caption{Results for the I--N transition of the HSC fluid
as obtained from the FMT of Hansen and Mecke \cite{Hansen2}. Symbols indicate simulation results from 
\cite{FLS}. Solid and dashed lines were calculated using different choices of the parameter $\zeta$ (see Ref. \cite{Hansen2} for details).}
\label{Hansen_fig}
\end{figure}

\subsection{Further studies on spatially nonuniform phases}
\label{further}

Hard models have been used to analyse other nonuniform liquid-crystalline phases different from the smectic, and also some
effects and phenomena present in general liquid-crystal phases, especially concerning peculiar structural and dynamical effects. 
One of the latter concerns the presence of transverse 
order in smectic phases made of hard rods. In studies of smectics, one assumes that the orientational order parameter is high and that,
as a result, the orientational distribution function should be highly peaked in the direction of the layer normal. The sucess (or not)
of the parallel-particle approximation is based on this expectation. Corrections to the free energy due to orientational freedom 
is expected to be quantitatively small, and therefore the coupling between orientations and positions should be effectively weak,
i.e. $\rho({\bm r},\hat{\bm\Omega})\simeq\rho({\bm r})h({\bm\Omega})$. This means that the extremely small fraction of rods located
close to the interstitial region should also be highly oriented along the layer normal. However, already at the level of Onsager 
density-functional theory, it was found by Rooij et al. \cite{Rooij_trans} that the orientational distribution function 
$h({\bm r},{\bm\Omega})$ is strongly modulated. Theory and simulations indicated that the interstitial particles are oriented parallel, 
instead of perpendicular, to the layers. 
\begin{figure}[h]
\includegraphics[width=7cm,angle=0]{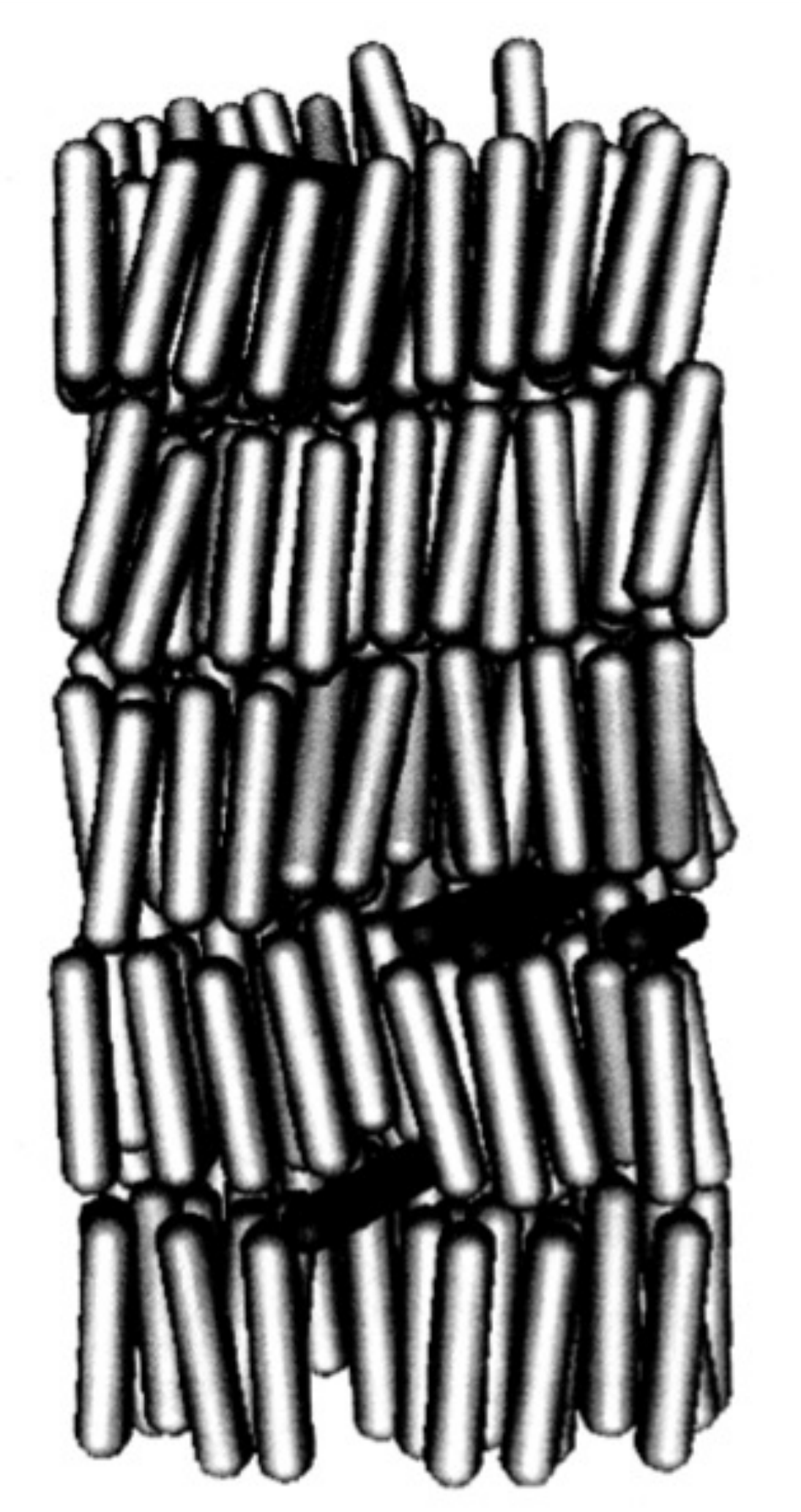}
\caption{Configuration of hard spherocylinders in a smectic arrangement showing interstitial particles with
transverse orientation (from simulations by van Rooij et al. \cite{Rooij_trans}). Reprinted with permission from \cite{Rooij_trans}. 
Copyright (1995) by the American Physical Society.}
\end{figure}

In more recent work, the related effect of interlayer diffusion has been studied. 
It has long been known that, in a nematic
phase, diffusion constants along and perpendicular to the director are very different \cite{Allen_Hess}, with some peculiar 
effects found in simulations of the nematic phase of hard ellipsoids \cite{Allen_diffusion}, where it was found that the 
longitudinal diffusion component increases with density in some density interval just above the isotropic-nematic transition. 
Longitudinal diffusion, which gives rise to interlayer diffusion or permeation in smectic phases, has been studied more
recently. Lamellar phases made of rod-like viruses were seen to exhibit single particle hops which occur by quasi-quantised
steps of one rod length; this one-particle diffusion can be explained on the basis of a simple model based on the
longitudinal diffusion of a nematic phase in a periodic potential that simulates the lamellar ordering. Diffusion mechanisms
in smectics has been studied in detail, using computer simulation, by Cinacchi and de Gaetani \cite{Giorgio_diffusion}.

The theories for nonuniform liquid-crystalline phases discussed so far were devised with a view to describing the transition
from the nematic to the smectic phase since the latter is the simplest nonuniform phase, exhibiting spatial order along only one coordinate. 
An additional level of nonuniformity is presented by the columnar phase, where the local density depends on two spatial coordinates.
Using computer simulation, Stroobants et al. \cite{Stroobants2} studied a fluid of hard parallel spherocylinders, and observed the
formation of a columnar phase (consisting of a two-dimensional hexagonal lattice of fluid columns) between the smectic and 
crystal phases $L/D>5$. This phase was later shown to be an artifact of the small system size of the samples \cite{Veerman}.
In fact, there are good reasons to expect that the columnar phase cannot be stable: in the large $L/D$ regime, hard parallel spherocylinders 
resemble hard parallel cylinders, and these behave exactly as hard discs if scaled along the long particle axis. Since hard discs directly freeze from the fluid 
into the crystal, there should be no intermediate phase with partial order in the system of parallel cylinders. In the corresponding freely-rotating model,
rotational entropy plays against columnar ordering, and the columnar phase is not stabilised either.

\begin{figure}[h]
\includegraphics[width=10cm,angle=0]{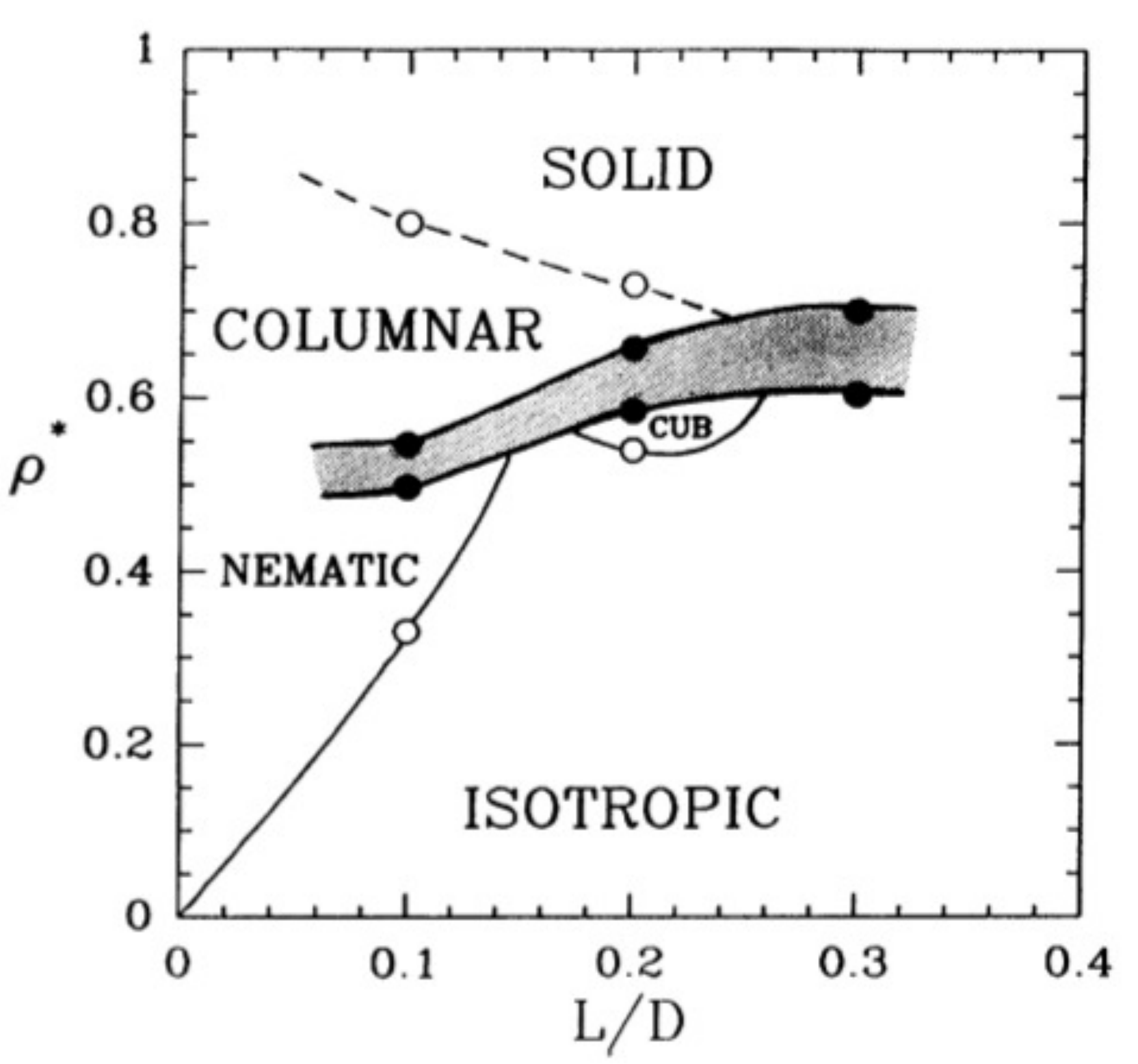}
\caption{Phase diagram of the hard cut-sphere model from computer simulation (from Veerman and Frenkel \cite{Frenkel_columnar1}).
Reprinted with permission from \cite{Frenkel_columnar1}. 
Copyright (1992) by the American Physical Society.}
\label{cutsphere}
\end{figure}

Therefore, except in mixtures of rods with different lengths \cite{Stroobants-2} (see Sec. \ref{binary_mixtures}) where
columnar ordering (generated by length bidispersity) was found, we do not expect to find columnar phases in fluids of (prolate) 
hard rods. This is why columnar phases were searched for in fluids of oblate particles. However, for this purpose the choice
of particle is not simple. The first model of oblate or discotic particle, the infinitely thin circular platelet discussed 
by Onsager, exhibits nematic ordering \cite{Frenkel-Eppenga0,Frenkel-Eppenga1}, but it possesses a vanishing excluded volume as particles become 
increasingly ordered and, consequently, there is no mechanism to generate spatially nonuniform phases beyond the nematic.
Another model considered, that of oblate hard ellipsoids, again shows no mesogenic phases different from the nematic 
\cite{Frenkel0,Allen_diffusion}, since already the parallel model scales to the hard-sphere model. To circumvent these problems,
Frenkel \cite{Frenkel_columnar} studied a particle model that favours the piling-up of particles to form columns. This is
the hard cut-sphere model, consisting of a sphere of diameter $D$ where two opposite parts have been removed, leaving a particle of 
thickness $L<D$ with two flat surfaces; when $L=D$ a full sphere is recovered.
Frenkel \cite{Frenkel_columnar} and Veerman and Frenkel \cite{Frenkel_columnar1} studied this model in detail by computer
simulation. Their complete phase diagram is shown in Fig. \ref{cutsphere}. As expected, a columnar phase, intermediate
between the nematic and the crystal phases, was formed for $L/D\alt 0.25$. Interestingly, an exotic nematic phase, the 
{\it cubatic} phase, was stabilised in a small region of density and aspect ratio. This phase possesses no positional
order but exhibits orientational order with cubic symmetry.

\begin{figure}[h]
\includegraphics[width=10cm]{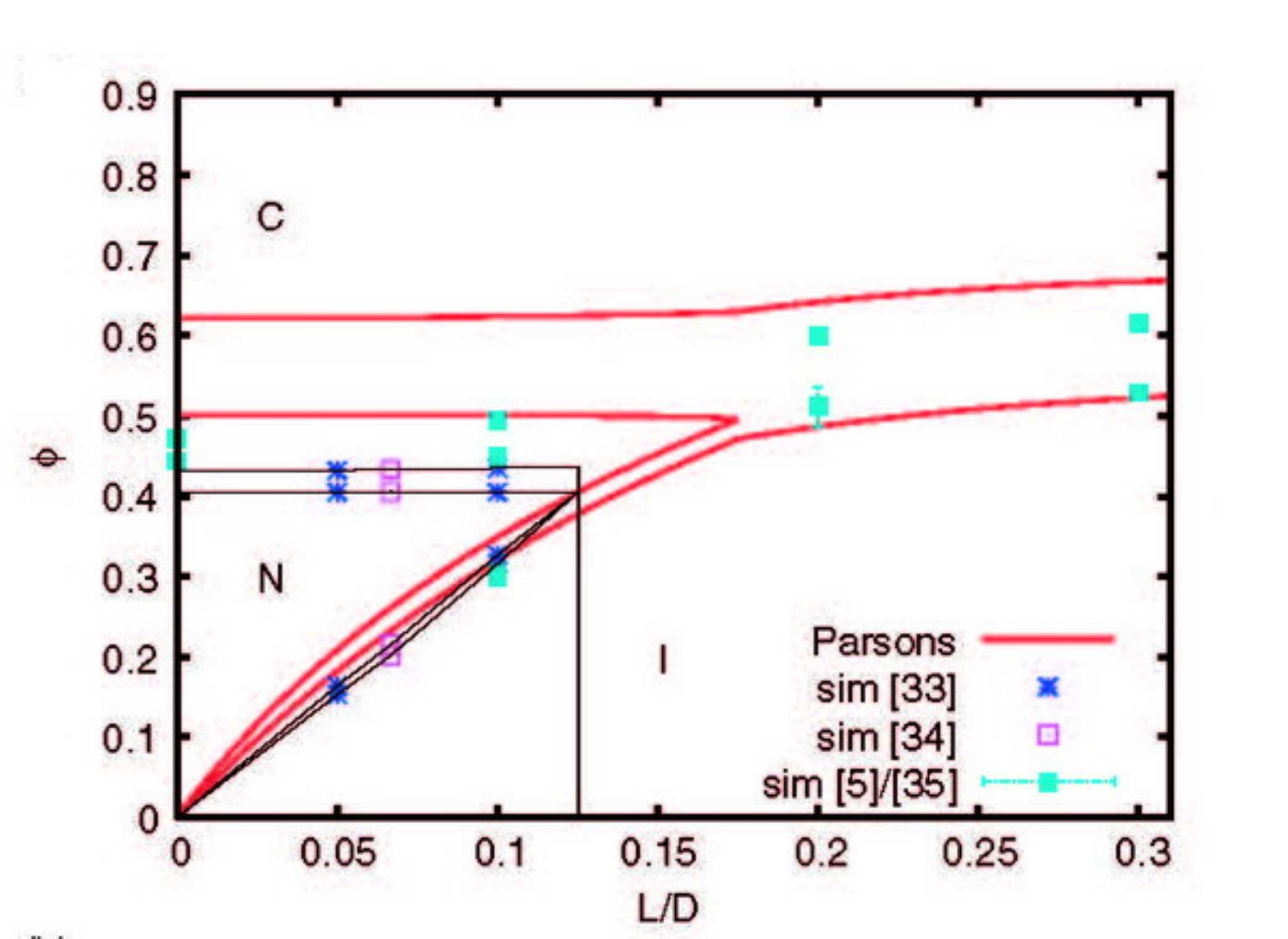}
\caption{\label{fig:wensink} Phase diagram of oblate cylinders of aspect ratio
$L/D$ as obtained in \cite{Wensink-1}. The packing fraction of the
system is represented in the vertical axis. Red lines are results
from the theoretical approximation. Symbols correspond to different
computer simulation studies (see original paper for details).
Reprinted by permission of the publisher (Taylor \& Francis Ltd., 
http://www.tandf.co.uk/journals).}
\end{figure}

More recently, further studies on the columnar phase stability in
systems of oblate hard particles have been carried out by Wensink and
Lekkerkerker \cite{Wensink-1} and by Marechal et al. \cite{Marechal-1}.
Wensink and Lekkerkerker mapped out the complete liquid-crystal
phase diagram of oblate cylinders of different aspect ratios $L/D<1$.
They used PL theory to describe isotropic and nematic phases, and a
Lennard-Jones-Devonshire cell model to address the question of the
stability of nonuniform phases (namely, columnar and solid) at high
packing fractions. At low enough aspect ratio ($L/D<\kappa_{t}\simeq 0.175$)
the system shows I--N and N--C transitions upon increasing its packing
fraction. For aspects ratios $L/D>\kappa_{t}$ the anisometry of the
particles is too small to produce a stable nematic phase and the system
shows a direct transition from the isotropic to the columnar phase.
Results compare satisfactorily with computer simulations, particularly
in the region $L/D>\kappa_{t}$, as can be seen in Fig.
\ref{fig:wensink}.

It is interesting to note that, within this model, the N--C transition
does not depend appreciably on the aspect ratio. By using a Gaussian
ansatz for the orientational distribution function, and taking advantage
of the fact that nematic order is very strong close to the N--C transition,
Wensink and Lekkerkerker were able to show that the nematic-columnar
transition is, in fact, universal and independent of particle shape.

\begin{figure}[h]
\includegraphics[width=10cm]{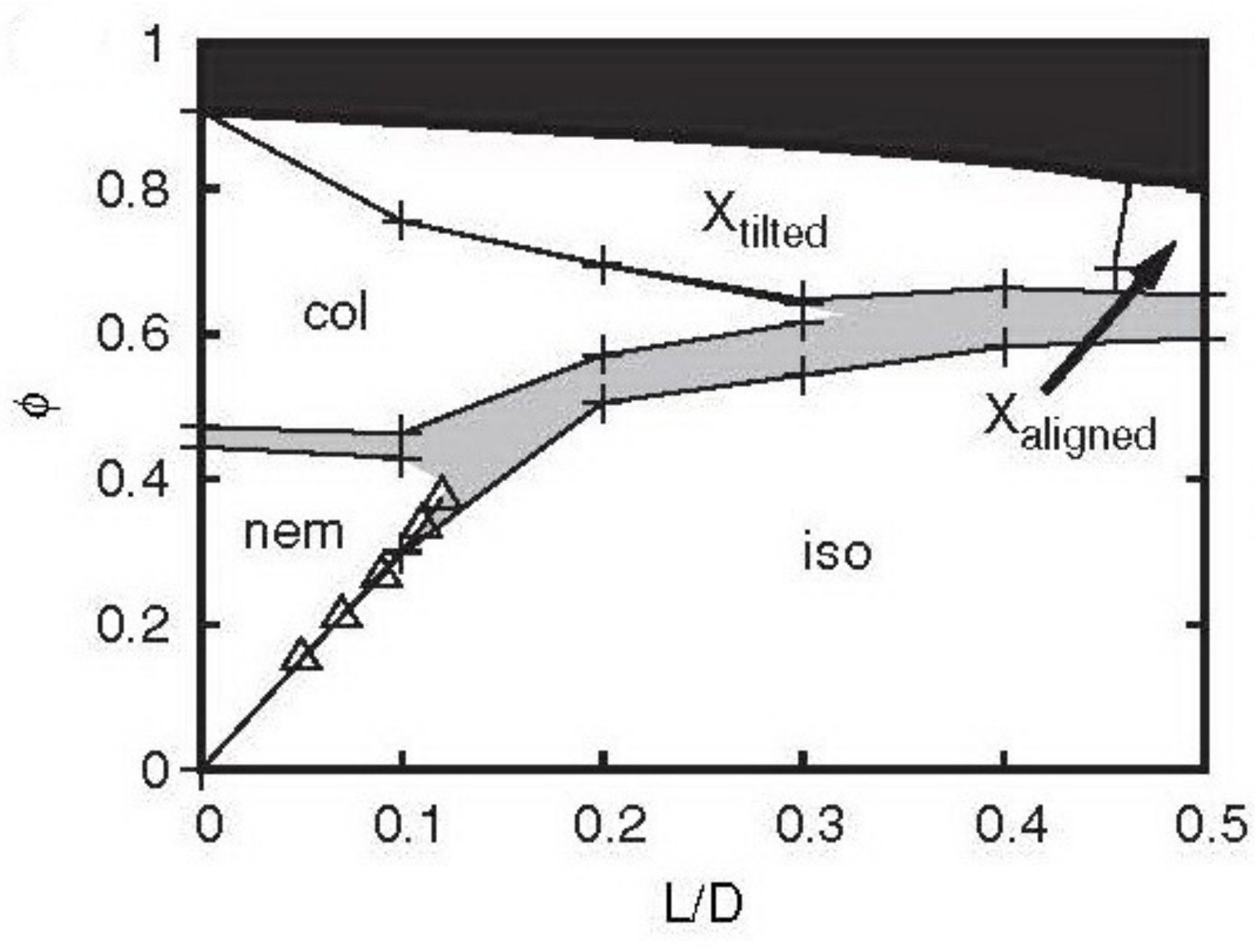}
\caption{\label{fig:marechal} Phase diagram of oblate hard spherocylinders
in the aspect ratio-packing fraction plane as obtained by Marechal
et al. \cite{Marechal-1}. The shaded region at high packing fraction
are states above the close packing limit. Crystalline phases $X_{\rm tilted}$
and $X_{\rm aligned}$, N and C mesophases and the I phase are stable.
Symbols are coexistence points obtained from the simulations, while
lines are only guides to the eye. Reprinted with permission from \cite{Marechal-1}. Copyright (2011), AIP Publishing LLC.}
\end{figure}

Marechal et al. \cite{Marechal-1} used Monte Carlo free energy calculations
to obtain the phase diagram of oblate hard spherocylinders (OHSC)
as a function of their aspect ratio. Their result has been reproduced
in Fig. \ref{fig:marechal}. Apart from the crystal phases, the topology
of the phase diagram is qualitatively similar to that obtained by
Wensink and Lekkerkerker for oblate cylinders using PL theory (Fig.
\ref{fig:wensink}). At low aspect ratios the system undergoes I--N
and N--C phase transitions as packing fraction is increased. Above a particular
value of aspect ratio, the N phase is no longer stable
and a direct I--C phase transition takes place. Moreover, the similarity
between both phase diagrams applies also to the N--C transition at
low aspect ratios: Like in the case of oblate cylinders, this transition
seems to be almost independent of $L/D$. 

\begin{figure}
\includegraphics[width=13cm]{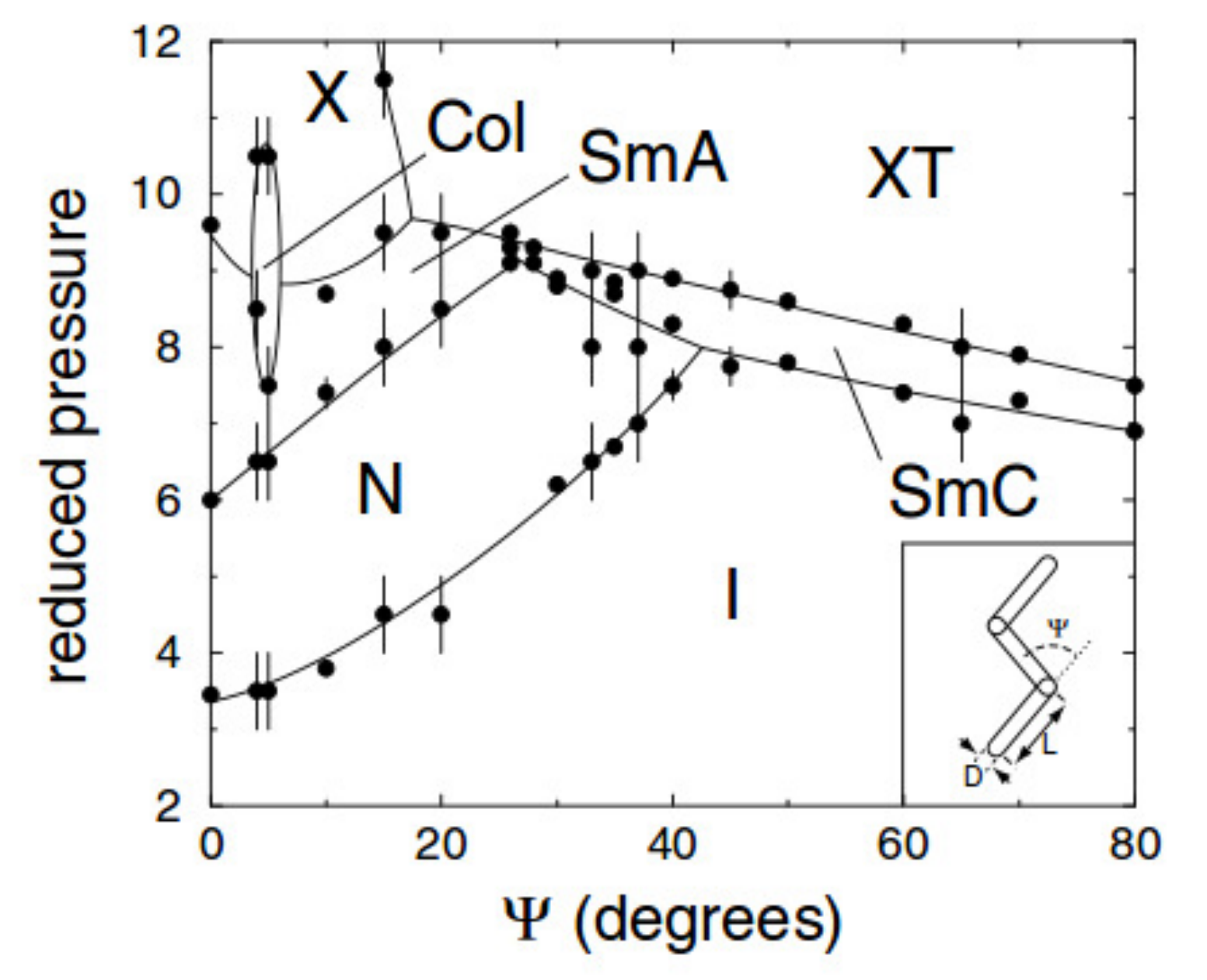}
\caption{Phase diagram of zigzag molecules made out of hard spherocylindrical sectors with aspect
ratio $L/D=2$ as a function of angle between the sectors $\Psi$, from \cite{Clark}. Labelled phases are:
isotropic (I), nematic (N), smectic A (Sm A), smectic C (Sm C), columnar (C), tilted crystal (XT) and
crystal (X). Reprinted with permission from \cite{Clark}. 
Copyright (2004) by the American Physical Society.}
\label{Clark_fig}
\end{figure}

\begin{figure}
\includegraphics[width=8.65cm]{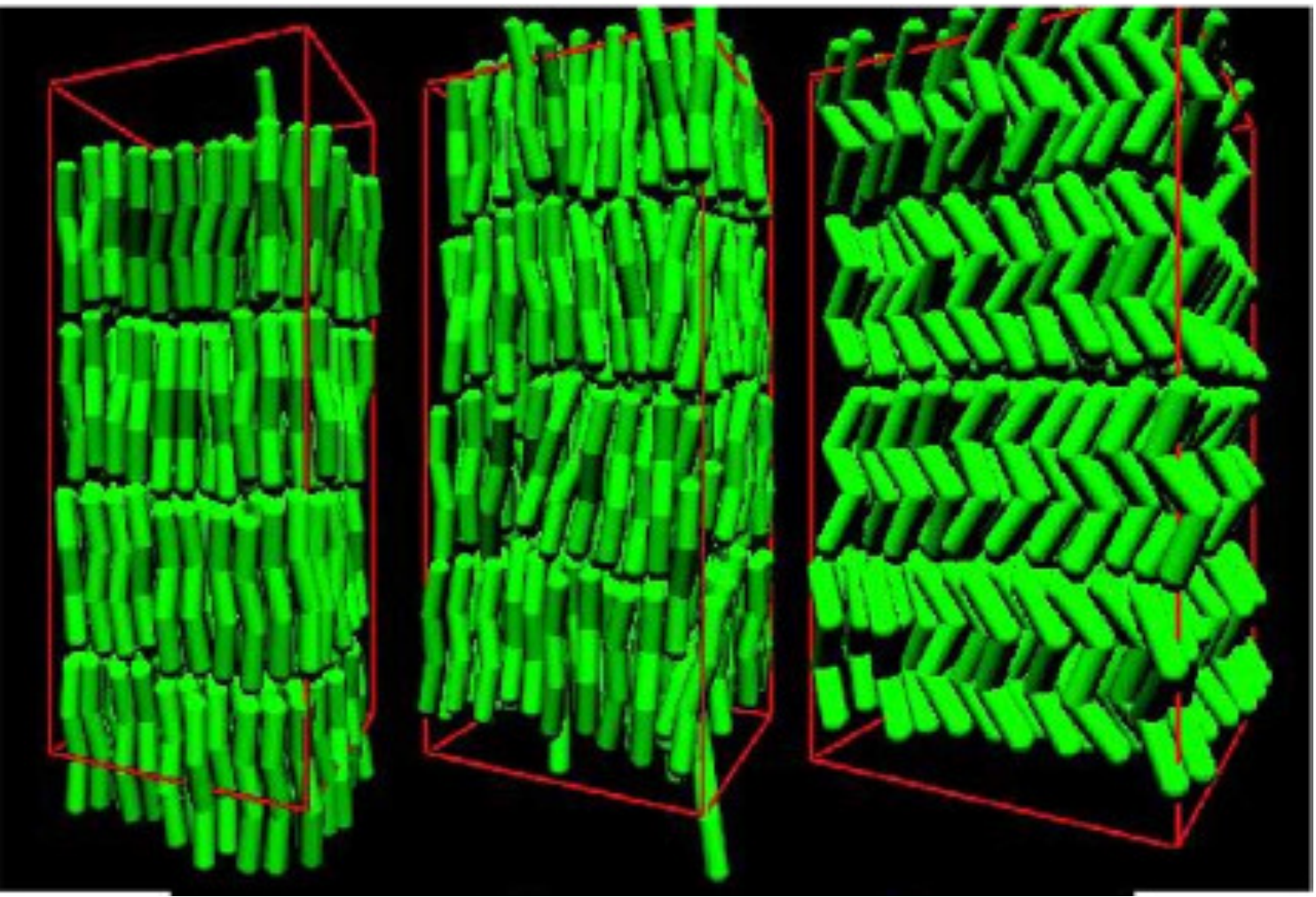}
\includegraphics[width=6cm]{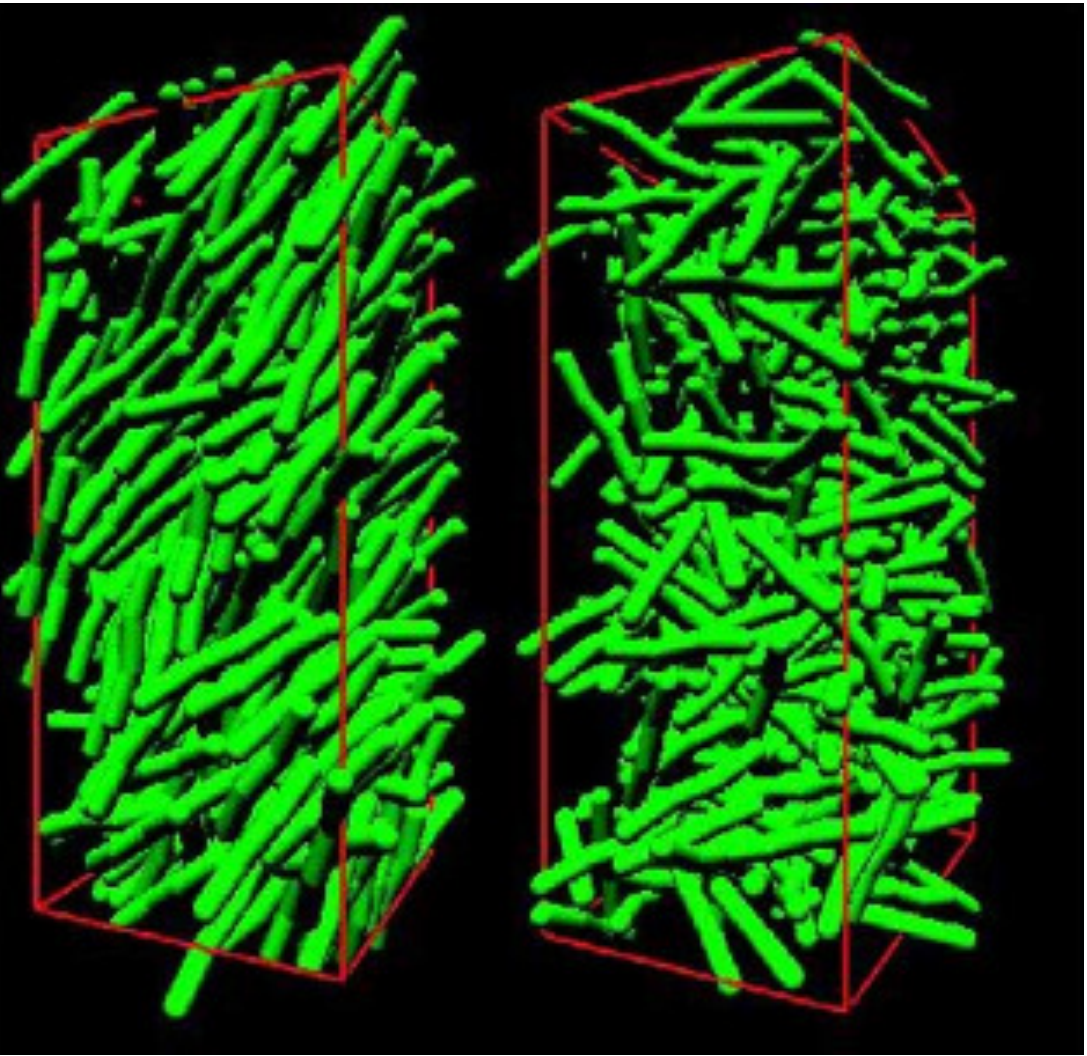}
\caption{Typical configurations in different from Monte Carlo simulations of a system of $400$ zigzag molecules, 
from \cite{Clark}. From left to right and top to bottom:
crystal (reduced $p^*=11$ and angle between consecutive sectors $\Psi=15^{\circ}$),  smectic A 
($p^*=9$ and $\Psi=15^{\circ}$), smectic C ($p^*=8$ and $\Psi=65^{\circ}$), nematic ($p^*=5$ and $\Psi=15^{\circ}$)
and isotropic ($p^*=1$ and $\Psi=15^{\circ}$) phases. Reprinted with permission from \cite{Clark}. 
Copyright (2004) by the American Physical Society.}
\label{Clark_fig1}
\end{figure}

There have been some studies on phase behaviour in fluids of rigid particles formed by linked rods.  The resulting
particles possess non-convex shapes, and nonstandard liquid-crystalline phases may be stabilised due to nontrivial
packing restrictions. For example, banana-shaped or V-shaped particles are biaxial bodies, and consequently they 
have been used to investigate the stability of the biaxial
nematic phases (see next section). However, these particles have been conjectured \cite{Bisi} to be able to also form smectic 
phases with antiferromagnetic arrangement, an effect originating in their polar character (here polarity has a steric 
origin associated to the shape). A very detailed simulation study of zigzag particles was performed by Maiti et al. 
\cite{Clark}, who considered particles with three sectors
made from three hard spherocylinders attached by their ends. An amazingly large number of stable
phases were found, including isotropic, nematic, smectic A, smectic C, columnar and two crystal phases (see Fig. 
\ref{Clark_fig}, where the phase diagram is represented in the plane pressure vs. angle between the sectors $\Psi$).
Fig. \ref{Clark_fig1} shows typical configurations of these phases.
The presence of the smectic C phase and a transition from the smectic-A to the smectic-C phase
is particularly interesting since it shows, as already demonstrated by Somoza and
Tarazona using density-functional theory \cite{ST,ST1}, that hard-body interactions alone can stabilise different types
of smectic phases and drive transitions between them.

\subsection{Biaxial particles}
\label{biaxial_section}

\begin{figure}
\includegraphics[width=15cm]{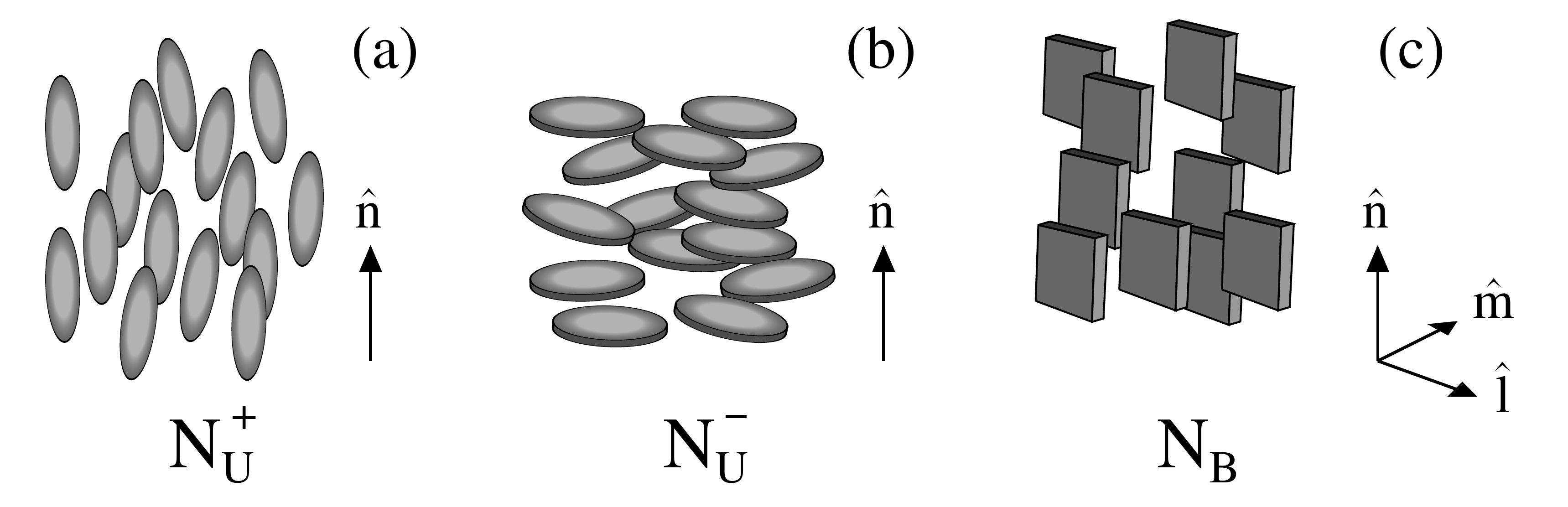}
\caption{Uniaxial nematic phases of (a) prolate and (b) oblate particles, and (c) biaxial nematic phase. 
See text for discussion.}
\label{uni-bi}
\end{figure}

Up to now we have discussed the appearance of liquid-crystalline order
as the direct consequence of the anisotropy along a single axis of common nematogens.
Simple entropic arguments rapidly lead to the conclusion that the
equilibrium uniform configuration of the system should be a nematic phase
when the density is high enough, as it was simply and elegantly shown
by Onsager \cite{Onsager}. But it is clear that molecules of most
liquid-crystal forming substances are not cylindrically symmetric
and, therefore, it should be possible, as first predicted by
Freiser \cite{Freiser-1} in 1970 in the context of an extended Maier-Saupe
model for thermotropic liquid crystals, to find a stable biaxial nematic
(N$_{\rm B}$) phase, i.e. a nematic phase that shows not only long-range
orientational order of the long molecular axis along one direction
(the director), as it is the case in common uniaxial nematics, but
also shows long-range order of the shorter molecular axes along two
other mutually perpendicular directors. In other words, a nematic
phase with three directors about which three molecular axes tend to
align. In Fig. \ref{uni-bi} we have shown common uniaxial nematic phases
N$_{\rm U}^{+}$ and N$_{\rm U}^{^{-}}$ which are usually formed by rod-like
molecules and disc-like molecules, respectively, and a biaxial nematic
phase formed by board-like molecules.

In the case of the uniaxial nematic phases (a) and (b) shown in Fig. \ref{uni-bi},
all the directions perpendicular to $\hat{\mathbf{n}}$ are equivalent.
Therefore, if we look at the system from a direction parallel to the
director, we will find that the molecules have, in average, a circular
cross section while they will show an elliptical cross section if we
look at the system in a direction perpendicular to $\hat{\mathbf{n}}$.
This leads to an axially symmetric ellipsoid and, hence, this kind
of nematic phases are optically uniaxial. In (a) the largest polarizability
is parallel to $\hat{\mathbf{n}}$ and, as a consequence, this uniaxial
nematic phase has positive birefringence. In case (b), however, the
largest polarizability is perpendicular to $\hat{\mathbf{n}}$, resulting
in a negative birefringence. But in the case (c) there is another
possibility: the existence of a secondary director $\hat{\mathbf{m}}$
perpendicular to $\hat{\mathbf{n}}$ (and, therefore, a third one $\hat{\mathbf{l}}$
perpendicular to the other two). In this case molecules will show
a non-circular (elliptical) averaged cross section if they are viewed
either parallel to the main director $\hat{\mathbf{n}}$ or parallel
to the secondary one $\hat{\mathbf{m}}$ (or $\hat{\mathbf{l}}$).
Therefore, this nematic phase has reduced symmetry and is optically
biaxial. In other words, molecules in the uniaxial phase are rotationally
disordered around the long axis while in the biaxial phase this rotation
is restricted in such a way that the molecules show a time averaged
board-like shape. From the macroscopic point of view, the biaxial
nematic has two optical axes, i.e. directions along which the optical
properties appear to be cylindrically symmetric \cite{Luckrust-1,Luckrust-1a,Tschierske-1}.

The theoretical prediction of the N$_{\rm B}$ phase in 1970 triggered
its search since it was immediately recognized that the existence
of N$_{\rm B}$ phases is a problem of great interest, not only theoretical
but also practical, due to the fact that these phases could be used
in the development of fast electro-optical devices (see the excellent
review by Tschierske and Photinos \cite{Tschierske-1,Berardi-1}). Consequently, after
its prediction by Freiser, a great effort has been devoted to their
theoretical analysis, experimental observation, and computer simulation.
The first experimental observation of the N$_{\rm B}$ phase, in a lyotropic system, 
was done in 1980 by Yu and Saupe \cite{Yu-1}, although their system was a ternary
mixture rather than a pure compound. The occurrence of the N$_{\rm B}$
phase in mixtures will be reviewed in Sec. \ref{binary_mixtures}. More recently
biaxial nematics have been observed in thermotropic liquid crystals
made of bent-core organic molecules \cite{Madsen-1,Acharya-1,Galerne-1,Madsen-2};
however, in this case controversial issues about the correct identification
of the biaxial phase still remain \cite{Luckrust-1,Luckrust-1a,Galerne-1,Madsen-2,VanLe-1}.
Direct experimental evidence of the existence of the biaxial nematic
phase in a system of particles that can be considered hard bodies
is that found in 2009 by van den Pol et al. \cite{van den Pol-1} in
a systems of board-like particles. These authors found both biaxial
nematic and biaxial smectic phases in a colloidal dispersion of goethite
particles. The particles are so big ($254\times83\times28$ nm$^3$
on average, with a polydispersity of $20\%-25\%$ in all directions)
that they can be regarded, to a very good approximation, as hard particles.
Small angle X-ray scattering was used to study the structure of the macroscopic
domains formed. Biaxiality is found in a system whose particles have
a shape that is almost exactly in between rod-like and plate-like,
in agreement, as we shall review below, with the predictions of theoretical
models. The results of this experimental study suggest that biaxial
phases can be indeed easily obtained by a proper choice of the particle
shape. Molecular asymmetry is also expected to have some other influences
on the macroscopic phase behaviour of the system, involving transitions
between non-biaxial phases as it has been shown, for example, by Somoza
and Tarazona \cite{Somoza-3} who used the decoupling approximation
to investigate the isotropic-nematic transition as a function of particle
biaxiality in a system of hard spheroplatelets. Their conclusion is
that the first-order character of this phase transition is significantly
weakened by the molecular asymmetry.

To our knowledge, the first hard-body model approximation to the biaxial
nematic phase is that of Shih and Alben \cite{Shih-1,Alben-2}. They
used a generalization of Flory's lattice model to compute the phase
diagram of rectangular plate-like particles of any length and width.
The system consists in a simple cubic lattice formed by $M$ cells.
$N$ rigid rectangular plates are then dispersed through the lattice.
Each one of these plate-like particles is $w$ cells wide, $l$ cells
long and one cell thick and it is required that the two sides of the
plates lie along mutually perpendicular base-vector directions. Therefore,
there are six possible molecular orientations in this model. There
are no interactions between particles except for the hard-core interaction
needed in order to avoid overlaping between plates. The thermodynamic
behaviour of the system is then obtained by considering the number
of distinct ways $g\left(\left\{ \alpha\right\} ,N,M\right)$ of packing
$N$ molecules into $M$ cells under the restriction that given fractions
$\alpha_{i}$ ($i=1,...,6$) of the plates have each of the six possible
orientations. In the limit of very large $N$ the Helmholtz free energy
of the system is related to $g$ evaluated at the set $\left\{ \overline{\alpha}\right\}$
for which it takes its maximum value:
\begin{eqnarray}
F=-kT\ln g\left(\left\{ \overline{\alpha}\right\} ,N,M\right)
\end{eqnarray}
\noindent Details of the calculation of $g\left(\left\{ \overline{\alpha}\right\} ,N,M\right)$
are to be found in the original paper of Shih and Alben \cite{Shih-1}.
The results obtained from the model are that a biaxial nematic phase
is obtained at high pressure for plates which are neither very square
nor very rod-like in shape. In this phase both the long axes and the
flat faces of the particles tend to be parallel. At lower pressure
the biaxial phase undergoes a second order phase transition to a uniaxial
phase. The model shows two different uniaxial nematic: prolate uniaxial
states (N$_{\rm U}^+$) in which the long axes tend to be parallel, and oblate uniaxial
states (N$_{\rm U}^-$) in which the flat faces tend to be parallel. Prolate
uniaxial nematics are stable for rod-like particles (length/width
$\gtrsim0.3$) at intermediate densities, and oblate uniaxial states
take place for plate-like particles (length/width $\lesssim0.3$)
also at intermediate densities. At even lower pressure the uniaxial
nematic phase undergoes a first order transition to an isotropic phase
showing discontinuities in the density and the orientational order
in going through the transition. The size of these discontinuities
depends strongly on the aspect ratio of the plates. 

\begin{figure}
\includegraphics[width=6cm]{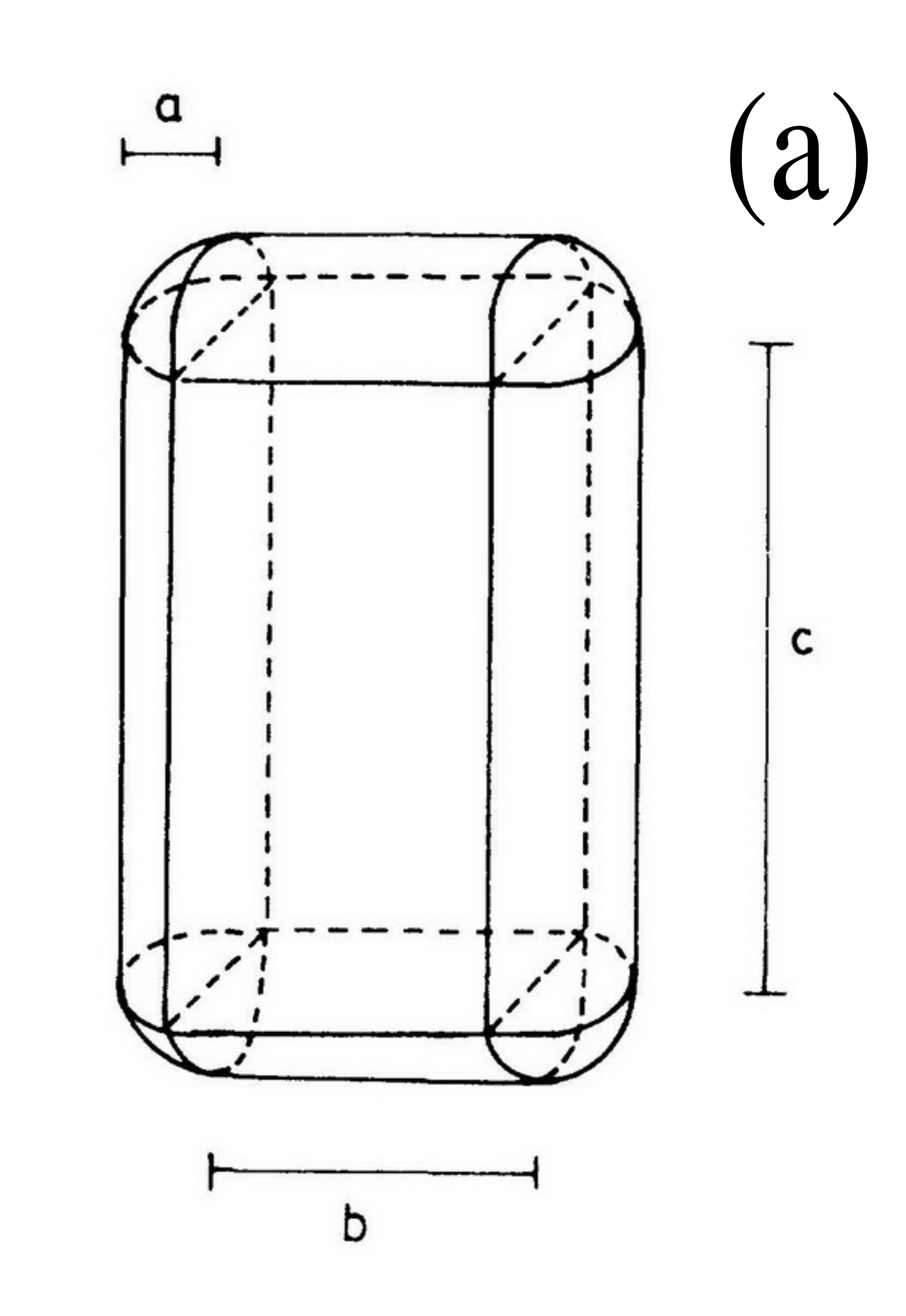}
\includegraphics[width=9cm]{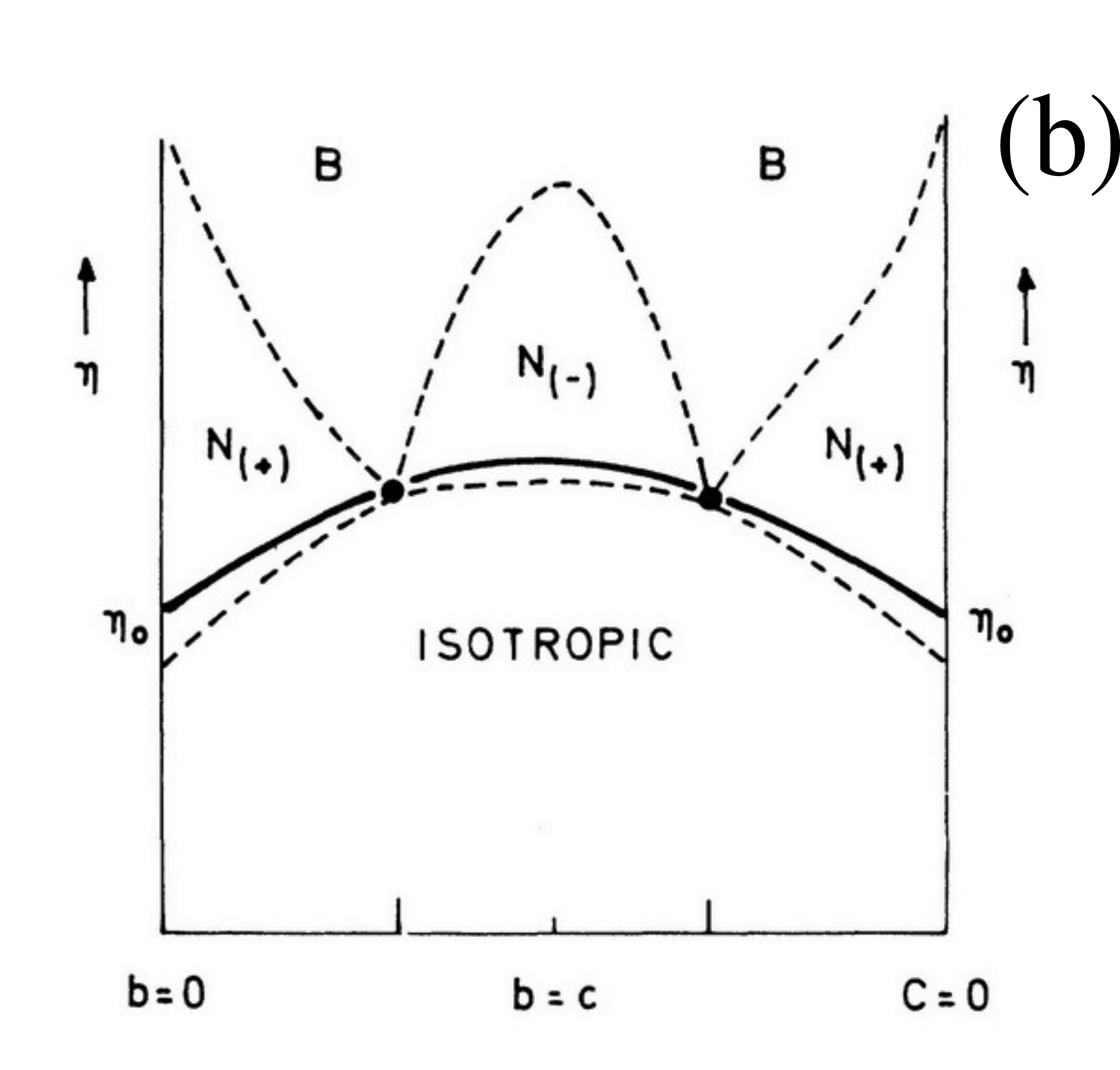}
\caption{(a) Spheroplatelet. (b) Conjectured phase diagram for a system of
spheroplatelets from Onsager's theory as a function particle geometry
and system's packing fraction. Reprinted with permission from \cite{Mulder-1}. 
Copyright (1989) by the American Physical Society.}
\label{Mulder_fig}
\end{figure}

A different kind of plate-like particles, namely spheroplatelets [see
Fig. \ref{Mulder_fig}(a)], was considered in 1989 by Mulder \cite{Mulder-1} who used,
taking advantage of the knowledge of the explicit expression for the excluded
volume between two spheroplatelets, Onsager's approximation to investigate
the phase behaviour of a system of such particles. His results are
not conclusive with respect to the precise topology of the phase diagram
since a complete free energy minimization was not carried out.
Instead, a bifurcation analysis with respect to particle geometry was performed and a conjectured phase
diagram was presented [see Fig. \ref{Mulder_fig}(b)], which showed a biaxial nematic
phase for high enough packing fraction. For moderate packing fractions, the biaxial phase was
conjectured to be stable for particles shapes not too rod- or plate-like.

Spheroplatelets of all possible values
of their elongation $c$ were also analysed the following year, in
1990, by Ho\l{}yst and Poniewierski \cite{Holyst-1} using the smoothed density approximation.
A Landau bicritical point,
at which a direct transition from the isotropic phase to the biaxial
phase occurs (and as conjectured by Mulder, Fig. \ref{Mulder_fig}),
was found. In fact, a whole line of these points is obtained when
the spheroplatelets elongation is allowed to vary. A line of bicritical
points was also found for a system of hard biaxial
ellipsoids of axes $a<b<c$ for elongations $c/a\leq 7$. Both spheroplatelets and biaxial
ellipsoids show similar scaling at the Landau bicritical point: $b/a\sim (c/a)^{1/2}$,
where $b/a$ is the breadth of the ellipsoid or the spheroplatelet.

\begin{figure}
\includegraphics[width=8.0cm]{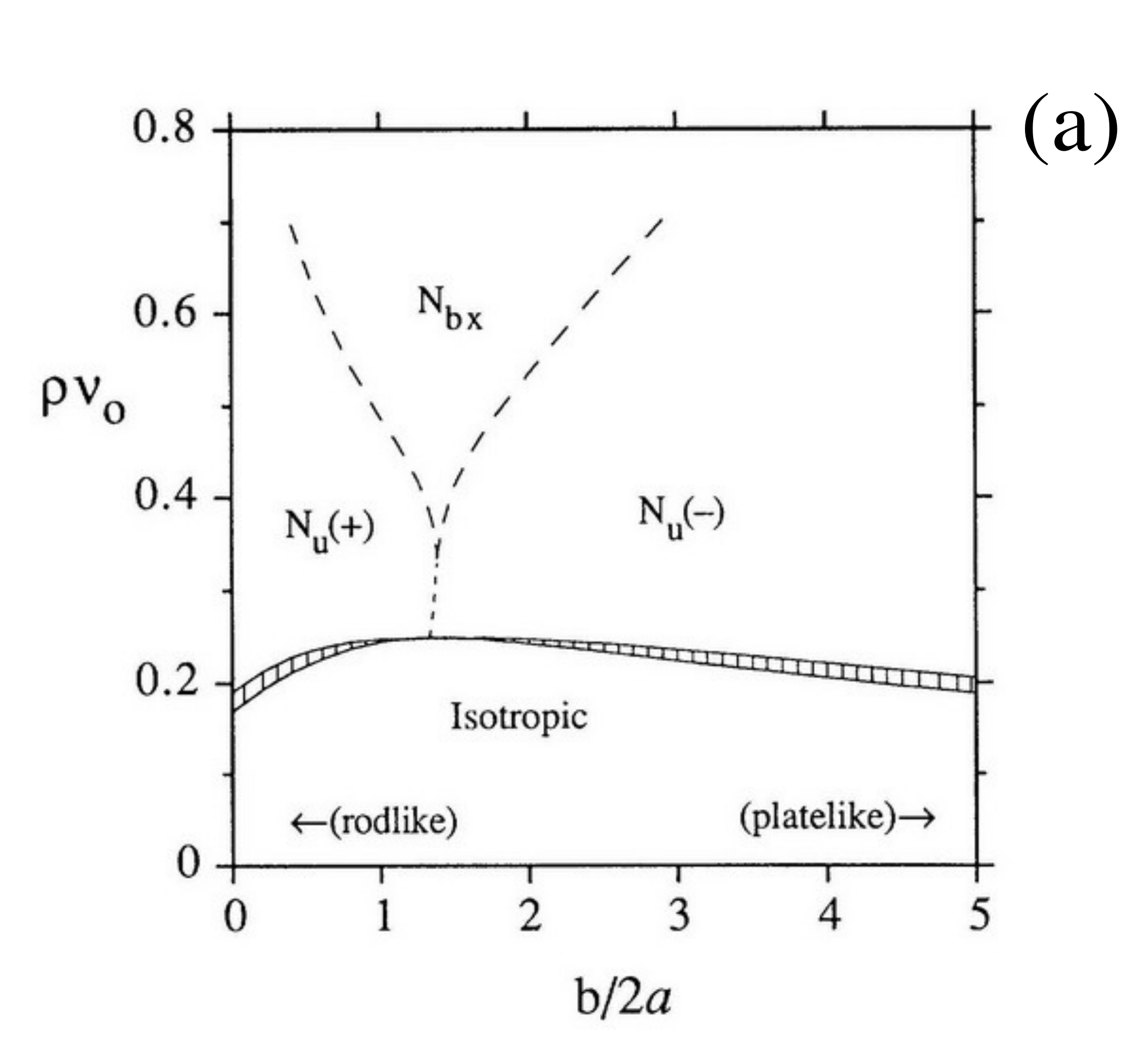}
\includegraphics[width=8.0cm]{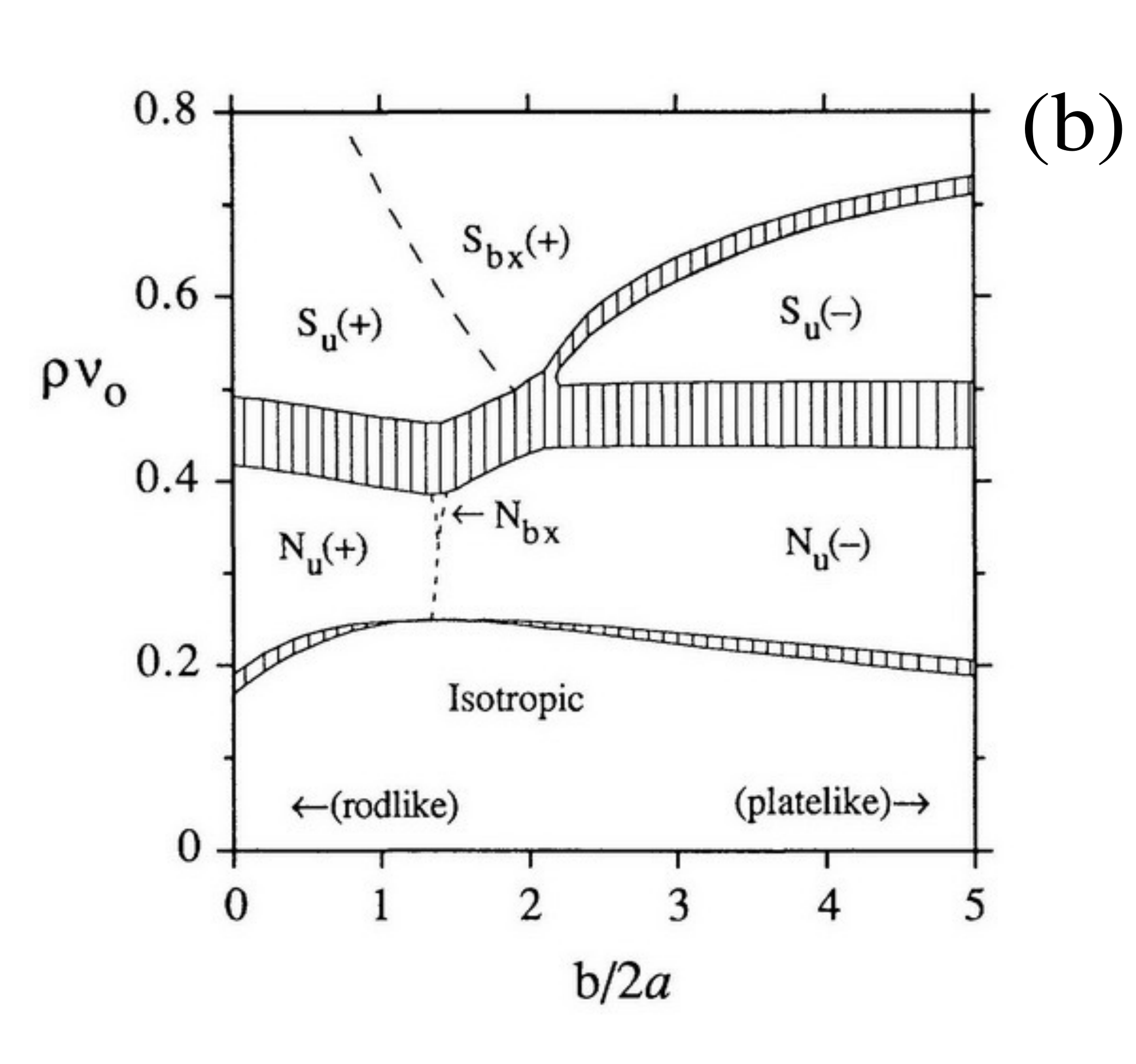}
\caption{(a) Phase diagram of spheroplatelets with $c/2a=5$ from the scaled
particle of Taylor and Herzfeld \cite{Taylor1}. (b) Same as (a) but
including translationally ordered smectic phases (see text for
discussion). Reprinted with permission from \cite{Taylor1}. 
Copyright (1991) by the American Physical Society.}
\label{Taylor_fig}
\end{figure}

The phase diagram of a system of spheroplatelets
was calculated later, in 1991, by Taylor and Herzfeld \cite{Taylor1},
this time making use of the scaled particle approximation. In order
to further simplify the calculations, the resulting scaled particle
equations were solved in the Zwanzig approximation, as in the previous work 
of Shih and Alben discussed above. 
Fig. \ref{Taylor_fig} shows the phase diagram obtained by Taylor and
Herzfeld for a system of spheroplatelets with $c/2a=5$. In the vertical
axis the packing fraction of the system is shown, while the horizontal
axis represents the particle geometry. Panel (a) shows the result obtained from the scaled particle
theory solved in the Zwanzig approximation, considering only the possible
formation of nematic phases. The shaded areas correspond to first-order
phase transitions while dashed lines correspond to second-order transitions.
Two uniaxial nematics are stable, namely the rod-like uniaxial nematic N$_{\rm U}(+)$
and the plate-like uniaxial nematic N$_{\rm U}(-)$. For particles that are not too
rod- or plate-like a biaxial nematic, N$_{\rm bx}$, is found
when the packing fraction is high enough, in agreement with previous
models. 

The elusiveness of the biaxial nematic phase led Taylor and
Herzfeld to wonder about the possibility that the biaxial nematic were
pre-empted by translationally ordered phases, namely by the smectic
phase. In order to shed some light on that issue, they considered
the stability of a smectic by using a cell model to treat the translational
order. In that model, smectic order was imposed by introducing impenetrable
cell walls which divide the system into smectic layers. Each layer
is then treated as a two-dimensional liquid confined within the one-dimensional
cell corresponding to the thickness of the smectic layer, the two-dimensional
density of this fluid being coupled to the one-dimensional cell entropy
through their mutual dependence on the smectic layer spacing. Three
possible smectic-A phases are then possible, depending on which of
the three spheroplatelet principal axes is associated with the smectic
ordering axis. Fig. \ref{Taylor_fig}(b) shows the phase diagram
obtained when this description of the smectic-A ordering is combined
with the scaled particle theory for the isotropic and nematic phases.
Stable smectic-A phases are found in the full range of particle biaxiality
(as measured by the value of $b/2a$) for packing fractions $\gtrsim0.5$.
In the rod-like limit ($b\rightarrow0$) there is a first-order transition
from the uniaxial nematic to the uniaxial smectic S$_{\rm U}(+)$ in which
the rod axes are aligned perpendicular to the smectic layers. In the
other extreme, for plate-like particles in the limit $b\rightarrow c$,
a first order transition between the uniaxial nematic N$_{\rm U}(-)$
and a uniaxial smectic-A phase S$_{\rm U}(-)$ is found. In this smectic
phase the short molecular axes are oriented perpendicular to the smectic
layers. In an intermediate molecular biaxility region of the phase
diagram, a stable biaxial smectic phase, S$_{\rm BX}(+)$, is obtained.
In this phase the long molecular axes are aligned perpendicular to
the smectic layers, like in the uniaxial S$_{\rm U}(+)$ phase, but an
in-layer second-order transition as the molecular biaxility is increased
leads to the ordering of the short molecular axes, resulting in the
biaxial smectic phase. This in-layer order of the short molecular
axes is lost again, via a first-order phase transition, upon increasing
even further the molecular biaxility, giving rise to the stable uniaxial
smectic S$_{\rm U}(-)$. There are two remarkable aspects in the phase
diagram of Fig. \ref{Taylor_fig}(b). The first is the almost
complete disappearance of the stable biaxial nematic island predicted
by this model when no translationally ordered phases are taken into
account (as shown in panel (a) of the figure). The second
is the strong first-order character of all the nematic-smectic
transitions as shown by the wide shaded areas that separate those
phases in the phase diagram. Although it is expected that the phase
diagram predicted in this paper could be qualitatively correct, one
must bear in mind that two different theories are used
to describe the translationally disordered phases and the smectic
one. Moreover, the approximation used to describe smectic phases will
probably overestimate the stability of the smectic phase since the
translational order is imposed by hand. 

\begin{figure}
\includegraphics[width=13cm]{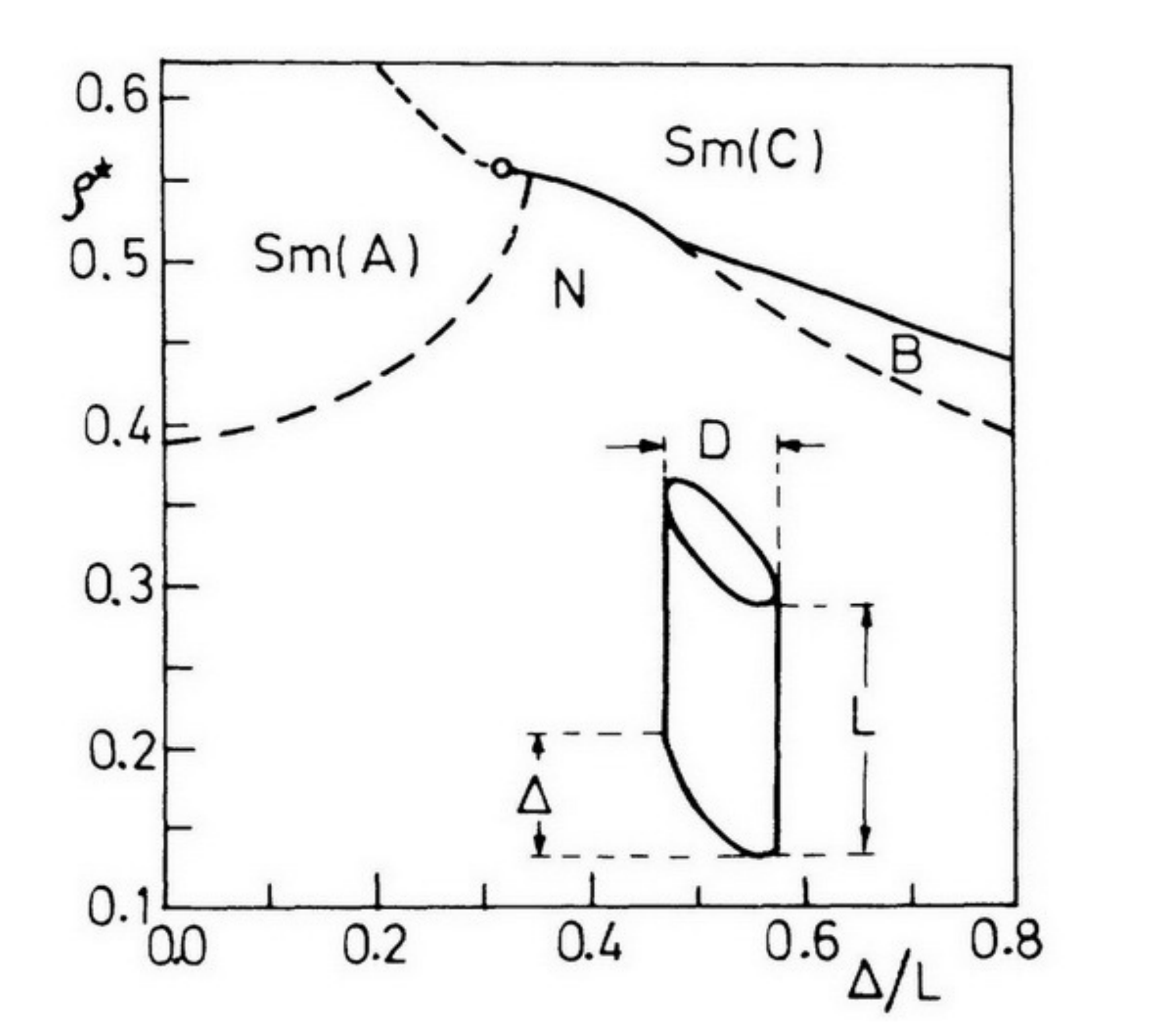}
\caption{Phase diagram of a system of parallel oblique hard cylinders as obtained
from the weighted-density theory of Somoza and Tarazona \cite{ST}. Densities in the vertical axis
are in units of the complete packing density. Reprinted with permission from \cite{ST}. 
Copyright (1988) by the American Physical Society.}
\label{Somoza_fig}
\end{figure}

A weighted-density theory was
employed by Somoza and Tarazona \cite{ST,ST1} to investigate
the phase diagram of a system of hard particles with a different biaxial
shape, namely oblique cylinders. The phase diagram of the system was
computed, for a system of parallel particles, as a function of the
molecular asymmetry as measured by the parameter $\Delta$ (see Fig. \ref{Somoza_fig}). 
The transition densities, $\rho^{*}$, shown in the vertical axis of the
figure are expressed in units of the close packing density 
$\rho_{cp}=2/\left(3^{1/2}D^{2}L\right)$. Apart from a
continuous nematic-smectic A phase transition for small values of
$\Delta/L$, i.e., for particles with small biaxiality, and a first-order
nematic-smectic C at intermediate molecular biaxialities, a stable
biaxial nematic (region $B$ in the figure) is obtained for large
values of $\Delta/L$. The predicted existence of a biaxial nematic
phase in a system of asymmetric hard particles is in agreement with
the results from previous theoretical models but, as pointed out by
the authors in their conclusions, in this particular model the biaxial
nematic phase could be an artifact due to the imposed perfect alignment
of the cylinders' axes which may favour the biaxial order.

\begin{figure}
\includegraphics[width=13cm]{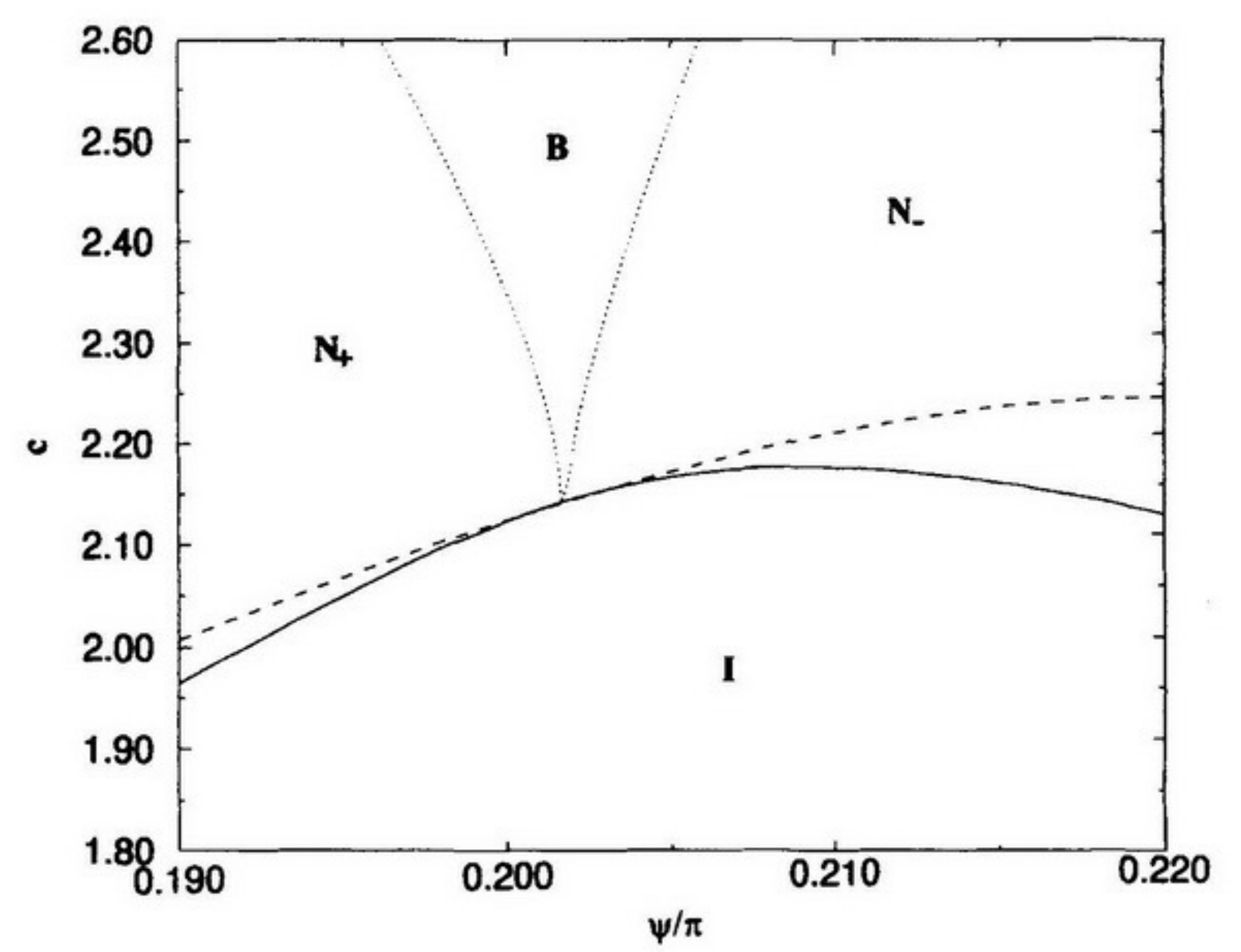}
\caption{Phase diagram of a system of hard-boomerang particles as obtained
from the Onsager theory by Teixeira et al. \cite{banana_shaped}.
Reprinted by permission of the publisher (Taylor \& Francis Ltd., 
http://www.tandf.co.uk/journals).}
\label{Teix_fig}
\end{figure}

A hard-boomerang fluid, formed by hard particles each one made of
two hard spherocylinders joined at their ends at an angle $\psi$,
has been studied by Teixeira et al. \cite{banana_shaped}. In this way
a biaxial particle is constructed, the value of $\psi$ being a measure
of its biaxility. The Onsager approximation was then
used to obtain the free energy of the system and bifurcation analysis
was employed in order to study the relative stability of the isotropic
and nematic phases as a function of the molecular biaxiality. The
phase diagram of the system, in the vicinity of the Landau point where
isotropic, uniaxial nematics and biaxial nematic meet, is shown in
Fig. \ref{Teix_fig}. In that figure $c=\left(\pi/4\right)DL^{2}\rho$ is the reduced
density of the system while in the horizontal axis molecular biaxiality
increases as we move to the right toward lager values of $\psi$.
The solid line is the locus of the lowest densities at which the model
has uniaxial nematic solutions. The dotted lines are the loci of second-order
uniaxial- biaxial transitions, and the dashed line is the limit of
stability of the isotropic phase with respect to nematic fluctuations.
Therefore, at both sides of the Landau point the model predicts a
first order isotropic-unixial nematic transition, although coexistence
was not calculated. At the Landau point the isotropic phase continuously
transforms into the biaxial nematic upon increasing the density of
the system. The phase diagram qualitatively resembles closely that
obtained by Taylor and Herzfeld for hard spheroplatelets (left side
of Fig. \ref{Taylor_fig}) and, as in that case, a uniaxial $N_{+}$ phase is obtained
in the limit of rod-like particles and a uniaxial nematic $N_{-}$
is stable in the opposite limit of plate-like particles. A simulation study of a bent-core
molecule made from HSC sectors was performed by Lansac et al. \cite{Lansac}. No biaxial nematic phase was found in
this study, although other interesting features of the model were discovered, in particular the formation of
a new class of smectic phase characterized by the spontaneous formation of macroscopic
chiral domains (even though the particles are achiral).

A closed model for biaxial particle, composed of hard spherocylindrical sectors attached by their
ends at an angle and forming a zigzag molecule, was studied by Maiti et al. \cite{Clark} by means of
Monte Carlo simulations. This study, already discussed at the end of Sec. \ref{further}, revealed the
presence of isotropic, nematic, smectic A, smectic C, columnar and two crystal phases. Of interest here
is the character of the smectic and crystal phases, which are biaxial, but not the nematic phase.
Simulations have shown that a biaxial nematic phase made of hard biaxial ellipsoids require a high degree of
biaxiality \cite{Allen_biax,Camp}. Therefore, it is possible that the zigzag model may have a stable
biaxial nematic phase for some specific molecular configuration.

\begin{figure}
\includegraphics[width=15cm]{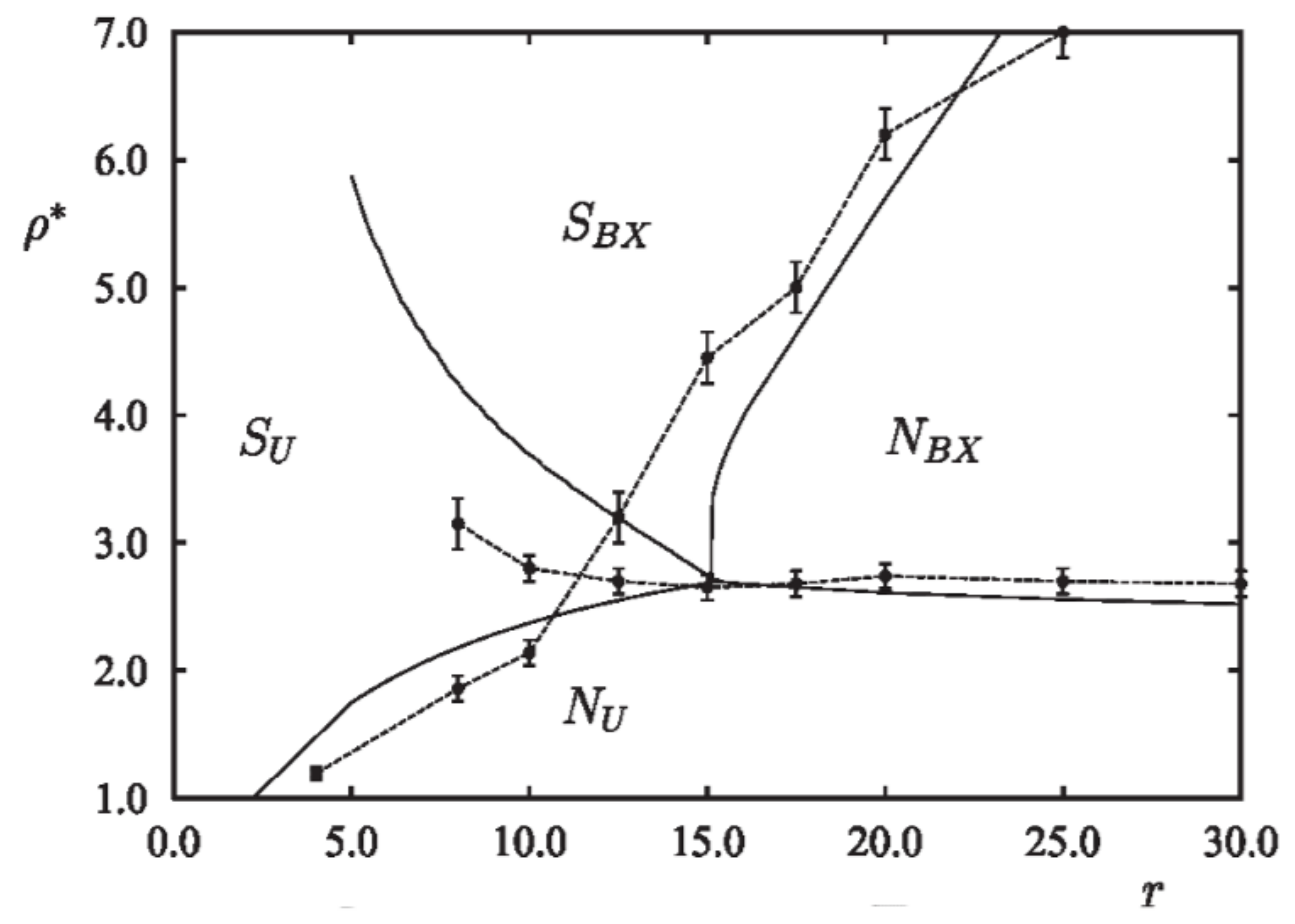}
\caption{Phase diagram of the one-component fluid of hard biaxial board-like particles studied by
Vanakaras et al. \cite{Photinos-1} in the packing fraction versus particle aspect ratio plane.
See text for a discussion. Reproduced from Ref. \cite{Photinos-1} with permission of the PCCP Owner Societies.}
\label{vana_fig}
\end{figure}

Vanakaras et al. \cite{Photinos-1} used Monte Carlo simulations and Onsager theory of
biaxial board-like particles with frozen long axis and a freely-rotating secondary axis.
They calculated the phase diagram of the one-component fluid using simulation,
and compared the results with Onsager theory. The phase diagram, in the plane packing fraction
versus aspect ratio of the rectangular cross section perpendicular to the frozen axis (see
Fig. \ref{vana_fig})
exhibited both uniaxial and biaxial nematic and smectic phases with phase boundaries all meeting 
at a single point. As discussed in Sec. \ref{binary_mixtures}, they also studied a binary
mixture, which was found to stabilise the biaxial nematic phase.

\begin{figure}
\includegraphics[width=8cm]{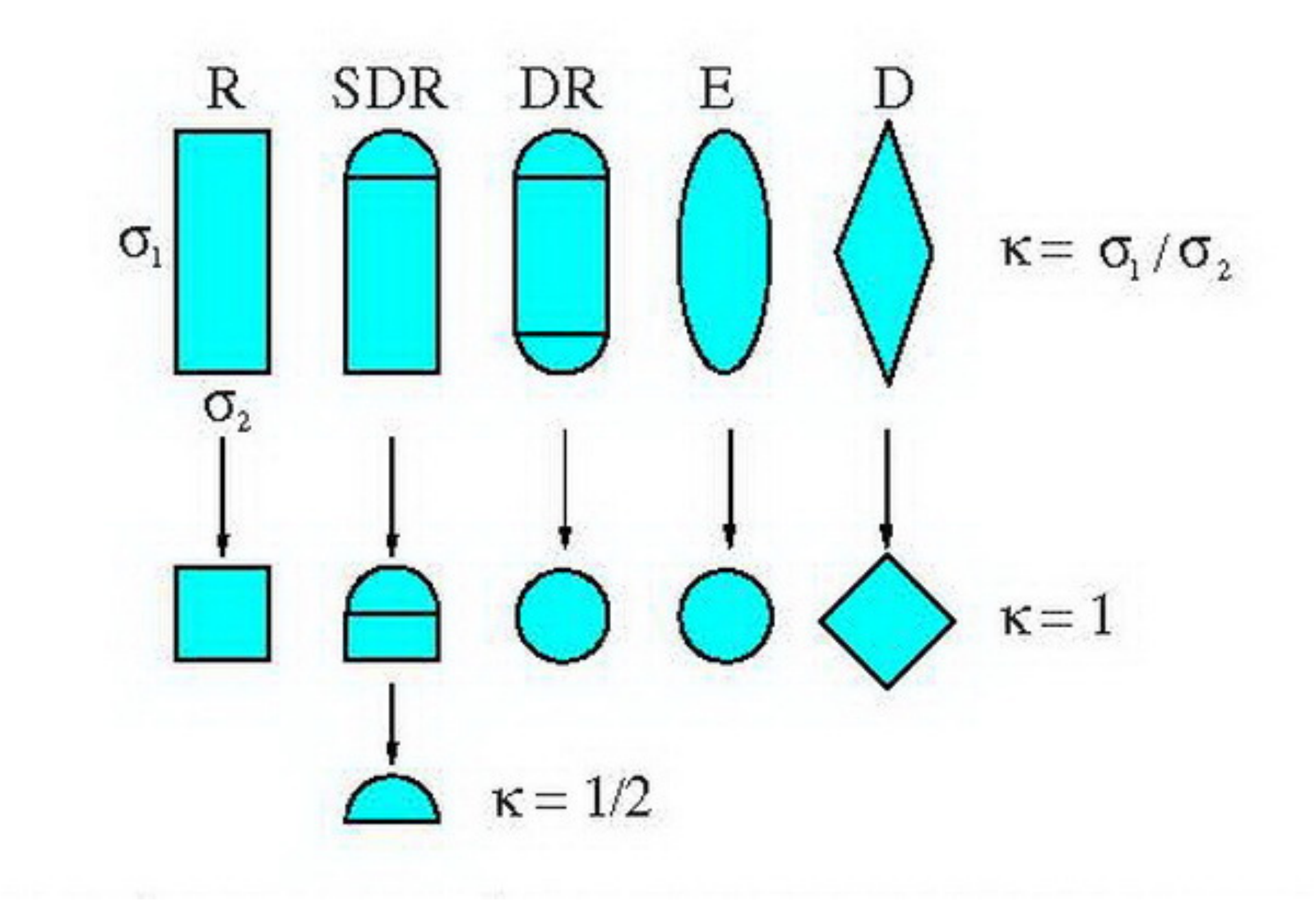}
\includegraphics[width=8cm]{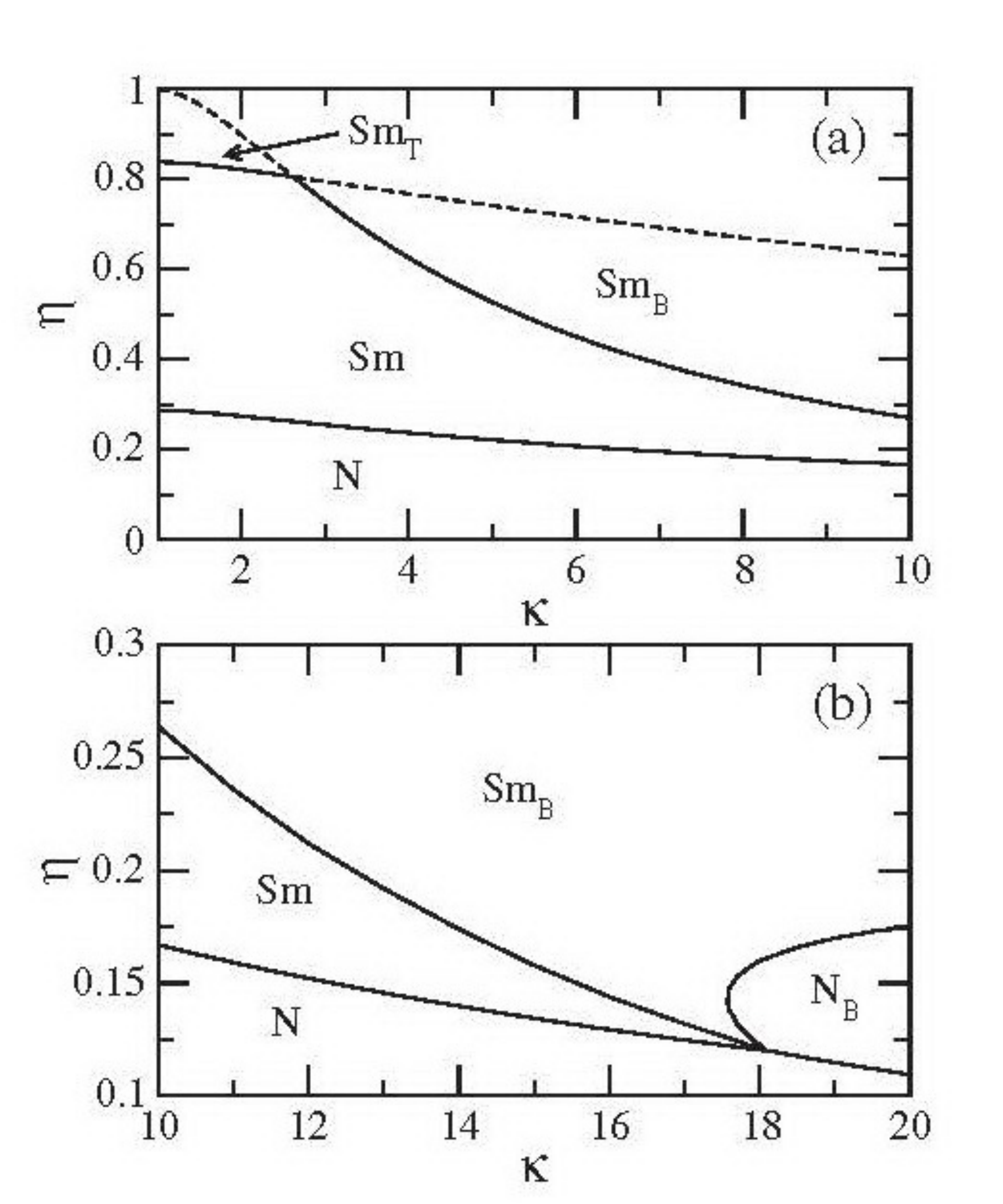}
\caption{Left: Cross section of biaxial particles studied in Ref. \cite{Yuri-3}.
Right: Phase diagram of a system of particles with R cross section.
See text for a discussion. Reprinted with permission from \cite{Yuri-3}. 
Copyright (2008) by the American Physical Society.}
\label{Yuri_fig}
\end{figure}

A more extensive theoretical study of the phase diagram of system made of five different
biaxial particles, differing in their cross sections, was done by
Mart\'{\i}nez-Rat\'on et al. \cite{Yuri-3}. The orientation of the principal
molecular axis was kept fixed along the $z$ axis while the secondary
axis can rotate freely. In this way, the system is able to show a
uniaxial nematic phase with its director along the $z$ axis, a biaxial
nematic, with its secondary director in the $xy$ plane, when long-range
orientational order of the short molecular axes is present, and a
tetratic nematic phase reminiscent of the corresponding tetratic nematic
phase observed in two-dimensional fluids of hard rectangles. Uniaxial,
biaxial, and tetratic smectic phases are also possible within this
restricted-orientations model, their orientational properties being
those of their nematic counterparts but showing long-range one-dimensional
translational order in such a way that particles arrange in layers
perpendicular to the $z$ direction. The theoretical approximation
used was a fundamental-measure density functional theory applied to
parallel hard bodies, as reviewed in Sec. \ref{FMT}. The model
consists of hard biaxial particles with characteristic length $L$
along the $z$ axis and $\sigma_{1}$ and $\sigma_{2}$ in $xy$ plane.
Five different transverse sections (see Fig. \ref{Yuri_fig}) were considered:
rectangle (R), semidiscorectangle (SDR, consisting of a rectangle
capped with only one semidisc in one of its ends), discorectangle
(DR, obtained from the previous one by adding another semidisc at
the other end), ellipse (E), and deltoid (D, composed of an isosceles
triangle and its reflection through its common base). Phase diagrams
of one-component systems made of these particles where obtained as
a function of particle biaxiality as measured by their cross section
aspect ratio $\kappa=\sigma_{1}/\sigma_{2}$ ($\sigma_{1}$ being
the largest size) 
and the system packing fraction $\eta$. The phase diagram of
a system of particles with R cross sections has been reproduced in
Fig. \ref{Yuri_fig} (a complete report of results is available in the original
paper \cite{Yuri-3}), cases (a) and (b) corresponding to two different
numerical minimizations of the free energy (a Gaussian parameterisation
in (a) and a Fourier expansion in (b) for the density profile). 
Continuous curves correspond to second-order
phase transitions while dashed lines show the continuation of the
Sm$-$Sm$_{\rm B}$ and Sm$-$Sm$_{\rm T}$ spinodals. For $\kappa=18.101$ four
second-order transition lines meet in a single point. The existence
of this four-phase point at high enough values of $\kappa$ is a common
feature to all the systems of particles with different cross sections
that have been studied. For aspect ratios of the cross section below
the four-point, the system undergoes second-order N$-$Sm and Sm$-$Sm$_{\rm B}$
phase transitions upon increasing the packing fraction of the system.
Beyond the four-phase point, for larger values of $\kappa$, the sequence
of second-order phase transitions is N$-$N$_{\rm B}$ and N$_{B}-$S$_{\rm B}$
when packing fraction increases. In the case of the R cross section
shown in Fig. \ref{Yuri_fig}, there is a region of $\kappa$ just below the four-phase
point where a re-entrant biaxial nematic is observed: the sequence
of transitions in that region is N$-$Sm, Sm$-$Sm$_{\rm B}$, Sm$_{\rm B}-$N$_{\rm B}$,
and N$_{\rm B}-$Sm$_{\rm B}$ as $\eta$ increases. The existence of this re-entrant
behaviour of the biaxial nematic phase (also seen for some other cross
sections studied, but not for all of them) is a genuine prediction
of this fundamental-measure approximation since Onsager theory predicts
a monotonic phase boundary between N$_{\rm B}$ and Sm$_{\rm B}$ phases \cite{Photinos-1}.

An additional conclusion from this work is that the location of the
four-phase point is highly sensitive to the shape of the particles
cross section. As concluded by the authors, this fact suggests that
the optimization of the particle geometry could be a useful criterion
in the design of colloidal particles that can exhibit an increased
stability of the biaxial nematic phase with respect to other competing
phases with spatial order. Other relevant difference between different
cross sections is the phase behaviour in the small $\kappa$ and high
$\eta$ region where, as shown in Fig. \ref{Yuri_fig} for the case of R particles,
the Sm$_{\rm T}$ phase could be stable. Not surprisingly, the region
of stability of the tetratic smectic phase is found to strongly depend
on the particles cross section, the reason being that the excluded
volume involved in a T-configuration (two particles in a perpendicular
configuration, needed in a stable tetratic phase) crucially depends
on the particle cross section. In any case, the tetratic smectic phase
is pre-empted by the crystalline phase, as the Monte Carlo simulations
seem to show \cite{Photinos-1}.

\begin{figure}
\includegraphics[width=16cm]{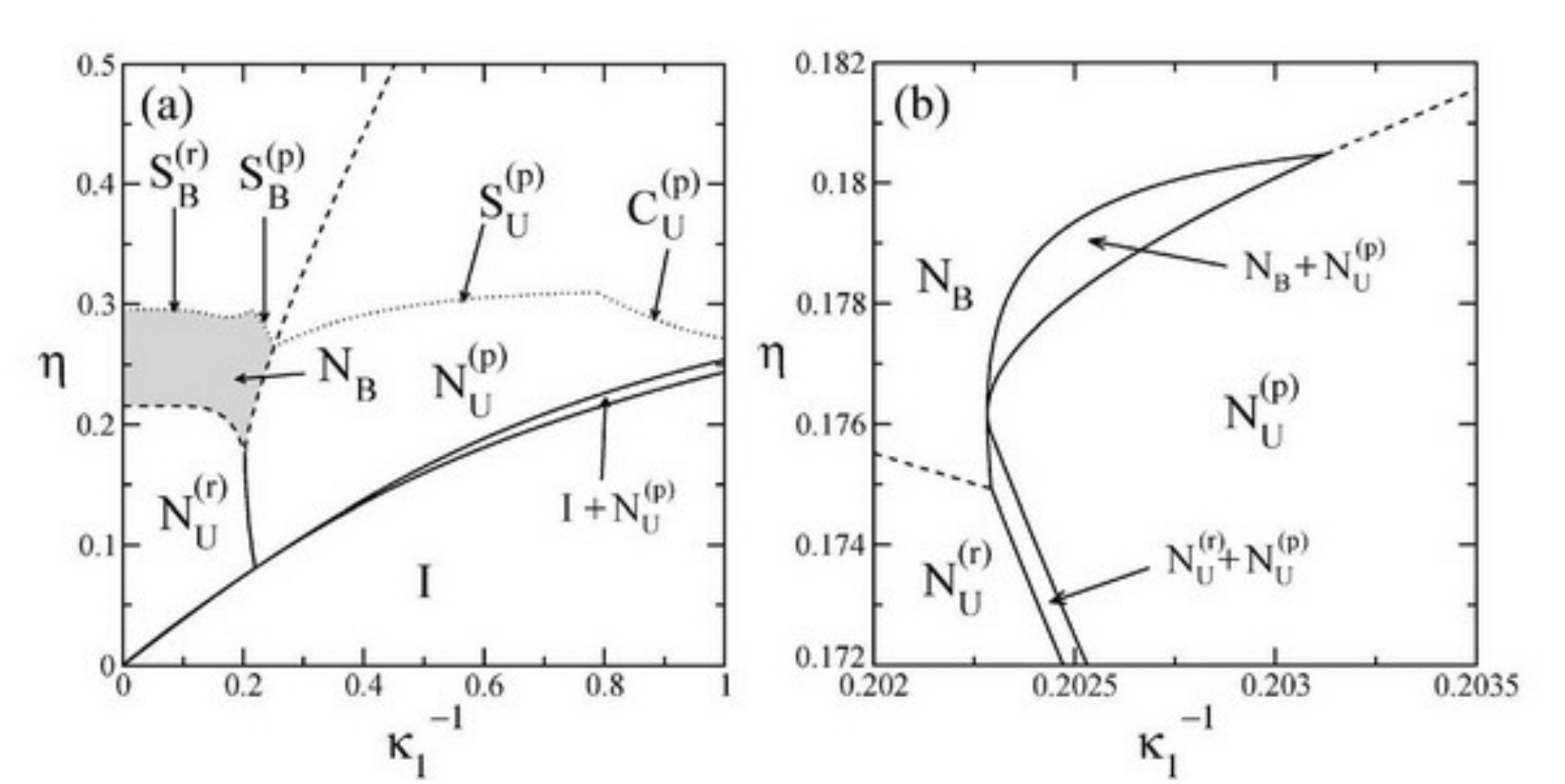}
\caption{(a) Phase diagram of a system of hard board-like particles with $\kappa_{2}=5$
as obtained by Mart\'{\i}nez-Rat\'on et al. \cite{Yuri-4}.
(b) A zoom around $\kappa_{1}^{-1}\sim0.2$. Labels indicate regions
of stability of the different phases, or the type of ordering with
respect to which bifurcation curves are computed. Solid and dashed
curves indicate first and second order phase transition, respectively.
Dotted curves are bifurcation lines to the corresponding non-uniform
phase from the stable bulk uniform phases. The shaded region in (a)
corresponds to the estimated maximum region of stability of the biaxial
nematic phase. Reproduced from Ref. \cite{Yuri-4} with permission of the PCCP Owner Societies.}
\label{Yuribis_fig}
\end{figure}

Inspired by the experimental finding by van den Pol et al. \cite{van den Pol-1}
of a biaxial nematic phase in a system of goethite particles (see
some details above), Mart\'{\i}nez-Rat\'on et al. \cite{Yuri-4} considered
recently a system of board-like particles of sizes $\sigma_{1}>\sigma_{2}>\sigma_{3}$.
As it has been pointed out in the preceding paragraphs, there exists
a general consensus among different approaches, and also computer
simulations, in that biaxial nematic phase in systems of board-like
particles (including spheroplatelets among them) could be stable
for particles with $\sigma_{1}/\sigma_{2}\approx\sigma_{2}/\sigma_{3}$
in such a way that they are not too rod-like or plate-like. In
fact, this is also the case in the van den Pol et al. experimental
study: They found a stable biaxial nematic, as compared to the
competing smectic and columnar phases, for board-like particles with
$\sigma_{1}/\sigma_{2}\approx\sigma_{2}/\sigma_{3}=3$. Mart\'{\i}nez-Rat\'on
et al. used a fundamental-measure density functional approximation,
solved in the restricted-orientation approximation (the so-called
Zwanzig approximation) in which only a discrete set of possible particles
orientations are considered. The stability of the biaxial nematic
with respect to smectic and columnar phases is studied as a function
of the particle shape as measured by the two aspect ratios $\kappa_{1}=\sigma_{1}/\sigma_{2}$
and $\kappa_{2}=\sigma_{2}/\sigma_{3}$. A systematic study of the
phase diagram of the system has been conducted by varying $\kappa_{1}$
while $\kappa_{2}$ is kept fixed. They have found phase diagrams
which include all the uniform phases: isotropic, uniaxial rod- and
plate-like nematics, and biaxial nematic (see Fig. \ref{Yuribis_fig}). Bifurcation analysis
of the uniform phases with respect to fluctuations of the smectic,
columnar and plastic-solid types were also included in this study. 
According to their results for different values of $\kappa_{2}$ (see
the original paper for other phase diagrams apart from that reproduced
here), the biaxial nematic phase begins to be stable for $\kappa_{2}\gtrsim2.5$,
in agreement with the experimental results of van den Pol et al. \cite{van den Pol-1}.
Another feature of the phase diagram obtained from this theory, in
agreement also with previous theories and simulations on biaxial hard
particles, is the existence of a region of biaxiality centred at $\kappa_{1}\sim\kappa_{2}$
which widens as $\kappa_{2}$ increases. The study of Mart\'{\i}nez-Rat\'on
et al. was the first to apply fundamental-measure theory to biaxial
particles, going in this way beyond the second-order virial approximation.
That allowed them to obtain a genuine result, not accounted for by
previous studies based on second-order theories: the prediction that
the phase diagram must be asymmetric in the neighbourhood of $\kappa_{1}\sim\kappa_{2}$.
Very recent computer simulations by Peroukidis and Vanakaras \cite{Peroukidis-1}
for systems of board-like particles (namely, spheroplatelets) have
confirmed the main conclusions of the work of Mart\'{\i}nez-Rat\'on et al.

We end this section by noting that a big computer-simulation effort has been done by different
research groups to shed some light on the problem of biaxial
liquid-crystal phases. Some of these studies use hard-body models \cite{Allen_biax,Camp}.
An excellent and up-to-date review
has been recently published in this journal by Berardi et al. \cite{Berardi-1}.
We refer the reader to this work for a more complete view on
the current state of the art in the field of biaxial phases in hard-body models.

\subsection{Elasticity}

Elasticity is a genuine property of liquid crystals and a crucial factor determining the response of the material to
external fields. A recent, general overview of the attempts to investigate liquid-crystal elasticity from the theoretical front has been
written by Ferrarini \cite{Ferrarini}. Here we discuss the use of hard-body models to obtain qualitative and quantitative information
on elastic constants of nematic and smectic materials.

Since the nematic phase has a broken rotational symmetry with respect to the isotropic phase, it presents elasticity. In practical
terms this means that a spatial distortion of the local director field, $\hat{\bm n}=\hat{\bm n}({\bm r})$ involves a cost in free energy.
At the macroscopic level one expects this free energy cost to be related to the square gradient of the director field components,
$\partial_{\alpha}\hat{\bm n}_{\beta}$, where $\alpha,\beta=1,2,3$ denote spatial coordinates. In an apolar
nematic, i.e. one that does not distinguish between $\hat{\bm n}$ and $-\hat{\bm n}$ (which is realised in practise when
particles have head--tail symmetry), a general square--gradient expansion reduces, on applying the relevant symmetries,
to the terms $\left(\nabla\cdot\hat{\bm n}\right)^2$, $\left[\left(\nabla\times\hat{\bm n}\right)\cdot\hat{\bm n}\right]^2$
and $\left[\left(\nabla\times\hat{\bm n}\right)\times\hat{\bm n}\right]^2$,
meaning that there are three independent elastic distortion modes, splay, bend and twist, respectively, each associated to a 
particular elastic constant. The general expression for the elastic free energy, due to Frank \cite{Frank}, is
\begin{eqnarray}
F_e=\frac{1}{2}\int_V d^3r \left\{K_1\left(\nabla\cdot\hat{\bm n}\right)^2+K_2
\left[\left(\nabla\times\hat{\bm n}\right)\cdot\hat{\bm n}\right]^2+K_3
\left[\left(\nabla\times\hat{\bm n}\right)\times\hat{\bm n}\right]^2\right\}
\label{Frank}
\end{eqnarray}
This is a purely macroscopic expression, giving the free-energy that has to be added to the non-distorted nematic to obtain
the complete free energy of the distorted nematic. It does not include any relaxation of the order parameter $Q$ involved in 
the distortion, which will certainly exist. The study of this effects requires a microscopic theory.

The first microscopic theoretical calculations of elastic constants for model nematics were due to Priest \cite{Priest},
who used an Onsager theory for hard rods that incorporate a rotated director field in the long-wavelength limit and an expansion in rotational
invariants. Despite using hard rods as a model particle and Onsager theory (meant to reproduce hard needles), Priest obtained good agreement 
with experiment. The order-of-magnitude agreement with experimental values
is remarkable, considering that the Onsager model strictly applies to very long rods. 
Shortly after this work, Straley \cite{Straley} derived a more general 
expression for the elastic free energy directly from the Onsager density functional for hard rods, assuming a
functional form for the distribution function but without any expansion. 
He obtained numerical values for the three constants close to 
the I--N phase transition, in agreement with the results of Priest. 

In seminal work, Poniewierski and Stecki derived general expressions for the elastic constants of a nematic in terms of the
direct correlation function \cite{Stecki,Stecki1}, $c^{(2)}({\bm r},\hat{\bm\Omega},\hat{\bm\Omega}')$ of the
undistorted nematic. For splay, bend and twist distortions the elastic constants are given by:
\begin{eqnarray}
K_1&=&\frac{1}{2}\int d{\bm r} x^2\int d\hat{\bm\Omega}\int d\hat{\bm\Omega}' 
\rho'(\theta) c^{(2)}({\bm r},\hat{\bm\Omega},\hat{\bm\Omega}') 
\rho'(\theta') \hat{\Omega}_x\hat{\Omega}_x',\nonumber\\\nonumber\\
K_2&=&\frac{1}{2}\int d{\bm r} x^2\int d\hat{\bm\Omega}\int d\hat{\bm\Omega}' 
\rho'(\theta) c^{(2)}({\bm r},\hat{\bm\Omega},\hat{\bm\Omega}') 
\rho'(\theta') \hat{\Omega}_y\hat{\Omega}_y',\nonumber\\\nonumber\\
K_3&=&\frac{1}{2}\int d{\bm r} z^2\int d\hat{\bm\Omega}\int d\hat{\bm\Omega}' 
\rho'(\theta) c^{(2)}({\bm r},\hat{\bm\Omega},\hat{\bm\Omega}') 
\rho'(\theta') \hat{\Omega}_x\hat{\Omega}_x'.
\label{Ks}
\end{eqnarray}
Here $\rho({\bm r},\hat{\bm\Omega})=\rho(\theta)$ is the local density distribution of the undistorted nematic, with 
$\cos{\theta}=\hat{\bm n}\cdot\hat{\bm\Omega}$. Numerical values for the elastic constants were obtained for hard rods, using
a direct correlation function from Onsager theory (i.e. the Mayer function, which is exact in the low-density limit).
Equations (\ref{Ks}) have later been rederived in more direct ways 
\cite{Singh_rederiv,Lipkin,ST_elas}. Also, the contribution from the relaxation of the distribution function was later evaluated \cite{ST_elas}, 
and the relative contributions from repulsive and attractive interactions have been assessed \cite{Gel,Singh_uniaxial}.
The derivation of general expressions for $K_i$ in terms of microscopic correlations was an important step, 
since it allowed the calculation of elastic constants from microscopic grounds: given a theory for $c^{(2)}$ for some particular molecular 
theory (e.g. a density-functional theory, from which the direct correlation function can be obtained by functional differentiation), 
one can evaluate the constants $K_i$, which in turn can be plugged in the macroscopic Frank expression to give
predictions about the behaviour of realistic models of distorted liquid crystals, something which is very important in 
technological applications. 

\begin{figure}
\epsfig{file=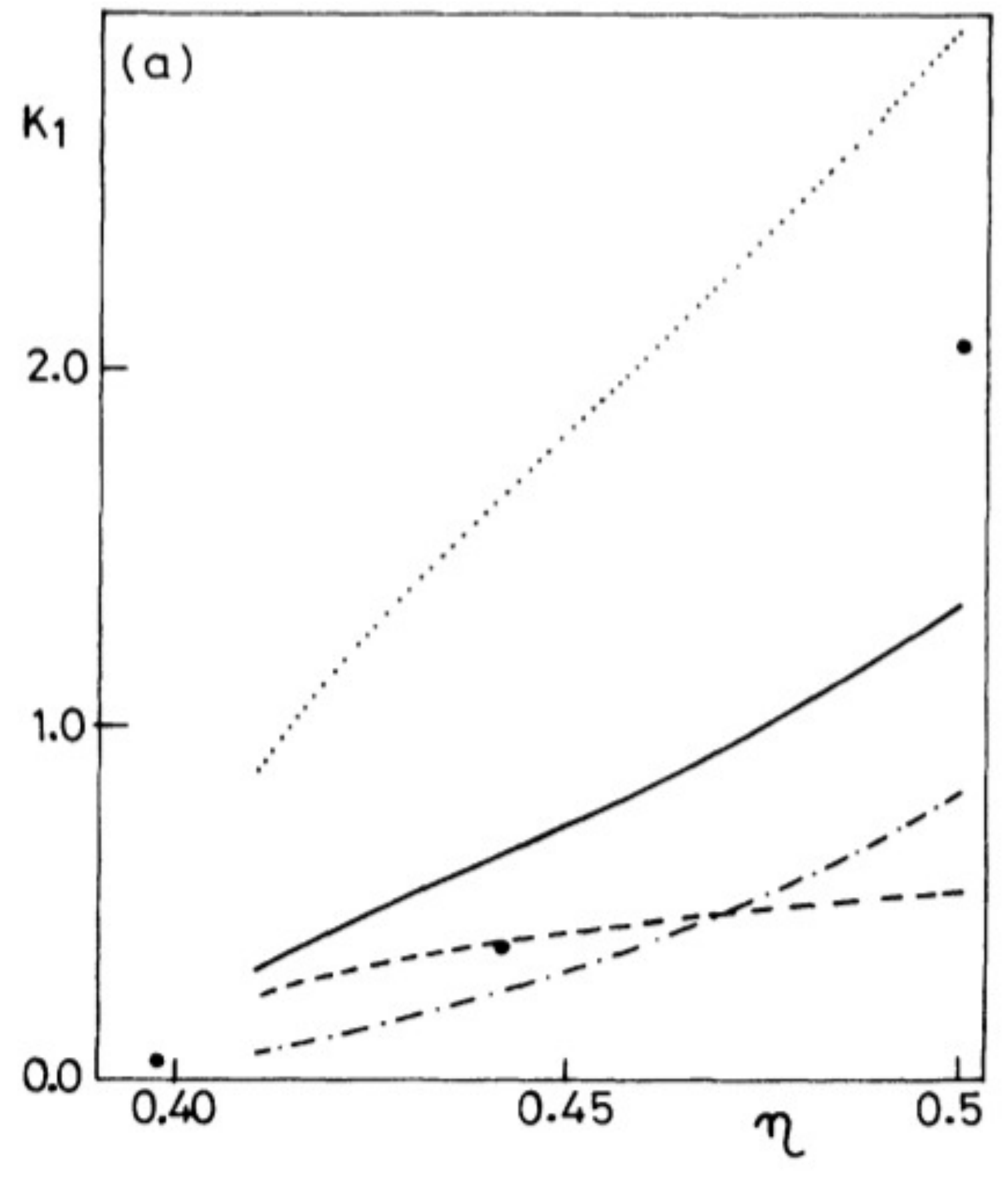,width=2.0in}
\epsfig{file=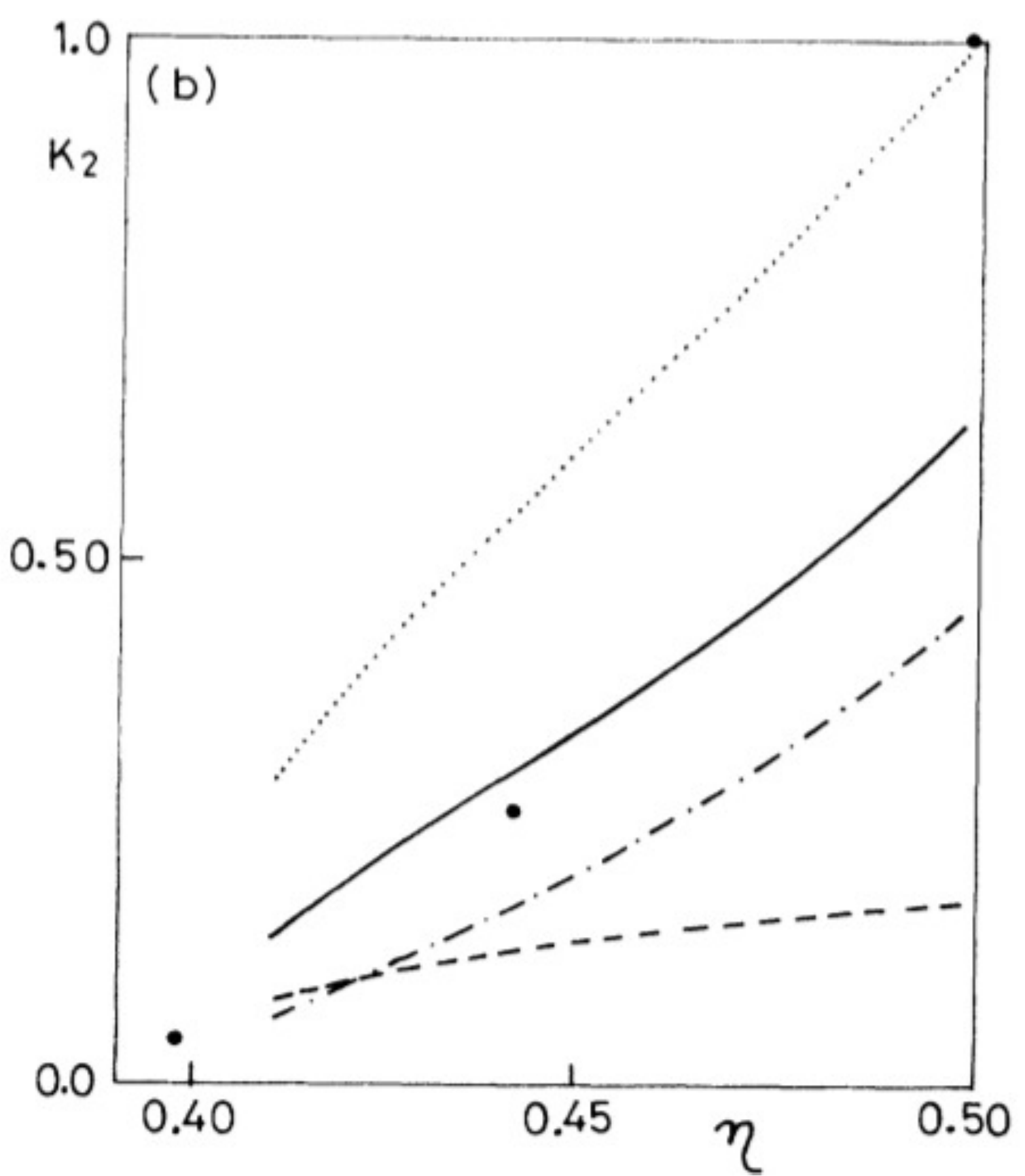,width=2.06in}
\epsfig{file=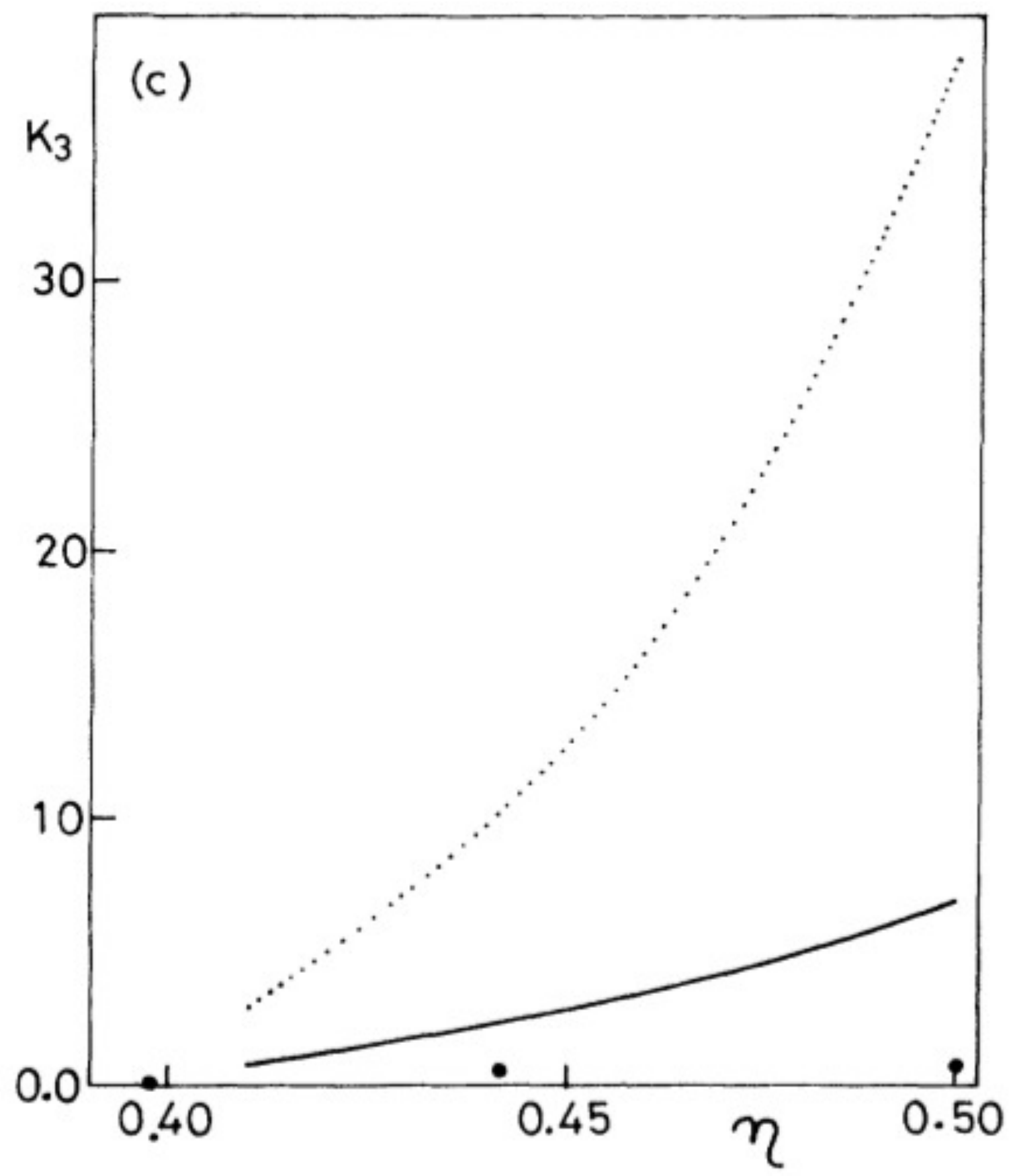,width=2.0in}
\caption{Results for the nematic elastic constants from the Somoza-Tarazona theory \cite{ST_elas_0} (continuous line) and
the Sin-Doo Lee theory \cite{SinDooLee_elas} (dotted lines). Simulation results from Ref. \cite{Allen_Ks_simul1} are given by circles.
Reprinted with permission from \cite{ST_elas_0}. 
Copyright (1989) by the American Physical Society.}
\label{elas1}
\end{figure}

Singh and Singh \cite{Singh_elas_ellip} calculated the elastic constants of hard ellipsoids, using series expansions for the direct correlation
function. Bad convergence properties with respect to the number of terms in the expansion
were found. Lee and Meyer \cite{Lee-Meyer} recalculated the elastic constants for 
hard rods using Straley's method and Onsager theory (hard needles), and found general agreement with Poniewierski and Stecki. 
They used a more accurate orientational 
distribution function than previous authors. Also, they noted that reasonable
predictions were obtained when comparing with real data for virus suspensions, but results were not so reasonable for 
solutions of more flexible colloidal particles. Shortly after, Lee \cite{SinDooLee_elas} and Somoza and Tarazona \cite{ST_elas_0} made calculations 
using different choices for the direct correlation function. Lee used PL theory to calculate the uniform nematic, 
and obtained the direct correlation function not from the PL density functional, but from an independent prescription. Also, 
he neglected terms of order $D/L$ and higher in the excluded-volume expansion, which produced the equality $K_1=3K_2$, 
a result not supported by simulations. In contrast, Somoza and Tarazona obtained the direct correlation function in a consistent way, i.e. 
from the second functional derivative of the free-energy density functional, and the same theory was used to obtain the reference 
nematic state. Including contributions to all orders, they obtained much better agreement with simulations (Fig. \ref{elas1}).

Poniewierski and Ho\l{}yst \cite{Ponie_more} used their proposed version of DFT for hard spherocylinders to derive the 
necessary direct correlation function for the nematic fluid and to calculate the elastic constants, also in a consistent
way. Their results were similar to those of Somoza and Tarazona and compared equally well with simulations.
Finally, Somoza and Tarazona incorporated the relaxation of the nematic order parameter in the distorted nematic 
\cite{ST_elas}, and found that this effect reduces the values of the elastic constants.

In the simulation front, 
Allen et al. used expressions for the elastic constants obtained from fluctuations of the order tensor \cite{Forster} to compute 
values from simulations of the hard ellipsoid and hard spherocylinder models 
\cite{Allen_Ks_simul1} (note the factor $9/4$ missing in the results, corrected in Ref. \cite{Allen_Ks_simul2}), finding moderate agreement
with the then existing theories \cite{Singh_elas_ellip}. Tjipto-Margo et al. \cite{Allen_Ks_simul3} made a detailed analysis of the comparison 
between simulation results and the
early theories for the direct correlation function; although order-of-magnitude agreement was found, discrepancies 
with simulation existed, which were attributed to the poor description of the function at long distances. 

Later, Singh et al. \cite{Singh_elas0} presented a more systematic study of their expansion method to calculate numerical values for the elastic constants, and
included attractive contributions to the interactions via perturbation, concluding that the repulsive contribution of the pair interaction 
is dominant. These authors also used the same method to calculate values for the twelve elastic constants of a biaxial nematic fluid \cite{Singh_elas1}. The
relative values of the constants and their behaviour as the biaxial phase goes to the uniaxial phase were discussed. An interesting result is that
the effect of biaxiality in the orientational ordering on the elastic constants is small.

The elastic behaviour of smectic phases has also been studied, although not so thoroughly as for nematic phases. 
In a smectic-A phase distortions of the director field must keep the integrity of the smectic layers intact, which means that
bend and twist distortions are not allowed. Due to the broken translational symmetry of the smectic-A along
the director, there exists an elastic modulus $B$, associated to layer compressibility, adding a term 
$\displaystyle\frac{1}{2}B\int_V d^3r\left(\partial u/\partial z\right)^2$ to the elastic free energy. $u$ is the local displacement field of 
the layers (similar to the displacement field in 3D solids). Two derivations of expressions for the smectic elastic
constants have appeared in the literature. Lipkin et al. \cite{Lipkin} 
derived expressions for the elastic constants of a smectic phase as a 
generalisation of the corresponding espressions for the nematic, and including the layer compressibility modulus $B$. Numerical values
for the constants were not estimated. More recently,
Singh et al. \cite{Singh_elas} derived expressions for the constants in terms of order parameters, and estimated their magnitude.
The formalism included tilted smectic phases.
However, in neither of these studies has a serious evaluation, related to a particle model, been made. 

In this section we have mentioned works on elastic constants that make use of hard-body models to represent particle
interactions. Many recent studies have been devoted to the effect of attractive interactions on the elastic properties of
nematic liquids, mainly through perturbative treatments. In line with the spirit of the present review, these studies,
more appropriate to establish contact with real materials, 
are not mentioned here, and we refer the reader to the review works referenced in the introduction.


\subsection{Multicomponent fluids}

\subsubsection{Binary mixtures}
\label{binary_mixtures}

When hard-body particles of different shapes are mixed together in
different concentrations, new phenomenology arises in the macroscopic
phase behaviour of the system. Again, since particles are hard bodies,
the new phenomena are due entirely to entropy. In the last 30 years
or so there has been considerable interest in the study of entropy-driven
transitions in mixtures.

Probably one of the first studies was due to Asakura and Oosawa \cite{Asakura},
who showed that the addition of a small amount of non-adsorbing polymer
to a dispersion of (hard) colloidal particles leads to an attractive
interaction between these particles. The origin of this attraction
is, of course, entropic: polymer molecules maximize their free volume
when the large colloidal particles are close to each other, the so-called
\emph{depletion} effect. The mixture can be theoretically modelled
in a simplified manner by getting rid of the polymer and considering
this attraction as an effective depletion potential between the colloidal
particles, and one can then use all the theoretical tools available for
simple fluids.

This route was followed by some authors in the last quarter of the
last century to analyse the phase behaviour of different binary mixtures.
For example, in 1999 Lekkerkerker and co-workers used an extension
of the free-volume approximation to investigate the depletion-induced
phase separation in mixed suspensions of colloidal spheres of diameter
$\sigma$ and colloidal rods of length $L$ and diameter $D$, in
the range $L\le\sigma$ \cite{Lekkerker-1}. Later, the same group
used the depletion approximation to conclude that the addition of a small
amount of nonadsorbing polymer significantly modifies the I--N phase
boundaries of a system of sterically stabilized colloidal platelets
\cite{Lekkerker-2}.

Mixtures of colloidal particles with nonadsorbing polymers, and also
mixtures of colloidal particles of different shapes, have been extensively
studied not only from the theoretical point of view, but also experimentally,
as shown, for example, in \cite{Poon-1,Fraden-1} and references therein. 
Experimental studies have probably triggered the interest of many
theoreticians in modelling mixtures in an effort to predict their macroscopic
behaviour. Models based on the depletion approximation, like those
mentioned above, have been applied to other mixtures of colloidal
particles, but those models will not be treated in this review since
they reduce the problem, at the end, to a simple system whose particles
interact via an effective potential that is not a hard-body interaction,
a subject not covered in this review.

\begin{figure}
\includegraphics[width=16cm]{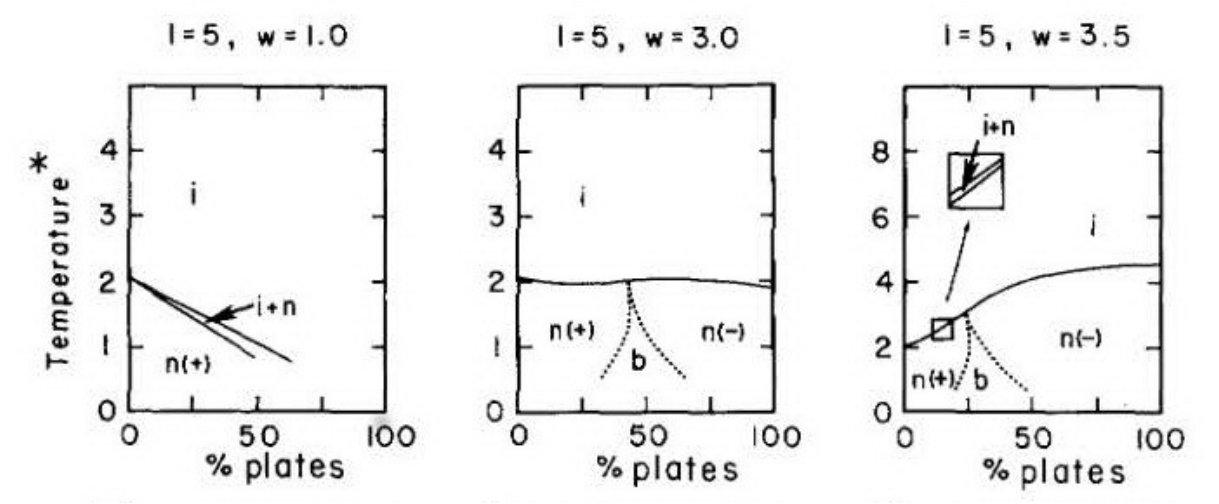} \caption{Phase diagrams of a mixture of rods and plates as obtained by Alben
\cite{Alben-1} using his mean-field lattice-model. The effective temperature shown
in the vertical axis is proportional to the temperature
divided by the pressure, as it corresponds to a system with hard-body
interactions where the internal energy is, therefore, identically
zero. Labels indicate isotropic (i), uniaxial nematics (n(+)
and n(-)), and biaxial nematic (b). Continuous lines correspond to
first-order phase transitions while dotted lines are second-order
phase transitions. Reprinted with permission from \cite{Alben-1}. Copyright (1973), AIP Publishing LLC.}
\label{Alben}
\end{figure}

One of the earliest approaches to describe theoretically binary
mixtures of anisotropic particles with different shapes was presented
in 1973 by Alben \cite{Alben-1}. He made mean-field lattice-model
calculations of a mixture of rod- and plate-like molecules. The
model is a modification of that originally proposed by Flory to describe
steric interaction in polymers. Space is represented by a cubic lattice
of $M$ cells. Dispersed on this lattice there are a total of $N$
molecules, a fraction $1-c$ of which are rods, each occupying $l$
consecutive cells along one of the lattice axes. A fraction $c$ of
the $N$ particles are square plates, each occupying $w\times w$
coplanar cells in a plane defined by two of the lattice axes. Both types 
of particles are one-cell thick. In this
way, there are three possible orientations for each type of particle,
making a total of six orientations. It is assumed that there
are no interactions between particles, except for the hard core interactions
that prevent two particles from occupying the same cell. The thermodynamic
properties of the system follow by approximating the configurational
partition function corresponding to the ways of placing the $N$ molecules
into the $M$ cells under the restriction that a fraction $c$ 
are plates and a fraction $1-c$ are rods. Details can be found in
Alben's paper. The results indicate that the I--N transition can be
continuously changed from first to second order and, also, that a
biaxial nematic phase can be stabilized for an appropiate choice of
the dimensions and concentrations of the two components of the mixture.
Some of the results of Alben are reproduced in Fig. \ref{Alben}.

A formally exact theoretical approximation for a HS-HE mixture was developed in 1994 by Samborski and Evans \cite{Samborski}. They followed
the standard route to express, in the framework of DFT, the free energy of the system in terms of a
functional integral of the direct correlation function. Since there is no way to determine the direct correlation function of a system
of hard anisotropic particles, even in the one-component case, Samborski and Evans approximated the direct correlation function from
existing equations of state of fluid mixtures of hard convex bodies: Integrating the pressure, the Helmholtz free energy of the system is obtained
and from this the direct correlation function can be calculated upon functional differentiation. The theory was applied to different HS-HE
mixtures, changing the HE elongation and the ratio between the HS and HE volumes. No demixing was found, even in cases where the HS volume
was considerably smaller than the HE volume. At high enough density, a I--N transition was found, accompanied by phase segregation such that
the N phase is richer in ellipsoids than the I phase. It was also found that the N phase has a higher order parameter and a higher pressure
than the N phase of a one-component HE fluid.

Many of the hard-body models applied to binary mixtures of colloidal
particles are extensions, at different degrees of
sophistication, of the Onsager theory for one-component fluids. Each
component $i$ of a binary mixture ($i=1,2$) made of uniaxial particles
is described in terms of a local density $\rho_{i}({\bm{r}},\hat{\bm{\Omega}})$.
For a uniform fluid $\rho_{i}({\bm{r}},\hat{\bm{\Omega}})=\rho_{0}x_{i}h_{i}(\hat{\bm{\Omega}})$,
where $x_{i}=N_{i}/N$ is the fraction of species $i$, $N_{i}$ its
number of particles, and $N=N_{1}+N_{2}$ the total number of particles.
$h_{i}(\hat{\bm{\Omega}})$ are the orientational distribution functions.
By substituting $\rho({\bm r},\hat{\bm\Omega})\to\rho_i({\bm r},\hat{\bm\Omega})$ in the expression for the ideal free energy
of a one-component system (\ref{fid}), summing over all species $i$, and using Onsager excess free-energy approximation
(\ref{Ons}) extended to all pairs of species $ij$, Lekkerkerker and coworkers
\cite{new_Lekker,Stroobants-1} extended Onsager's second-virial theory to
a mixture of hard particle. The resulting expresion
for the Helmholtz free energy per particle is 
\begin{eqnarray}
\frac{\beta F}{N} & = & \log{\left(\rho_{0}\Lambda_{i}^{3}\right)}+\sum_{i=1}^{2}x_{i}\left(\log{x_{i}}-s_{{\rm or}}^{(i)}\right)+\rho_{0}\sum_{i=1}^{2}\sum_{j=1}^{2}x_{i}x_{j}B_{2}^{(ij)}[h_{i},h_{j}],\label{eq:Lekker-1}
\end{eqnarray}
where $\Lambda_{i}$ is the thermal wavelength of species $i$, $s_{{\rm or}}^{(i)}=-\left<\log{\left[4\pi h_{i}(\hat{\bm{\Omega}})\right]}\right>_{h_{i}}$
are the orientational entropies per particle of species $i$ in units
of the Boltzmann constant $k$ {[}which is a functional of the orientational
distribution function $h_{i}(\hat{\bm{\Omega}})${]}, while 
\begin{eqnarray}
B_{2}^{(ij)}[h_{i},h_{j}]=
\frac{1}{2}\left<\left<v_{{\hbox{\tiny excl}}}^{(ij)}(\hat{\bm{\Omega}}_{i},
\hat{\bm{\Omega}}_{j})\right>\right>_{h_{i},h_{j}}=
\frac{1}{2}\int_{V}d\hat{\bm{\Omega}}_{i}\int_{V}d\hat{\bm{\Omega}}_{j}
v_{\hbox{\tiny exc}}^{(ij)}(\hat{\bm{\Omega}}_{i},\hat{\bm{\Omega}}_{j})h_{i}(\hat{\bm{\Omega}}_{i})
h_{j}(\hat{\bm{\Omega}}_{j}).
\end{eqnarray}
represents orientation-averaged excluded volumes of two particles
of species $ij$. Three different entropy contributions are apparent
in Eqn. (\ref{eq:Lekker-1}): mixing, orientation, and excluded volume
entropy terms, respectively. The phase behaviour of the system is
the result of a delicate balance between the three entropic terms.
The theory was first applied to a binary mixture of rods of different
lengths \cite{new_Lekker}. A I--N transition was found, with the N phase
being significantly richer in long rods than the I phase. The order 
parameter of the long rods was found to be larger in the coexisting N phase
than in the one-component fluid made of equally long rods. The order 
parameter of the short rods first increased and then decreased with an
increasing mole fraction of the long rods.

\begin{figure}
\includegraphics[height=10cm]{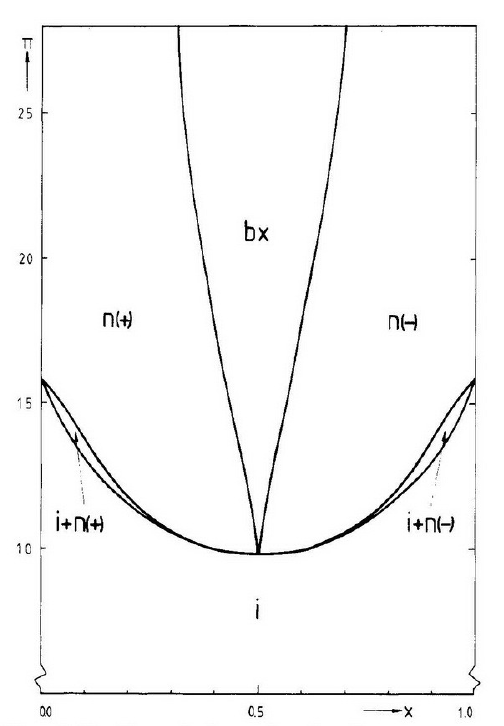} \caption{Osmotic pressure ($\Pi$) - fraction of discs ($x$) phase diagram
of a mixture of rods and discs for the case $B_{2}^{(11)}=B_{2}^{(22)}=B_{2}^{(12)}$
obtained from the second virial Onsager theory of Stroobants and Lekkerkerker \cite{Stroobants-1}.
Reprinted with permission from Ref. \cite{Stroobants-1}. Copyright (1984) American Chemical Society.}
\label{Stroobants-1fig}
\end{figure}

In another application, Stroobants and Lekkerkerker \cite{Stroobants-1}
used the theory to mixtures of rods and plates, modelled as hard cylinders 
of appropiate dimensions. The results obtained
indicated that, depending on the total concentration of particles
and the fraction of rods and disks, the mixture exhibits, in addition
to the I phase, two uniaxial N phases and a biaxial N$_{\rm B}$ phase
(see Fig. \ref{Stroobants-1fig}). One of the uniaxial N phases corresponds
to a rod-rich mixture, with the rod director oriented along some direction
and the plate director situated in the plane perpendicular to the
rod director (n$(+)$ in the figure). The other uniaxial N phase is plate-rich
with the rod director along the plane perpendicular to the plate director
(n$(-)$ in the figure). In the N$_{\rm B}$ phase the two perpendicular
directors are kept at a fixed orientation. If we take into account
that in Fig. \ref{Stroobants-1fig} the vertical axis is essentially the
inverse of the effective temperature plotted in the vertical axis
of Fig. \ref{Alben}, a remarkable qualitative agreement between results
from the lattice-model of Alben and the Onsager theory of Stroobants
and Lekkerkerker is apparent.

In 1994 van Roij and Mulder \cite{vanRoij-Mulder} studied the relative stability of the
N$_{\rm B}$ phase against N--N demixing as a function of aspect ratio in a
symmetric binary mixture of plates and rods (meaning that the rod aspect ratio
is the inverse of the plate aspect ratio), using the Onsager second-virial theory in the
Zwanzig approximation. They found that, for an aspect ratio equal to 5, the mixture 
separates into two uniaxial nematic phases, but with no biaxial nematic phase present. In contrast,
for an aspect ratio equal to 15, a stable biaxial phase was found.

The same problem was analyzed, also using Onsager theory, by Vanakaras
et al. \cite{Photinos-2} (see also \cite{Photinos-1} where the
N$_{\rm B}$ phase in mixtures of biaxial board-like particles is investigated
using Onsager theory and MC simulations). They concluded that no stable
N$_{\rm B}$ phase is found in mixtures of hard rods and plates for small
aspect ratios (about 5, the typical value for usual nematogens).
Stabilisation was predicted, however, at much higher values of the
aspect ratio (30 or higher).

The prediction of a biaxial N$_{\rm B}$ phase in fluids of uniaxial particles
called the attention of other researchers \cite{Camp-1,Camp-2}. In
these investigations the rod-plate mixture was modelled by prolate
HE of finite aspect ratio $\kappa=e$ (rods) and oblate hard ellipsoids
of aspect ratio $\kappa=1/e$ (plates), and equal volume. Computer
simulations where performed for $e=10,15$, and $20$. The same four
distinct phases predicted by the theory were found: I, two uniaxial
N and the biaxial N$_{\rm B}$ phases. However, the region of stability
of the N$_{\rm B}$ phase was found to be severely limited by demixing into
two coexisting uniaxial phases, N$_{\rm U}^+$ and N$_{\rm U}^-$, the N$_{\rm B}$
phase being stable only in a narrow region of elongation and mixture
composition.

An extension of Onsager theory to include, in the spirit of the PL
approximation, the contribution from higher-order virial terms, was
also carried out by Varga et al. \cite{Jackson-1}. The Helmholtz free energy excess
per particle of the mixture, giving the contribution from excluded-volume
interactions, is written as: 
\begin{eqnarray}
\frac{\beta F_{{\rm ex}}[h_{1},h_{2}]}{N}=\Psi_{{\rm HS}}(\eta)\frac{{\displaystyle \sum_{i=1}^{2}\sum_{j=1}^{2}x_{i}x_{j}B_{2}^{(ij)}[h_{1},h_{2}]}}{B_{2}^{{\rm (HS)}}},\label{PL_mezclas}
\end{eqnarray}
The HS volume of the reference system, $v_{0}$, is obtained as a
mole-fraction-weighted sum of the particle volumes of each species,
$v_{0}=x_{1}v_{1}+x_{2}v_{2}$. This is, in essence, the generalisation
to mixtures of the procedure that improves Onsager theory to give
PL theory for a one-component fluid. Again the density prefactor accounts for
the contribution to the free energy arising from higer-order virial
terms. Compared with the simpler Onsager's second-virial
approach, the application of this theory to the analysis of HE mixtures improved the theoretical
predictions as compared to simulations \cite{Jackson-1}. However,
in contrast to simulation results, the rescaled Onsager theory predicts
a symmetric phase diagram with respect to the equally-molar mixture
of prolate and oblate ellipsoids, as a result of the prolate-oblate
symmetry of the excluded volume.

The possible existence of the biaxial N$_{\rm B}$ phase in mixtures of
uniaxial particles was further analyzed by other authors, and not
only for rod-plate mixtures. In 2000 Kooij and Lekkerkerker \cite{Lekkerker-3}
came back to study the problem, this time experimentally. They used
a combination of two systems, rod-like boehmite ($\textrm{AlOOH}$)
and plate-like gibbsite ($\textrm{Al(OH\ensuremath{)_{3}}}$) colloids,
which can be approximated as hard bodies. For the particular
particle sizes used (see original reference for details), the study ruled out
the existence of a N$_{\rm B}$ phase with rods and plates orientationally
ordered in mutually perpendicular directions. It was found that the
biaxial N$_{\rm B}$ phase is unstable with respect to demixing into an
isotropic and two uniaxial nematic phases, in the line of the findings
of Refs. \cite{Camp-1,Camp-2}. The origin of demixing is the larger
excluded volume of a rod-plate pair compared to the rod-rod and
plate-plate excluded volumes, an effect that compensates the mixing
entropy, which favours the biaxial phase. As concentrations increase,
the excess excluded volume becomes more important and the system demixes.
Nevertherless, a word of caution is needed, because experimental systems are
always polydisperse, and its phase behaviour may be dramatically different
from that of strictly binary mixtures (see Sec. \ref{polidispersidad}).

The PL theory for binary mixtures was applied in 2001 by Wensink et
al. \cite{Lekkerker-4} to study the phase behaviour of a mixture
of thin and thick platelets. I--N density inversion (where the I phase
is heavier than the N phase) was found, in agreement with previous
experimental results.

\begin{figure}
\includegraphics[width=14cm]{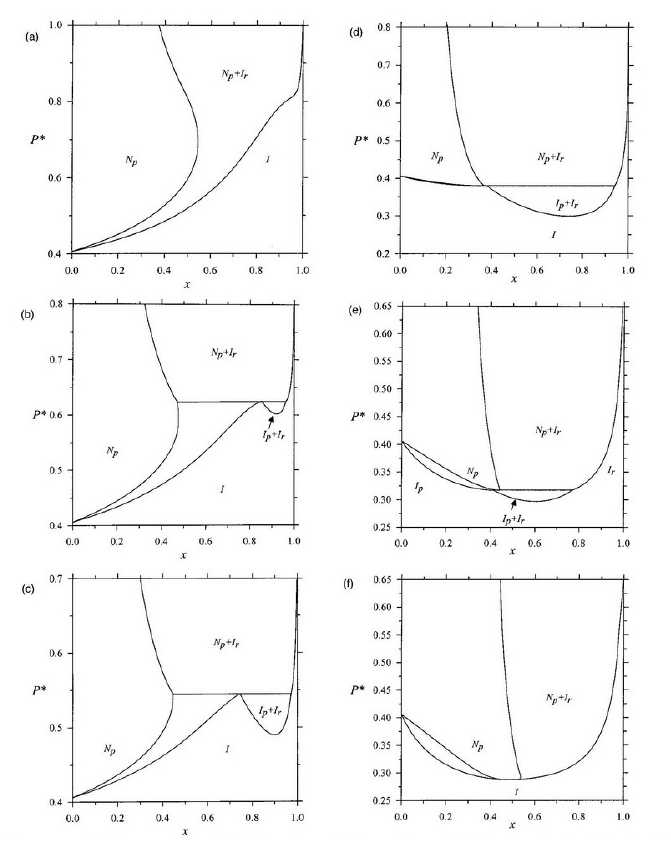} 
\caption{Pressure versus composition phase diagrams of mixtures of plates and
rods with aspect ratios 50 and 1/50 and different plate/rod diameter
ratios $d$, (a) $d=80$, (b) $d=60$, (e) $d=50$, (d) $d=25$, (e) $d=18$,
and (f) $d=6$) as obtained by Jackson and coworkers using the Parsons-Lee
theory \cite{Jackson-1}. The labels I$_{r}$, I$_{p}$, N$_{r}$, N$_{p}$ denote the
rod-rich isotropic, plate-rich isotropic, rod-rich nematic, and plate-rich
nematic phases, respectively. $x$ is the mole fraction of rods. Reprinted with permission from \cite{Jackson-1}. Copyright (2002), AIP Publishing LLC.}
\label{Jackson}
\end{figure}

The same theory was applied to different kinds of mixtures by
Jackson and coworkers in an extensive series of works in the period
2002-2005 \cite{Jackson-1,Jackson-2,Jackson-3,Jackson-4}. The phase
behaviour of a mixture of rods an plates \cite{Jackson-1}, both approximated
by hard cylinders of appropiate dimensions, was studied. In all cases
the plate volume was always orders of magnitude larger than the rod
volume. The authors did not carry out systematic free-energy minimization
and phase-coexistence calculations for all the mixtures investigated;
instead, they used a combined analysis based on I--N bifurcation and spinodal-demixing 
calculations to determine the geometrical requirements for the occurrence
of a demixing transition involving two I phases. What they found (see Fig. \ref{Jackson}) was
that the I phase is unstable relative to demixing for a wide range
of molecular parameters. The reason is the large excluded volume associated
with the mixing of unlike particles. However, using stability analysis
they found, for certain aspect ratios, that the I--N transition always
preempts I--I demixing, irrespective of the particle diameters. On
the other hand, when I--I demixing was found, it was accompanied by
the existence of an upper bound at large size ratios (Asakura and
Oosawa limit), and a lower bound at small size ratios (Onsager limit),
beyond which the system showed a mixed I phase. The results of the
stability analysis were confirmed by full phase-diagram calculations
for selected values of the particle geometrical parameters.

Mixtures of rod-like and disc-like particles were studied also by
Jackson and coworkers \cite{Jackson-3} using MC computer simulations
in the canonical ensemble and PL theory. Particles were modelled as
HSC of aspect ratio 5 and hard cut spheres of aspect ratio 0.12, the
ratio of diameters being chosen so that both particles had the same
volume. Simulations starting from a mixed isotropic state showed that, at low
total density, the I phase is stable with respect to ordered states.
Coexistence between two uniaxial N phases and the stabilization of a disc-rich I 
phase was found. At high densities the mixture exhibits N--C and S--C phase coexistence.
In agreement with simulations, no stable biaxial nematic phases were found.
The PL theory used in this work is not suitable for the study of translationally 
ordered states, such as C or S phases and, consequently, the study was restricted 
to the I and N phases.

The last kind of mixtures studied by Jackson and coworkers were mixtures
of hard rod-like particles. In a first paper \cite{Jackson-2}, a
mixture of HC of aspect ratios 15 and 150 was considered. The phase
behaviour of the mixture was explored as a function of the ratio $d$
between the diameters of the two cylinders using again the PL theory.
The results obtained can be summarised as follows: For large enough
values of the diameter ratio, for example $d=50$, the mixture shows
only an I--N phase transition driven by the excluded-volume interaction
between the large particles. In this case of very different sizes,
big particles can be considered as large colloidal particles. Since
exclusion interactions between particles of the other component and
between unlike particles are significantly small compared to interactions
between large particles, small particles play the role of a weakly
interacting solvent and I--I or N--N demixing is not to be expected.
However, when the diameter ratio is reduced, excluded interactions
involving small particles become more important, particularly
those between unlike particles. As a consequence, I--I and N--N demixing
transitions are indeed observed. The precise topology of the phase
diagram of this particular mixture can be rather complicated and,
in fact, it changes significantly with variations in the diameter
ratio $d$. This is not surprising since the three excluded volume
interactions become more and more comparable as $d$ is reduced, and
the final topology of the phase diagram is the result of a delicate
free-energy balance from these contributions. Detailed results can
be found in the original paper \cite{Jackson-2}.

In the last paper of their series \cite{Jackson-4}, Jackson and
coworkers again applied the PL theory of hard-rod mixtures to model
an experimental mixture of thin and thick viral particles, consisting
of charged semiflexible fd-virus (thin rods) and fd-virus irreversibly
coated with the neutral polymer polyethylene glycol, fd-PEG (thick
rods). The geometrical parameters of the hard rods were chosen to
mimic the experimental system. Thin and thick particles were modelled
as rods of the same length $L=24D_{{\rm thick}}$, where $D_{{\rm thick}}$
is the diameter of the thick rods, and different diameters. This particular
choice of $L$ corresponds to the experimental aspect ratio of the
thick particles. Phase diagrams of the mixture for diameter ratios
$d=D_{{\rm thick}}/D_{{\rm thin}}$ ranging from 3.7 to 1.1 were considered
in order to reproduce the experimental systems. The theory predicts
a region of I--I coexistence which is not observed experimentally.
Also, for small values of $d$, an I--N phase transition is found at
low concentrations, while at high values of $d$ a N--N coexistence region,
ending in a lower critical point, is predicted. As $d$ increases,
the N--N lower critical point moves to lower rod concentrations and,
upon increasing $d$ even further, a parameter region is entered
where the phase diagram shows an I--N--N coexistence region capped by
a region of N--N coexistence bounded by an upper critical point, and by
an additional region of N--N coexistence bounded by a lower critical
point. Finally, for very large $d$, these two regions of N--N coexistence
coalesce to form a single N--N coexistence region (see Fig. 3 in \cite{Jackson-4}
for details).

The mixture of thin and thick hard rods of the same length had been
previously analyzed by van Roij et al. in 1998 \cite{van Roij-1}
using the simple Onsager theory. They found that the mixtures not
only show an I--N transition and the previously-predicted depletion-driven
I--I demixing transition, but also a N--N demixing transition driven
by the orientational entropy of the thin rods. Several cases, corresponding
to different values of the diameter ratio $d$, were studied, and phase
diagrams exhibiting I--N, I--I and N--N coexistence, I--N--N and I--I--N
triple points, and I--I and N--N critical points, were obtained. When
compared with experiments \cite{Jackson-4}, these results showed that
Onsager second-virial theory qualitatively reproduces the main features
of the experimental phase diagram for large values of $d$. However, the theory 
is not able to account for the phase-behaviour evolution
from a totally miscible N phase to a demixed N--N
state upon increasing $d$. The use of PL rescaling shows that an I--N--N
coexistence region is not required for a region of N--N coexistence
to exist, in contrast to the predictions of the second virial theory.
But not everything is fully understood yet. For example, the N--N upper
critical point, predicted for very long rods by both theories, has
not been observed in the experiments. The effect of a small degree of flexibility was not taken 
into account in the theories, but fd-virus particles are not completely rigid.
Therefore, direct comparison of theoretical predictions
with experiments on mixtures of fd-virus has to be done with care.

All of the above theoretical studies on nematic ordering in mixtures of hard bodies
use Onsager theory or variations thereof. In contrast, Schmidt \cite{Schmidt3} used
his fundamental-measure-based DFT to calculate
the isotropic phase of a hard needle-hard sphere mixture for different values of the aspect
ratio $L/D$, where $D$ is the diameter of the spheres. The phase diagrams produced by the theory 
were similar to those obtained from free-volume theory \cite{free_volume}, which predicts a
demixing transition ending in a critical point, with each of the
demixed phases rich in one of the components. The sphere-sphere radial distribution 
function, as obtained from the theory, was in good agreement
with the Monte Carlo simulation results performed by the same author \cite{Schmidt3}.

\begin{figure}
\includegraphics[width=9cm]{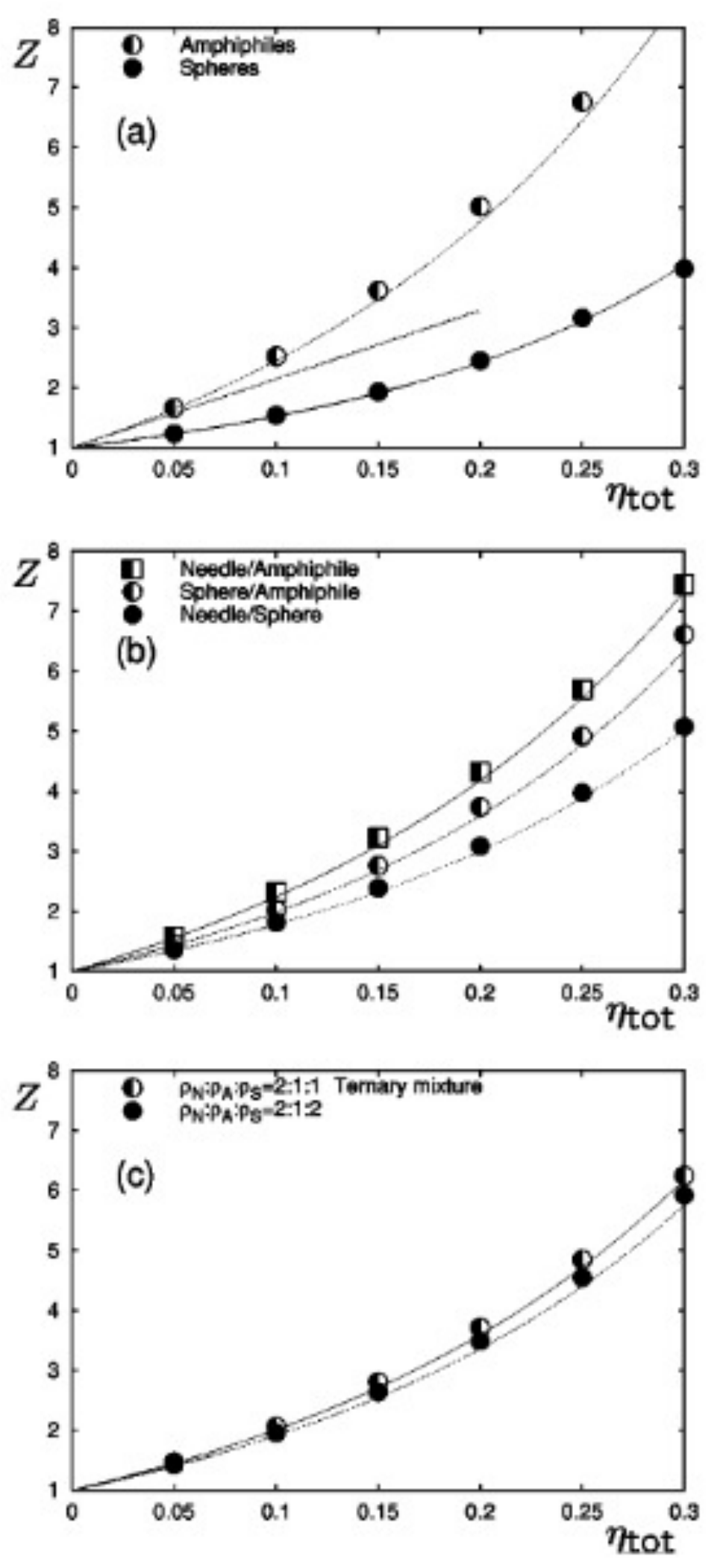} \caption{Compressibility factor $Z$ as a function of total packing 
fraction $\eta_{\rm tot}$ of an amphiphilic mixture, according to the theory of Schmidt and von Ferber \cite{Ferber}.
MC simulations (symbols) are compared with the theory (lines). Straight lines show the low-density limit governed by the 
second virial theory. (a) Pure systems; (b) binary mixtures; (c) ternary mixtures. Reprinted with permission from \cite{Ferber}. 
Copyright (2001) by the American Physical Society.}
\label{amph}
\end{figure}

This functional was further extended by Schmidt and von Ferber \cite{Ferber} to incorporate an amphiphilic-like 
hard-body particle formed by a HS with a radially attached hard needle. The resulting ternary mixture constitutes a 
simple model for a water (HS)-oil (needle)-amphiphilic (HS+needle) mixture where particles interact only 
via non-overlapping excluded volumes. While the Mayer functions for a pair of HS or a pair of hard needles are exactly 
obtained by convolutions of weight functions, for a HS and amphiphilic-particle or two of the latter the Mayer
functions are only approximate, due to their non-convex excluded volumes. However, the authors showed that the deviations 
of the the second virial coefficients with respect to the exact result are relatively small. 
Equations of state for the pure amphiphilic fluid, and for the amphiphilic-needle, amphiphilic-HS and HS-needle binary 
mixtures (all of them in the isotropic state) were calculated and compared with canonical MC simulations carried out by 
the same authors. The theory is in remarkable agreement with simulations for total packing fractions less than 0.3. Also, 
the equations of state for a ternary mixture of a particular composition compare very reasonably, Fig. \ref{amph}.
The authors calculated analytically the demixing spinodals of all possible binary mixtures and of the ternary mixture,
and showed that the amphiphilic-needle and amphiphilic-HS binary mixtures are more miscible than the HS-hard needle mixture, 
a property exhibited by real amphiphilic molecules due to their simultaneous hydrophilic and hydrophobic tendencies.

The same formalism was used by Schmidt and Denton to treat a ternary mixture of colloidal HS-hard needle-ideal polymer 
mixture \cite{Denton}. The ideal polymer was approximated by a sphere that interacts via excluded volume with the 
colloidal spheres, but there is no interaction between the polymers. The effect of the 
polymer-needle interaction on the phase behaviour of the ternary mixtures was studied, and
different phase diagrams were calculated by solving the conditions for mechanical and chemical equilibrium of the isotropic
coexisting phases. Calculations were carried out for the case where the HS diameter, the polymer diameter and the needle 
length are equal. In the case of no polymer-needle interaction, demixing between colloid-rich and colloid-poor phases is 
found. When the hard needle-polymer interaction is included, rich phase diagrams, exhibiting three-phase 
coexistence and reentrant demixing behaviour, were obtained. The addition of needles to the HS-polymer mixture stabilises the 
ternary mixture, an effect due to the competition between the polymer- or needle-mediated depletion effect and the 
interaction between them. 

More recently, FMT functionals have been devised and applied to more complicated mixtures.
A functional developed for a ternary mixture of HS, hard needles and hard platelets of
vanishing thickness obtained by Esztermann et al. \cite{Schmidt6},
was later generalized by Philips and Schmidt \cite{Schmidt7} to study
the phase behaviour of binary mixtures of infinitely thin platelets
with different diameters $D_1$ and $D_2$. Different phase diagrams were
produced for $\kappa=D_1/D_2$ ranging from 1 to 5.  For
$\kappa=D_1/D_2<1.7$ the phase diagram includes the usual I--N
transition, which becomes wider as $\kappa$ is increased. For $\kappa>2$ there
appears N--N demixing and an associated I--N--N triple point. No I--I demixing was
found. The results from the present theory were compared with those
from Onsager theory, and three major differences were found:
(i) the FMT predicts a smaller I--N biphasic region, (ii) the transitions
obtained by FMT are located at lower densities as compared to the Onsager model, 
and (iii) for the same $\kappa$, the N--N demixing region obtained from FMT spans a larger range of
compositions. However, the general phase diagram topologies predicted
by both theories were similar. Fractionation between
coexisting phases increases with $\kappa$, and big particles in each 
of the coexisting phases are more orientationaly ordered.

de las Heras and Schmidt \cite{de_las_Heras_x} reformulated the FMT functional of Esztermann et al. 
\cite{Schmidt6} for a binary mixture of HS and hard platelets of vanishing thickness, calculating the
phase behaviour of the mixture. They found a strong broadening of the I--N biphasic
region upon increasing the pressure. Also, for large values of the
aspect ratio $\kappa=D_{\rm P}/D_{\rm HS}$, N--N
demixing with different platelets concentrations was found. In the same work, the authors
formulate a FMT-based functional for mixtures of
hard platelets of vanishing thickness and overlapping hard spheres as an
approximate model to study the phase behaviour of platelet-polymer
mixtures. For low platelet-polymer size ratios, aside from I--N and
N--N phase coexistences, the mixture exhibits I--I demixing (not found
in hard-core platelet-sphere mixtures).

In addition to the more or less systematic studies on positionally
disordered phases of mixtures reviewed in the previous paragraphs,
there have been studies of very particular mixtures using both computer
simulations and theoretical models. For example, Seara et al. \cite{Seara-1}
used Onsager theory to study the phase behaviour of a binary mixture
of rods with different aspect ratios. By choosing particle sizes adequately,
the values of second-order virial coefficients were adjusted to be
equal. Since virial coefficients are the same, phase diagrams of the
one-component fluids are, at least in the framework of Onsager theory,
identical and independent of the particle aspect ratio in the hard-needle
limit. Therefore, the paper focuses on the case of mixtures where
one component is longer and thinner than the other, and the evolution
of the phase-diagram topology is studied as a function of the shape
difference between the two components. The main result is the occurrence
of I--I demixing, with an associated critical point, and N--N demixing
is also obtained. There is a N--N critical point that shifts as the
shape difference increases, eventually reaching the I--N transition
and giving rise to a four-phase region (for binary mixtures Gibbs
phase rule permits a maximum of three phases in simultaneous coexistence,
except in special cases of symmetric particles like the present one).

So far we have discussed the uniform I and N phases of binary mixtures.
When positionally-ordered phases, in particular S and C phases, are
considered, is easy to understand that the already complicated phase-diagram
topologies previously discussed will become even more complex. In
particular, part of the phase equilibria involving I and N phases
could be preempted by nonuniform phases and become metastable.
Moreover, Koda et al. \cite{Koda_depletion} performed constant-pressure 
Monte Carlo simulations of binary mixtures of hard disc-like particles 
with diameter-to-thickness ratios 2.5 and 5.0, and found strong tendency of
the large discs to form clusters with columnar ordering, induced by
strong depletion effects due to the small discs.

\begin{figure}
\includegraphics[width=14cm]{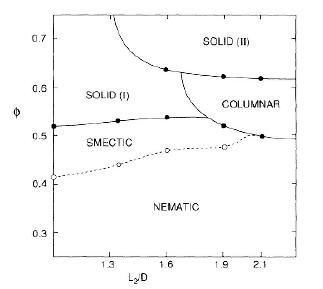} 
\caption{Monte Carlo results of Stroobants 
\cite{Stroobants-2} for the phase
diagram of a mixture of HSC. One of the species has fixed aspect ratio
$\kappa_{1}=L_1/D=1$. The aspect ratio $\kappa_{2}=L_2/D$ of the second species
of the mixture is shown in the horizontal axis. Mixture's concentration
is kept fixed at the equivalence point (i. e. the point at which partial
volume fractions of both components are equal). The total packing
fraction is represented in the vertical axis. Labels are as follows:
Solid (I): substitutionally disordered binary crystal. Solid (II):
phase separated pure component crystals. Open circles: continuous
transition. Solid circles: first-order transition. Lines are guides to the eyes. Reprinted with permission from \cite{Stroobants-2}. 
Copyright (1992) by the American Physical Society.}
\label{Stroobants-2}
\end{figure}

To simplify the analysis, the first attempts to deal with positionally
ordered phases in hard-body binary mixtures assumed perfect orientational
order (i. e. systems of parallel molecules). In 1992 Stroobants \cite{Stroobants-2}
performed MC simulations of binary mixtures of parallel HSC. Different
mixtures were considered, keeping the aspect ratio of one of the species
fixed, $\kappa_{1}=L_1/D=1$, and varying the aspect ratio of the other
in the range $\kappa_{2}=L_2/D=1.3-2.1$. The diameter of both particle
species was the same. In all cases the composition of the mixture
was adjusted so that the partial volume fractions of both components
were equal. A stable C phase was observed when $\kappa_{2}>1.6$.
Moreover, for $\kappa_{2}<1.9$, the N--S phase transition was found
to be preempted by a N--C transition (see Fig. \ref{Stroobants-2})
Since no stable C phase exists in the one-component systems, the conclusion
was that bidispersity favours C over S order, at least for these particular
mixtures.

Two years later, Cui and Chen \cite{Cui} extended the third-virial
coefficient free-energy functional approximation of Mulder for parallel
one-component cylinders \cite{Mulder_smec} to mixtures of the same
diameter and different lengths. The theory is trivially extended by
considering two local densities $\rho_{i}(z)$, where the $z$ axis
is along the director. The Helmholtz free-energy functional is written
as $F[\rho_{1},\rho_{2}]=F_{{\rm id}}[\rho_{1},\rho_{2}]+F_{{\rm ex}}[\rho_{1},\rho_{2}]$,
with 
\begin{eqnarray}
\frac{\beta F_{{\rm id}}[\rho_{1},\rho_{2}]}{A}=\sum_{i=1}^{2}\int dz\rho_{i}\left(z\right)\left[\log{\left(\rho_{i}(z)\Lambda_{i}^{3}\right)}-1\right].\label{eq:mul0}
\end{eqnarray}
The excess free-energy part can be written as a virial expansion:
\begin{eqnarray}
\frac{\beta F_{{\rm ex}}[\rho_{1},\rho_{2}]}{A} & = & \sum_{i=1}^{2}\sum_{j=1}^{2}\int dz\int dz'\rho_{i}\left(z\right)\rho_{j}\left(z'\right)B_{2}^{(ij)}(z,z')\nonumber \\
\nonumber \\
 & + & \sum_{i=1}^{2}\sum_{j=1}^{2}\sum_{k=1}^{2}\int dz\int dz'\int dz''\rho_{i}\left(z\right)\rho_{j}\left(z'\right)\rho_{k}(z'')B_{3}^{(ijk)}(z,z',z'')+\cdots\label{eq:mul}
\end{eqnarray}
where $A$ is the area of the layers. The first virial coefficients
are given by 
\begin{eqnarray}
 &  & B_{2}^{(ij)}(z,z')=-\frac{1}{2}\int d{\bm{r}}_{\perp}'f_{ij}\left({\bm{r}}-{\bm{r}}'\right),\nonumber \\
\nonumber \\
 &  & B_{3}^{(ijk)}(z,z,z'')=-\frac{1}{6}\int d{\bm{r}}_{\perp}'\int d{\bm{r}}_{\perp}''f_{ij}\left({\bm{r}}'\right)f_{jk}\left({\bm{r}}'-{\bm{r}}''\right)f_{ik}\left({\bm{r}}''\right).
\end{eqnarray}
Cui and Chen \cite{Cui} truncated the expansion beyond the third
term, and calculated $B_{2}^{(ij)}(z,z')$ and $B_{3}^{(ijk)}(z,z',z'')$ explicitely
for parallel HC. Conclusions similar to those of Stroobants \cite{Stroobants-2}
were obtained.

At the same time, the simpler Onsager theory, containing only the
$B_{2}^{(ij)}(z,z')$ function, was used by Koda and Kimura \cite{Koda-1}
to study the N--S phase transition in the same mixture. Different phase
diagrams, in the packing fraction-composition plane and for different
values of the length of the long cylinders, were obtained. The main
findings of this work were: (i) the region of S stability is suppressed
or enlarged depending on the length of the long cylinders and, (ii)
two types of smectic phases exist. In one, smectic layers
consist of uniform mixtures of large and short particles. In the other,
microsegregation occurs with alternate layers of short and long particles.

The stability of the N phase of mixtures of long and short parallel
cylinders against S or C phase formation was also studied, using Onsager
theory, by Sear and Jackson in 1995 \cite{Jackson_estab}. In agreement
with the previous studies summarized above, they found that the addition
of cylinders of a different length decreases the tendency towards
S ordering, to the extent that the N phase can coexist directly with
the C phase without forming a S phase first.

The study of the interplay between N and S phase stability and the
possible microphase segregation was continued with studies on mixtures
of parallel HSC and HS in the second half of the '90. In 1996 Koda
et al. \cite{Koda-2} again used Onsager theory, together with constant-pressure
MC simulations, to conclude that the addition of spherical molecules
to a system of parallel HSC induces the formation of a micro-segregated
S phase with alternating layers of HSC and HS. Later on, in 2000,
Dogic et al. \cite{Dogic-1} revisited the problem using computer
simulations. The conclusions of this study were the following: (i)
Entropy-driven micro-segregation occurs in mixtures of parallel
rods and spheres. (ii) Adding spheres smaller than the rod width decreases
the total packing fraction needed for the formation of the S phase,
and therefore small spheres effectively stabilize the S phase; the
opposite is true for large spheres. (iii) The degree of stabilization
increases with increasing rod length.

An extra element was included in the contribution made by Mart\'{\i}nez-Rat\'on
et al. \cite{Yuri-2} in 2006. In contrast to other works, these
authors incorporated orientational order and carried out a more systematic
study. Mixtures of rods of different aspect ratios and spheres of
different diameters were considered and treated within Onsager theory
(see below). The study concluded that depletion effects, and consequently
S stability, decrease significantly as a result of orientational disorder
in the S phase when compared with the corresponding data based on the
frozen-orientation approximation.

\begin{figure}
\includegraphics[width=14cm]{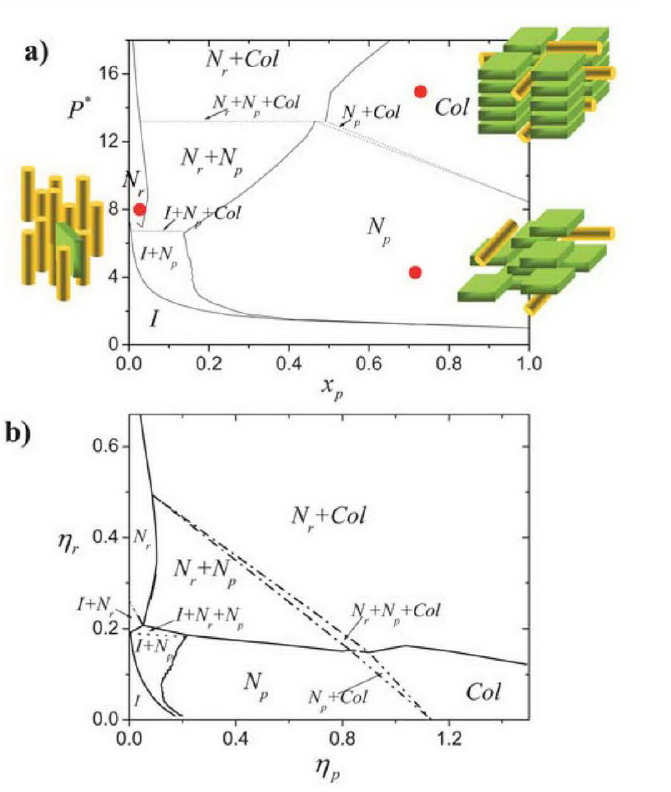} 
\caption{(a) Phase diagram of plate\textendash{}rod binary mixtures of molecular
sizes $L_{p}=9$ and $L_{r}=13$ in the plane pressure-plate concentration
as calculated by Peroukidis et al. \cite{Photinos-3}. Cartoons of
molecular organization are shown for the marked points on the
phase diagram. (b) Phase diagram in the plane packing fraction of
rods ($\eta_{r}$) - packing fraction of plates ($\eta_{p}$) corresponding
to the same case shown in (a). Reproduced from Ref. \cite{Photinos-3} with permission of 
the Royal Society of Chemistry.}
\label{photinos-fig}
\end{figure}

The work of Koda et al. on mixtures of parallel HSC and spheres was
later extended by Vesely \cite{Vesely-1} to mixtures of other linear
particles (fused spheres, ellipsoids and sphero-ellipsoids) and HS.
Mixtures of HS with shape-anisometric colloids, namely rod-like and
plate-like particles, have been studied using a lattice-model approach
by Peroukidis et al. \cite{Photinos-3} (see Fig. \ref{photinos-fig}). 
Also, Varga and coworkers
published results on further extensions of Koda's theory to mixtures
of equally long but differently wide cylinders \cite{Varga-1}, and
HC of equal diameters but different lengths \cite{Varga-2}.

In 2004 the authors of this review and their coworkers published the
first of a series of papers devoted to the phase behaviour of binary
mixtures of freely rotating hard-body particles, paying special attention
to the stability and properties of the S phase \cite{Cinacchi-1}.
The approach used was a version of the PL theory suitable for mixtures,
but extended to deal with spatially ordered (nonuniform) systems.
The local densities are now a function of the particle orientations,
$\rho_{i}({\bm{r}},\hat{\bm{\Omega}})$. Let us formulate
the theory without assuming any special symmetry. 
The Helmholtz free-energy functional is given by 
\begin{eqnarray}
\beta F_{\rm id}[\rho_{1},\rho_{2}]=\sum_{i=1}^{2}\int_{V}d{\bm{r}}
\int d\hat{\bm{\Omega}}\rho_{i}({\bm{r}},\hat{\bm{\Omega}})\left[\log{\left(\rho_{i}({\bm{r}},\hat{\bm{\Omega}})\Lambda_{i}^{3}\right)}-1\right],
\end{eqnarray}
and 
\begin{eqnarray}
&&\beta F_{\rm ex}[\rho_{1},\rho_{2}]=-\frac{\Psi_{\hbox{\tiny HS}}(\eta)}{2B_{2}^{(\hbox{\tiny HS})}}
\sum_{i=1}^{2}\sum_{j=1}^{2}\int_{V}d{\bm{r}}\int d\hat{\bm{\Omega}}\int_{V}d{\bm{r}}\int d\hat{\bm{\Omega}}'
\rho_{i}({\bm{r}},\hat{\bm{\Omega}})\rho_{j}({\bm{r}},\hat{\bm{\Omega}})
\nonumber\\\nonumber\\&&\hspace{6cm}\times\hspace{0.1cm}
f_{ij}({\bm{r}}-{\bm{r}}',\hat{\bm{\Omega}},\hat{\bm{\Omega}}'),\label{eq:nos-1}
\end{eqnarray}
where, as usual, ${\displaystyle \eta=\rho_{0}\sum_{i=1}^{2}x_{i}v_{i}}$
is the total packing fraction of the mixture (which is equal to that
of the reference HS fluid). For uniform phases, $\rho_{i}({\bm{r}},\hat{\bm{\Omega}})=
\rho_{0}x_{i}h_{i}(\hat{\bm{\Omega}})$,
this theory reduces exactly to the PL approximation for mixtures and,
consequently, all results pertaining to the I and N phases of previous
sections apply here. Also, in the low-density limit and for strictly
parallel particles with smectic symmetry, $\rho_{i}({\bm{r}},\hat{\bm{\Omega}})=\rho_{i}(z)
\delta(\hat{\bm{\Omega}}-\hat{\bm{z}})$,
the expression for $F_{\hbox{\tiny ex}}[\rho_{1},\rho_{2}]$ reduces
to Eqn. (\ref{eq:mul}) truncated at second order. However, the generalised
form of the theory can describe spatially ordered phases of any symmetry,
including the S and C phases. In \cite{Cinacchi-1}, the theory was
applied to the S phase by making the simplification $\rho_{i}({\bm{r}},\hat{\bm{\Omega}})=
\rho_{i}(z,\hat{\bm{\Omega}})$.
Computational details on free-energy minimisation to obtain the equilibrium
density profiles and orientational distribution functions (from which
N and S order parameters, and thermodynamic and phase transition properties,
can be easily derived) can be found in \cite{Cinacchi-1}. In this
paper binary mixtures of HSC of the same diameter but different lengths
(and, therefore, different aspect ratios), were considered. Phase
diagrams were found to depend strongly on the aspect ratio of each
component, and also on their length ratios. When the mole fraction
of long rods is larger, it was found that layered phases present a
S structure with short rods located at the layers, mixed with long
rods. However, in the opposite case (short rods more abundant that
long rods), a S$_{2}$ phase is obtained consisting of layers of
short rods with long rods located parallel to the latter but in the
interlayer region.

The same kind of mixture was analyzed by Cinacchi et al. in Ref. \cite{Cinacchi-2}.
The special case where one of the components is a HS was also considered.
Particular emphasis was put on the interplay between S formation versus
S-S segregation. It was found that, in general, S-S segregation occurs
in a wide range of compositions and pressures, except when particles
of both components have similar lengths in which case segregation
appears only at high pressure. When lengths are very different
and the mixture is poor in long molecules, a micro-segregated (but
macroscopically homogeneous) phase, where the minority species is
expelled to the interlayer regions, was found to be stable.

\begin{figure}
\includegraphics[width=8cm]{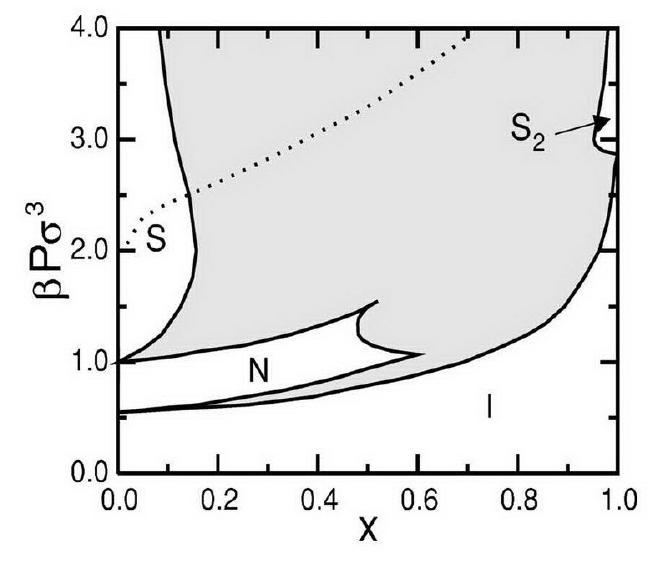}
\includegraphics[width=8cm]{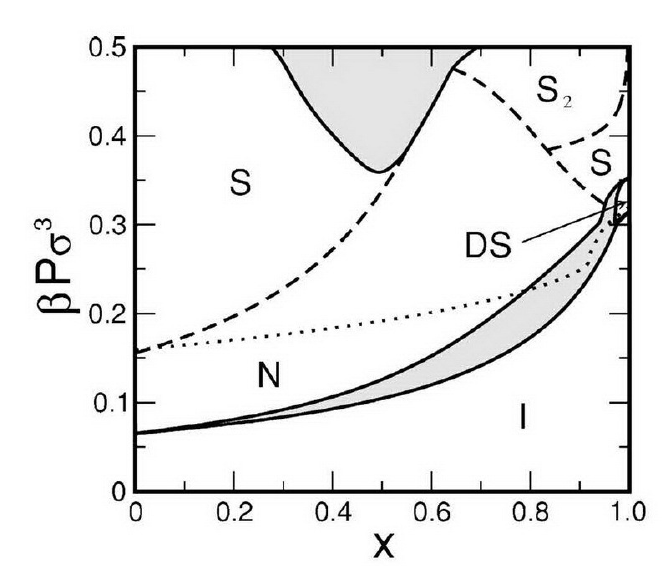}
\caption{Phase diagrams in the pressure - composition plane ($x$ being the fraction
of shortest particles) for different mixtures, from \cite{Yuri-1}. (a) Mixture of HSC with the same diameter
$\sigma$ and aspect ratios $\kappa_{1}=4.5$ and $\kappa_{2}=8.0$,
as obtained from an extended PL approach. (b) Phase diagram of a mixture
of hard parallelepipeds of aspect ratios $\kappa_{1}=4.5$ and $\kappa_{2}=8.0$
and cross section $\sigma^{2}$, as obtained from a FMT approach with a Zwanzig approximation. 
The continuous lines indicate first-order phase transitions. The shaded regions are two-phase regions
of phase coexistence. The regions of stability are labeled by S
(standard smectic formed by layers identical in composition), N
(nematic), I (isotropic), S$_{2}$ (microsegregated smectic phase
with long particles located in the interlayer space), and DS (discotic
smectic). The dotted line is the spinodal line corresponding to the
instability of the nematic phase with respect to columnar-type fluctuations. Reprinted with permission from \cite{Yuri-1}. Copyright (2005), AIP Publishing LLC.}
\label{Yuri-mezclas}
\end{figure}

A more general analysis of the phase behaviour of binary mixtures of
hard rods of different lengths and diameters was published by Mart\'{\i}nez-Rat\'on
et al. in 2005 \cite{Yuri-1}. Attention was focused on the formation
of C phases and the relative stability between S and C phases. Once
again the extended PL theory was used, this time complemented with
a restricted-orientation (Zwanzig) approximation of the FMT for hard
parallelepipeds. This strategy provided a complete picture of the
problem: while the PL theory fully incorporates the orientational
degrees of freedom, but treats spatial and orientational correlations
in an approximate manner, the Zwanzig approximation of FMT theory
considers only orientations along three mutually orthogonal axes,
but correlations are more faithfully represented. Results for the
phase diagram of a mixture of HSC using the extended PL theory have
been reproduced in the left side of Fig. \ref{Yuri-mezclas}, results
for a mixture of hard parallelepipeds obtained using FMT within the
restricted-orientation Zwanzig approximation are shown on the right
side of the same figure.

The main feature of the phase diagram obtained using PL theory is
the very strong segregation of the smectic phase; it is in fact so strong that it
preempts the I--N transition. Apart from the standard smectic phase, a second
kind of smectic phase, S$_{2}$, is found to be stable in the region of the phase diagram
corresponding to mixtures rich in the short particles. In the micro-segregated
S$_{2}$ phase, layers of different composition alternate in such a
way that the density distributions of the two species are shifted
one with respect to the other by half a smectic period. In the case
of the Zwanzig model (right side of the figure), the smectic segregation
takes place at a much higher pressure (as compared, for example, with
the location of the isotropic-nematic transition). As a result, the
I--N transition is not preempted. There is no direct I--S phase transition
like the one resulting from the PL theory for a mixture rich in short particles.
Instead, we have continuous N--S transitions and a first-order S--S$_{2}$
transition at higher pressure. However, it has to be noted that when
the stability of the nematic phase against columnar-type fluctuations
is taken into account, these results have to be modified considerably.
The nematic-columnar spinodal line has been represented as a dotted
line in Fig. \ref{Yuri-mezclas}. As it is evident from the figure,
the conclusion is that the FMT-Zwanzig approximation greatly enhances
the stability of the C phase, preempting smectic order completely.
By contrast, in the PL model, the C phase may preempt a large region
of S-phase stability in some mixtures, but some regions where the
S phase is stable remain. Note that the stability against C-type
fluctuations in the PL theory was estimated by computing spinodal
lines; a proper calculation including binodal lines would probably
result in enhanced stability of the C phase.

In a final study using the PL theory, Cinacchi et al. \cite{Cinacchi-3}
calculated the phase diagrams of a collection of binary mixtures of
thin and thick HSC. Attention was paid to two cases: (i) binary mixtures
where the two components have the same length and, (ii) binary mixtures
where the two components have the same volume. Spherocylinders of
the same total length and different diameter tend to demix considerably
as soon as the diameter ratio deviates from unity, especially at (high)
pressures such that at least the phase richer in the thicker component
is S. In the case where the two components have equal volumes, demixing
is further increased due to the disparity not only in particle diameter
but also in particle length.

Many of the theoretical studies on mixtures of hard anisotropic bodies have been inspired in,
or have inspired, experiments on colloidal suspensions of particles. When comparing the
experimental results with the theories as regards the stable phases and the different
regions of coexistence, it is important to pay attention to the possible effects of
gravity, since the sedimentation profile is a cross result of phase behaviour and
gravity effects. The sedimentation-diffusive equilibrium of colloidal liquid-crystal suspensions,
either mixtures or polydisperse systems, gives rise to a stack of distinct
layers with different properties (such as particle composition). This problem has recently been
discussed by de las Heras and Schmidt \cite{de_las_Heras_xx}. They used a Legendre
transform on the FMT-based functional introduced by Esztermann et al. \cite{Schmidt6} to study sedimentation
of a binary mixture of hard platelets with vanishing thickness and to account for the effect of gravity on the
number of stacking diagrams. The authors found that even simple binary
(not necessarily polydisperse) mixtures produce a stacking diagram containing six
types of stacks with up to four distinct layers. The extended Gibbs
phase rule that determines the maximum number ($N_{\rm max}$) of
sedimented layers reads: $N_{\rm max}=3+2(n_b-1)+n_i$, where $n_b$ is
the number of binodals present in the bulk phase diagram, while $n_i$ is
the number of their inflection points.

\subsubsection{Polydisperse systems}
\label{polidispersidad}

In this section we focus on continuously polydisperse mixtures, i.e. mixtures beyond binary and ternary that contain an arbitrarily large 
number of species, so large that in fact sizes can be characterised by a continuous distribution.
Particle size polydispersity is now recognised as an important actor in the phase behaviour of real colloidal suspensions of
anisotropic particles. Already in binary mixtures (Sec. \ref{binary_mixtures}) interesting effects can be identified: 
(i) broadening of the coexistence gap, 
(ii) enrichement of coexisting phases in some species (an effect predicted to occur by Onsager in mixtures with a discrete number of 
species), and (iii) exotic phase equilibria, such as reentrant nematic phases,
nematic-nematic coexistence or even three-phase, isotropic-nematic-nematic coexistence. 
These effects have experimental evidence. Size polydispersity is a realistic feature of real fluids, and has to be taken into account in theoretical treatments.
Polydispersity adds an extra element of complexity which may entail richer phase
diagrams. In particular, as the number of components increases without limit,
Gibbs' phase rule allows for the possibility of an unlimited number of phases 
to coexist. However, the theoretical analysis of polydisperse anisotropic 
fluids is not easy, as subtle questions arise concerning
particle size distributions, lack of accuracy of perturbation treatments, and numerical issues.

The experimental realisation of polydisperse mixtures of hard particles consists of suspensions of colloidal 
synthetic particles (natural particles such as viruses are usually monodisperse) in an aqueous environment
at high Coulomb screening conditions. The latter condition ensures that particle interactions will faithfully be
represented by overlap or hard interactions. Many experimental procedures have been devised to synthesise
colloidal particles with different geometries and shapes (a topic not covered in the present review). All of these
procedures lead to samples containing particles of different sizes in all particle dimensions. Theoretical
treatments of the resulting suspensions necessarily have to incorporate polydispersity. The polydispersities
in each dimension are typically poorly controlled in the experiments and have to be measured by different
means (dynamical light scattering, direct optical visualisation, etc.)

Polydisperse fluids of hard rods were first considered in the context of Onsager theory. 
Onsager already outlined the possible generalisation of his theory to discretely polydisperse systems in length,
and advanced the broadening effect on the coexisting gap for the isotropic-nematic transition. Later, 
several experimental studies motivated the analysis of these
systems. McMullen et al. \cite{McMullen} generalised Onsager theory to polydisperse micelles 
(approximated by hard rods), considering a simple aggregation model where all rods are in chemical equilibrium and
the distribution is obtained from the model itself. Later 
Odijk et al. \cite{Odijk_poly}, and Sluckin \cite{Sluckin_poly} and Chen \cite{Chen} further used it with slight differences, considering an expansion 
valid for narrow size distributions. 
The first attempt to study continuously polydisperse fluids forming liquid-crystalline phases for general (not only small)
polydispersity coefficient was made by Clarke et al. \cite{Cuesta_poly}.
More recently, an approximate but consistently more accurate procedure, the moment 
method, was used by Speranza and Sollich \cite{Sollich_i-n1}. Before reviewing the results, we present the theory and then discuss the different implementations. 
Finally, we will review other works on polydispersity focused on particle shapes other than hard rods; these works use different versions of fundamental-measure
density-functional in the Zwanzig approximation.

We consider hard rods of length $L$ and breadth $D$, and introduce the
variables $\sigma_1=L/\left<L\right>$ and $\sigma_2=D/\left<D\right>$, denoted collectively by $\boldsymbol{\sigma}\equiv (\sigma_1,\sigma_2)$. Here 
$\left<L\right>$ and $\left<D\right>$ are the averaged particle sizes 
(note that ${\bm\sigma}$ is a dimensionless quantity, but the definition of 
polydisperse variables is not universal, and some authors use different
definitions).
If one or more of these variables are continuously distributed, the local density distribution $\rho({\bm r},\hat{\bm\Omega})$
has to be generalised to $\rho({\bm r},\boldsymbol{\sigma},\hat{\boldsymbol{\Omega}})$. 
The present definition of polydisperse variables guarantees
that the local distribution is a density, in the sense that its normalisation
is $\displaystyle\int d{\bm\sigma}\int d\hat{\bm\Omega}\rho({\bm r},\boldsymbol{\sigma},\hat{\boldsymbol{\Omega}})=N$. Assuming for the moment that there is
no spatial dependence, we have $\rho({\bm r},\boldsymbol{\sigma},\hat{\boldsymbol{\Omega}})=
\rho(\boldsymbol{\sigma},\hat{\boldsymbol{\Omega}})=
\rho(\boldsymbol{\sigma}) h(\boldsymbol{\sigma},\hat{\boldsymbol{\Omega}})$. These functions satisfy the normalisation 
conditions $\int d\boldsymbol{\sigma}\rho(\boldsymbol{\sigma})=\rho_0$ (with $\rho_0$ the total mean density of 
the polydisperse mixture) and $\displaystyle\int d\hat{\boldsymbol{\Omega}} h(\boldsymbol{\sigma},\hat{\boldsymbol{\Omega}})=1$.  
It is also convenient to introduce $\rho_0(\boldsymbol{\sigma})=\rho_0f(\boldsymbol{\sigma})$, the total density distribution 
function, where $f(\boldsymbol{\sigma})$ is the so-called {\it parent} distribution function, which reflects 
the particle size distribution obtained from the particular experimental procedure used to synthesize particles.
Polydispersity is usually quantified in terms of polydispersity coefficients $\Delta_i$ which give the standard deviation
of the size distribution function, $\Delta_i=\sqrt{\langle\sigma_i^2\rangle/\langle\sigma_i\rangle^2-1}$ 
with $\displaystyle\langle\sigma_i^{n}\rangle=\int d{\bm\sigma} \sigma_i^{n}f({\bm\sigma})$.

The extended Onsager theory for the polydisperse mixture is \cite{Odijk_poly,Sluckin_poly}
\begin{eqnarray}
\frac{\beta {\cal F}[\rho]}{V}&=&\int d\boldsymbol{\sigma}
\int d\hat{\boldsymbol{\Omega}}\rho(\boldsymbol{\sigma},\hat{\boldsymbol{\Omega}})
\left\{\log{\left[\rho(\boldsymbol{\sigma},\hat{\bm\Omega})\Lambda({\bm\sigma})\right]}-1\right\}\nonumber\\\nonumber\\
&+&\frac{1}{2}
\int d\boldsymbol{\sigma}\int d\hat{\boldsymbol{\Omega}} \rho(\boldsymbol{\sigma},\hat{\boldsymbol{\Omega}}) 
\int d\boldsymbol{\sigma}'\int d\hat{\boldsymbol{\Omega}}' \rho(\boldsymbol{\sigma}',\hat{\boldsymbol{\Omega}}')
V_{\rm exc}(\boldsymbol{\sigma},\boldsymbol{\sigma}',\hat{\boldsymbol{\Omega}},\hat{\boldsymbol{\Omega}}'),
\label{Onsager_poly}
\end{eqnarray}
where $V_{\rm exc}(\boldsymbol{\sigma},\boldsymbol{\sigma}',\hat{\boldsymbol{\Omega}},\hat{\boldsymbol{\Omega}}')$ 
is the excluded volume between a pair of particles with dimensions 
$\boldsymbol{\sigma}$, $\boldsymbol{\sigma}'$ and orientations $\hat{\boldsymbol{\Omega}}$, $\hat{\boldsymbol{\Omega}}'$, 
respectively, and $\Lambda({\bm\sigma})$ is the thermal wavelength of the species with dimensions ${\bm\sigma}$. 
Minimisation with respect to the function $h(\boldsymbol{\sigma},\boldsymbol{\Omega})$ provides the self-consistent
equation
\begin{eqnarray}
h(\boldsymbol{\sigma},\hat{\boldsymbol{\Omega}})=
\frac{\displaystyle\exp\left[-\int d\boldsymbol{\sigma}'\rho(\boldsymbol{\sigma}')
\int d\hat{\boldsymbol{\Omega}}'h(\boldsymbol{\sigma}',\boldsymbol{\Omega}')
V_{\rm exc}(\boldsymbol{\sigma},\boldsymbol{\sigma}',\hat{\boldsymbol{\Omega}},\hat{\boldsymbol{\Omega}}')\right]}
{\displaystyle\int d \boldsymbol{\Omega} \exp\left[-\int d\boldsymbol{\sigma}'\rho(\boldsymbol{\sigma}')
\int d\hat{\boldsymbol{\Omega}}'h(\boldsymbol{\sigma}',\boldsymbol{\Omega}')
V_{\rm exc}(\boldsymbol{\sigma},\boldsymbol{\sigma}',\hat{\boldsymbol{\Omega}},\hat{\boldsymbol{\Omega}}')\right]},
\label{tt}
\end{eqnarray}
from which the degree of orientational order for species $\boldsymbol{\sigma}$ can be obtained:
\begin{eqnarray}
Q(\boldsymbol{\sigma})=\int d\hat{\boldsymbol{\Omega}} P_2(\hat{\bm\Omega}\cdot\hat{\bm n}) 
h(\boldsymbol{\sigma},\hat{\boldsymbol{\Omega}}).
\end{eqnarray}
Here $\hat{\bm n}$ is, as usual, the nematic director. Isotropic, $Q(\boldsymbol{\sigma})=0$, and nematic,
$Q(\boldsymbol{\sigma})\ne 0$, branches can be obtained from these equations. 

The calculation of phase equilibrium for polydisperse mixtures is not a trivial task, since there is in general a
whole region in the thermodynamic phase diagram where two or more phases can coexist, and the density of particles of
a given size changes depending on the coexistence point (giving rise to
fractionation). Let $ \rho_{\alpha}(\boldsymbol{\sigma})$ be the density of
particles of size $\boldsymbol{\sigma}$ in phase $\alpha$, and $\mu_{\alpha}(\boldsymbol{\sigma})$ the corresponding
chemical potential. If $m$ phases are in chemical equilibrium, all chemical potentials of a given species $\boldsymbol{\sigma}$ 
should be equal to $\mu_0(\boldsymbol{\sigma})$, a function which is determined by the conservation of the number of 
particles (lever rule): 
\begin{eqnarray}
\rho_0(\boldsymbol{\sigma})=\sum_{\alpha=1}^m \gamma_{\alpha} \rho_{\alpha}(\boldsymbol{\sigma}),
\hspace{0.4cm}\sum_{\alpha=1}^m\gamma_{\alpha}=1,
\label{lever}
\end{eqnarray}

The coefficients $\gamma_{\alpha}$, with $0\leq \gamma_{\alpha}\leq 1$, represent the fraction 
of volume occupied by the phase $\alpha$. Calculating the chemical potentials from Eqn. (\ref{Onsager_poly})
and using (\ref{tt}) and (\ref{lever}) gives
\begin{eqnarray}
\rho_{\alpha}(\boldsymbol{\sigma})=
\frac{\displaystyle\rho_0(\boldsymbol{\sigma})
\int d\hat{\boldsymbol{\Omega}}\exp{\left[-
\int d\boldsymbol{\sigma}'\rho_{\alpha}(\boldsymbol{\sigma}') 
\int d\hat{\boldsymbol{\Omega}}'h_{\alpha}(\boldsymbol{\sigma}',
\hat{\boldsymbol{\Omega}}')V_{\rm exc}(\boldsymbol{\sigma},
\boldsymbol{\sigma}',\hat{\boldsymbol{\Omega}},\hat{\boldsymbol{\Omega}}')
\right]}}{\displaystyle
\sum_{\beta=1}^m\gamma_{\beta} 
\int d\hat{\boldsymbol{\Omega}}\exp{\left[-
\int d\boldsymbol{\sigma}'\rho_{\beta}(\boldsymbol{\sigma}') 
\int d\hat{\boldsymbol{\Omega}}'h_{\beta}(\boldsymbol{\sigma}',
\hat{\boldsymbol{\Omega}}')V_{\rm exc}(\boldsymbol{\sigma},
\boldsymbol{\sigma}',\hat{\boldsymbol{\Omega}},\hat{\boldsymbol{\Omega}}')
\right]}}.
\label{la_rho_de_sigma}
\end{eqnarray}
Fixing $\rho_0$ and solving (\ref{la_rho_de_sigma}) and  (\ref{tt}) for each phase $\alpha$ we can find
all the functions $h_{\alpha}(\boldsymbol{\sigma},\hat{\boldsymbol{\Omega}})$ and $\rho_{\alpha}(\boldsymbol{\sigma})$.
The latter gives information on particle fractionation. 
The $m-1$ independent coefficients $\gamma_{\alpha}$ 
are calculated through the mechanical equilibrium conditions
\begin{eqnarray}
p_1[\rho_1;\{\gamma_{\tau}\}]=p_2[\rho_2;\{\gamma_{\tau}\}]=\cdots=p_m[\rho_m;\{\gamma_{\tau}\}],
\end{eqnarray}
with $\beta p_{\alpha}[\rho_{\alpha};\{\gamma_{\tau}\}]=\rho_{\alpha}+\beta F_{\rm ex}[\rho_{\alpha}]/V$ (valid for Onsager second-virial theory). Here 
$\rho_{\alpha}=\int d\boldsymbol{\sigma}\rho_{\alpha}(\boldsymbol{\sigma})$ is the number density of phase $\alpha$.
For the case of two-phase coexistence, say between I and N phases, the onset of order is calculated by fixing 
$\gamma_{\rm N}=0$ (a vanishingly small amount of N phase, with the I phase occupying the whole volume). 
Solving the equations for chemical and mechanical equilibrium, the so-called I-cloud and N-shadow densities,
$\rho_{\rm I}^{(\rm c)}$ and $\rho_{\rm N}^{(\rm s)}$ respectively, can be found.
In the other limit, $\gamma_{\rm N}=1$ (vanishingly small amount of I phase), the same coexistence equations
provide the N-cloud and I-shadow densities, $\rho_{\rm N}^{(\rm c)}$ and $\rho_{\rm I}^{(\rm s)}$, respectively. 
Usually phase diagrams are presented plotting the total number density $\rho_0$ as a function of the polydisperse 
coefficients. In the case where only the particle length $L$ is polydisperse, the
I-cloud and N-cloud curves in the $\rho_0-\Delta_L$ plane define the  
boundaries of the two-phase I--N coexistence. 

The calculation of coexistence parameters, even for the simplest two-phase transition, is a daunting task:
Eqns. (\ref{tt}) and (\ref{la_rho_de_sigma}), which include integrations on angular and polydispersity variables,
have to be discretised and a huge number of grid points (and, therefore, of unknowns) have to be used. 
Therefore, some approximations are needed.
Sluckin \cite{Sluckin_poly} and later Chen \cite{Chen} were the first to study the effect of a small amount of
polydispersity (i.e. the limit $\Delta\to 0$) on the coexistence densities of the I--N transition and the
pressure values at the cloud and shadow points, using the Onsager model for length-polydisperse hard rods, i.e.
rods with $L_0/D_0\to\infty$,
fixed $D_0$ and polydisperse length $L$, with $L_0=\left<L\right>$ the mean length. Sluckin \cite{Sluckin_poly}
used the Gaussian parameterisation for the orientational distribution function proposed by Odijk,
\begin{eqnarray}
h({\bm\sigma},\hat{\bm\Omega})=
\left\{
\begin{array}{ll}
\displaystyle{\frac{\alpha(l)}{4\pi}\exp\left[-\frac{\alpha(l)}{2}\theta^2\right]}, & 
\displaystyle{0\leq\theta\leq \frac{\pi}{2}}\\\\
\displaystyle{\frac{\alpha(l)}{4\pi}\exp\left[-\frac{\alpha(l)}{2}(\pi-\theta)^2\right]}, 
& \displaystyle{\frac{\pi}{2}\leq\theta\leq \pi},
\end{array}
\right.
\end{eqnarray}
with $l\equiv L/L_0$, 
and considered the polydispersity coefficient $\Delta$ as a perturbation parameter, obtaining transition densities,
orientational order parameter, and fractionation in terms of $\Delta$. As in binary mixtures, he obtained a
broadened density gap (more evident in the I-cloud-N-shadow boundary) and an enrichment of long rods in the nematic phase. 
Chen \cite{Chen} developed a second-order
perturbation theory without assuming any particular parameterisation. Results were somewhat in disagreement
with those of Sluckin, in that, for small $\Delta$, the biphasic gap was predicted to be narrower than in the
monodisperse fluid, a result attributed to the fact that the theory was valid only for small polydispersity. It might be that this apparent disagreement
is due to the fact that Chen focused on the I-shadow-N-cloud boundary.

Speranza and Sollich \cite{Sollich_i-n1} reconsidered the problem of length-polydisperse hard rods in the Onsager
limit, in an effort to generalise the theory to general size distributions (not necessarily narrow). Also, they
calculated the whole phase diagram and discussed the conditions for the appearance of exotic features such as
three-phase I$-$N$-$N coexistence regions and N--N transitions.
In order to make the problem tractable, Speranza and Sollich
used a spherical harmonic expansion, truncated to second order, of the excluded volume
$V_{\rm exc}(l,l',\hat{\boldsymbol{\Omega}},\hat{\boldsymbol{\Omega}}')=2L_0D_0^2 l l'|\sin\gamma|$,
[with $\gamma$ the angle between the unit vectors $\hat{\boldsymbol{\Omega}}$ and $\hat{\boldsymbol{\Omega}}'$]. 
Within this approximation, the free energy depends only on two generalised one-particle {\it moments}, $n_0$ and $n_2$, 
where
\begin{eqnarray}
n_k=\int_0^{\infty} dl\hspace{0.1cm} l\int_0^{\pi} d\theta\sin\theta  P_k(\cos\theta)\rho(l) h(l,\theta),
\end{eqnarray}
and $\rho(l)$ is the density distribution function scaled by the factor $\pi L_0^2D_0/4$. Speranza and Sollich
used the {\it moment method} to calculate the I--N phase coexistence. In this method the ideal free energy
is projected onto a subspace generated by a finite set of moments. For a truncatable (with respect to the moments) 
excess free-energy density $\Phi_{\rm ex}(\{n_k\})$, the ideal part of free-energy density in reduced thermal units 
\begin{eqnarray}
\Phi_{\rm id}=\int_0^{\infty} dl \int_0^{\pi} d\theta\sin\theta \rho(l,\theta)\left[
\log \frac{\rho(l,\theta)}{h(l)}-1\right],
\end{eqnarray}
is minimized with respect to $\rho(l,\theta)$, with the constraint of 
having fixed values for the generalized moments $n_{k}$. Note that the factor $h(l)$ (the parent size distribution) inside 
the logarithm does not affect the phase behaviour but it is useful to derive the coexistence equations. 
The constrained minimization results in the following expression for the total free-energy density
\begin{eqnarray}
\Phi(\{n_k\})=\sum_{k} \lambda_{k} n_k-n_0
+\Phi_{\rm ex}(\{n_k\}),
\label{moments}
\end{eqnarray}
where $\lambda_k$ are the Lagrange multipliers that guarantee the 
constraints. As can be seen from (\ref{moments}), the total free energy depends 
on the generalized moments $n_{k}$, which can be viewed as 
densities of \emph{quasiparticles} corresponding to a multicomponent mixture. 
Thus, if the excess part of the free-energy density depends on two moments,
the usual thermodynamic formalism for binary mixtures can be applied to calculate the phase coexistence 
between the I and N phases. We should note that the method is exact for the calculation
of coexistence between a phase that occupies a vanishingly small part of the volume (shadow phase)
and another that spans the whole volume (cloud phase). 
For a more formal discussion on the method of moments see Refs. 
\cite{Warren,Sollich_0,Sollich_1,Sollich_2,Sollich_3}.

Speranza and Sollich analysed the consequences of assuming both unimodal and bimodal length distribution functions
in the phase behaviour. The latter was written as a combination of two Schultz distributions of variance $\Delta_0$
centred at two lengths $L_1$ and $L_2$, with $r=L_2/L_1$. The phase diagram was calculated in the 
$\rho_0-\Delta_0$ plane.
For a unimodal distribution, broadening of the density gap between the I and N cloud curves as $\Delta_0$ is increased
was obtained. Also strong fractionation was observed, with the N-shadow phase being more populated by long rods (first moment 
$n_1$ much higher than that in the I-cloud phase). For high enough $\Delta_0$, the zeroth moment (number density) 
of the N-shadow phase becomes lower than that of the I-cloud phase, i.e. the vanishingly small coexisting N phase 
has less rods (although they are longer). Also, three-phase I$-$N$-$N coexistence did not occur in fluids with unimodal (Schultz) length distributions,
according to the truncated Onsager theory.

The results for a bimodal length distribution were compared with those for a binary mixture (for which the distribution function is
a sum of two delta functions) of the same length asymmetry $r$. When the polydispersity $\Delta_0$ is large enough,
the I$-$N$-$N and N$-$N coexistencies disappear, while the reentrant N phase is still present.
The authors concluded that the bimodal distribution function should be sufficiently asymmetric and with two clearly visible peaks for the
three-phase coexistence to exist \cite{Sollich_i-n1}. 

In another work, Speranza and Sollich \cite{Sollich_i-n2} used the same model but with a parent distribution function exhibiting 
a so-called fat tail (i.e. a lower-than-exponential decay at large lengths, which needs to be truncated at a cut-off $l_{\rm max}$ 
to avoid divergences of the mean length in the N-shadow distribution function).

Now the phase behaviour in the $\rho_0-\Delta$ plane is different:
for small $\Delta$ the fluid exhibits the usual I$-$N phase transition, but beyond some value $\Delta^*$ the I-cloud curve 
possesses a kink. At this point the N-shadow curve has a discontinuity, since the coexisting N becomes bimodal with 
a second maximum at $l_{\rm max}$ that moves to lower values as $\gamma_{\rm N}$ is increased. 
Above but close to $\Delta^*$, the fluid exhibits a narrow region of three-phase I$-$N$-$N coexistence,
limited above by I$-$N coexistence. When the cutoff $l_{\rm max}$ is increased, the value of $\Delta^*$ decreases \cite{Sollich_i-n2}.   

\begin{figure}
\epsfig{file=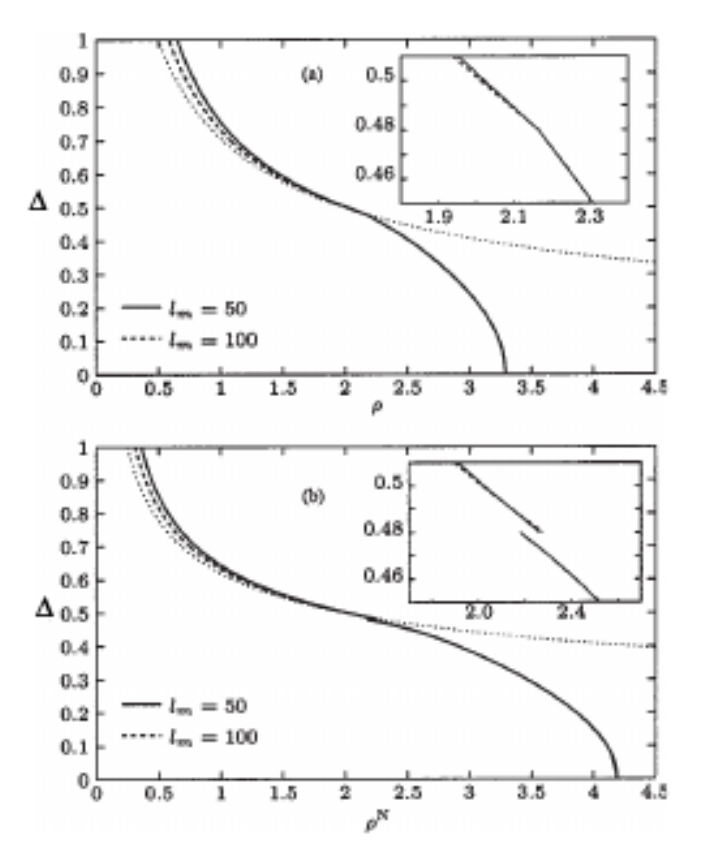,width=3.5in}
\caption{(a) I-cloud and (b) N-shadow curves in the polydispersity vs. scaled density
plane for length-polydisperse rods with Schultz distribution function truncated at $l_{\rm max}=50$ (solid) and 100 (dashed), from \cite{new_Sollich}.
The dotted lines in both figures represent the limiting case $l_{\rm max}\to\infty$. Insets in (a) and (b) show details of 
the main figures about the kink (a) and the discontinuity (b) of the cloud and shadow curves respectively. Reprinted with permission from \cite{new_Sollich}. 
Copyright (2003) by the American Physical Society.}
\label{sollich_fig}
\end{figure}

To confirm the results given by the truncated Onsager theory, Speranza and Sollich \cite{new_Sollich} also 
obtained the exact numerical solution to the I-cloud--N-shadow coexistence equations. The other set, 
the N-cloud--I-shadow equations, 
was not solved due to its inherent numerical complexity. Only unimodal (fat-tailed 
and Schultz) length-distributions were used. The existence of kinks and discontinuities in the cloud and shadow curves
confirmed the existence of three-phase coexistence in a narrow interval of polydispersities for both distributions, while the cloud-point density was seen to decrease to zero as the cutoff length tends to infinity in the case of the fat-tail distribution. Contrary to the 
truncated Onsager model, the Schultz distribution predicts both a kink in the
I-cloud curve (implying I--N--N coexistence), 
and a finite value for the cloud-point density as the cutoff length
increases. Fig. \ref{sollich_fig} is a plot of the I-cloud and N-shadow curves in the 
$\rho_0-\Delta$ plane resulting for the truncated Schultz distributions with two different cut-off $l_{\rm max}=50,100$. 
These results indicate that exponentially-decaying distributions 
separate the functional space of size distribution functions in those that give 
three-phase I--N--N coexistence for finite $l_{\rm max}$ and those which do not. 

\begin{figure}
\epsfig{file=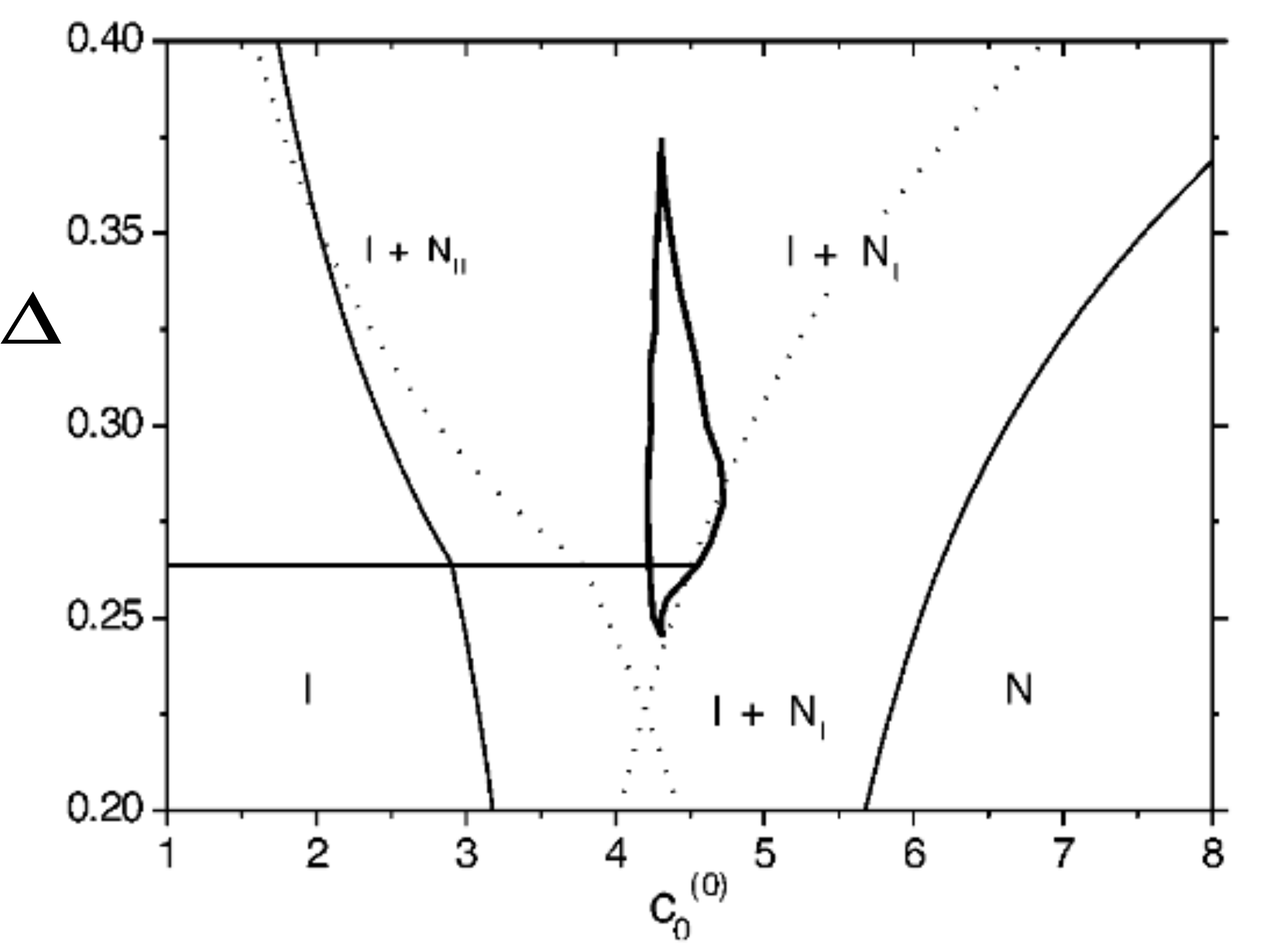,angle=0,width=4.5in}
\caption{Phase diagram in the scaled density $c_0^{(0)}=\rho_0 \pi D_0 L_0^2/4$--polydispersity $\sigma$ plane for length-polydisperse rods with a
log-normal density distribution function truncated at $l_{\rm min}=0.01$ and $l_{\rm max}=10$, from \cite{Wensink_poly}. Thin-solid and dotted
lines correspond to the I--N-cloud and the I--N-shadow curves, respectively. Thick lines enclose the region where one of 
the nematic phases loses stability with respect to the second nematic 
(so that the three-phase I--N$_1$--N$_2$ coexistence should be inside this region). 
The horizontal line represents the polydispersity at which the I-cloud curve exhibits a kink. 
Reprinted with permission from \cite{Wensink_poly}. Copyright (2003), AIP Publishing LLC.}
\label{phd_wensink}
\end{figure}

Although three-phase I--N--N coexistence was confirmed for fat-tail distributions, explicit calculation of coexistence boundaries
was not carried out in the above works. In order to do this, the three-phase coexistence equations have to be solved
well inside the I--N coexistence region. Taking $\gamma_{N_1}=\gamma\neq 0$ ($\gamma_{\rm I}=1-\gamma$) 
and taking $\gamma_{N_2}=0$, the point in the $\rho_0-\Delta$ phase diagram for which there appears a 
vanishingly small fraction of the coexisting N$_2$ phase in the already phase separated I--N$_1$ 
mixture can be calculated. The other point is calculated by setting $\gamma_{N_2}=\gamma\neq 0$ 
and $\gamma_{N_1}=0$. This region was approximately obtained by Wensink and Vroege \cite{Wensink_poly} 
by plotting the pressure $p$ at the two-phase I--N coexistence as a function of $\gamma_N$, searching for the two values 
of $\gamma_{\rm N}^{(i)}$ ($i=1,2$) for which $d p/d\gamma_{\rm N}=0$. When $\gamma_{\rm N}$ is between these values,
the pressure is a decreasing 
function of $\gamma_{\rm N}$, indicating the instability of the original N phase with respect to phase separation 
between two different nematics. The I-cloud and N-cloud 
curves and the phase diagram calculated by Wensink and Vroege is shown in Fig. \ref{phd_wensink}. 
A Gaussian function for the orientational distribution was used in the I--N phase-coexistence calculations, with 
an analytic expression obtained for the free energy in the limit of high order.
The approximate three-phase-coexistence is delimited by two consulate points, the lower one located at a value of polydispersity 
below the kink of the I-cloud curve. Use of a Schulz distribution did not produce any indication for
three-phase I--N--N coexistence. It can be concluded that the calculations with exponential distribution functions
are very sensitive to the approximation used for the angular distribution function.

The study of freely-rotating polydisperse particles other than rods has not been carried out yet due to the inherent numerical
difficulties involved in solving the coexisting equations in the context of the presently-available DFT approaches.
To circumvent this problem, restricted-orientation (Zwanzig) approximations have been used for particles with different shapes.
For example, the FMT-based DFT can be recast into a form where the weighted densities depend only on one-particle weights, which
makes the problem numerically tractable. Within this approximation, 
the bulk behaviour of polydisperse plate-like particles has been analysed \cite{chino,Yuri5,Velasco1}.
Also, bimodal polydisperse mixtures of rods and plates \cite{Yuri6,Yuri7}, 
polydisperse fluids of biaxial board-like particles \cite{Roij1} and parallel hard cylinders \cite{Holyst,Yuri8} have been studied. 
In general, studies of this type aim at calculating the limits of two- or three-phase coexistences between uniform phases, 
and in some cases also their instabilities against spatial density fluctuations with different liquid-crystal symmetries.

The FMT formalism for hard board like, biaxial particles, in the Zwanzig approximation, can be easily generalized
for the polydisperse case. The theory is written in terms of 
the density profiles $\rho_{\mu\nu}^{(\rm i)}({\bm r},\boldsymbol{\sigma})$ ($\mu\neq\nu=x,y,z$), which gives the density 
of particles in the $i$th phase at point ${\bm r}$ with main axis pointing along the Cartesian axis $\mu$ and
secondary axis pointing along $\nu$. The edge-lengths ${\bm \sigma}=(\sigma_1,\sigma_2,\sigma_3)$ are
continuously polydisperse. The particle weights have to be redefined in order to consider the particle biaxiality. 
The constrained minimization of the total free-energy per unit of volume $\beta{\cal F}^{(i)}/V$ corresponding to the 
phase $i$ (in coexistence with the other $m-1$ phases) with respect to the density profiles 
$\rho^{(i)}_{\mu\nu}({\bm r},\boldsymbol{\sigma})$, together with the lever-rule constraint 
(\ref{lever}), provides an equation for these profiles, which is solved numerically. Use of Fourier transforms usually
simplifies the problem. To find the spinodal instability of a uniform phase, say a biaxial nematic phase, with respect to 
nonuniform periodic modulation of a given wave vector ${\bm q}$, one may apply a standard bifurcation analysis. 
Onsager theory has also been used in the Zwanzig approximation to treat hard-board particles in the polydisperse case.

\begin{figure}
\epsfig{file=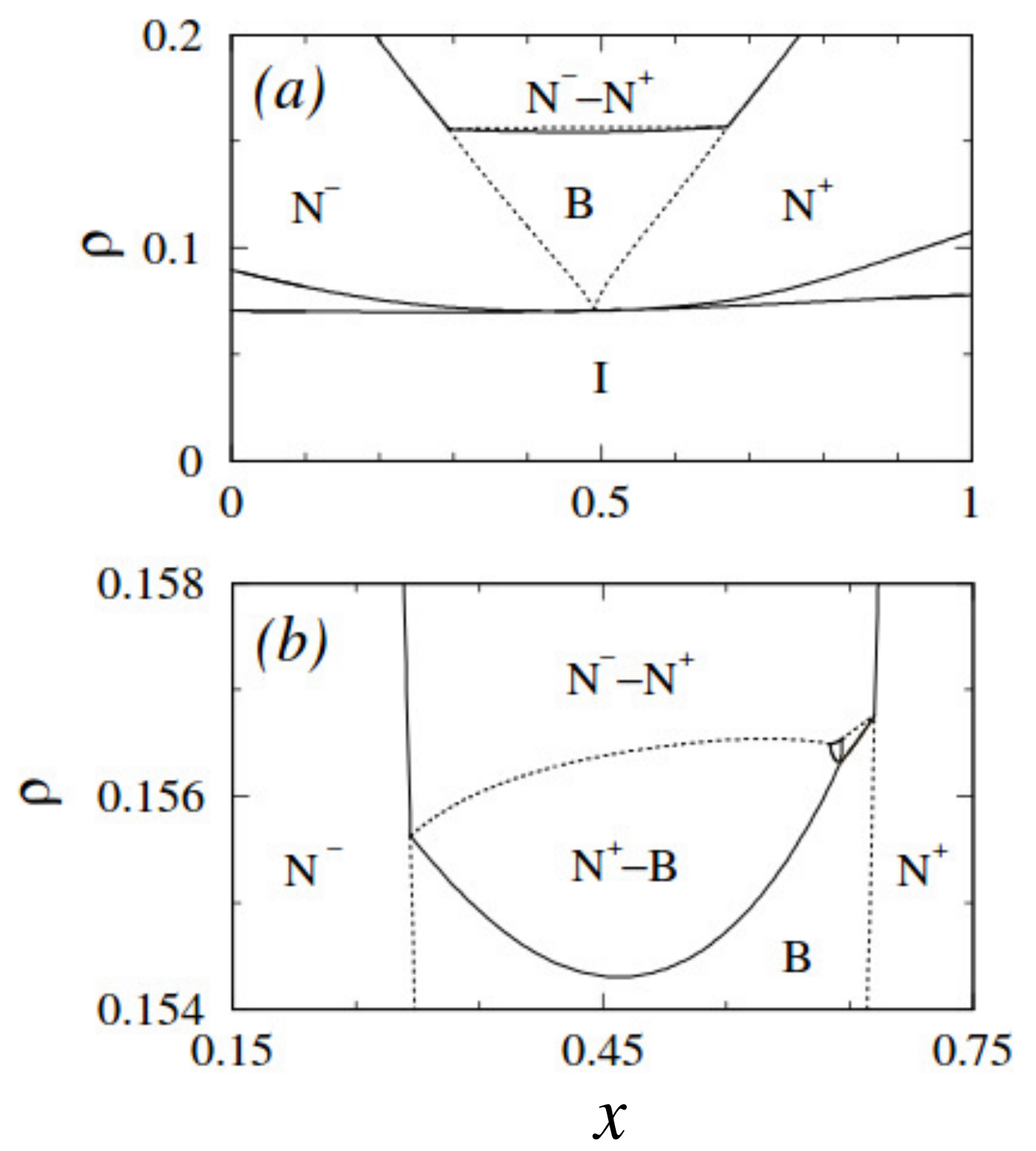,angle=0,width=3.5in}
\caption{(a) $x-\rho$ phase diagram of a polydisperse mixture of Zwanzig rods and plates for $\kappa_0=5$, 
with $x$ the fraction of rods \cite{Yuri6}.
The thickness and width polydispersities are $\Delta_{2/3}=0.610$ and $\Delta_{-1/3}=0.302$ respectively. Different stable phases 
and two-phase coexistences are correspondingly labelled. (b) A detail of the phase diagram shown in (a).
Reprinted with permission from \cite{Yuri6}. 
Copyright (2002) by the American Physical Society.} 
\label{enhanced}
\end{figure}

Using these techniques, some effort has been devoted to the study of the biaxial nematic phase N$_{\rm B}$ in polydisperse fluids.
As mentioned in Sec. \ref{biaxial_section}, binary mixtures of uniaxial hard rods and plates with high enough asymmetry can stabilise
the biaxial nematic phase with respect to the occurrence of N$^+_{\rm U}-$N$^-_{\rm U}$ demixing. The N$_{\rm B}$ phase has not been
observed in experimental rod-plate colloidal suspensions. Although particles can be prepared to closely resemble hard bodies, they are
inevitably polydisperse, not only in size, but also in geometry. In a series of two papers, the 
effect of polydispersity on the stability of the N$_{\rm B}$ phase
with respect to N$^+_{\rm U}-$N$^-_{\rm U}$ demixing has been investigated by Mart\'{\i}nez-Rat\'on and Cuesta \cite{Yuri6,Yuri7}, 
using hard board-like particles of square cross section and dimensions $(L,\sigma,\sigma)$. The volume of particles was set to
unity and the polydispersity variable was the aspect ratio $\kappa=L/\sigma$, which accounts for both size and 
geometry polydispersity, and a bimodal parent probability distribution function was chosen, with peaks at $\kappa_0>1$ (rod sector)
and $\kappa_0^{-1}$ (plate sector). The phase behaviour of the mixture was studied as a 
function of the aspect ratio $\kappa_0$ and polydispersity. The main result of this model is the enhanced stability of the N$_{\rm B}$ 
phase with polydispersity. For example, for $\kappa_0=5$ in the bidisperse limit,
the binary mixture, just above the Landau point, exhibits a N$_{\rm U}^+$--N$_{\rm U}^-$ demixing. However, when some polydispersity 
is added, the N$_{\rm B}$ becomes stable against N$_{\rm U}^+$-N${\rm_U}^-$ demixing \cite{Yuri6} (see Fig. \ref{enhanced}). 
For $\kappa_0=15$, for which the 
binary mixture exhibits a region of N$_{\rm B}$ phase stability, polydispersity gives rise to a rather complex phase diagram.
Above the N$_{\rm B}$ there appears a region of three-phase N$_{\rm U}^+$--N$_{\rm U}^-$--N$_{\rm B}$ coexistence surrounded by a stable 
N$_{\rm B}$ phase (below), a two-phase N$_{\rm U}^-$--N$_{\rm U}^+$ coexistence region (above), a N$_{\rm U}^+$--N$_{\rm B}$ two-phase region
(left) and finally a N$^-$--N$_{\rm B}$ coexistence region (right) \cite{Yuri7}. 
The coexisting phases also exhibit strong fractionation, more pronounced in the shadow coexisting phases. In both studies the 
instability of the uniform phases with respect to density modulations of given symmetries (like C and S) were estimated via 
the calculations of spinodal curves in the $x-\rho$ plane \cite{Yuri6,Yuri7}.

\begin{figure}
\includegraphics[width=8cm,angle=-90]{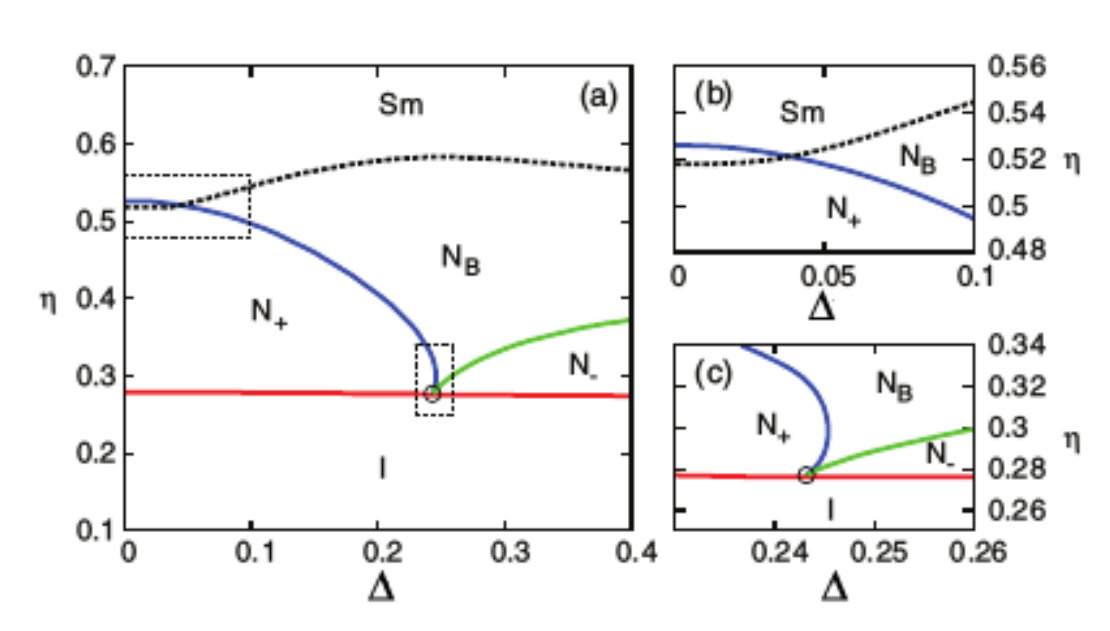}
\caption{(a) Phase diagram in the packing fraction-polydispersity plane of a polydisperse fluid of hard boardlike 
biaxial particles with $L/T=9.07$, $W/T=2.96$ and polydispersity in the length $T$, according to Onsager theory
in the Zwanzig approximation \cite{Roij1}. The solid lines show the boundaries limiting the 
stability regions of uniform phases (I, N$_{\rm U}^{\pm}$ and N$_{\rm B}$, 
accordingly labelled in the figure). The dashed line shows the 
instability of uniform phases with respect to periodic modulations with smectic
symmetry (the smectic phase is denoted Sm in the figure). 
(b) and (c) details of (a) around the crossing 
point between the N$^*$-N$_{\rm B}$ transition curve and the N--S spinodal curve, and (b) around the tetracritical point shown 
with circle. Reprinted with permission from \cite{Roij1}. 
Copyright (2011) by the American Physical Society.}
\label{board_like_biaxial}
\end{figure}

van den Pol et al. \cite{Pol} have recently managed to obtain experimentally the
elusive biaxial nematic phase in colloidal suspensions of mineral board-like biaxial particles, 
polydisperse in the three lengths $L>W>T$ (length, width and thickness, respectively) but with approximately the same 
shape. The effect of polydispersity on the stability of the N$_{\rm B}$ phase has been studied theoretically by Belli 
et al. \cite{Roij1}, focusing on the previous experimental results and using Onsager theory in the Zwanzig approximation.
Belli et al. showed that, under certain conditions, polydispersity can enhance N$_{\rm B}$ stability. 
The conditions are: (i) the shape parameter $\nu=L/W-W/T$ for all particles should be approximately the same and 
close to zero (the value for a perfect particle biaxiality), and (ii) polydispersity is taken on the particle volume. 
To match the experimental conditions, the authors chose $L/T=9.07$ and $W/T=2.96$ (which results in $\nu=0.1$) for all species.
21 species with different Gaussian-distributed lengths $T$, characterised by mean value $\langle T\rangle$ and 
polydispersity coefficient $\Delta$ (again fixed in accordance with experiments), were chosen, and
the phase diagram in the total packing fraction-polydispersity plane was calculated, neglecting the fractionation of 
particles between different coexisting phases (this is justified if phase transitions are continuous or 
of weakly first order). While the one-component fluid exhibits the sequence I--N$_{\rm U}^+$--S (where N$_{\rm U}^+$ is a uniaxial
nematic phase with the longest side of particles pointing on average along the nematic director), polydispersity (see phase
diagram in Fig. \ref{board_like_biaxial}) stabilizes 
the N$_{\rm B}$ phase in a region bounded below and above by the N$_{\rm U}^+$ and S phases, respectively. 
At some particular polydispersity, a tetracritical point appears, where I, N$_{\rm U}^+$, N$_{\rm B}$ and a new N$_{\rm U}^-$ phase   
(with the shortest particle axes pointing in the direction of the nematic director) meet. For larger values of
$\Delta$, the N$_{\rm B}$ phase is bounded below and above by the N$_{\rm U}^-$ and S phases, respectively. 

\begin{figure}
\epsfig{file=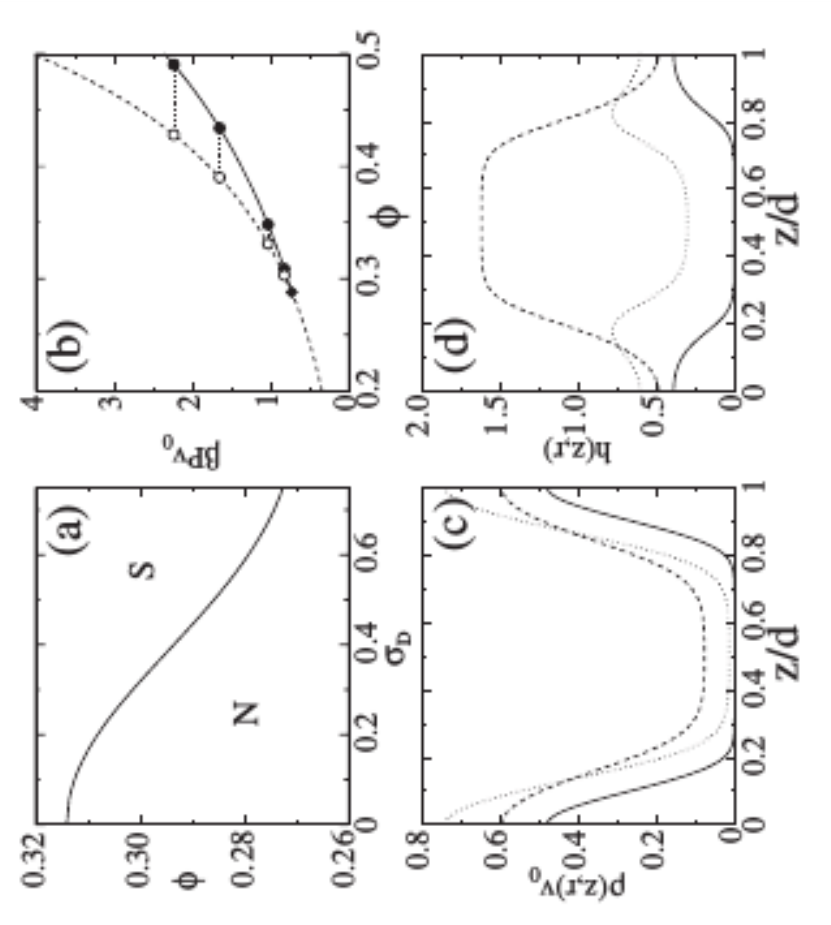,angle=-90,width=4.in}
\caption{Theoretical results for a suspension of length-polydisperse hard cylinders of the same thickness \cite{chino}.
(a) N--S spinodal packing fraction $\eta$ as a function of diameter polydispersity $\Delta$. (b) Equation of state 
(reduced pressure vs. packing fraction) of the N and S phases for $\Delta=0.52$. Open circles represent the relative, 
experimental values of the packing fractions with respect to the bifurcation point, while black circles correspond 
to the S phase with the same osmotic pressure. (c) Density profiles $\rho(z,r)$ as a function of $z$ in reduced units 
for $r=1.5$ (solid line), 0.4 (dashed line) and 0.8 (dotted line). 
(d) Normalized density distribution function, defined in the text, as a function of $z/d$ for the same 
values of $r$ as in (c) and with lines having the same meaning. In (c) and (d) the packing fraction and the period of 
the S phase are $\eta=0.452$ and $d/L_{\rm eff}=1.211$, respectively. Reprinted with permission from \cite{chino}. 
Copyright (2009) by the American Physical Society.} 
\label{las_cuatro}
\end{figure}

The smectic phase is usually found to be stable at high density in colloidal suspensions of close-to-monodisperse 
rod-like particles. It is known from theory and experiment that length-polydispersity in suspensions of particles
with close-to-monodisperse widths can destroy smectic ordering and favour columnar ordering \cite{Lekker_Nature}. 
Intuitively one would expect that
plate-like particles of approximately constant thickness will not form the columnar phase if the polydispersity in diameter
is sufficiently high, and that smectic ordering will result instead. The elusive smectic phase was recently obtained in
colloidal suspensions of mineral plate-like particles made of $\alpha$-Zirconium phosphate \cite{chino}. These particles are
completely monodisperse in thickness $L$, while their polydispersity in diameter $\sigma$ is high, and
interact through a complex long-range repulsive interaction due to the surface charges and solvent effects. 
The suspensions
exhibit a first-order I--N transition with a wide density gap at low densities and a continuous or possibly
weak first-order N--S transition at high density \cite{chino}. A DFT in the Zwanzig approximation was used by 
the same authors to 
rationalize these experimental findings, considering polydisperse hard cylinders of constant effective thickness $L_{\rm eff}$ 
and mean diameter $\sigma_0$. A second virial approximation was used to study the I--N transition, as the experimental
mean aspect ratio $L/\sigma_0$ of the particles was very small. The effective thickness $L_{\rm eff}$ was chosen to
match the volume fraction $\eta$ of the experimental I-cloud point.
For diameter polydispersities $\Delta$ between 0.3 and 0.5, both the density gap and the curvature of the 
function $\gamma_{\rm N}(\eta)$ (the fraction of the sample volume occupied by the N phase as a function of the packing fraction) 
were correctly estimated \cite{chino}. As regards the N--S transition, a parallel hard-cylinder model treated 
with a FMT approach, suitably extended to consider polydispersity in diameter, was used.
The parallel alignment approximation is justified here due 
to the high nematic ordering of colloidal particles at packing fractions close to that of the N--S transition. Setting $\Delta$   
to the experimental value, equations of state for nematic and smectic phases were calculated. The N--S transition was found 
to be continuous, while the packing fraction at the transition turned out to be a decreasing function of $\Delta$ [see Fig. 
\ref{las_cuatro} (a)]. The experiment provided the value of packing fraction
$\eta_0^*$ at which there first appears a small amount of
smectic. Slow kinetics prevents from observing fully developed smectic domains when the sample packing fraction $\eta_0$ 
is increased, and the smectic packing fraction $\eta_S$ as a function of the sample packing fraction could not be directly
measured. However, it was indirectly estimated from the theoretical equations of state for the nematic and smectic phases,
Fig. \ref{las_cuatro}(b). The ratio $L_{\rm eff}/L$ was determined as a function of $\eta_0$ from the 
knowledge of $L/d$ from experiments (with $d$ the 
smectic period) and $d/L_{\rm eff}$ given from the theory. The extrapolation of $L_{\rm eff}/L$ as a function 
of $\eta_0$ to the value for the I--N transition provides a value almost identical to the
one giving the best fit of the curve $\gamma_{\rm N}(\eta_0)$. 
Finally, the theory predicted a strong microsegregation in the S phase, 
with maxima of the local fraction of wider particles located at the layers and maxima of small particles located at interstitials. 
This is shown in Fig. \ref{las_cuatro}(d), where this effect is shown through the function
$\displaystyle h(z,r)=\rho(z,r)/\int_0^{\infty} dr \rho(z,r)$,
where $r=\sigma/\sigma_0$ and $\rho(z,r)$ is the smectic density profile of species with relative diameter $r$. 

Mejia et al. \cite{Mejia} performed further studies on the above experimental system in order to study more systematically the  
effect of polydispersity on the two-phase I--N coexistence. Samples with different polydispersities and mean aspect ratios 
were prepared, and phase diagram in the total packing fraction, $\eta_0$, vs. fraction of the sample volume occupied by the N 
phase, $\gamma_{\rm N}$, plane were calculated. The more polydisperse sample exhibits a huge coexistence gap and strong non-linearity 
in the function $\gamma_{\rm N}(\eta_0)$, results which were rationalised by Mart\'{\i}nez-Rat\'on and Velasco \cite{Yuri5} 
using a FMT-based theory for polydisperse oblate parallelepipeds in the Zwanzig approximation. A value for the effective 
aspect ratio $L_{\rm eff}/\langle\sigma\rangle$ was chosen to account for the effective repulsive interactions 
between platelets, and the parent size-distribution function was chosen to be unimodal or bimodal. 
While samples with small polydispersity are well described by a unimodal distribution, large polydispersities require
\emph{marginally} bimodal distributions (i.e. distributions made from two overlapping peaks such that no second maximum
is visible). Bimodal distributions with two clearly separated maxima
can give rise to a loop in the function $\gamma_{\rm N}(\eta_0)$. Thus the fraction of total 
volume occupied by nematic exhibits a non-monotonic behaviour, not related to the presence of three-phase I--N--N coexistence,
a genuine prediction of the model which is expected to be confirmed in future experiments.  

\begin{figure}
\epsfig{file=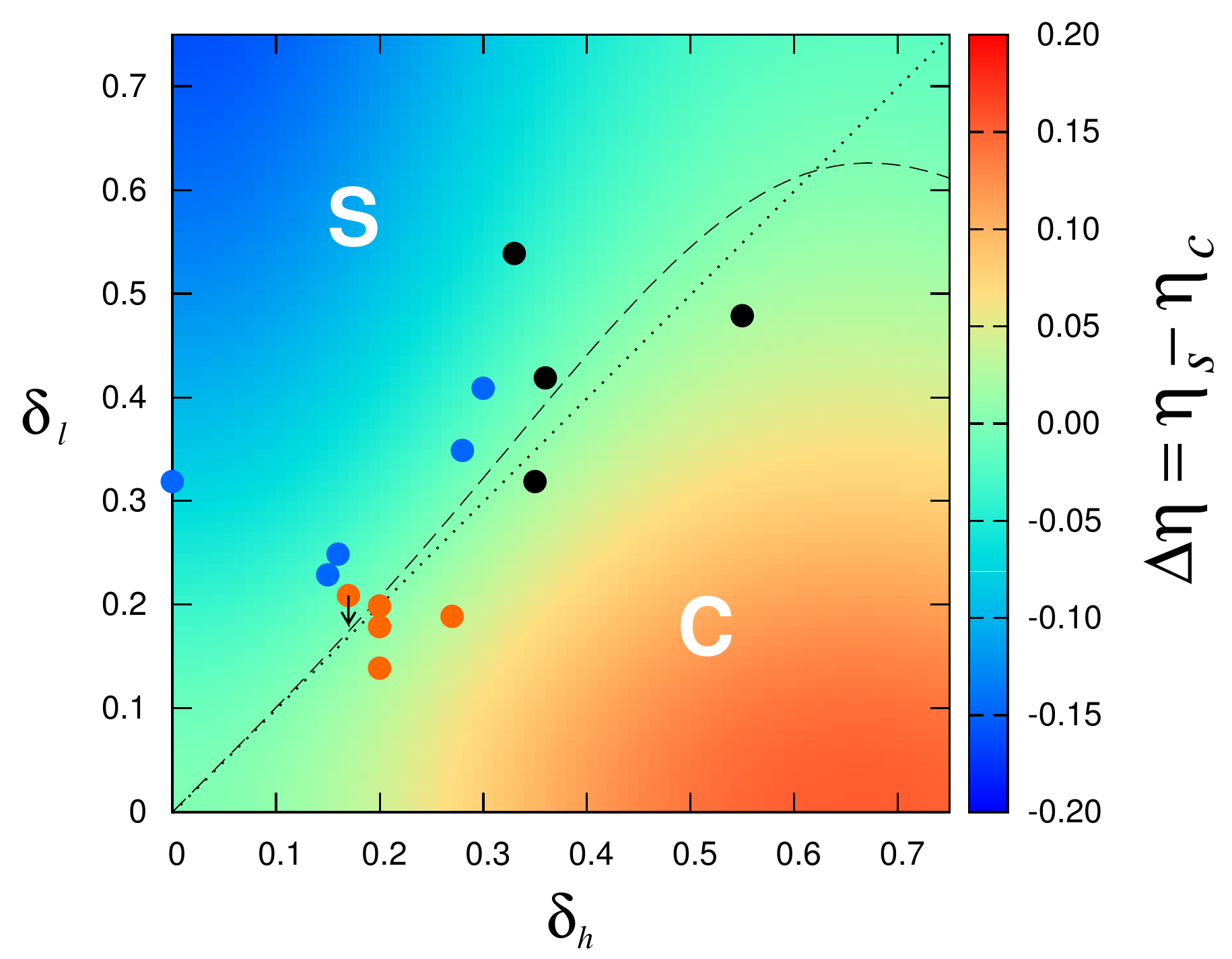,width=5.5in}
\caption{Regions of stability of the S (blue) and C (orange) phases as obtained from a bifurcation analysis of
a fundamental-measure theory of polydisperse hard platelets \cite{Velasco1}. The 
difference $\Delta\eta=\eta_s-\eta_c$ between the packing fractions of S and C bifurcations from the N phase is plotted
as a function of polydispersities in side-length $\delta_l$ and thickness 
$\delta_h$ in colour code (scale bar at right). Blue, orange and black 
circles represent the values of polydispersities corresponding to experimental samples in the S (blue), C (orange) and S--C (black) 
phases. Reproduced from Ref. \cite{Velasco1} with permission of the PCCP Owner Societies.} 
\label{colors}
\end{figure}

As mentioned before, synthetic mineral particles cannot be made completely monodisperse in all their characteristic lengths. For example,
in studies on colloidal platelets \cite{Lekker_Nature}, particles are usually highly polydisperse in both diameter and
thickness and, to our knowledge, only in one case \cite{chino} has the platelet thickness been perfectly controlled to exactly
one monolayer. Diameter-polydispersity in platelets of equal thickness can destabilise the columnar phase
with respect to the smectic phase, while rod-like particles exhibit the opposite behaviour. The 
effect of length-polydispersity on the relative stability of smectic and columnar phases in fluids of hard parallel
cylinders was studied through coexistence calculations \cite{Holyst} and bifurcation analysis \cite{Yuri8}. 
However, realistic models meant to reproduce experimental results must take into account both polydispersities, a programme
recently followed by Velasco and Mart\'{\i}nez-Rat\'on \cite{Velasco1}. These authors considered hard board-like particles
of square section $l\times l$ and thickness $h$, with main axes along a common direction, and polydisperse in both lengths 
but with decoupled size distribution. A FMT theory, extended for a general polydisperse fluid, was adopted.
Fixing the two polydispersity coefficients $(\delta_l,\delta_h)$, bifurcation theory was used to obtain the N--(C,S) bifurcation 
packing fractions $\eta_{\rm C,S}$, complemented with free-energy calculations. Results are plotted in Fig. \ref{colors}, with
symbols representing experimental data collected from the literature. 
Colours of symbols represent the nature of the different phases obtained in experiments (blue for S, orange for 
C and black for S--C coexistence). The model correctly describes the interplay between C and S phase 
stability as a function of both polydispersities. Free-energy relative differences between smectic and columnar
phases confirm the scenario obtained from bifurcation theory: at high enough density 
and for $\delta_h\gg \delta_l$, the C is more stable than the S phase, while for $\delta_l\ll \delta_h$   
their relative stability changes. There are values of $(\delta_l,\delta_h)$ 
for which the two branches cross at a density  $\rho_0^*$ larger than the two bifurcation points: for $\rho_0<\rho_0^*$ the 
smectic is more stable, with the opposite behaviour for $\rho_0>\rho_0^*$. Finally, the microsegregation mechanism already
observed in the smectic phase was also confirmed for the stable C phases: particles located at the sites of the
square lattice have a side-length distribution similar to the global one, while particles at interstitial points 
exhibit a size distribution centred at much smaller values.

\begin{figure}
\includegraphics[width=11cm,angle=0]{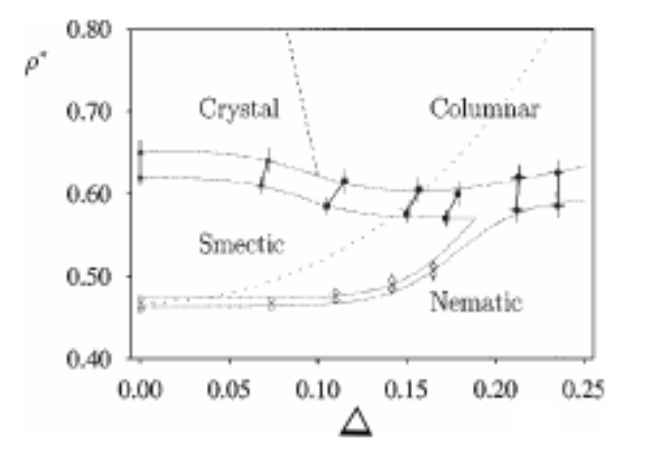}
\caption{Phase diagram in the scaled density $\rho^*$ versus polydispersity coefficient $\Delta$ for the
free-rotating HSC model, as obtained from simulation \cite{Bates1}. Reprinted with permission from \cite{Bates1}. Copyright (1998), AIP Publishing LLC.}
\label{fig_Frenkel_poly}
\end{figure}

Computer simulations of polydisperse systems are technically difficult to perform, and consequently there are very few studies.
In 1998 Bates and Frenkel \cite{Bates1} studied the effect of
length-polydispersity on the phase behaviour of freely-rotating HSC in
the limit of infinite aspect ratio. Their phase diagram in the density versus polydispersity coefficient
plane is represented in Fig. \ref{fig_Frenkel_poly}. The authors found that when the
polydispersity parameter $\Delta$ is less than 0.08, the
phase behaviour is unchanged with respect to the one-component case,
namely the phase transition sequence N--S--K is obtained. For
polydispersities in the range $0.08<\Delta<0.18$ the window of stability of
the S phase shrinks, while those of the N and C phases (bracketing the
smectic region) become wider. Finally, for
$\Delta>0.18$, the S phase is not longer stable, since rods have a high dispersion in length and
they are unable to arrange into periodic layers, and the N phase exhibits a
direct first-order transition to the C phase, which seems to be stable
up to close packing. The same authors studied the I--N transition in a
fluid of hard freely-rotating platelets of vanishing thickness \cite{Bates2}.
Semi-grand canonical Gibbs ensemble simulations were performed to study particle
fractionation as a function of diameter-polydispersity. Larger
particles tend to populate the coexisting N phase, while the I phase is richer
in small platelets. Also, large platelets are more orientationally
ordered than small platelets in the N phase.

\subsection{2D fluids}

Two-dimensional systems have attracted a lot of interest for a long time, mainly because of their peculiar properties
relating to their low dimensionality. Also, these systems are a simple realisation of real materials forming
monolayers adsorbed on a substrate or self-assembled structures on fluid interfaces. 2D fluids may also serve as
models for biological or synthetic membranes and the understanding of the assembling properties of molecules of
different shapes on a surface is becoming an increasingly important topic in condensed-matter physics and
cross-disciplines such as physical chemistry and biophysics.

The peculiar properties of 2D crystals and the associated liquid-crystal freezing transition (existence of bond-orientational 
order and the hexatic phase, absence of true long-range positional order, etc.) for fluids made of
isometric or effectively isometric particles
have their counterpart in mesophases possessing orientational order. There are indications that the I--N transition
in 2D fluids might be governed by a Kosterlitz-Thouless mechanism, similar to the one thought to operate in the liquid-hexatic
transition of an isotropic fluid. Since the order parameter of a nematic is
a tensor quantity, there are subtle questions on the existence of order and its dependence on the dimensionalities
of the order parameter matrix and of the physical space. A good introduction to these topics can be found in the book by 
Lubensky and Chaikin \cite{Lubensky}. Another important difference with respect to the 3D case is the apparent nonexistence 
of smectic order in 2D, which is probably associated with the effect of two-dimensional fluctuations on the emergence of 
spatial order along one direction.

The question on the existence and nature of orientational order in 2D fluids is an old one \cite{Straley_2D}. 
Evidence based on simulations \cite{Frenkel-Eppenga_2D} and experiments \cite{Winkle} indicate that, similar to spatial 
order in 2D crystals, true long-range orientational order does not 
exist in 2D nematics and that $D=2$ is the lower critical dimensionality in nematics. 
As many other 2D systems, the nematic state presents anomalously large thermal fluctuations which result in a highly
fluctuating nematic director. Assuming that 
the free energy can be described by a Frank-type elastic model in the one-elastic-constant approximation,
\begin{eqnarray}
F_e=\frac{1}{2}\int_A d^2r \left\{K_1\left(\nabla\cdot\hat{\bm n}\right)^2+K_3\left|\nabla\times\hat{\bm n}\right|^2\right\}=
\frac{1}{2}K\int_A d^2r\left|\nabla\theta\right|^2,
\end{eqnarray}
[where $\hat{\bm n}=(\cos{\theta},\sin{\theta})$], the fluctuations in $\theta$, the director tilt angle with respect to a
fixed reference direction, would depend on the number of particles $N$ as $\left<\theta^2\right>\sim \log{N}$, with a vanishing order
parameter in the thermodynamic limit, $q_1=\left<\cos{2\theta}\right>\sim N^{-kT/2\pi K}$, and an
orientational correlation function $g_n(r)=\left<\cos{[n\theta(r)]}\right>\sim r^{-n^2kT/2\pi K}$ that would
decay algebraically rather than presenting long-range order. All of these results imply that, strictly speaking, the ordered nematic
phase does not exist in the thermodynamic limit $N\to\infty$, although the dependence is slow and even large nematic samples, or confined nematics,
will be well ordered. This property, stemming from fluctuations, puts into question the very existence of the usual Frank elastic constants 
\cite{Straley_2D}, which should be regarded to adopt values renormalised by fluctuations \cite{Nelson}. 

Concerning the nature of the I--N transition, an interesting property of 2D nematics is that, for apolar particles exhibiting head-tail symmetry, 
a general Landau expansion in the order parameter $Q$ cannot depend on its sign, so that the I--N transition can be of first or
second order; this is in contrast with 3D nematics, for which the Landau expansion contains cubic terms which drive the transition
to be first-order (although weak as seen experimentally). Simulations of hard ellipses \cite{Cuesta_Frenkel} have indeed found the presence of a tricritical
point separating first-order from continuous I--N transition as a function of aspect ratio.

\begin{figure}[h]
\includegraphics[width=13cm,angle=0]{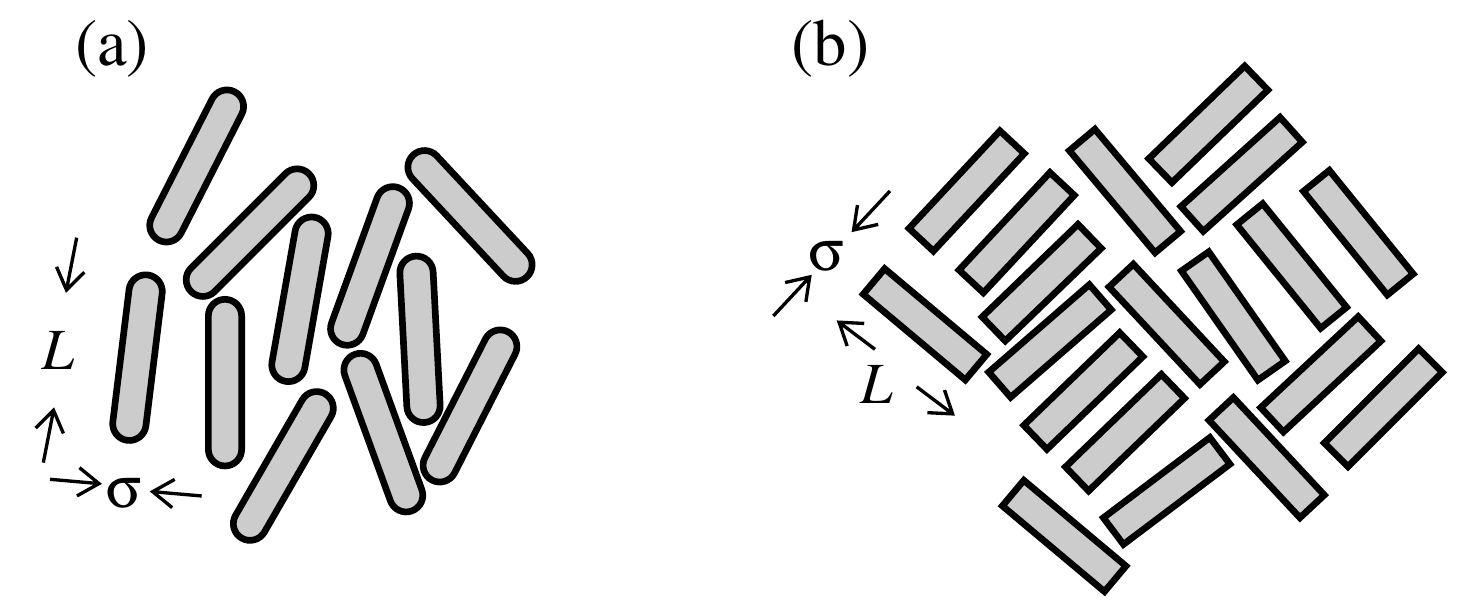}
\caption{Typical models for two-dimensional fluids made of anisotropic
particles. (a) Hard discorectangles, with $L$ the cylinder length and
$\sigma$ the width (note that the total lenght is $L+\sigma$). (b)
Hard rectangles of length $L$ and width $\sigma$.}
\label{tetratic}
\end{figure}

As in many other ordering phenomena in condensed matter physics, the role played by hard-body models in the investigations of orientational 
order in 2D has been crucial. Fig. \ref{tetratic} shows two of the more popular particle models.
The orientational transition in 2D was initially explained using entropy arguments. 
Straley \cite{Straley_2D} investigated Onsager theory for a 2D nematic and found it to predict an I--N transition. However, there is
an important subtlety. The argument 
Onsager used to truncate the virial expansion at second order for 3D rods is not valid in 2D since virial coefficients higher
than the second are much more important in 2D than in 3D. Rigby \cite{Rigby} calculated higher-order virial coefficients for
2D isotropic fluids made of elliptical and discorectangular particles with a wide range of aspect ratios, including the hard-needle
limit. He reported negative values for $B_4$ and $B_5$ for large elongations but, in the case of ellipses, $B_5$ becomes negative
already for a length-to-breadth ratio of 1:6. Therefore, Onsager theory is not strictly valid even in the long-particle limit, and
results should always be taken with caution.

The first evidence by simulation of nematic ordering, due to Vieillard-Baron \cite{VB}, was in fact in a 2D fluid. But the results were not
conclusive. Later Frenkel and Eppenga \cite{Frenkel-Eppenga_2D} studied the same system and concluded that
the nematic phase exhibits quasi-long-range order (with algebraically-decaying orientational correlations). 
Cuesta et al. \cite{Cuesta_Baus} used density-functional theory to describe the I--N and the freezing transitions. 
The theory was an extension of one used previously for hard ellipsoids. For uniform phases it has the same structure as the 
PL theory for 2D fluids, but at low densities it does not reproduce the exact second-virial coefficient, which is especially
problematic for ellipses with large aspect ratio. 

\begin{figure}[h]
\includegraphics[width=12cm,angle=0]{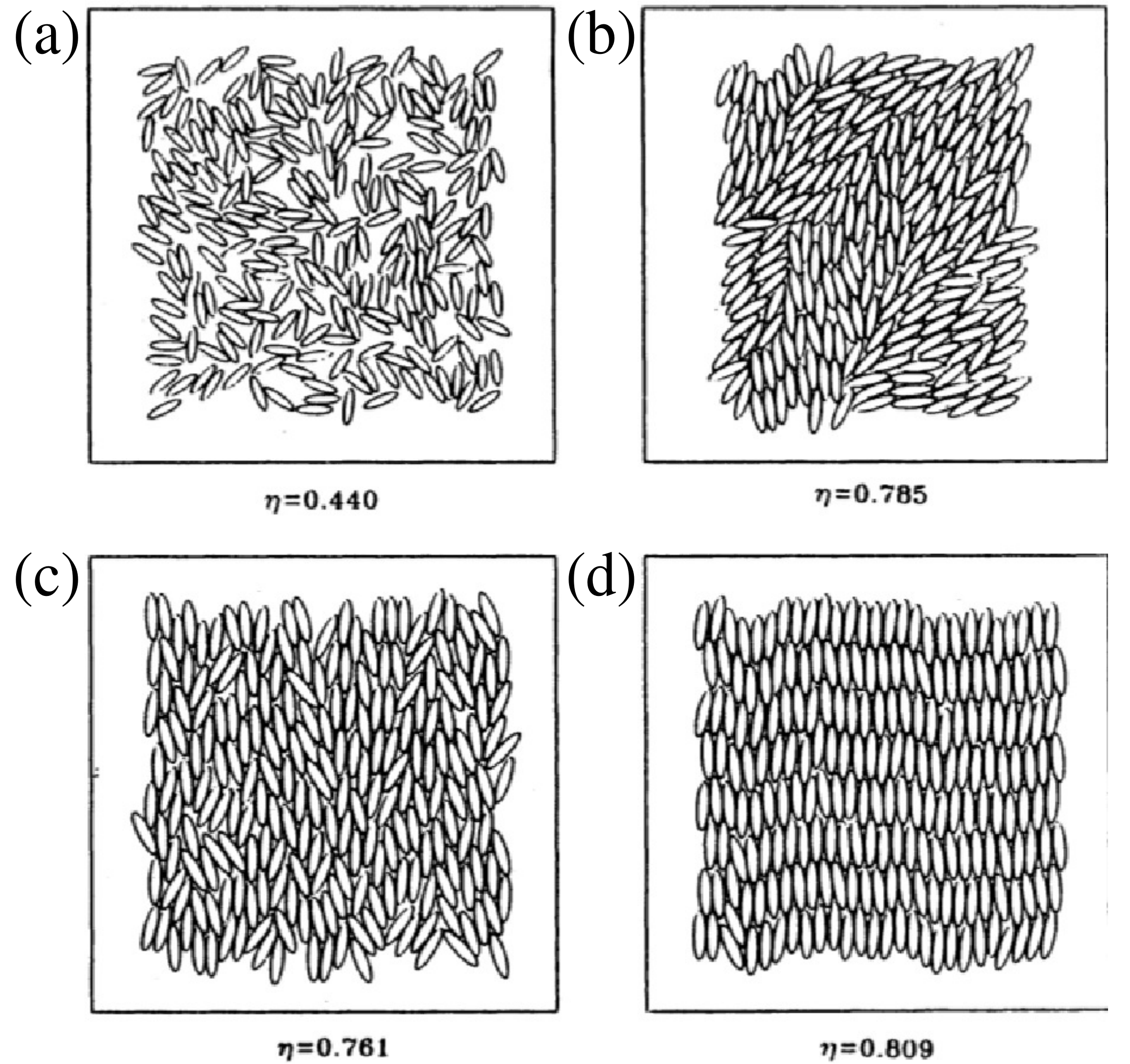}
\caption{Snapshots of typical particle configurations of hard ellipses in different phases from simulations by Cuesta and 
Frenkel \cite{Cuesta_Frenkel} for aspect ratio $\kappa=4$. (a) Isotropic phase, at packing fraction $\eta=0.440$; 
(b) unstable nematic phase, $\eta=0.785$, obtained by compression from the isotropic phase; 
(c) stable nematic phase, $\eta=0.761$; (d) solid phase, $\eta=0.809$. The last two states were obtained by expanding 
a solid phase. Reprinted with permission from \cite{Cuesta_Frenkel}. 
Copyright (1990) by the American Physical Society.}
\label{Cuesta_Frenkel}
\end{figure}

In more extensive work, Cuesta and Frenkel \cite{Cuesta_Frenkel}
investigated the transition by constant-pressure MC simulation. In these simulations the N phase appeared to be stable for 
aspect ratios $\kappa>2$, with a crystal phase at high density. The order of the transition seemed to change from first-order to continuous 
as the aspect ratio increased,
pointing to the existence of a tricritical point. For $\kappa=6$ the transition proceeded via disclination unbinding. Cuesta and Frenkel
also proposed a rescaling of the excluded volume to remedy the deficiencies of the Cuesta et al. theory \cite{Cuesta_Baus}, and obtained a 
correct second-virial coefficient. Comparison with the MC results was satisfactory for both the equation of state and the location of
the transition. Fig. \ref{Cuesta_Frenkel} shows particle configurations of the system at 
different packing fractions for the case $\kappa=4$, where isotropic, nematic and solid phases are stabilised. The 
configuration corresponding to the unstable nematic phase, panel (b), shows the presence of nematic domains, which herald
the formation of a stable nematic phase at higher densities. It is also
clear from the configuration pertaining to the solid phase that particles are not very localised and large positional
fluctuations exist. One important point is that the computer simulations have not found a smectic phase in fluids
of hard convex bodies, \cite{Frenkel-Eppenga_2D,Cuesta_Frenkel,HDR_Bates}, which is an important difference with respect 
to the 3D case 
and could be associated with the reduced dimensionality and the amplification of spatial fluctuations that destroy smectic order. 
2D smectics have been observed, though, in adsorbed monolayers of molecules on graphite \cite{graphite}.

Bates and Frenkel \cite{HDR_Bates} investigated in more detail the nature of the continuous isotropic-nematic transition using
Monte Carlo simulation of a fluid of hard discorectangles. Analysis of the
orientational correlations, which decay algebraically (quasi-long-range order) with distance, seemed to confirm a 
Kosterlitz-Thouless scenario for the continuous transition, governed by a disclination-unbinding mechanism. For rods of 
low aspect ratio the isotropic phase changes to a crystal phase via a first-order transition. The isotropic phase exhibits 
strong short-ranged particle correlations in position and orientation. More recently, 
Khandkar and Barma \cite{Khandkar} performed  grand canonical Monte Carlo of hard needles, and confirmed the quasi-long-range 
nature of the nematic phase.

The Onsager theory for 2D hard needles was investigated in more detail by Chrzanowska \cite{Chrzanowska}. She made a careful
analysis of the theory based on bifurcation theory, and a detailed comparison of the order parameter and the thermodynamics with the 
available computer simulations 
of Frenkel and Eppenga \cite{Frenkel-Eppenga_2D} and Chrzanowska and Ehrentraut \cite{Chrzanowska0}, and from her new MC data
\cite{Chrzanowska}. The agreement is very good, which is attributed to two causes: the small system size of the simulations, which
prevents the appearance of director distortion and point defects, and the fortuitous cancellation of virial coefficients due to
some of them being negative. Also, it was conjectured that the disclination unbinding mechanism associated with the
Kosterlitz-Thouless transition observed in the simulations takes place after the transition, i.e. in the nematic region.

Extended Onsager theories à la Parsons-Lee have been proposed for hard rods in 2D. Varga and Szalai \cite{Varga_Szalai} studied the
isotropic-nematic transition for hard ellipses using two versions of PL theory. In one, the reference fluid was taken to be the
hard-disc fluid, whereas in the other the isotropic phase of hard ellipses was used. In general, PL theory gives reasonable agreement
with MC data, but the order of the transition is not correct in the narrow range of aspect ratio about $\kappa=4$.
A PL theory for hard discorectangles was also proposed by de las Heras et al. \cite{elas_2D}, using parallel hard ellipses
as a reference fluid. Results were compared with the MC simulations of \cite{HDR_Bates} for $\kappa=L/D=15$, and very good agreement was found 
for the equation of state. In the case of the transition density the results were not in such an agreement, since the theory predicted
an isotropic-nematic transition at a packing fraction $\eta=0.257$, compared with $0.363$ from the simulations. However, there is an
interesting observation. de las Heras et al. also calculated the two elastic constants of a 2D nematic (splay $K_1$ and bend $K_3$), using the
same theory and expressions for the elastic constants similar to those for the 3D nematic. For $\kappa=15$, the ratio $K_1/K_3$ decreases
very quickly with density as the system goes into the nematic phase, which shed doubts about results based on the common one-constant
approximation $K_1=K_3$ (note that there are no estimations of elastic constants in 2D based on simulations). Also, Bates and Frenkel 
\cite{HDR_Bates} estimated the transition density by assuming a Kosterlitz-Thouless-type
transition and applying the condition $K_c=8kT/\pi$ for the value of the renormalised constant. Taking $K=(K_1+K_3)/2$ and values from the
DFT calculation, one obtains $\eta=0.36$, in good agreement with the simulation value. 

Hard ellipses and hard discorectangles have topologically similar phase diagrams, with I, N and K phases. The character of
the I--N transition seems to depend delicately on the particle geometry and particle aspect ratio. However, beyond the
type of transition, there are important subtleties
associated with the particle shape. One example is the formation of oriented phases with different symmetries, and 
recently interest has focused on the phase behaviour of particles with shapes different from the simple
ellipsoidal or discorectangular shape. One of the most studied is the rectangular shape, which may be considered as
a deformation of the ellipsoidal shape, given on the $xy$ plane by $(x/a)^n+(y/b)^n=1$ with $n=2$, in the limit when 
the exponent $n\to\infty$. The hard-rectangle model has received a lot of attention due to the possibility that the so-called
{\it tetratic} phase might be stabilised. The tetratic phase is a nonstandard nematic phase possessing two, instead of
one, equivalent and mutually orthogonal directors, and it is equivalent to the cubatic phase in 3D. Therefore, the tetratic
phase has four-fold symmetry, even though the particles possess two-fold symmetry.
In the tetratic phase the angular distribution function $f(\varphi)$,
with $0\le\varphi\le 2\pi$, exhibits peaks not only at $\varphi=0$ and $\pi$, but also at $\pi/2$ and $3\pi/2$, and
all of these directions are equally probable. The formation of these nontrivial arrangements in self-assembled 
suspensions of colloidal rectangular particles could be important in relation with the creation of patterned templates
which could give rise to new layered materials of technnological importance \cite{Fichthorn}. Therefore, the 
theoretical description of spontaneous ordering derived from packing (entropic) problems in nontrivial structures is a very 
relevant problem and the use of hard models is very adequate. 

The first indication for the existence of a tetratic phase in fluids of rectangular particles came in 1998 from the work 
by Schlacken et al. \cite{Schlacken}, who used SPT theory to study the orientational transition in fluids of 
hard ellipses and rectangles. For the case of ellipses they discussed the results for the equation of state and the
transition densities and compared with the existing Monte Carlo results for ellipses of Cuesta and Frenkel \cite{Cuesta_Frenkel}.
In the case of hard rectangles they used the same theory, but with a different excluded area, and a tetratic phase was
found to be stable for low aspect ratios, in a small region is the density-aspect ratio plane. The phase diagram was found to 
depend on the value of aspect ratio. For $\kappa\equiv L/\sigma>5.44$ (with $L$ the length and $\sigma$ the width of the rectangle) the 
isotropic fluid was predicted to change to a uniaxial nematic phase via a continuous phase transition. However, 
for $2.62<\kappa<5.44$ the transition changed to first-order. Finally, for $\kappa<2.62$ a direct (continuous) transition from the isotropic
phase to the tetratic phase was predicted to occur at a rather high density. At even higher densities the tetratic phase was conjectured 
to exhibit a second continuous transition to the uniaxial nematic phase (no explicit spinodal calculations were performed). 

The reason why the tetratic phase can be stabilised lies in the existence of sharp corners in the particle shape.
We do not expect to find tetratic phases for particles such as ellipses or discorectangles (the 2D version of spherocylinders).
The reason is that there is a fundamental difference between rectangles and models with smooth corners. This difference can be 
appreciated by looking at the angular dependences of the excluded area, $v_{\rm exc}(\varphi)$, of the two models, 
with $\varphi$ the relative angle between the long axes of two particles. In the case of hard discorectangles there are
minima at $\varphi=0^{\circ}$ and $180^{\circ}$, but for hard rectangles there are additional, relative minima at $\varphi=90^{\circ}$
and $270^{\circ}$, which the fluid can use in order to compensate for the loss of orientational entropy. 
The existence of tetratic order was ruled out for hard discorectangles in a Monte Carlo study by Bates and Frenkel \cite{HDR_Bates}.

Wojciechowski and Frenkel \cite{Frenkel_Polaco} studied
the hard-square fluid in 2004. The fluid did not appear to freeze from the isotropic phase, but
an intermediate phase with quasi-long-ranged (algebraic) orientational order and no positional (faster than 
algebraic) order appeared; this phase corresponds to a tetratic phase, although it cannot be considered a proper
liquid-crystal phase since the particle does not differentiate the usual nematic phase from the tetratic phase.

\begin{figure}[h]
\includegraphics[width=9.5cm,angle=0]{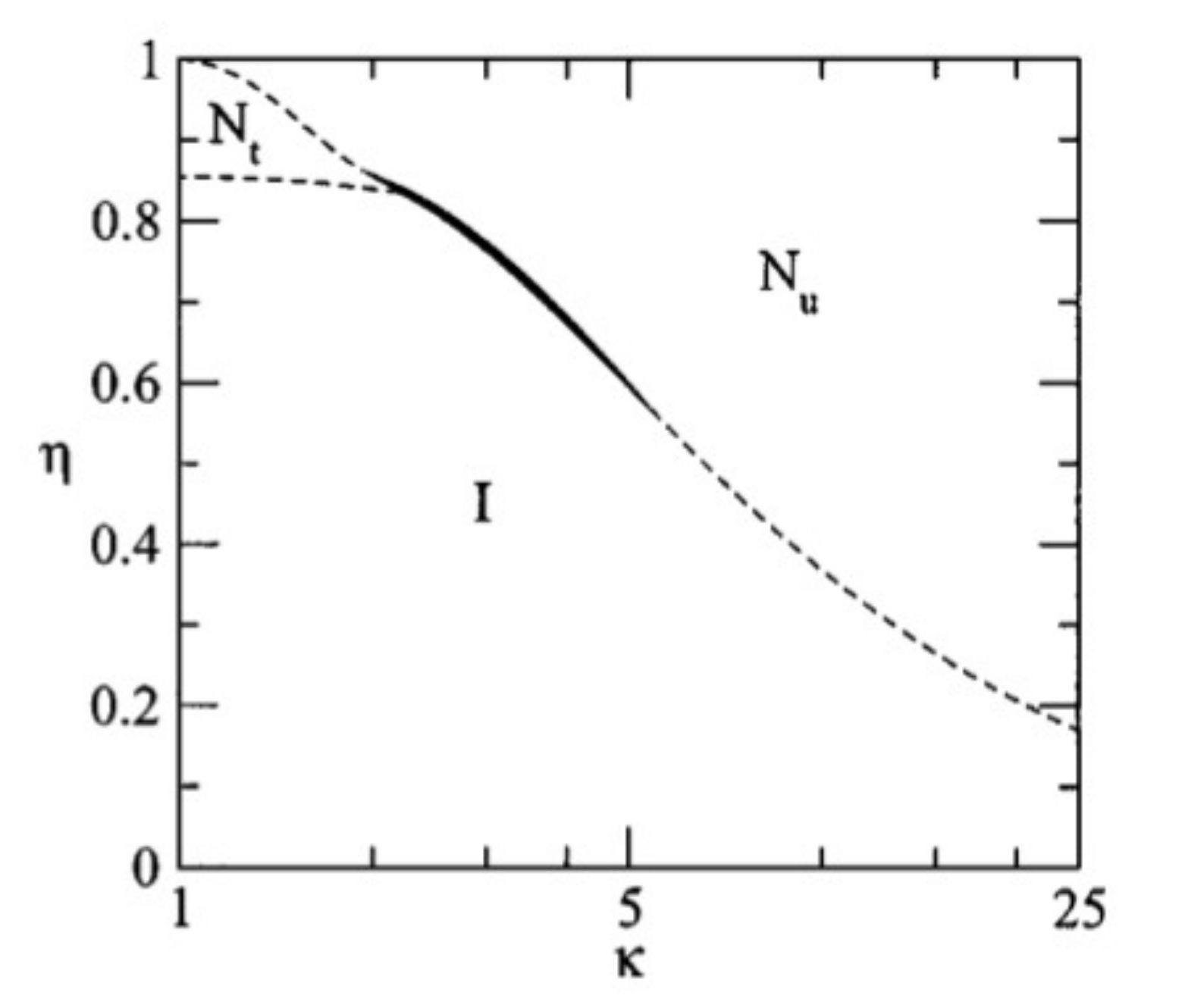}
\caption{Phase diagram of the hard rectangle system in the packing 
fraction-aspect ratio plane as obtained by Mart\'{\i}nez-Rat\'on et al. 
\cite{Tetratic_MRVM} using scaled-particle theory. Reprinted with permission from \cite{Tetratic_MRVM}. Copyright (2005), AIP Publishing LLC.}
\label{Phase_diag_2D}
\end{figure}

A year later, Mart\'{\i}nez-Rat\'on et al. \cite{Tetratic_MRVM} performed a detailed study of the ordering properties of
the hard discorectangle and the hard rectangle models using, as in the work of Schlacken et al. \cite{Schlacken},
the SPT theory to describe spatially uniform phases, but extending and correcting the work of the latter authors.
However, this theory predicts tetratic ordering at rather high values of packing fraction, $\eta\simeq 0.85$, in a region 
where phases with spatial order should be stable. Therefore, Mart\'{\i}nez-Rat\'on et al. also studied the 
formation of spatially nonuniform phases using an extended SPT theory that incorporated elements from fundamental-measure
theory, giving a reasonable account of spatial correlations. The results are shown in the left panel of
Fig. \ref{Phase_diag_2D}, which represents the SPT phase diagram of the model including only uniform phases. 
These results complement those of Schlaken et al. \cite{Schlacken} in that not only instability lines were calculated, but 
also the full phase equilibria. A complex phase behaviour in the region where the tetratic
phase begins to be stable was found. In the interval $2.21<\kappa<5.44$ the isotropic-uniaxial nematic transition
is of first order. For $\kappa<2.21$ the tetratic phase begins to be stable, with a stability region bounded
by a continuous isotropic-tetratic transition below, and by a tetratic-uniaxial nematic transition above;
this transition is of first order for $1.94<\kappa<2.21$, and continuous for $1<\kappa<1.94$. This means that there
are two tricritical points: one at $\kappa=5.44$, already predicted by
Schlacken et al., and a second one at $\kappa=1.94$ whose calculation requires a proper bifurcation study from the tetratic
to the uniaxial nematic phase. The spinodal line corresponding to the instability 
of the isotropic phase against spatial fluctuations was also calculated from this extended SPT theory. The line
is located in the range $0.5\alt\eta\alt 0.6$, a long way below the predicted stability for the tetratic phase, which
would mean that, even though tetratic correlations may be large, they are not sufficient to stabilise the tetratic phase.

The hard-rectangle system was explored in detail for the case $\kappa=3$ by Martínez-Ratón \cite{Yuri2D}. His aim was
to investigate layering and commensuration transitions in a confined setup (hence the choice of an aspect ratio that
ensured that the columnar phase was the stable phase at high density), but the bulk phase sequence was calculated as
a preliminary step. The theory used was a FMT for hard rectangles in the Zwanzig approximation, considering 
orientations along two perpendicular axes. A number of phases were found: isotropic, nematic, two different smectic
phases, a plastic solid, a perfectly oriented crystal and a columnar phase (note that this theory cannot describe a
tetratic phase). However, the equilibrium sequence of phases was isotropic and columnar, with a direct first-order
transition between the two. 

Parallel with these developments, experiments on vibrated quasi-2D monolayers made of (macroscopically-sized) 
granular rods have been performed. Vibration induces motion of the grains which is similar to thermal motion
in statistical systems. Although the nature of vibrated granular matter is very different from that of thermal systems,
they are both partially controlled by overlap (entropic or packing) interactions, and in some regimes they are expected to 
generate similar types of structures, either steady (granular matter) or equilibrium (thermal matter) states.
In fact, Narayan et al. \cite{Narayan} performed experiments on vibrated monolayers of anisotropic grains, such
as basmati rice, pinrolls and steel cylinders. Pinrolls exhibit nematic phases, while basmati-rice grains have smectic ordering. 
But cylinders, which have sharp corners, form nematic phases with strong tetratic correlations. These conclusions were 
obtained directly from visible inspection of real images, but also from the proper orientational correlation functions.
In particular, defining correlation functions by $g_n(r)=\left<\cos{n\phi(r)}\right>$ (where $\phi(r)$ is the angle
of the long axis of a particle at distance $r$ relative to another particle fixed at the origin), nematic 
ordering is measured by
$g_2(r)$, while $g_4(r)$ controls the presence of tetratic order. One interesting thing is that tetratic correlations were 
seen to be substantial for rather large aspect ratios ($\kappa=12.6$). In contrast, SPT predicts that tetratic order should 
be small for aspect ratios $\kappa\agt 2.21$.

At high packing fractions hard rectangles would be expected to crystallise with particles pointing in the same direction
and with their centres of mass arranged on a rectangular lattice (the metastable phase obtained by Martínez-Ratón
\cite{Yuri2D} in his density-functional study mentioned above). But in the case $\kappa=2$ two
such rectangles can form a dimer with square shape and a definite axis (the long sides of the rectangles). A collection of such 
square dimers 
can arrange to form a crystal with dimers lying at the sites of a square lattice, and with close-packing limit at $\eta=1$ 
(perfect packing), but with residual
entropy associated with the random orientation of the dimer axis. Therefore, a nonperiodic degenerate solid phase could be the lowest
free-energy phase. This is an interesting scenario explored by Wojciechowski et al. \cite{Wo2} in 1991 using hard-disc dimers. In effect, they 
identified a nonperiodic crystal at high density, but the underlying lattice was triangular. 

In 2006 Donev et al. investigated hard rectangles
of aspect ratio $\kappa=2$, i.e. particles that can form square 
dimers \cite{Donev_Stillinger}. They demonstrated that this system
exhibits both phases with tetratic order and nonperiodic solids at high density. The latter is a nonperiodic tetratic phase
made of a random tiling of dimers forming a square lattice, with a residual entropy of $1.79k$ per
particle and a possibly glassy character. The result of Donev et al. are consistent with a 
two-stage transition scenario of the Kosterlitz-Thouless type, with an isotropic-tetratic liquid transition followed by a 
tetratic-solid transition. The robustness of these results was checked by using two complementary approaches: a Monte Carlo
simulation on hard rectangles and a Molecular Dynamics simulation using rectangles with rounded corners.
These results suggest that hard-rectangle fluids show a strong tendency to form clusters, and that theories based on 
two-body correlations \cite{Schlacken,Tetratic_MRVM} should not give accurate results for their structure and thermodynamics.

Therefore, Mart\'{\i}nez-Rat\'on, Velasco and Mederos \cite{3-body} revisited the problem of the stability of the tetratic phase using a 
modified SPT theory that incorporates the third virial coefficient, thus including three-body correlations that should be important to describe
clustering of hard rectangles. A phase diagram was calculated using a variational 
procedure. The range of stability of the tetratic phase was found to
be increased with respect to the standard SPT approach based solely on two-body correlations, both in density and particle
aspect ratio. The same group studied the problem with a focus on clustering \cite{HR_cluster}, using Monte Carlo
simulation as a guide to construct a model based on stable polydisperse clusters of rectangles. Clustering was seen to
greatly enhance the relative stability of the tetratic phase with respect to the standard, uniaxial nematic phase.
In particular, it was predicted that square clusters have a dominant contribution to the free energy.

An important experimental study on the issue of the tetratic phase was performed, using colloidal particles, by Zhao et al. \cite{Chaikin0}.
A solution of PMMA discs, with $\kappa\simeq 6.4$, were prepared standing 
on edge on a planar surface, and their ordering properties were analysed. 
The system should be similar to a fluid of hard rectangles. Almost smectic behaviour was observed at high densities but, 
more interestingly, a single Kosterlitz-Thouless transition from the isotropic to the nematic phase was found.
Nematic order seems to be destroyed by wall defects, which lead to strong short-range tetratic order on the isotropic
side of the transition.

More recently, Geng and Selinger \cite{Selinger} have investigated the conditions for tetratic order to appear.
They used a soft-potential, Maier-Saupe-type model, analysed with a mean-field theory supplemented with Monte Carlo simulations. 
The model included an anisotropy coefficient that reflects the amount of two-fold symmetry breaking. Although a soft-potential
model, the results are interesting also to interpret hard-body studies. An important finding is that the tetratic phase can
exist up to a relatively high value of the anisotropy coefficient which, when transferred to a hard-particle view, points to
rather high particle aspect ratios. Also, the phase diagram obtained is surprisingly close to that predicted by
Mart\'{\i}nez-Rat\'on et al. \cite{Tetratic_MRVM} for the hard rectangle fluid, including the different phases, the
nature of the phase transitions and even the existence of tricritical points. This result indicates that the phase
behaviour is very generic and independent of the particular interactions.

In a very complete MC simulation work, Triplett and Fichthorn \cite{Fichthorn} studied the hard-rectangle fluid with
a range of aspect ratios and densities. For large aspect ratios the results compared even quantitatively with the results
of Ref. \cite{Tetratic_MRVM} for the isotropic to uniaxial nematic transition. For a moderate value, $\kappa=7.5$, a detailed
study of angular correlations showed strong tetratic ordering, but not sufficiently large to give a stable tetratic phase
with long- or quasi-long-range order.

The ordering properties of 2D hard squares have recently been revisited by Avenda\~no and Escobedo \cite{Escobedo1},
who went a step further from the simulations of Wojciechowski and Frenkel \cite{Frenkel_Polaco} for hard squares. 
Avenda\~no and Escobedo considered a fluid of rounded hard squares and investigated the evolution of phase behaviour
with respect to a roundness parameter, using rather large system sizes. For this system there is still some debate on
the true mechanism of orientational and positional ordering, which in some sense parallels that around the freezing
transition in the hard-disc fluid (a debate not completely closed, see \cite{Krauth}). Experimental studies on hard colloidal
platelets of square and pentagonal shapes \cite{Zhao,Zhao1} in part motivated this study, together with simulation work on 
pentagons \cite{Schilling}. In general, the crystal phases predicted by the simulations are not observed in the experiments, 
a result that has been associated with
roundness effects and periodic boundary conditions. Avenda\~no and Escobedo found that the shape roundness in a very
important factor and that the experimental phases can indeed be observed in the simulations. For a particular range of
roundness parameter, the tetratic mesophase mediates the phase evolution from the isotropic to the crystal phase.

\begin{figure}[h]
\includegraphics[width=10cm,angle=0]{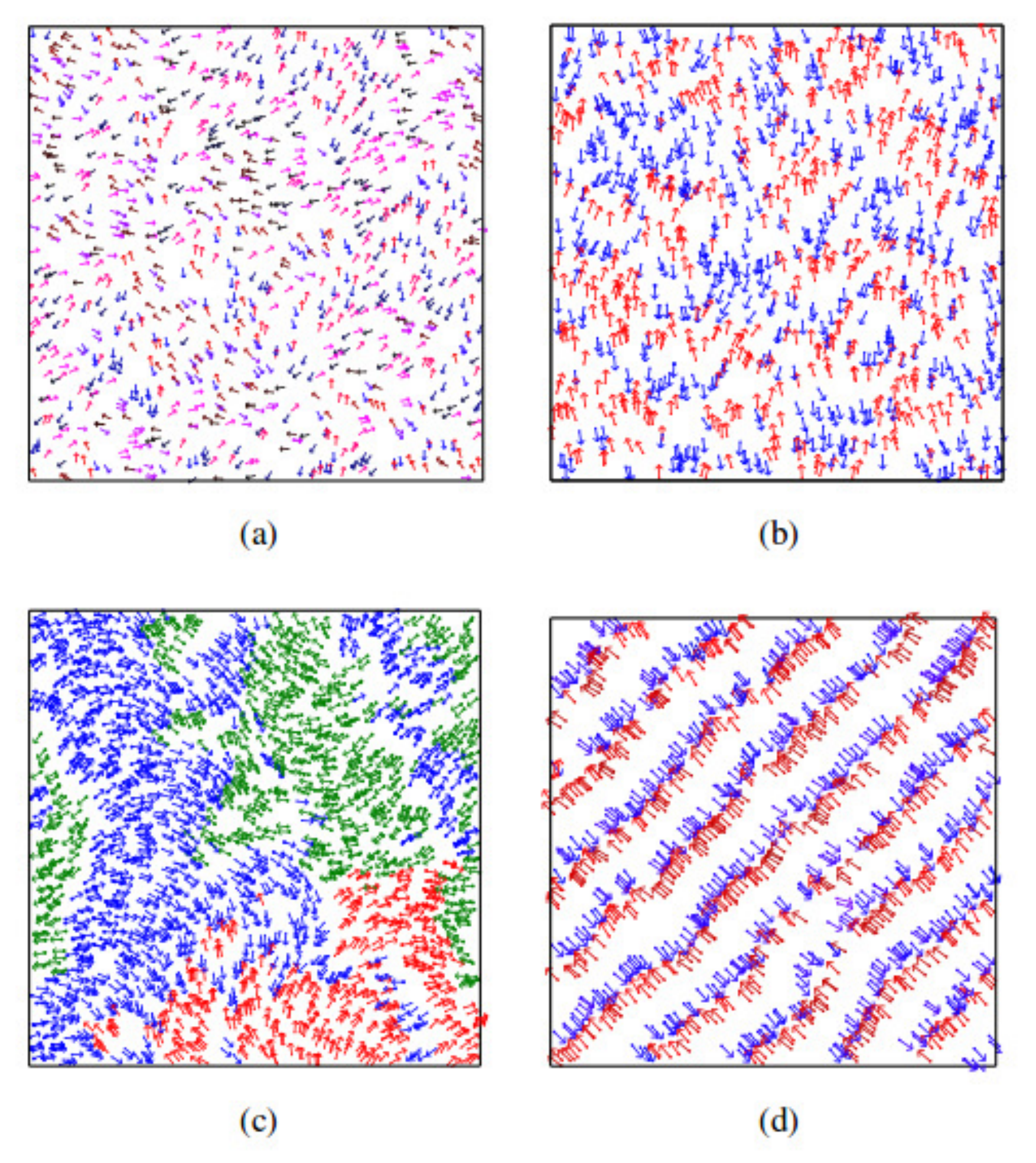}
\caption{Typical configurations of V-shaped hard-needle particles in 2D as obtained from simulations \cite{Gurin_Mex}.
(a) Isotropic (reduced pressure $p^*=15$, number of particles in the simulations $N=1000$);
(b) nematic ($p^*=40$, $N=1000$). 
(c) deformed nematic ($p^*=60$, $N=2000$), all for a bending angle between the needles of $\alpha=\pi/8$ (linear configuration
would have $\alpha=0$). (d) Antiferroelectric smectic ($p^*=55$, $N=1000$) for $\alpha=3\pi/8$.
See original article for details.}
\label{Anti}
\end{figure}

As mentioned before, there is evidence, based on computer simulations, that the smectic phase is not stable in 2D fluids
of hard ellipses and hard discorectangles \cite{Frenkel-Eppenga_2D,Cuesta_Frenkel,HDR_Bates} and that, in these fluids, the only
mesophase between the isotropic and crystal phases is the nematic. In an effort to search for other mesophases in different, 
not necessarily convex, bodies, Varga et al. \cite{Sabi_Mex} considered the fluid of zigzag needles introduced by \cite{Perusquia},
and applied Onsager theory. The study was motivated by computer simulations of Peón at al. \cite{Peon}, who collected evidence
for the stabilisation of the smectic phase and the existence of isotropic-nematic, nematic-smectic and isotropic-smectic transitions.
Crystal structures cannot be formed in this system, as in any other system of particles without volume. Varga et al. included
the rotational freedom of particles to describe the isotropic and nematic phases, but considered a parallel particle approximation
for the smectic phase. The results were seen to agree reasonably with the simulations. The stabilisation of the smectic phase was 
associated with the effect of increased particle terminal line segments or bent angle of the zigzag. Its structure is peculiar,
since the central core of the zigzag is tilted with respect to the layer normal, an optimised arrangement that maximises the free volume 
inside the layers.

Bisi et al. \cite{Bisi}, based on packing arguments, have speculated on the formation of antiferroelectric smectic phases in V-shaped 
particles due to a shape-polarity effect similar to that taking place in real dipolar molecules.
More recently, Martínez-González et al. \cite{Gurin_Mex} have considered the simpler V-shaped needles and performed Monte Carlo
simulation and calculations based on Onsager theory. They observed the formation of deformed, stable nematic phases consisting of
orientationally ordered polar domains with bent director orientation, but with zero overall polarisation. Using Onsager theory,
it can be shown that the V-shaped particle geometry favours bend deformations due to a free-energy reducing
bend torque, while splay deformations have an associated free-energy cost. As pressure is increased, these polar domains become
smaller and transform into inear arrays with alternating polarity, i.e. an antiferroelectric smectic phase. 
The theoretical results were seen to be in good agreement with the simulations.
A sequence of typical configurations can be seen in Fig. \ref{Anti}.

\begin{figure}[h]
\includegraphics[width=10cm,angle=0]{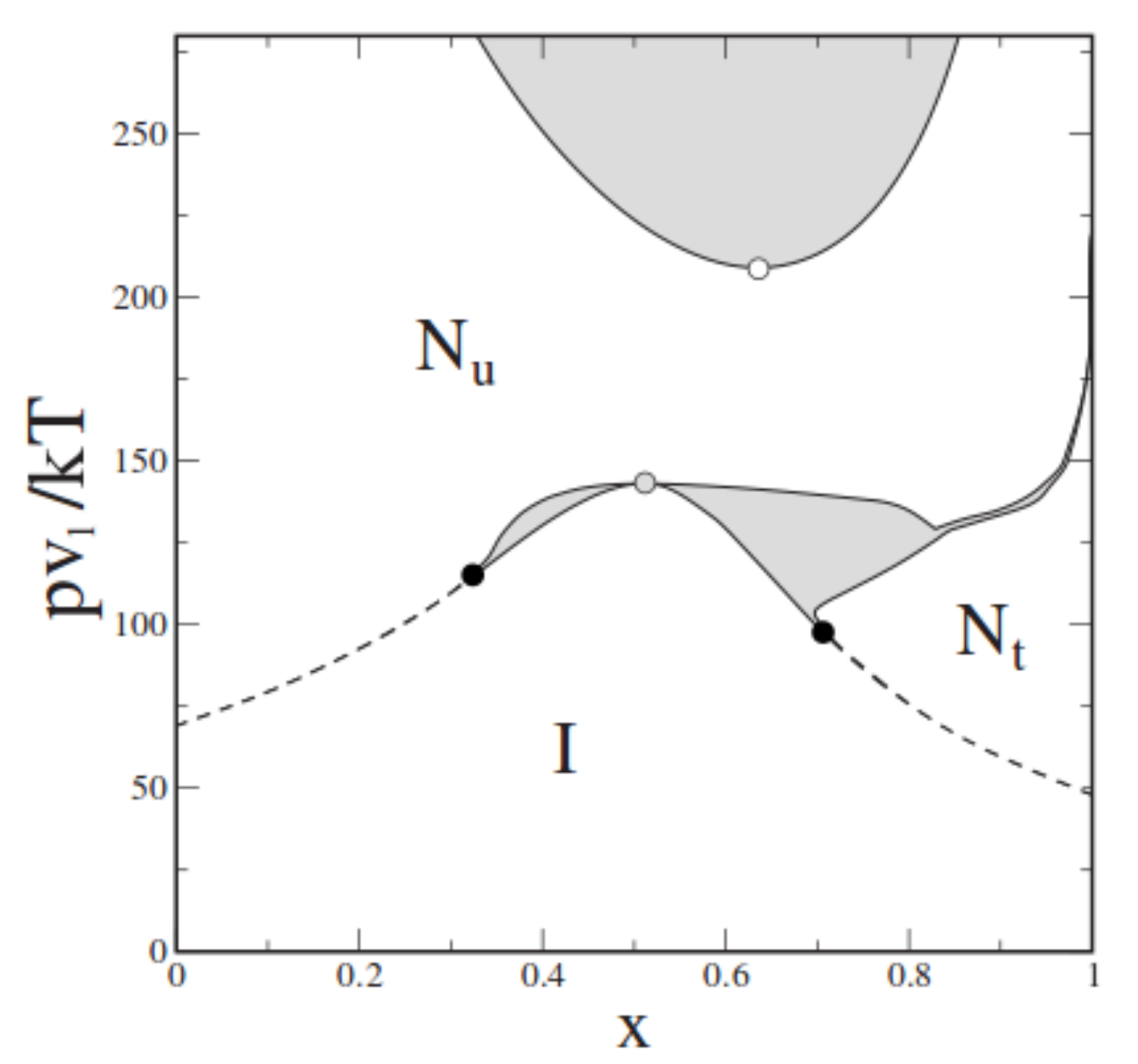}
\caption{Phase diagram for a hard rectangle/hard discorectangle mixture in the scaled-pressure versus composition of
the hard-rectangle component, from \cite{Us_2D_Tetratic_mix}. Values of the shape parameters are:
aspect ratio $\kappa_1=1.5$ and width $\sigma_1=1$ for the rectangles, and
aspect ratio $\kappa_2=2$ and same particle
area as a rectangle of aspect ratio equal to $2$ and unit width for the discorectangles.
The open circle indicates the critical point, the
shaded circle denotes an azeotropic point, while the filled circles
indicate tricritical points. Two-phase regions are indicated by the grey areas. Reprinted with permission from \cite{Us_2D_Tetratic_mix}. 
Copyright (2007) by the American Physical Society.}
\label{mix_Dani}
\end{figure}

Mixtures of anisometric particles in 2D have also been investigated. One of the important issues is the possibility
of nematic-nematic demixing phenomena. In 2005, Mart\'{\i}nez-Rat\'on et al. \cite{Us_mix_2D} used 
scaled particle theory to study binary mixtures of hard discorectangles and hard rectangles, with a view to
exploring possible liquid-crystal demixing scenarios in two dimensions. Using a bifurcation analysis
from the isotropic phase, tested against calculations based on a theoretical restricted-orientation model,
it was shown that both isotropic-nematic phase separation and nematic-nematic demixing ending in a critical point
was possible in a binary mixture of hard discorectangles and also in mixtures of hard discs and 
discorectangles. In addition, an isotropic-nematic-nematic triple point for a mixture of hard disks and hard discorectangles
was found. Similar demixing phase diagrams were obtained when one or two of the species have an elliptical shape, as 
shown in  \cite{Yuri_mix}. The elliptical shape enhances the demixing gap since, for a given aspect ratio, the ellipse
is the more convex body in two dimensions.

de las Heras et al. \cite{Us_2D_Tetratic_mix} used scaled-particle theory to study binary mixtures composed
of hard rectangles and other particles not possessing stable tetratic order by themselves (either hard discorectangles
or hard discs). Due to packing frustration associated with particle shape, the tetratic phase in hard rectangles of
low aspect ratio is destabilised when the second component is added, leading to demixing involving a second phase
(uniaxial nematic or isotropic). The effect is minimised for hard squares. The effect is also observed when the
second component is a hard rectangle of different aspect ratio but the same particle area or 
different particle area but the same aspect ratio. When the size ratio is sufficiently large, isotropic-tetratic
or tetratic-tetratic demixing was obtained in mixtures of hard squares. Fig. \ref{mix_Dani} shows a pressure-composition
phase diagram for a mixture of hard rectangles with aspect ratio $\kappa_1=1.5$ and width $\sigma_1=1$, and hard
discorectangles of aspect ratio $\kappa_2=2$ and same particle
area as a rectangle of aspect ratio equal to $2$ and unit width. The phase diagram includes a lower critical point 
terminating a region of uniaxial nematic demixing, an azeotropic point, and two
tricritical points separating first-order from continuous phase transitions. Isotropic I, uniaxial nematic N$_{\rm U}$ and 
tetratic $N_{\rm T}$ phases were found, with corresponding regions of stability shown in the figure.

Finally, we mention some applications of hard-body models to systems whose properties are 
in the region between two and three dimensions and that share some characteristics with
2D and 3D fluids. A very important example is that of Langmuir monolayers. These systems are
composed of liquid-crystal-forming molecules located at an approximately two-dimensional
surface (usually flat but with possibly substantial thermal fluctuations in case the surface is a liquid interface), 
and with molecular axes that can freely rotate in 3D. Experimentally these monolayers are observed to undergo
an amazingly complex variety of phase transitions \cite{Kaganer}, including positional ordering, chain freezing and
expanded-to-condensed phase transitions, and tilt transitions. These transitions are associated with orientational 
order of the molecules, conformational changes in the molecular chains, and positional order of the centres of mass. 
Some of these transitions are possibly explained, partially or totally, in terms of excluded volume. 

A model of hard rods (in the Onsager limit) grafted to a flat surface was analysed by various authors
\cite{Halperin_mono0,Moore_mono,Halperin_mono,Chen_mono}. Contrary to the full 3D case, this system does not possess 
a discontinuous nematic-like transition. Somoza and Desai \cite{Somoza} used the same 
model to examine the possibility that the fluid may exhibit a transition to a tilted nematic phase. In agreement 
with previous findings
the totally repulsive model was not shown to exhibit such a tilt transition, although a bifurcation analysis to a
phase with a broken azimuthal symmetry showed a tendency towards the tilted phase. In order to induce
a phase transition, an external field \cite{Chen_mono} or interparticle attractive interactions had 
to be incorporated, and Somoza and Desai \cite{Somoza} obtained a rich phase diagram using simple model attractions
between the rods.

More recently, Martínez-Ratón et al.
\cite{monolayers} used a fundamental-measure density functional to study a monolayer of
particles consisting of hard uniaxial boards of sizes $L\times\sigma\times\sigma$ with centres lying
on a flat surface. The aim was to study the existence of a phase transition from a uniaxial to a
biaxial nematic phase for both prolate ($L>\sigma$) and oblate ($L<\sigma$) geometries.
Since the Zwanzig approximation (with three possible orientations of the distinct axis) was used,
the three-dimensional one-component fluid can be mapped onto a two-dimensional fluid mixture
of three components consisting of hard squares and two hard rectangles pointing along 
perpendicular directions (these 2D bodies are the projections of the 3D particles on the surface).
The authors found that, for oblate particles and $\kappa^{-1}\equiv\sigma/L>3.5$, there exists a
planar N$_{\rm U}$ (with the same number of rectangles pointing along the
two perpendicular directions) to N$_{\rm B}$ (with more rectangles pointing along one direction) 
continuous phase transition, which is bounded above by a stability region of a nonuniform phase (with 
some kind of spatial order, be it columnar, smectic or crystal). In contrast, for $\kappa<3.5$, there is a 
direct transition from the N$_{\rm U}$ to a nonuniform phase. For prolate particles, but only for
$\kappa=L/\sigma>21.34$, is there a (reentrant) N$_{\rm B}$ phase, surrounded by regions of
N$_{\rm U}$-phase stability.

\section{Conclusions}

For more than a century, hard-core models have been playing a crucial role in understanding the structure
and phase behaviour of condensed phases of matter. In particular, since the pioneering work of Onsager,
hard bodies have been used to qualitatively understand the formation of orientationally ordered phases (see Table \ref{tablita}
for a summary of models and applications). 
Even though anisotropic hard interactions are enough to explain the stability of nematic, smectic,
columnar and many other mesophases, this does not mean that they are essentially responsible for the stability of 
liquid-crystalline phases in real materials. However, their study is very important insofar as they
play an essential role in theories that incorporate more realistic molecular interactions. Also, many
real colloidal systems (e.g. naturally occurring virus particles) have interactions which can be approximated 
by hard interactions. The actual possibility to fabricate tailor-made colloidal particles of different shapes 
has recently revitalised the attention on hard-particle models, and research on theories that quantitatively 
solve their statistical 
mechanics is still very active. Coupling between orientational and positional order, crucial in mesophases that
exhibit partial spatial order,
gives rise to nontrivial issues that these theories have to face. Most theories proposed so far rely on Onsager theory, which
only considers two-body correlations, and different schemes have been proposed to go beyond this approximation,
with limited success. Perturbative treatments on hard-body models have reached some predictive power in some
instances, but lack of knowledge on correlations functions of hard-particle fluids is a serious drawback in these
applications. The recently developed FMT version of density-functional theory as applied to freely rotating anisotropic 
particles opens up a promising avenue of research 
on hard-body models, and in the future we expect an important activity both in the use of new hard-body models and in the
development of new theoretical tools.

\begin{table}
\begin{center}
\tiny{
\begin{tabular}{|c|c|c|}
\hline\hline
{\bf Models} & {\bf Theories} & {\bf Applications} \\
\hline\hline
\begin{tabular}{c}Hard ellipsoids\\Hard Gaussian overlap\end{tabular} & 
\begin{tabular}{c}Onsager\\ Extended Onsager\\ Simulation\end{tabular}  & \begin{tabular}{c}Uniaxial \& biaxial nematics\\\end{tabular}\\
\hline
Hard cylinders & \begin{tabular}{c}Onsager\\ Scaled-particle theory\\ Extended Onsager\\ Bifurcation \\
FMT (parallel and freely \\
rotating of vanishing thickness) \\ Simulation
\end{tabular} & \begin{tabular}{c}Nematics\\ Smectics\\ Mixtures\\ Columnar \\\end{tabular} \\
\hline
Hard spherocylinders & \begin{tabular}{c}Onsager\\ Scaled-particle theory\\ Bifurcation\\ Extended Onsager\\ FMT\\ Simulation\end{tabular} &
\begin{tabular}{c}I-N-S-K transitions\\Plastic phase\\Elastic constants\\ Mixtures\\ Polydisperse fluids\end{tabular}\\
\hline
Hard cut spheres & \begin{tabular}{c}Extended Onsager\\ Simulation\end{tabular} &
\begin{tabular}{c}Columnar\\Cubatic\\ Mixtures\end{tabular}\\
\hline
Hard boards & \begin{tabular}{c}Onsager\\ FMT (Zwanzig approx.)\\ Cell theory\\Simulation\end{tabular} & 
\begin{tabular}{c}Uniaxial \& biaxial nematics\\ Smectics\\ Exotic phases \\
Polydisperse fluids\end{tabular}\\
\hline
Hard ellipses (2D) & \begin{tabular}{c}Scaled-particle theory \\ Extended Onsager \\ Simulation\end{tabular} & \begin{tabular}{c}I-N-K transitions\\Nature of I-N transition\\
Effect of particle geometry
\end{tabular}\\
\hline
Hard discorectangles (2D) & \begin{tabular}{c}Scaled-particle theory\\ Extended Onsager\\ Simulation \end{tabular}& 
\begin{tabular}{c}I-N-K transitions\\Nature of I-N transition\\ Mixtures\\
Effect of particle geometry
\end{tabular}\\
\hline
Hard rectangles (2D) & \begin{tabular}{c}Scaled-particle theory\\Extended Onsager\\ Simulation \end{tabular}& 
\begin{tabular}{c}Tetratic nematics\\I-N-T-K transitions\\ Clustering\\ Mixtures\\
Effect of particle geometry
\end{tabular}\\
\hline\hline
\end{tabular}
}
\end{center}
\caption{\label{tablita} Summary of hard-body models and applications.}
\end{table}

\newpage

\acknowledgments

This work was partially supported by grants MODELICO-CM/S2009ESP-1691 (Comunidad
Autónoma de Madrid, Spain), and FIS2010-22047-C05-01, FIS2010-22047-C05-02 and 
FIS2010-22047-C05-04 (MINECO, Spain).

\end{document}